\documentclass{elsart}
\journal{Physics Reports}
\usepackage{amsfonts}
\usepackage{amsmath}
\usepackage{rotating}
\usepackage{amssymb}
\usepackage{latexsym}
\usepackage{graphicx}
\usepackage{psfig}
\usepackage[active]{srcltx}

\newtheorem{THEOREM}{Theorem}[section]
\newtheorem{LEMMA}{Lemma}[section]

\newcommand {\dx} {\frac{d}{dx}}
\newcommand {\dxx} {\frac{d^2}{dx^2}}
\newcommand {\dy} {\frac{d}{dy}}
\newcommand {\dyy} {\frac{d^2}{dy^2}}
\newcommand {\dz} {\frac{d}{dz}}
\newcommand {\dzz} {\frac{d^2}{dz^2}}
\newcommand {\dt} {\frac{d}{dt}}
\newcommand {\dtt} {\frac{d^2}{dt^2}}
\newcommand{\pa}{\partial}

\newcommand{\al}{\alpha}
\newcommand{\bet}{\beta}

\newcommand{\ep}{\epsilon}
\newcommand{\veps}{\varepsilon}

\newcommand{\la}{\lambda}
\newcommand{\ta}{\tau}

\newcommand{\om}{\omega}
\newcommand{\Om}{\Omega}
\newcommand{\de}{\delta}

\newcommand{\ka}{\kappa}
\newcommand{\De}{\Delta}

\newcommand{\vphi}{\varphi}
\newcommand{\vrho}{\varrho}

\newcommand{\rar}{\rightarrow}
\newcommand{\lrar}{\leftrightarrow}
\newcommand{\non}{\nonumber}

\newcommand{\D}{\mathcal{D}}
\newcommand{\p}{\mathcal{P}}
\newcommand{\F}{\mathcal{F}}
\newcommand{\lc}{\mathcal{L}}
\newcommand{\h}{\mathcal{H}}

\begin{document}
\setcounter{tocdepth}{5}
\setcounter{secnumdepth}{5}
%
%
%
\begin{frontmatter}
 \title{One-Dimensional Quasi-Exactly Solvable Schr\"odinger Equations}
 \author[UNAM]{Alexander V.~Turbiner}
 \ead{turbiner@nucleares.unam.mx, alexander.turbiner@stonybrook.edu}

\address[UNAM]{Instituto de Ciencias Nucleares,
Universidad Nacional Aut\'onoma de M\'exico,
Apartado Postal 70-543, 04510 M\'exico\\ and\\
Department of Physics and Astronomy, Stony Brook University,
Stony Brook, NY 11794-3800, USA}


\begin{abstract}
Quasi-Exactly Solvable Schr\"odinger Equations occupy an
intermediate place between exactly-solvable (e.g. the harmonic
oscillator and Coulomb problems etc) and non-solvable ones. Mainly, they were
discovered in the 1980ies. Their major property is
an explicit knowledge of several eigenstates while the remaining ones
are unknown. Many of these problems are of the
anharmonic oscillator type with a special type of anharmonicity. The
Hamiltonians of quasi-exactly-solvable problems are characterized
by the existence of a hidden algebraic structure but do not have
any hidden symmetry properties. In particular, all known
one-dimensional (quasi)-exactly-solvable problems possess a
hidden $\mathfrak{sl}(2,\bf{R})-$ Lie algebra. They are equivalent to the
$\mathfrak{sl}(2,\bf{R})$ Euler-Arnold quantum top in a constant magnetic field.

Quasi-Exactly Solvable problems are highly non-trivial, they shed light on the delicate analytic properties of the Schr\"odinger Equations in coupling constant, they lead to a non-trivial class of potentials with the property of Energy-Reflection Symmetry. The Lie-algebraic formalism allows us to make a link between the
Schr\"odinger Equations and finite-difference equations on uniform
and/or exponential lattices, it implies that the spectra is preserved. This
link takes the form of quantum canonical transformation. The
corresponding isospectral spectral problems for finite-difference
operators are described. The underlying Fock space formalism giving
rise to this correspondence is uncovered. For a quite general class of
perturbations of unperturbed problems with the hidden Lie algebra
property we can construct an algebraic perturbation theory,
where the wavefunction corrections are of polynomial nature, thus, 
can be found by algebraic means.

In general, Quasi-Exact-Solvability points to the existence of a
hidden algebra formalism which ranges from quantum mechanics to
2-dimensional conformal field theories.
\end{abstract}

%

\end{frontmatter}

\tableofcontents
\newpage

\begin{center}
{\it INTRODUCTION }
\end{center}

\addcontentsline{toc}{section}{Introduction \protect \hfill}

\vskip 1.5cm

\renewcommand{\theequation}{0.{\arabic{equation}}}
\setcounter{equation}{0}

\renewcommand{\thefigure}{0.{\arabic{figure}}}
\setcounter{figure}{0}

\renewcommand{\thetable}{0.{\arabic{table}}}
\setcounter{table}{0}

Exact solutions of non-trivial problems
can provide valuable information about the real and hidden properties of the problem.
Very often such solutions lead to the discovery of unexpected features.
The Schr\"odinger equation is the basic object of quantum mechanics and thus it is quite
important to find as much as possible exact information about it.
This motivates the search for exact solutions. Although
everybody has its natural, intrinsic definition what does {\it exact
solution} mean, the general, a widely accepted definition of ``exact solution"
(which would be rigorous enough from a mathematical point of view) was
lacking until recently. Such a definition should
possess a certain heuristic value giving us a chance to apply a
mathematical formalism. This definition should also have a
clear physical meaning.

Knowledge of the exact solutions often allows us to develop a
constructive perturbation theory, where concrete feasible
calculations of relevant quantities can be done.

In recent times a renewed interest in exact solutions was greatly
inspired by discovery of a new class of quantum mechanical spectral
problems, the so-called {\it quasi-exactly-solvable} problems.
In these spectral problems a finite number of eigenstates can be found
explicitly, by algebraic means. It has led to a certain definition
of both {\it the exact solution}, and {\it a solvable problem}
admitting an exact solution(s).

Perhaps, one of the most important and natural philosophical ideas
for the present author is the idea about impossibility to learn everything
about Nature~
 \footnote{In one of the most famous Russian
 folklore books of the XIXth century "Koz'ma Prutkov Thoughts"
 written anonymously by several well-known Russian writers of 19th
 century it was repeated about hundred times `Nobody can embrace
 everything, do not trust anyone who claims that he can embrace
 everything'}.
The idea of {\it quasi-exact-solvability} can be considered as a
manifestation of the limits of exact (algebraic) knowledge at a given moment. One
can find several eigenstates exactly (algebraically) but not all.
It can be considered as a certain hint to the relevance of
quasi-exactly-solvable problems to Nature.
The remaining eigenstates are of transcendental nature and can be found only
approximately.

One of the simplest examples of how to find an exact solution of a
non-trivial Schr\"odinger equation is to reverse the standard logic.
Namely, instead of looking for a solution of the Schr\"odinger
equation with a given potential, we
\begin{quote}
 {\it  take any normalizable (square-integrable) function
 $\Psi_0(x)$ and find a potential in the Schr\"odinger operator
 for which this function is an eigenfunction
\begin{equation}
\label{V0}
 \frac{\De \Psi_0(x)}{\Psi_0(x)}\ =\ V_0(x)-E_0\ ,
\end{equation}
 where without loss of generality one can place $E_0=0$.}
\end{quote}
Using this method one can generate a zillion potentials, where a
single eigenstate is know explicitly.

A first example of this ``non-trivial" procedure was proposed to
the author by Felix~A.~Berezin in the mid-1970ies. Take
$$\Psi_0(x)= e^{-\al x^4/4}\ .$$
From (\ref{V0}) we find $$V_0(x)=\al^2 x^6 - 3 \al x^2\ ,\ E_0=0\ .$$
Hence, if $\al>0$ we know the ground state eigenfunction exactly in the potential $V_0(x)$.
Even on first sight, it is quite a non-trivial result, since we
know an exact eigenfunction of the ground state for some
anharmonic oscillator. In reality, this result is even
more non-trivial. The Schr\"odinger operator with potential
$V_0(x)$ is Hermitian for any real $\al$, and has infinitely-many
bound states if $\al \neq 0$ and is analytic in $\al$. For any
$\al>0$ the function $\Psi_0$ is the ground state eigenfunction
with zero energy, but in the domain $\al<0$ the function $\Psi_0(x)$ is neither
normalizable, nor the energy can be continued analytically from $\al >0$.
It can not be explained in the framework of the standard Stokes phenomenon
approach. We do have two analytically disconnected spectral problems:
one which is defined at $\al>0$ and another one which is defined at $\al<0$, see
\cite{Bender-Turbiner:1993} and \cite{Turbiner:1992}.
Chapter 2 will be devoted to this phenomenon. However, the
non-triviality of this potential is not exhausted by the above-mentioned
observation. The potential $V_0(x)$ is a double-well potential
with degenerate minima but $\Psi_0(x)$ has a single maximum
corresponding to a position of unstable equilibrium. Thus, the particle
prefers to be near the maximum of potential contrary to our intuition
\footnote{In classical mechanics, the motion at $E=0$ is not periodic:
for particle it takes infinite time to reach the tip of the barrier.}.
So, it is an explicit example for which neither a standard semiclassical
analysis, nor instanton calculus can be applied straightforwardly.

Using the above procedure we can create as many potentials as we like for which
one eigenstate can be found explicitly. This reasoning leads to a natural question:
Can we find a potential in which we know two, three ... , a finite number
of eigenstates constructively?
In essence, the attempt to find an answer to this question has led to the
discovery of a new class of spectral problems -- quasi-exactly-solvable
problems. A straightforward attempt to solve this problem failed and
a solution led to another discovery -- a non-trivial link between
the spectral theory and the representation theory of Lie algebras.

One of the goals of this Review is to describe this new connection
between the representation theory of Lie algebras and
linear differential equations.

The main idea of this connection is surprisingly simply. Let us
consider a certain set of differential operators of the first
order
\begin{equation}
\label{diff}
 J^{\al}(x)\ =\ a^{\al , \mu}(x) \pa _{\mu} \ + \
 b^{\al}(x) \quad,\quad  \pa _{\mu} \equiv \frac{d}{dx^{\mu}}\ ,
\end{equation}
where $\al =1,2,\dots ,k \ , x \in {\bf R}^{n} \ , \mu = 1,2,\ldots, n$
and \ $a^{\al , \mu}(x) , \ b^{\al}(x)$ are certain functions on
${\bf R}^{n}$. Assume that the operators form a basis of some Lie
algebra $g$ of dimension $k=\dim g$. Now let us take a polynomial $h$
in generators $J^{\al}(x)$ and ask the question:
\begin{quote}
{\em Does the differential operator $h(J^{\al}(x))$ have some
specific properties, which distinguish this operator from a
general linear differential operator?}
\end{quote}
Generically, the answer is {\it negative} -- nothing special appears.
Perhaps, it may be helpful to study the integrability of $h$, to find
operators commuting with $h$ using the Lie-algebra properties.
For instance, the Casimir operators (for the case of a reducible
representation) are evident integrals.
However, the situation becomes totally different if the algebra
$g$ is taken in a finite-dimensional representation. The answer
becomes {\it affirmative} :
\begin{quote}
{\em The differential operator $h(J^{\al}(x))$ begins to possess a
finite-dimensional invariant subspace. This  finite-dimensional
invariant subspace coincides with a finite-dimensional
representation space of the algebra $g$  of the first order
differential operators. If a basis of this
finite-dimensional representation space can be constructed
explicitly, the operator $h$ can be presented in a block-diagonal
form explicitly.}
\end{quote}
Such a differential operator having a finite-dimensional invariant
subspace with an explicit basis in functions is called {\it
quasi-exactly-solvable}. In general, a finite-dimensional
invariant subspace with an explicit basis can not be connected
with a representation space of some finite-dimensional Lie
algebra.

The first {\it explicit} examples of quasi-exactly-solvable problems
were published by Razavy \cite{Razavy:1980,Razavy:1981} and by
Singh-Rampal-Biswas-Datta \cite{Singh:1980}. In explicit form the general idea of
quasi-exact-solvability had been formulated for the first time in
\cite{Turbiner:1988f,Turbiner:1988z}, which led to a complete catalogue
of one-dimensional, quasi-exactly-solvable Schr\"odinger operators based on
spaces of polynomials \cite{Turbiner:1988qes}. The term
\textit{quasi-exact-solvability} has been suggested in
\cite{Turbiner:1987}~\footnote{In translation
of \cite{Turbiner:1988z} from Russian into English it appeared as
\textit{quasi-solvability} which later was also used in different articles.}.
The connection between quasi-exact-solvability and the finite-dimensional representations
of the $sl_2$ algebra was mentioned for the first time by
Zaslavskii-Ulyanov \cite{Ulyanov:1984}. Later, the idea of
quasi-exact-solvability was generalized to multidimensional differential operators,
matrix differential operators \cite{Shifman:1989},
finite-difference operators of different types (on different
lattices \cite{Ogievetsky:1991}), \textit{mixed}
operators containing differential operators, permutation
operators \cite{Turbiner:1994a}, and Dunkl operators.

In \cite{Morozov:1990} a connection was described between
quasi-exact-solvability and two-dimensional conformal quantum field theories
(see also Halpern-Kiritsis \cite{Halpern:1989} and Tseytlin \cite{Tseytlin:1994,Tseytlin:1994last},
and for a review \cite{Shifman:1994,Halpern:1996}). A relationship with solid-state physics is given in \cite{Ulyanov:1992} and very exciting results in this direction were found by Wiegmann-Zabrodin \cite{Wiegmann:1992,Wiegmann:1994}.
Survey of quasi-exact-solvability together
with a formalism of invariant subspaces in functions of
differential, finite-difference, matrix-differential operators
and differential operators containing with reflection
operators was done in \cite{Turbiner:1994,Turbiner:1995}.
Following this formalism, in the last 20 years, the (Lie)-algebraic nature of
the Calogero-Moser-Sutherland models, see e.g. \cite{Olshanetsky:1983},
was uncovered and explored \cite{Ruhl:1995,Brink:1997,Boreskov:2011,Sokolov-Turbiner:2015}
(see \cite{Turbiner:2011rat,Turbiner:2013trig} for a review and references therein).

Since 1988 \cite{Turbiner:1988qes} hundreds of papers were published on
quasi-exact-solvability - certainly, it is impossible to embrace all obtained results
in a single review paper of a limited volume. Thus, the presentation will be
limited to the basic facts and results which follow the taste
of the present author. Chapter 1 contains a mathematical introduction, the general theory of the second order QES differential equations and a description of twelve (major) QES Schr\"odinger operators.
Chapter 2 gives a description of one of the most important particular cases - the Quasi-Exactly-Solvable Anharmonic Oscillator.
In Chapter 3 the Algebraic Perturbations of Exactly-Solvable Problems which are closely
related to QES problems is briefly introduced.
Chapter 4 is devoted to the Fock space formalism, the Lie-algebraic discretization,
and the QES finite-difference equations on uniform and exponential lattices.

%
\graphicspath{{./}
{Chapter_1/}
{Chapter_2/}
{Chapter_3/}
{Chapter_4/}
}

\clearpage









%
%
%
%
\newpage

\begin{center}
{\bf \large Chapter\ 1.}\\[20pt]
 {\Large Quasi-Exact Solvability - a general consideration }
\end{center}
\addcontentsline{toc}{toc}{Chapter 1. Quasi-Exact Solvability -
general consideration \hfill}

\vskip 1cm


\renewcommand{\theequation}{1.{\arabic{equation}}}
 \setcounter{equation}{0}

 \renewcommand{\thefigure}{1.{\arabic{figure}}}
 \setcounter{figure}{0}

 \renewcommand{\thetable}{1.{\arabic{table}}}
 \setcounter{table}{0}

\addtocounter{equation}{-1}

%

The goal of this Chapter is to describe a connection between
the representation theory of Lie algebras and linear
differential equations. Specifically, it will be taken the algebra
$\mathfrak{sl}(2,\bf{R})$ realized by the first order differential operators
in one variable and demonstrated a connection to the
ordinary differential equations.


Thus, this Chapter is devoted to a general description of
quasi-exactly-solvable operators acting on functions
in one real (complex) variable.

\renewcommand{\theequation}{1.1.{\arabic{equation}}}
\setcounter{equation}{0}
\section{Generalities}

Let us take $n$ linearly-independent functions $f_1(y), f_2(y)
\ldots f_n(y)$ in one variable and form a linear space
\begin{equation}
\label{e1.1.1}
 {\F}_n^{(1)} \ = \ \langle f_1(y), f_2(y), \dots ,
 f_n(y) \rangle ,
\end{equation}
over complex (real) numbers. By construction the space ${\F}_n$
has the dimension $n$, $dim ({\F}_n) = n$. This space is
defined ``ambiguously" in a sense that functions $f_i(y)$ admit a
change of variable and multiplication of all of them by an
arbitrary function. Thus, we have to introduce a notion of {\it
equivalence}:

\begin{quote}
 {\em Two functional spaces ${\F}_n^{(1)}$ and
${\F}_n^{(2)}$ are called {\bf equivalent}, if one space can be
transformed into another through a change of variable and/or the
multiplication by a function,
\begin{equation}
\label{e1.1.2}
 {\F}_n^{(2)} \ =\ g(y){\F}_n^{(1)} |_{y=y(x)}\ ,\quad
 g(y) \neq 0 \ .
\end{equation}
Such a transformation is called {\bf gauge (similarity)}
transformation and the function $g(y)$ is called {\bf gauge
factor}.}
\end{quote}

Choosing a space (\ref{e1.1.1}) and considering all possible
changes of variable, $y=y(x)$, and functions $g(y)$, one can
describe a family of equivalent spaces (orbit). In general, in the
family there is a certain space which contains, as a subspace,
the space of linear functions
\begin{equation}
\label{e1.1.3}
 {\F}_n \ = \ \langle 1, x, \phi_1(x), \phi_2(x),
\dots , \phi_{n-2}(x) \rangle .
\end{equation}
\begin{quote}
{\em A linear functional space containing a subspace of linear
functions is called {\bf basic linear space}.}
\end{quote}

If a space (\ref{e1.1.1}) is given, then in order to construct a
basic linear space (\ref{e1.1.3}) we have to choose the gauge
factor as $g(y)=f_1(y)^{-1}$ and the variable $x=f_2(y)/f_1(y)$.
It allows us to introduce a {\bf standardization}: hereafter, if it is
not explicitly mentioned, only spaces of the form
(\ref{e1.1.3}) are considered. In general, any other space of the
family can be constructed by making a change of variable and by
multiplication by  an appropriate factor.

It is easy to see that if a linear differential operator
$h(y,\dy)$ acts on a space ${\F}_n^{(1)}$, then a similarity-transformed,
linear differential operator
\begin{equation}
\label{e1.1.4}
 \bar h(y,\dy)= g(x) h(x,\dx) g^{-1}(x) |_{x=x(y)} \ ,
\end{equation}
acts on an equivalent space ${\F}_n^{(2)}$. Therefore, the
operators $\bar h(y,\dy)$ and $h(x,\dx)$ are {\bf gauge-equivalent}.

Through this Chapter we focus on the only particular basic linear
space -- the linear space of polynomials of order not higher
than $n$ with real coefficients,
\begin{equation}
\label{e1.1.5}
  {\p}_{n+1} \ = \ \langle 1, x, x^2, \dots , x^n
\rangle \equiv (x^k|\ 0\leq k \leq n)\ ,
\end{equation}
where $n$ is a non-negative integer and $x \in {\bf R}$. Let
\[
{\tilde \p}_{n+1} \ = \ \bigg (\ \frac{1}{x^k}\ \vert \ \ 0\leq k
\leq n \ \bigg ) \ .
\]
It is worth noting an important property of the invariance:
\begin{equation}
\label{e1.1.6}
 x^n \bigg ({\p}_{n+1}\vert_{x \rar \frac{1}{x}}\bigg ) \
 =\ x^n {\tilde \p}_{n+1} \
 = {\p}_{n+1} \ .
\end{equation}
The following stems from an evident feature of polynomials: if $p_k(x) \in
{\p}_{n+1} $ is a polynomial of degree $k$, $k \leq n$, then $x^k
p_k(1/x) \in {\p}_{n+1}$ is a polynomial of degree $k$ with
inverse order of the coefficients in comparison to $p_k(x)$.
In general, in the space of polynomials ${\p}_{n+1}$, the M\"obius
transformation acts,
\begin{equation}
\label{e1.1.6mob}
 x\ \rar\ \frac{ax+b}{cx+d}\ ,\ ad-bc \neq 0\ .
\end{equation}
It maps the real line to itself. The 2 x 2 matrix $[a,b;c,d]$ is an element
of the $SL(2,\bf{R})$ group. The inversion (\ref{e1.1.6}) is a particular
case of the M\"obius transformation which appears if $a=d=0$ and $b=c=1$.
Another important transformation which acts in ${\p}_{n+1}$ is a linear
transformation
\begin{equation}
\label{e1.1.6lin}
 x\ \rar\ ax+b\ ,\ a \neq 0\ .
\end{equation}
It corresponds to the $SL(2,\bf{R})$ matrix $[a,b;0,1]$.

The spaces of polynomials can be ordered: $ {\p}_1 \subset  {\p}_2
\subset {\p}_3 \subset \dots \subset {\p}_n \subset \dots $. They
form an object which is called {\bf infinite flag (filtration)},
\begin{equation}
\label{e1.1.7}
    {\p} \equiv {\p}_1 \subset  {\p}_2 \subset {\p}_3 \subset \ldots
    \subset {\p}_n \subset \ldots \quad .
\end{equation}
It is worth mentioning that the linear transformation (\ref{e1.1.6lin})
preserves each particular subspace ${\p}_n$ of the flag. Hence, it preserves
the flag ${\p}$. It is evident that the M\"obius transformation does not
preserve the flag ${\p}$. It preserves a single subspace ${\p}_n$ only.

\renewcommand{\theequation}{1.2.{\arabic{equation}}}
\setcounter{equation}{0}
\section{Ordinary differential equations}
\subsection{General consideration}

Take the linear space  ${\p}_{n+1}$ of polynomials of degree not
higher than $n$ (\ref{e1.1.5}).

\begin{quote}

 {\em A linear differential operator of the $k$th order,
$T_k(x,\dx)$, is called {\bf quasi-exactly-solvable}, if it
preserves the linear space of polynomials ${\p}_{n+1}$,
\begin{equation}
\label{QES}
    T_k(x,\dx):{\p}_{n+1} \rar {\p}_{n+1}\ .
\end{equation}
This operator is block-triangular in the basis of monomials $x^m,
m=0,1,\ldots$. \linebreak The operator $E_k(x,\dx)$ is called {\bf
exactly-solvable}, if it preserves the infinite flag ${\p}$ of
spaces of polynomials (\ref{e1.1.7}), implying that each
individual space ${\p}_j$ is preserved
\begin{equation}
\label{ES}
    E_k(x,\dx): {\p}_j \rar {\p}_j \ ,\ j=0,1,\ldots \ .
\end{equation}
This operator is block-triangular in the basis of monomials $x^m,
m=0,1,\ldots $. }
\end{quote}

It has been known since Sophus Lie that the algebra
$\mathfrak{sl}(2,\bf{R})$ can be realized by first order
differential operators in one variable
\[
J^+_n = x^2 \dx - n x \ ,
\]
\begin{equation}
\label{sl2r}
 J^0_n = x \dx - \frac{n}{2} \  ,
\end{equation}
\[
J^-_n = \dx \ .
\]
where $n$ is the mark of the representation ($j=\frac{n}{2}$ is called the spin of representation).
It is easy to check that for any $n \in \bf{R}$ the generators (\ref{sl2r}) obey the
$\mathfrak{sl}(2,\bf{R})$ commutation relations:
\[
 [J^{\pm}_n,J^0_n] = \pm J^{\pm}_n \quad
 ,\quad [J^+_n,J^-_n] = -2 J^0_n\ .
\]
If the identity operator, $X=const$ is added to (\ref{sl2r}), then
the generators $J^{\pm,0}_n, X$ span the algebra
$\mathfrak{gl}(2,\bf{R})$
 \footnote{The representation (\ref{sl2r}) is one of the so-called `projectivized' representations, see \cite{Turbiner:1988z,Turbiner:1988f}.}.
If in (\ref{sl2r}) the parameter $n$ is an integer, $n \in
\mathbf{N}$, the representation becomes finite-dimensional and the operators
$J^{\pm,0}_n, X$ have the space ${\p}_{n+1}$, (\ref{e1.1.5}) as a common
invariant subspace, which is a finite-dimensional representation space
\[
   J^{\pm, 0}_n, X:\ {\p}_{n+1} \rar {\p}_{n+1} \ .
\]
They realize the irreducible finite-dimensional representation: they act on
${\p}_{n+1}$ irreducibly. Hence, the quadratic Casimir operator $C_2$,
\begin{equation}
\label{esl2C}
 C_2\ =\ {\half} \{J^+_n,J^-_n\}- J^0_n J^0_n\ =\ -\frac{n}{2}
\big (\frac{n}{2}+1 \big )\ ,
\end{equation}
which commutes with all generators, becomes constant. Here $\{ a,b
\}\equiv ab+ba$ is the anticommutator. Relation (\ref{esl2C}) holds for
any real value of $n$, hence, (\ref{esl2C}) can be considered as an artifact
of realization of the algebra $\mathfrak{sl}(2,\bf{R})$ by the first
order differential operators. It is worth mentioning that for
any $n$ the generators $J^{0,-}_n$ form the Borel subalgebra,
$\mathfrak{b}_2 \subset \mathfrak{sl}(2,\bf{R})$. It is
convenient to introduce a notation $J^{0,-} \equiv J^{0,-}_0$.

One can prove a classification

\begin{LEMMA} \cite{Turbiner:1994}
\label{Lams}

 (i) Suppose $k < (n+1)$.  Any quasi-exactly-solvable
operator $T_k$ can be represented by a $k$th degree polynomial of
the operators $J^{\pm,0}_n$. If $k \geq (n+1)$, the part of the
quasi-exactly-solvable operator $T_k$ containing derivatives up to
order $n$ can be represented by an $n$th degree polynomial in the
generators (\ref{sl2r}).

 (ii) Conversely, any polynomial in (\ref{sl2r}) is
quasi-exactly solvable.

 (iii) Among quasi-exactly-solvable operators there exist
exactly-solvable operators, $E_k$. Any $E_k$ can be
represented by $k$th degree polynomial in generators $J^{0,-}$.
Hence, $E_k$ is a degeneration of $T_k$ when terms which contain
$J^{+}_n$ are absent.
\end{LEMMA}

\noindent
 {\bf Proof.}\,
In order to prove the part (i) let us mention at first that the
operators $J^{\pm,0}_n$ act on the space ${\p}_{n+1}$
irreducibly. This means there is no invariant subspace in
${\p}_{n+1}$ under the action of $J^{\pm,0}_n$. Thus, one can apply
the Burnside theorem which claims that if a set of linear
operators acts on some finite-dimensional linear space irreducibly
then any operator acting on this space is a polynomial in these
operators \footnote {Burnside theorem (see e.g. the book
\cite{Weyl:1931}) is a particular case of a more general Jacobson
theorem (see e.g. the book \cite{Lang:1965}, Chapter
XVII.3)}\, . If the dimension $(n+1)$ of the space ${\p}_{n+1}$ is
less than the degree $k$ of the operator $T_k$, the operator
contains a part for which ${\p}_{n+1}$ is the kernel. This part should be
subtracted. Then the part (i) holds. The statement (ii) is
evident. For a proof of the part (iii) let us take, for example,
two spaces ${\p}_{{n}_1+1}$ and ${\p}_{{n}_2+1}$ with $n_1\neq n_2$
and apply the part (i) to classify the operators preserving both of them. We
assume that $k$ is larger than both $(n_1+1)$ and $(n_2+1)$.
From this we can conclude that the only way to preserve
both spaces is to exclude the operators $J^{+}_{n_1,n_2}$. From
another side, the operators $J^{0}_{n_1,n_2}$ differ from each other by
a constant proportional to $(n_1-n_2)$ and an identity operator.
Thus, the operator $E_k$ preserves the space ${\p}_{n+1}$ for any
$n$. $\blacksquare$

\begin{quote}  Let us call a  {\it universal enveloping algebra}
$U_{\mathfrak{sl}(2,\bf{R})}$ of a Lie algebra
$\mathfrak{sl}(2,\bf{R})$ the algebra of all ordered
polynomials in generators $J^{\pm,0}_n$. The notion {\it
`ordering'} means that in any monomial in $J^{\pm,0}_n$
the generator $J^{+}_n$ is always placed to the left and the generator
$J^{-}_n$ is placed to the right. Taking a realization of $\mathfrak{sl}(2,\bf{R})$
in terms of first order differential operators (\ref{sl2r}) we get
a realization of the universal enveloping algebra in differential
operators $U^d_{\mathfrak{sl}(2,\bf{R})}$. It is a subalgebra
of the algebra of differential operators in one variable,
$U^d_{\mathfrak{sl}(2,\bf{R})} \subset \mbox{diff}
(1,\bf{R})$. Any element of $U^d_{\mathfrak{sl}(2,\bf{R})}$
preserves ${\p}_{n+1}$. The converse is almost true: the algebra of
differential operators which preserves ${\p}_{n+1}$ is the infinite-dimensional
algebra of differential operators generated by
the operators (\ref{sl2r}) (which is $U^d_{\mathfrak{sl}(2,\bf{R})}$)
plus annihilator $B \frac{d^{n+1}}{dx^{n+1}}$, where $B$ is any linear differential
operator. The algebra $U^d_{\mathfrak{sl}(2,\bf{R})}$ depends on $n$.
It is evident that two algebras characterized by different $n$'s are not isomorphic.
\end{quote}

\noindent
 {\it Comment 2.1}  \  The notion - the universal enveloping
algebra - allows us to state that at $k < n+1$ a quasi-exactly-solvable
operator $T_k$ of the order $k$ is simply an element of the universal
enveloping algebra $T_k \in U^d_{\mathfrak{sl}(2,\bf{R})}$. However,
if $k \geq n+1$, then $T_k$ is represented as an element of
$U^d_{\mathfrak{sl}(2,\bf{R})}$ plus $B
\frac{d^{n+1}}{dx^{n+1}}$, where $B$ is any linear differential
operator of order not higher than $(k-n-1)$. Evidently, an
operator $B \frac{d^{n+1}}{dx^{n+1}}$ annihilates the space
(\ref{e1.1.5}). In other words, the algebra of differential
operators acting on the space ${\p}_{n+1}$ coincides with the
algebra $U^d_{\mathfrak{sl}(2,\bf{R})}$ plus the annihilators.

\noindent
 {\it Comment 2.2}  \  As a consequence of the invariance
(\ref{e1.1.6}) of the space ${\p}_{n+1}$ the algebra
${\mathfrak{sl}(2,\bf{R})}$ generated by (\ref{sl2r}) is
covariant with respect to conjugation
\begin{eqnarray}
\label{e1.2.2}
 &x^{-n}\ J^{+}_n(x, \dx)\ x^n \mid _ {x=\frac{1}{z}}\,
 = -J^{-}_n(z, \dz)\,, \non \\
 &x^{-n}\ J^{0}_n(x, \dx)\ x^n \mid _ {x=\frac{1}{z}}\,
 = -J^{0}_n(z, \dz)\,,  \non \\
 &x^{-n}\ J^{-}_n(x, \dx)\ x^n \mid _ {x=\frac{1}{z}}\,
 = -J^{+}_n(z, \dz)\, .
\end{eqnarray}

It is natural to introduce the notion of {\it grading} for the generators
(\ref{sl2r}). It is easy to see that any $\mathfrak{sl}(2,\bf{R})$-generator
from (\ref{sl2r}) maps a monomial into monomial, $J^{\al}_n x^p \mapsto x^{p+d_{\al}}$.
Therefore, let us call $d_{\al}$ the {\it grading} of the generator
$J^{\al}_n$: $\mbox{deg} ( J^{\al}_n ) = d_{\al}$. Following this definition
\begin{equation}
\label{e1.2.3}
 deg (J^+_n) = +1 \ , \ deg (J^0_n) = 0 \ , \ deg (J^-_n) = -1 ,
\end{equation}
and
\begin{equation}
\label{e1.2.4}
 deg [(J^+_n)^{n_+} (J^0_n)^{n_0}(J^-_n)^{n_-}] \  =
 \ n_+ - n_- .
\end{equation}
The notion of the grading allows us to classify the operators $T_k$ in
a Lie-algebraic sense.

\begin{LEMMA}
 {\em A quasi-exactly-solvable operator $T_k \subset
 U_{\mathfrak{sl}(2,\mathcal{R})}$ has no terms of positive
 grading, if and only if it is an exactly-solvable operator.}
\end{LEMMA}

\noindent
 {\it Comment 2.3} .\
After transformation (\ref{e1.2.2}) the terms of zero grading
remain of zero grading, ones of positive grading become of negative
grading and, correspondingly, the terms of negative grading become
of positive grading. Thus, any exactly-solvable operator
having terms of negative and zero grading converts to the operator
with terms of positive and zero grading under transformation
(\ref{e1.2.2}). A quasi-exactly-solvable operator, however, always
possesses terms of positive grading both in $x$-space
representation and in $z$-space representation.

\begin{THEOREM} \cite{Turbiner:1992}
\label{Turbiner:1992}
 Let $n$ be a non-negative integer. Consider the eigenvalue problem
 for a linear differential operator of the $k$th order in one variable
\begin{equation}
\label{e1.2.5}
 T_k(x, \dx)\, \vphi (x)\ = \ \veps\, \vphi (x)\ ,
\end{equation}
where $T_k$ is symmetric. The problem (\ref{e1.2.5}) has $(n+1)$
linearly independent eigenfunctions in the form of a polynomial in
variable $x$ of order not higher than $n$, if and only if $T_k$ is
quasi-exactly-solvable. The problem (\ref{e1.2.5}) has an infinite
sequence of polynomial eigenfunctions, if and only if, the operator
is exactly-solvable.
\end{THEOREM}

\noindent {\it Proof} .\ The ``if" part of the first and the
second statements is obvious. The ``only if" part is a direct
corollary of Lemma~\ref{Lams} $\blacksquare$

Theorem \ref{Turbiner:1992} gives a general classification of
ordinary differential equations
\begin{equation}
\label{e1.2.6}
 L_k \vphi(x) \equiv \sum_{j=0}^{k} a_j (x) \frac{d^j \vphi (x)}{dx^j}  \
 = \ \veps \vphi(x)\ ,
\end{equation}
having polynomial solution(s) in $x$. The problem of finding
spectra which corresponds to these polynomial eigenfunctions is
reduced to the algebraic procedure of solving the system of $(n+1)$
linear equations \footnote{We will call this part of the
spectra of the equation (\ref{e1.2.6}) the {\it algebraic
sector}.}. This leads to the solution of the secular (characteristic)
equation which is a polynomial in $\veps$ of degree $(n+1)$.
In general, the spectral problem (\ref{e1.2.5}) in addition to
polynomial eigenfunctions possesses infinitely-many non-polynomial ones.
Finding them is a non-algebraic procedure and is reduced to a diagonalization
of a generic infinite-dimensional matrix.

\begin{quote}
{\em A linear differential operator with polynomial coefficient
functions is called {\bf algebraic}. An operator which can be
represented in terms of generators of some Lie algebra called
{\bf Lie-algebraic}.}
\end{quote}

According to the necessary condition formulated in
Theorem~\ref{Turbiner:1992} the coefficient functions $a_j (x)$ in
(\ref{e1.2.6}) should be of the form
\begin{equation}
\label{e1.2.7}
 a_j (x) \ = \ \sum_{i=0}^{k+j} a_{j,i} x^i \ .
\end{equation}
Hence, the linear differential operator in the l.h.s. of
(\ref{e1.2.6}) should be algebraic of the Fuchs type, with Fuchs index $k$.
However, it is worth emphasizing that not every algebraic operator admits
polynomial eigenfunctions or, in other words, the existence of an algebraic
sector.

Sufficient condition provided by Theorem~\ref{Turbiner:1992}
requires that the coefficients $a_{j,i}$ in (\ref{e1.2.7}) for
coefficient functions in (\ref{e1.2.6}) are not arbitrary.
They have to be of such a form that the linear differential
operator in the l.h.s. of (\ref{e1.2.6}) can be rewritten as a $k$th
degree polynomial element of the universal enveloping algebra
$U^d_{\mathfrak{sl}(2,\bf{R})}$ taken in realization
(\ref{e1.2.3}) of spin $n$. Therefore, the operator $L$ should be
not only algebraic but also Lie-algebraic. This generates a set
of constraints and reduces the number of free parameters of the algebraic
operator $T$. The number of free parameters of the algebraic
operator $T$ defines the number of free parameters of the
polynomial solutions in (\ref{e1.2.6}). It can be obtained by
counting the number of parameters which characterize a general
$k$th degree polynomial element of the universal enveloping
algebra $U^d_{\mathfrak{sl}(2,\bf{R})}$ factorized over the
Casimir operator. Thus, all elements containing $J^0_n
J^0_n$ (see (\ref{esl2C})) should be excluded from counting. A
straightforward calculation leads to the following formula
\begin{equation}
\label{e1.2.8}
 par (T_k) = (k+1)^2 + 1\ ,
\end{equation}
where the number of free parameters of the quasi-exactly-solvable
operator $T_k$ is denoted $par(T_k)$ where the mark of representation
(\ref{e1.2.3}) $n$ is included. If the parameter $n$ is integer,
the representation (\ref{e1.2.3}) becomes finite-dimensional. In this case
the parameter $n$ defines a degree of a polynomial solution, it also
appears in coefficients of polynomial eigenfunctions.

\noindent
 {\it Comment 2.4} .\
We can see that in the basis of monomials, $x^p, p=0,1,\ldots$\,,
the Fuchs type operator $L_k$ with Fuchs index $k$, see (\ref{e1.2.6}) - (\ref{e1.2.7}),
has the form of a $(2k+1)$-diagonal infinite matrix with $k$ upper and $k$
lower sub-diagonals. In the quasi-exactly-solvable case the matrix $L_k$ additionally becomes upper (lower) block-triangular. In the exactly-solvable case the matrix $L_k$ additionally becomes upper (lower) triangular. By a suitable change of $x$-coordinate one sub-diagonal
can be removed.

For the case of second order differential operators, the number of
free parameters (\ref{e1.2.8}) is equal to $par(T_2)=10$, while
a generic second order differential operator $L_2$ in (\ref{e1.2.6})
with coefficients (\ref{e1.2.7}) is characterized by 15 free parameters
\footnote{It is worth mentioning that the generic Fuchs type second-order
differential operator with Fuchs index 2 is nothing but the celebrated Heun
operator (see e.g. Kamke \cite{Kamke:1959})}.
Imposing five conditions on the operator (\ref{e1.2.6}) with coefficients
(\ref{e1.2.7}) (said differently, on a generic Heun operator)
we get a second order differential operator which can be written in terms
of the ${\mathfrak{sl}(2,\bf{R})}$ generators with the mark of representation
equals to some integer $n$.
Thus, the second order differential operator has the meaning of the Hamiltonian
of the Euler-Arnold quantum top.
A condition of quasi-exact solvability implies the finite-dimensionality of the
representation of ${\mathfrak{sl}(2,\bf{R})}$.
Eventually, it leads to six constraints such that, five parameters of
the original Heun operator become fixed and the sixth parameter $n$ is restricted
to an integer value. It will be discussed in details below, in Section \ref{2nd}.

In the case when the number of polynomial solutions is infinite,
the expression (\ref{e1.2.7}) simplifies to
\begin{equation}
\label{e1.2.9}
 a_j (x) \ = \ \sum_{i=0}^{j} a_{j,i} x^i\ ,
\end{equation}
thus, $k=0$, in agreement with the results by Krall  \cite{Krall:1938} (see
also \cite{Littlejohn:1986}). The number of free parameters is equal to
\begin{equation}
\label{esl20}
 par (E_k)\ =\ \frac{(k+1)(k+2)}{2}\ .
\end{equation}
In this case the $n$th eigenfunction is a polynomial in $x$ of
degree $n$. One can easily find the eigenvalue which corresponds to
the eigenfunction given by the polynomial of degree $n$,
\begin{equation}
\label{esl20E}
 \veps_n\ =\ \sum_{j=0}^{k} a_{jj} \frac{n!}{(n-j)!}\ .
\end{equation}
Thus, this eigenvalue is the polynomial in $n$ of degree $k$.

It can be shown that the operator $T_k$ with the coefficients
(\ref{e1.2.9}) preserves a finite flag  $$ {\p}_0 \subset {\p}_1
\subset {\p}_2 \subset \dots \subset {\p}_k $$ of spaces of
polynomials. It can be easily verified that the preservation of
such a finite flag of spaces of polynomials implies the
preservation of an infinite flag of such spaces.

A class of spaces which are equivalent to the space of polynomials
(\ref{e1.1.5}) is presented by
\begin{equation}
\label{esl21}
 \langle \al , \al\,  \bet , \dots , \al\,
  \bet ^n \rangle \ ,
\end{equation}
where $\al=\al (z),\ \bet=\bet (z)$ are arbitrary functions. A linear
differential operator acting on the (\ref{esl21}) is easily
obtained from a quasi-exactly-solvable operator
(\ref{e1.2.5})--(\ref{e1.2.7}) (see Lemma~\ref{Lams}) by making
the change of variable
\[
x = \bet (z)\ ,
\]
and the gauge (similarity)
transformation
\[
 \tilde T = \al (z)\ T\ \al (z)^{-1}\ .
\]
It is worth noting that in the case of the M\"obius transformation (\ref{e1.1.6mob})
\[
x = \frac{az+b}{cz+d}\ ,\ \al (z) = (cz+d)^n\ ,
\]
the space (\ref{esl21}) remains the space of polynomials and the operator $\tilde T$ is
an algebraic quasi-exactly-solvable operator.
Explicitly, the operator of $k$th order has the form
\begin{equation}
\label{esl22}
 \bar{T}_k\  = \ \al (z) \sum_{j=0}^{k}
 \big(\sum_{i=0}^{k+j} a_{j,i} \bet (z)^i \big)
 \frac{d^j}{dx^j} \vert_{x = \bet (z)}\ \al (z) ^{-1} \ ,
\end{equation}
where the coefficients $a_{j,i}$ are the same as in
(\ref{e1.2.7}).

As a result the expression (\ref{esl22}) gives a general form of
the linear differential operator of $k$th order acting on a space
(\ref{esl21}) equivalent to (\ref{e1.1.5}). Since {\it any} one- or
two-dimensional invariant functional space can be presented in the
form (\ref{esl21}) and thus can be reduced to (\ref{e1.1.5}), the following
general statement holds:
\begin{THEOREM}
\label{T:1.2.2}
 There are no linear operators possessing one- or
 two-dimensional invariant sub-space with an explicit basis other
 than given by Lemma~\ref{Lams}.
\end{THEOREM}

Therefore, an eigenvalue problem (\ref{e1.2.5}) for which a single
eigenfunction can be found in an explicit form, is related to the
operator
\begin{equation}
\label{esl23}
 {\bf T^{(1)}}\ =\ B\big( x,\dx \big) \dx \ +\ q_0 \,,
\end{equation}
where $B(x, \dx)$ is an arbitrary linear differential operator.
The operator (\ref{esl23}) has a constant eigenfunction with an
eigenvalue $\ep = q_0$.

It is worth mentioning that one can introduce a useful notion
of {\it gauge} transformation of derivative. It corresponds to
replacing the derivative by the {\it covariant} derivative,
\begin{equation}
\label{e1.D1}
 e^{-A(x)} \dx\,  e^{A(x)} = \dx + A'(x)\equiv {\D}(A(x))\ ,
\end{equation}
where $A$ is called the gauge phase and $A'(x)$ is a connection. It is evident
that this transformation is canonical: the Lie bracket, $[\dx, x]$ remains unchanged.
The gauge transformation can be accompanied by a change of variable,
\begin{equation}
\label{e1.D2}
 e^{-A(x)} \dx \,e^{A(x)}|_{z=z(x)} = {\D}(A(x))|_{z=z(x)}\
 =\ z'(x(z)) \dz + A'_x(x(z))\ .
\end{equation}
Making the gauge transformation of (\ref{esl23}) we arrive at
the gauge-transformed operator
\begin{equation}
\label{esl23.D}
 {\bf \tilde T^{(1)}}\ =\ B(x,{\D}( A(x))) {\D}( A(x))
 \ +\ q_0 = \tilde B(x, \dx) (\dx + A'(x)) + q_0 \,,
\end{equation}
which has an eigenfunction $\vphi (x) = e^{-A(x)}$.

A general differential operator possessing two eigenfunctions is
the Lie-algebraic operator: following Theorem~\ref{T:1.2.2},
its explicit form is given by
\begin{equation}
\label{esl24}
 {\bf T^{(2)}}\ =\ B \big(x, \dx\big) \dxx \ +\ q_{2} (x) \dx \
 +\ q_{1}(x),
\end{equation}
where $B(x, \dx)$ is again an arbitrary linear differential
operator. The coefficients $q_{1,2} (x)$ are the first- and
second-order polynomials, respectively, with coefficients such
that  $q_{2}(x) \dx + q_{1}(x)$ can be expressed as a linear
combination of the generators  $ J_1^{\pm,0}$ (see (\ref{sl2r})).
In general, the operator (\ref{esl24}) has two eigenfunctions
in a form of linear function $\vphi(x)= x + c$. Performing gauge
transformation (\ref{e1.D2}) of (\ref{esl24}) we arrive at
\[
 {\bf \tilde T^{(2)}}\ =\ B(x,{\D}(A(x))) {\D}^2( A(x))+
 \ q_{2} (x){\D} \ +\ q_1 (x)
\]
\begin{equation}
\label{esl24.D}
 = \tilde B(x, \dx) (\dx + A'(x))^2\ +
 \ q_{2}(x) \left(\dx + A'(x)\right)+q_1(x) \,,
\end{equation}
where two``algebraic" eigenfunctions are of the form $(x+c) e^{-A(x)}$.

\subsection{Second-order differential equations}
\label{2nd}

Second-order differential equations play an exceptionally
important role in a vast majority of applications in different sciences.
Due to their importance they certainly deserve a detailed and careful consideration.
We aim to explore one particular aspect of a general theory: a description
of the second-order differential equations (\ref{e1.2.5}) possessing
polynomial solutions. This problem is reduced to studying the eigenvalue problem
for the Heun operator, see e.g. \cite{Ronveaux:1995}
\begin{equation}
\label{heun-he}
    h_e\ =\ -P_{4}(x) \dxx \ +\ P_{3}(x) \dx\ +\ P_{2}(x) \ ,
\end{equation}
where $P_{4,3,2}(x)$ are polynomials of degrees $4,3,2$ , respectively,
(hence, we study the Heun equation with constrained coefficients), which has
polynomial eigenfunctions. Saying differently, we reduce this problem
to a classification of the second order differential operators with
finite-dimensional invariant subspace in polynomials. This, it is further
reduced to a classification of the Heun operators
admitting finite-dimensional invariant subspace in polynomials, in other words,
operators which are quasi-exactly-solvable.

A general Heun operator with arbitrary coefficients in the basis of monomials is given
by five-diagonal matrix. For a quasi-exactly-solvable (QES) Heun operator this
five-diagonal matrix becomes block-triangular. However, in all known examples the
QES Heun operator in the basis of monomials can be reduced to a tri-diagonal (Jacobi),
block-triangular matrix.

Theorem~\ref{Turbiner:1992} says that the second order differential operator which is quasi-exactly-solvable should be of the form
\[
 T_2 (J^{\al}_n,c_{\al \bet}, c_{\al}) =
 c_{++} J^+_n J^+_n  + 2 c_{+0} J^+_n  J^0_n
 + 2 c_{+-} J^+_n  J^-_n  + 2 c_{0-} J^0_n  J^-_n
 + c_{--} J^-_n  J^-_n
\]
\begin{equation}
\label{QES-2}
 + c_+ J^+_n  + 2 c_0 J^0_n  + c_- J^-_n  + c \ ,
\end{equation}
where $c_{\al \bet}, c_{\al}, c$ are arbitrary constants, the
factor 2 in front of the crossterms is introduced for convenience, and $J^{\al}_n$
are the generators of the algebra $\mathfrak{sl}(2,\bf{R})$
(see (\ref{sl2r})) in $(n+1)$-dimensional representation. The number of free parameters
is $par (T_2) = 9$ with an extra (integer) parameter $n$. The operator $T_2$ (\ref{QES-2})
is simply the Hamiltonian of the $\mathfrak{sl}(2,\bf{R})$ Euler-Arnold quantum top
\[
 h_{E-A} =
 c_{++} J^+_n J^+_n  + c_{+0} \{J^+_n , J^0_n\}
 + c_{+-} \{J^+_n , J^-_n\}  + c_{0-} \{J^0_n , J^-_n\}
 + c_{--} J^-_n  J^-_n
\]
\begin{equation}
\label{EA-qes-2}
 + (c_+ + c_{+0}) J^+_n  + 2 (c_0 - c_{+-}) J^0_n  + (c_- + c_{0-}) J^-_n  + c \ ,
\end{equation}
in a constant (imaginary) magnetic field,
\[
     {\bf B}\ =\ ((c_+ + c_{+0})\ ,\ 2 (c_0 - c_{+-})\ ,\ (c_- + c_{0-}))\ .
\]
Here $c_{\al \bet}$ play the role of components of the tensor of inertia.
If $n$ takes an integer value, the operator $T_2$ has $(n+1)$-dimensional invariant
subspace: it preserves the space ${\p}_{n+1}$.

As a consequence of the invariance (\ref{e1.1.6}) of the space ${\p}_{n+1}$,
see also (\ref{e1.2.2}), the operator $T_2$ (\ref{QES-2}) is equivalent to
\[
 T_2^{(e)} (J^{\al}_n,c_{\al \bet}, c_{\al}) =
 c_{++} J^-_n J^-_n  + 2 c_{+0} J^-_n  J^0_n
 + 2 c_{+-} J^-_n  J^+_n  + 2 c_{0-} J^0_n  J^+_n
 + c_{--} J^+_n  J^+_n
\]
\begin{equation}
\label{QES-2e}
 - c_+ J^-_n  - 2 c_0 J^0_n  - c_- J^+_n  + c \ .
\end{equation}

Substituting the representation (\ref{sl2r}) into the operator
(\ref{QES-2}) we arrive at the Heun operator $h_e$ (\ref{heun-he}),
where the coefficient functions $P_{4,3,2}$ are polynomials of degrees
4,3,2, respectively. The eigenvalue problem $h_e \vphi \ =\ \veps \vphi $
(cf. (\ref{e1.2.5})) takes the form
\begin{equation}
\label{QES-2exp}
 - P_{4}(x) \dxx \vphi (x) \ +\ P_{3}(x) \dx
 \vphi (x) \ +\ P_{2}(x) \vphi (x) \ =\ \veps \vphi (x)\ ,
\end{equation}
where the coefficients of $P_{j}(x)$ are related to
$ c_{\al \bet}, c_{\al}$ and $n$. If the polynomial coefficients $P_{j}(x)$
have arbitrary coefficients this equation becomes the well-known {\it Heun}
equation (see e.g. \cite{Kamke:1959}). This equation is characterized
by four singular points and is considered to be of the next level of
complexity after hypergeometrical (Riemann) equation. The general theory
of the Heun equation is presented in the book by Ince \cite{Ince:1956} (see
also \cite{Kamke:1959,Ronveaux:1995,Maier:2005} for concrete examples).

Theorem~\ref{Turbiner:1992} provides the necessary and sufficient
conditions for the spectral problem with the operator
(\ref{QES-2})-(\ref{QES-2exp}) to have $(n+1)$ polynomial
solutions of the form of polynomials in $x$ of degree not higher
than $n$. It implies that the coefficient functions $P_{j}(x)$
must be of the form
\[
 -P_4 (x)\ =\ c_{++} x^4 + 2 c_{+0} x^3 + 2 c_{+-} x^2 + 2c_{0-} x + c_{--}
 \ ,
\]
\[
 P_3 (x)\ =\ 2(1-n)\, c_{++} x^3 + [(2-3n)\, c_{+0} + c_{+}]x^2 - 2(n\, c_{+-} -
 c_{0}) x - n\, c_{0-} + c_{-} \ ,
\]
\begin{equation}
\label{QES-2coef}
 P_2 (x)\ =\  n^2\, c_{++} x^2 +  n^2\, c_{+0} x - n\, c_{+} x - n\, c_{0} + c \ .
\end{equation}
In general, Theorem~\ref{Turbiner:1992} states that there are no
other algebraic operators of the second order beyond the Heun
operator which generically admit polynomial eigenfunctions.

\noindent
{\it Comment 2.5}\ .
It is important to note that the reference point for coordinate $x$
can be chosen by replacing $x \rar x +a$ in such a way that
\[
 -P_4 (a)=c_{++}a^4 + 2c_{+0}a^3 + 2 c_{+-}a^2 + 2c_{0-}a + c_{--}\ =\ 0\ ,
\]
hence the coefficient function $P_4$ has no constant term and one of its singular
points is located at the origin. It implies that in the operators $T_2$ (\ref{QES-2})
and $T_2^{(e)}$ (\ref{QES-2e}) the coefficient $c_{--}=0$, and the terms
$J^-_n  J^-_n$ and $J^+_n  J^+_n$ are absent in these operators, respectively.
In this case the eigenvalue problem for $T_2^{(e)}$ degenerates to
\begin{equation}
\label{QES-2expe}
 - {\tilde P}_{3}(z) \dzz \vphi (z) \ +\ {\tilde P}_{2}(z) \dz
 \vphi (z) \ +\ {\tilde P}_{1}(x) \vphi (z) \ =\ \veps \vphi (z)\ ,
\end{equation}
where ${\tilde P}_{3,2,1}$ are polynomials of degree 3,2,1, respectively. This is
the so-called polynomial form of the celebrated Heun equation.

Let us proceed by exploring a few particular cases of the Heun operator.



\begin{LEMMA}
\label{Lemma123}
 {\it If the operator (\ref{QES-2}) is such that
\begin{equation}
\label{es123}
 c_{++}=0 \quad \mbox{and}
 \quad c_{+} = (n - 2 m) c_{+0} \ ,
\end{equation}
where $m$ is a non-negative integer, then the operator $T_2$
preserves both ${\p}_{n+1}$ and ${\p}_{m+1}$. In this case the
number of free parameters is reduced, $par (T_2) = 7$.}
\end{LEMMA}

The proof of Lemma~\ref{Lemma123} is based on the fact that under
conditions (\ref{esl23}), the operator $T_2 (J^{\al}_n,c_{\al
\bet}, c_{\al})$ can be rewritten in terms of the generators
$J^{\al}_m$ like $T_2 (J^{\al}_m,\allowbreak c'_{\al
\bet},c'_{\al})$. As a consequence of Lemma~\ref{Lemma123} and
Theorem~\ref{Turbiner:1992}, in general, among polynomial
solutions of (\ref{e1.2.6}) there are some solutions given by polynomials
of order $n$ and others given by polynomials of order $m$.

{\bf Remark.}  From the Lie-algebraic point of view
Lemma~\ref{Lemma123} indicates the existence of representations of
second-degree polynomials in the generators (\ref{sl2r})
possessing two invariant sub-spaces. In general, if $n$ in
(\ref{sl2r}) is a non-negative integer, then among representations
of $k$th degree polynomials in the generators (\ref{sl2r}), lying
in the universal enveloping algebra, there exist representations
possessing $1,2,...,k$ invariant sub-spaces. Even starting from an
infinite-dimensional representation of the original algebra
\footnote{$n$ in (\ref{sl2r}) is {\it not} a non-negative
integer}, one can construct the elements of the universal
enveloping algebra having finite-dimensional representation (for
example, the parameter $n$ in (\ref{esl23}) is non-integer,
however, the operator $T_2$ has the invariant sub-space of
dimension $(m+1)$). Also this property implies the existence of
representations of the polynomial elements of the universal
enveloping algebra $U_{\mathfrak{sl}(2,\bf{R})}$, which can
be obtained starting from different representations of the
original algebra (\ref{sl2r}).

\begin{LEMMA}
\label{Lemma124}
 \cite{Shifman:1999} If the
operator (\ref{QES-2}) is such that
\begin{equation}
\label{e1.2.ER}
 c_{++} = c_{+-} = c_{--} = c_{0} = c = 0 \ ,
\end{equation}
then in the spectral problem (\ref{e1.2.5}) the $(n+1)$
eigenvalues corresponding to polynomial eigenfunctions have a
center of symmetry at $\veps = 0$.
\end{LEMMA}

The proof is based on the fact that under condition (\ref{e1.2.ER}) the
tridiagonal matrix ${\bf T}_2$, which describes the algebraic sector
of the operator (\ref{QES-2}), has vanishing diagonal matrix
elements. It can be easily verified that $\mbox{tr} ({\bf
T}_2^{2k+1})\ =\ 0$ for any $k=0,1,2\ldots$. Hence, the
characteristic polynomial has the form
\[
S_{n+1}(\veps)= \veps^p s_{[\frac{n+1}{2}]} (\veps^2) \ ,
\]
where $p=0$ or 1 depending on $(n+1)$ is even or odd,
respectively\,.$\blacksquare$

This phenomenon is called the {\it energy-reflection symmetry}
\cite{Shifman:1999,Dunne:2002at}. There are several interesting potentials
having this symmetry. The simplest of them is
\[
 V(x) = x^6 - (3 + 2n) x^2\ ,\ n=0,1,2,\ldots \ .
\]
It is worth mentioning that the presence of the
energy-reflection symmetry simplifies the finding of eigenvalues.

A special situation occurs if the parameter $n$ takes the value 0 or
1. For $n=1$ as a consequence of Theorem~\ref{T:1.2.2}, a spectral
problem for the operator which is more general than
(\ref{QES-2})-(\ref{QES-2exp}) arises
\begin{equation}
\label{esl28}
 - F_{3}(x) \dxx \vphi (x) \ +\ \tilde P_{2}(x) \dx
 \vphi (x) \ +\ \tilde P_{1}(x) \vphi (x) \ =\ \veps \vphi (x)\ ,
\end{equation}
where $F_3(x)$ is an arbitrary function and
\[
 \tilde P_2= c_+ x^2 + 2 c_0 x + c_-\ ,
 \ -\tilde P_{1}=c_+x+c_0-c \ ,
\]
with arbitrary coefficients $c$'s. The problem (\ref{esl28}) is
characterized by two polynomial solutions
\begin{equation}
\label{esl28E}
 \vphi_{\pm}=N_{\pm} (c_+x + c_0 \pm \sqrt{c_0^2-c_+ c_-} ) \quad ,
 \quad \veps_{\pm}=c \mp \sqrt{c_0^2-c_+ c_-}\ ,
\end{equation}
where $N_{\pm}$ is a normalization factor. It is evident if the
operator in the l.h.s. of (\ref{esl28}) is symmetric, then the
condition
\begin{equation}
\label{esl28ES}
 {c_0^2-c_+ c_-}\ >\ 0 \ ,
\end{equation}
will be fulfilled.

For the case $n=0$ (one polynomial solution, $\vphi_0 = const $)
the spectral problem (\ref{e1.2.5}) becomes
\begin{equation}
\label{esl29}
 - F_{2}(x) \dxx \vphi (x) \ +\ F_{1}(x) \dx
 \vphi (x) \ +\ c \vphi (x) \ =\ \veps \vphi (x) ,
\end{equation}
(cf.~(\ref{QES-2exp})) with two arbitrary functions $F_{2,1}(x)$
and $c \in \mathbb{R}$. Here the eigenvalue $\veps_0=c$ which
corresponds to $\vphi_0 = const $.

A very important particular case arises if in the operator
(\ref{QES-2}) some of the coefficients vanish: $c_{++} = c_{+0} = c_+ =
0$. The quasi-exactly-solvable operator $T_2$ becomes the
exactly-solvable operator $E_2$ (Lemma 1.2.2)
\begin{equation}
\label{ES-2}
 E_2\ =\   2 c_{+-} J^+_n  J^-_n  + 2c_{0-} J^0_n J^-_n +
 c_{--} J^-_n  J^-_n  + 2c_0 J^0_n  + c_- J^-_n  + c\ ,
\end{equation}
and the number of free parameters is reduced to $par (E_2) = 6$.
Substituting the generator (\ref{sl2r}) into (\ref{ES-2}) and then into
(\ref{e1.2.6}), we obtain the equation
\begin{equation}
\label{ES-2eq}
 - Q_{2}(x) \dxx \vphi (x) \ +\ Q_{1}(x) \dx
 \vphi (x) \ +\ Q_{0}(x) \vphi (x) \ =\ \veps \vphi (x)\ ,
\end{equation}
where $Q_{j}(x)$ are polynomials of $j$th order
\[
-Q_2 (x)\ =\  2\, c_{+-}\, x^2\, +\, 2\, c_{0-}\, x\,+\, c_{--}\ ,
\]
\[
Q_1 (x)\ =\ - 2\, (n\, c_{+-}\,-\, c_{0})\, x - n\, c_{0-}\, +\, c_{-} \ ,
\]
\begin{equation}
\label{ES-2coef}
 Q_0 (x)\ =\  -\, n\, c_{0}\ +\ c \ .
\end{equation}
Different values of $n$ correspond to a redefinition of parameters
$c_{0,-}$ and $c$, thus, without loss of generality, one can set $n=0$.
The equation (\ref{ES-2eq}) with coefficient functions
(\ref{ES-2coef}) coincides with the hypergeometrical equation. The
$c$-coefficients in (\ref{ES-2eq})-(\ref{ES-2coef}) are
arbitrary. However, effectively the number of free parameters has been reduced
by three:
(i) the parameter $c$ corresponds to a choice of the reference point for the
energy and can be set equal to zero: Since the equation (\ref{ES-2eq})
is defined up to a linear change of variable (\ref{e1.1.6lin}) without a
loss of generality, (ii) the parameter $c_{+-}$ or $c_0$ can be set
equal to one and also (iii) the parameter $c_{--}$ or $c_-$ can be set
equal to zero.

As for exact-solvability, there exists families of equivalent, isospectral
exactly-solvable operators $E_2$ whose spectrum coincide up to a reference point.
Effectively, the number of free parameters of the exactly-solvable operator
$E_2$ is equal to three, $par (E_2) = 3$.

The operator on the l.h.s. of (\ref{ES-2eq}) coincides with the
generic hypergeometrical (Riemann) operator. It leads to

\begin{quote}
  {\bf Corollary 2.1}. {\it The hypergeometrical operator
\[
q = (2 c_{+-}x^2 + 2c_{0-}x + c_{--})\dxx + (2c_{0} x+c_-) \dx + c \ ,
\]
where $c$'s are arbitrary numbers, is the only exactly-solvable operator among
second order differential operators.}
\end{quote}

In general, the eigenvalues of (\ref{ES-2eq}) are given by a quadratic
polynomial in an integer number $k=0,1,2,\ldots$ which enumerates eigenstates
\begin{equation}
\label{e1.2.21}
 \veps_{k}= 2 c_{+-} k^2 -2 (c_{+-} - c_0)k + c \ ,
\end{equation}
while the $(k+1)$th eigenfunction is the polynomial in $x$ of order $k$. The parameter $c$ has the meaning of the reference point for eigenvalues, it can be set equal to zero, without loss of generality.

\newpage

\renewcommand{\theequation}{1.3.{\arabic{equation}}}
\setcounter{equation}{0}

\section{(Quasi)-Exactly-Solvable Schr\"odinger Equations}
\subsection{Generalities}
\label{qesp}
The stationary Schr\"odinger equation
\begin{equation}
\label{e1.3.21}
 \big(- \dzz \ + \ V(z) \big) \Psi (z) \ = \ E \Psi(z),
\end{equation}
with $L^2$-boundary conditions is the central equation of quantum physics.
Hence, it is one of the most important types of second-order differential
equations. Here the spectral parameter $E$ is the energy
and $\Psi (z)$ must be a square-integrable (normalizable) function
in the domain where the problem is defined. The operator in the l.h.s. of
(\ref{e1.3.21}) is the Hamiltonian which is sometimes called the
Schr\"odinger operator. Needless to say that any exact solution of
(\ref{e1.3.21}) is of great importance revealing non-trivial
properties, serving as a starting point for developing
a perturbation theory, modeling physical phenomena, etc. Quasi-exact
solvability is one method of creating exact solutions.

There are three domains where the equation (\ref{e1.3.21}) can be
defined: a real line, semi-line or interval. Our task is
two-fold: first to find (and classify) the {\it
quasi-exactly-solvable} Schr\"odinger operators which preserve a
finite-dimensional subspace of a Hilbert space. Second,
to find the {\it exactly-solvable} Schr\"odinger operators which
preserve the infinite flag of finite-dimensional subspaces of
a Hilbert space.

One possible way to construct the (quasi)-exactly-solvable
Schr\"odinger operators is to take the Lie-algebraic
quasi-exactly-solvable $T_2$ (or exactly-solvable $E_2$) operator
acting on finite-dimensional space(s) of polynomials
(\ref{e1.1.5}) and try to transform it into the Schr\"odinger type
operator. This can always be done by making a change of a variable
and a gauge transformation (see (\ref{e1.1.2})) as a consequence
of the one-dimensional nature of the differential equations we are
studying. In practice, the realization of this transformation is nothing but a
conversion of (\ref{QES-2exp})-(\ref{ES-2eq}) into
(\ref{e1.3.21}). All obtained solutions following such a procedure
have a factorizable form of a polynomial in some variable multiplied
by some factor. Usually, this factor is an entire function of $z$ without
zeroes at real $z$ within the domain. For all the known quasi-exactly-solvable
problems it has the meaning of {\it the ground state eigenfunction of a
primitive quasi-exactly-solvable problem at $n=0$}. One open question
remains to be answered: whether obtained solutions of the equation
(\ref{e1.3.21}) belong to square-integrable ones or not? From a general point of view,
this question will be discussed in the end of this Section. In what follows,
we restrict our consideration to particular examples being limited to real
functions and real variables only.

Let us consider the eigenvalue problem for the Heun operator (\ref{heun-he}), $h_e\equiv T_2$,
\[
 T_2 \big(x, \dx \big)\, \vphi (x)\ = \ \veps\, \vphi (x)\ ,
\]
where $h_e$ is the Lie algebraic operator of second degree
(\ref{QES-2}) which is explicitly,
\[
 - P_{4}(x) \dxx \vphi (x) \ +\ P_{3}(x) \dx
 \vphi (x) \ +\ P_{2}(x) \vphi (x) \ =\ \veps \vphi (x)\ ,
\]
(see (\ref{QES-2exp})-(\ref{QES-2coef})). It  is worth noting that
at $n=0$ the coefficient function $P_{2}(x)$ becomes constant. Without
loss of generality, this constant can be placed equal to zero by
redefining $\veps$. By introduction of a new variable $x=x(z)$ and a new function
\begin{equation}
\label{e1.3.22}
 \Psi (z) \ =\ \vphi (x(z)) e^{-A(z)} ,
\end{equation}
one can always reduce the spectral problem for the Lie algebraic
operator $T_2$ (\ref{QES-2exp}) to (\ref{e1.3.21}) with the potential
\cite{Turbiner:1988qes}
\begin{equation}
\label{e1.3.23}
 V(z) = (A')^2 - A'' + P_2 (x(z)) \ .
\end{equation}
Sometimes, $A$ is called {\it prepotential.} Here
\begin{equation}
\label{e1.3.23.1}
 A = \int \left(\frac{P_3}{P_4}\right)dx - log z'\ ,
 \ z = \pm\int \frac{dx}{\sqrt{P_4}} \ .
\end{equation}
Let us mention that in $x$-variable the derivative of $A$ looks like a ratio
of polynomials,
\[
   A'(x) = \frac{P_3(x) - P_4'(x)}{2 P_4(x)}\ .
\]
It makes sense as the logarithmic derivative of a gauge factor
(see (\ref{e1.D1})) - it is a very important object of the theory.
The final form of the most general quasi-exactly-solvable potential
in $x$-variable is given by the rational function
\begin{equation}
\label{e1.3.23.2q}
 V(x)\ =\ \frac{\big(P_4'(x) + 2 P_3(x)\big)\big(3P_4'(x)+2 P_3(x)\big)}{16 P_4(x)}
 - \frac{P_4''(x)}{4} - \frac{P_3'(x)}{2} + P_2 (x) \ ,
\end{equation}
where $P_{4,3,2}$ are from (\ref{QES-2coef}). The Hamiltonian in $x$-variable is
\[
{\mathcal H}(x)\ =\ -\De_g\  +\
\]
\begin{equation}
\label{e1.3.23.3q}
 \frac{\big(P_4'(x) + 2 P_3(x)\big)\big(3P_4'(x)+2 P_3(x)\big)}{16\, P_4(x)}\
 -\ \frac{P_4''(x)}{4}\ -\ \frac{P_3'(x)}{2}\ +\ P_2 (x) \ ,
\end{equation}
where $\De_g$ is the one-dimensional Laplace-Beltrami operator with metric $g^{11}$,
\[
    \De_g = \frac{1}{\sqrt{g}}\dx g^{11} \sqrt{g} \dx\ =\ g^{11}\dxx + \frac{g^{11}_{,1}}{2}\dx \quad ,
    \ g^{11} = P_4(x)\ ,
\]
and determinant $g=g_{11}=\frac{1}{g^{11}}$. Changing the variable $x$ to $z$ (\ref{e1.3.23.1}) the Laplace-Beltrami operator in $x$ becomes the second derivative in $z$.

A similar procedure can be performed for the equation (\ref{esl28})
with two known eigenfunctions. It requires
\[
 A = \int \left({\tilde P_2 \over F_3}\right)dx - log z'\ ,
 \ z = \pm\int \frac{dx}{\sqrt{F_3}}\ ,
\]
and leads to a potential
\begin{equation}
\label{e1.3.23.2}
 V(z) = (A')^2 - A'' + {\tilde P}_1 (x(z)) \ ,
\end{equation}
(cf. (\ref{e1.3.23})). For the case of the equation (\ref{esl29})
with one known eigenfunction
\[
 A = \int \left(\frac{F_1}{F_2}\right)dx - log z'\ ,
 \ z = \pm\int \frac{dx}{\sqrt{F_2}}\ ,
\]
and
\begin{equation}
\label{e1.3.23.3}
 V(z) = (A')^2 - A'' + c \ .
\end{equation}
where a new variable $z$ in (\ref{e1.3.23.1})-(\ref{e1.3.23.3}) is found
from the condition that the coefficient function in front of the second
derivative in (\ref{e1.3.21}) is equal to one.
If the functions (\ref{e1.3.22}) obtained after transformation
belong to the ${\lc}_2(D)$-space\footnote{Depending on the change
of variable  $x = x(z)$, the space $D$ can be the real line, a
semi-line and a finite interval.}, we arrive at the
quasi-exactly-solvable Schr\"odinger equations
\cite{Turbiner:1988f,Turbiner:1988z,Turbiner:1988qes}, where a
finite number of eigenstates is found algebraically.
In this case the spectral parameter $\veps$ plays the role of energy $E$.
We call these quasi-exactly-solvable equations the {\it quasi-exactly-solvable
equations of the first type}.

There exists another type of quasi-exactly-solvable Schr\"odinger equations
related with the generalized eigenvalue problem,
\begin{equation}
\label{e1.3.24}
 (- \dzz \ + \ {\tilde V}(z) ) \Psi (z) \ = \
 \veps \vrho(x(z)) \Psi (z) \ ,
\end{equation}
where a weight function $\hat \vrho(z) = \vrho(x(z))$ is inserted in the r.h.s.
(cf. (\ref{e1.3.21})). Such a spectral problem occurs naturally if a new
variable is found following the requirement that
the coefficient in front of the second derivative in $z$ is equal to some
function
\[
  P_4 (z')^2 = \frac{1}{\vrho(x)}\ ,
\]
different from 1. In this case a new variable occurs
\begin{equation}
\label{e1.3.24.1}
z\ =\  \pm \int \frac{dx}{\sqrt{\vrho (x) P_4 (x)}}\ .
\end{equation}
(cf. (\ref{e1.3.23.1})). The equation which is obtained from
(\ref{QES-2exp}), (\ref{esl28}), (\ref{esl29}) by taking the same
gauge factor (\ref{e1.3.22}) has the form
\[
 \big\{ \frac{1}{\vrho(x(z))}\big[-\dzz \ + (A')^2 - A''\big] + P_2 (x(z))-
 \veps \big\} \Psi(z)=0 \ .
\]
Multiplying both sides of this equation by $\vrho(x(z))$ we arrive at
\begin{equation}
\label{e1.3.24.2}
 \big\{- \dzz \ + (A')^2 - A'' + \big[ P_2 (x(z))-
 \veps \big]\vrho (x(z))\big\} \Psi(z)=0 \ ,
\end{equation}
which can be rewritten in the form (\ref{e1.3.24}) with a potential
\begin{equation}
\label{e1.3.24.3}
\tilde V(z) = (A')^2 - A'' + P_2 (x(z)) \vrho (x(z)) \ ,
\end{equation}
this is a slight modification of (\ref{e1.3.23}). For all known
quasi-exactly-solvable problems of the second type the weight factor is a
monomial  $\vrho (x)=x^d$, where $d$ is equal to either $\pm 1$ or $\pm 2$.
In the case of those problems the spectral parameter $\veps$ looses its meaning
of energy, while the energy enters explicitly (or implicitly) to the
potential as a constant term. The meaning of the spectral problem (\ref{e1.3.24})
is rather unusual. We study a family of potentials such that the ground state
energy in the first potential is equal to the energy of the first excited state
in the second potential, which is itself equal to the energy of the second excited state
in the third potential, etc. The eigenfunctions of these states have the form of a
polynomial of a fixed degree multiplied by some factor. This procedure is
widely used in atomic physics in the case of the Coulomb problem. Namely,
instead of studying the spectral problem for the Coulomb Hamiltonian
$(-\De - \frac{\al}{r})\Psi \ =\ E \psi$, we explore a modified spectral
problem $(-\De - E)\Psi = \frac{\al}{r} \psi$, or equivalently,
$(-r\De - E r)\Psi = \al \psi$\,, where the energy $E$ is fixed
and quantization of the charge $\al$ is considered. It ends up with quantization
formula $\al_k= {\sqrt{-2 E}}\, k$\,, where $k=1,2,3, \ldots$. If we assume now that
the charge $\al$ is fixed this formula leads immediately to the familiar
formula for the quantization of energy: $E_k = -\frac{1}{2k^2}\,,\ k=1,2,3,\ldots$.
From a physical point of view such an approach means that we are looking for the Coulomb
problem of different charges which contains an eigenstate of a certain fixed energy $E$.
Such a representation of the Schr\"odinger equation as a spectral problem is called
the {\it Sturm} representation for the Coulomb problem. Our spectral problem is
a generalization of the Sturm representation to other potentials. We will continue to call it
the {\it Sturm representation}. This representation turned out to be very useful for studying the quasi-exactly-solvable problems \cite{Turbiner:1988z,Turbiner:1994PR}.

Below, we will follow the catalogue given in
\cite{Turbiner:1988qes}. A presentation of results is given the
following sequence: First, we display the quadratic element $T_2$ of the
universal enveloping algebra $\mathfrak{sl}(2,\bf{R})$ in the
representation (\ref{sl2r}) and then its equivalent form of
differential operator $T_2(x, \frac{d}{dx})$. Second, we  present the corresponding
potential $V(z)$ or $\tilde V(z)$ if the weight factor $\vrho (x) \neq 1$ and
an explicit expression for the change of the variable $x=x(z)$,
weight function $\vrho (z)$. Third, we present the explicit form of $y=-A'(z)$,
which is the logarithmic derivative of the ground state eigenfunction for the
potential at $n=0$ with negative sign. Finally, the functional form of
the eigenfunctions $\Psi (z)$ of the ``algebraized" part of the spectra
(the algebraic sector of the space of eigenfunctions) is
given~\footnote{The functions $p_n(x)$ occurring in the expressions
for $\Psi (z)$ denote polynomials of the $n$th order. They are nothing but
the polynomial eigenfunctions of the operator $T_2(x,d_x)$}.

All known one-dimensional quasi-exactly-solvable Schr\"odinger equations
degenerate either to well-known exactly-solvable ones, or the
non-solvable special ones  - like the Mathieu or Lame equations,
being associated with them. In total, there exists ten types of the
quasi-exactly-solvable, one-dimensional Schr\"odinger equations.

We begin our consideration with the quasi-exactly-solvable
equations associated with the exactly-solvable Morse oscillator. It
implies that at the limit, when the number of algebraic
eigenstates $(n+1)$ goes to infinity, the Morse oscillator is
recovered. There are three quasi-exactly-solvable problems of this
type which are given by Case I, II, III. Their potentials are finite
pieces of the Laurent series in variable $e^{-\al z}$.

\vskip .3cm

\subsection{Morse-type potentials}

The {\it Morse} oscillator is one of the well-known exactly-solvable quantum-mechanical
problems (see e.g. Landau and Lifschitz \cite{LL-QM:1977}). It is described by
the Hamiltonian with the potential
\begin{equation}
\label{e1.3.2-morse}
V(z) = A^2 e^{-2\al z} - 2 A e^{-\al z}  \ ,\  A >0 \ ,\ \al > 0\ ,
\end{equation}

\begin{figure}[tb]
\begin{center}
     {\includegraphics*[width=3in]{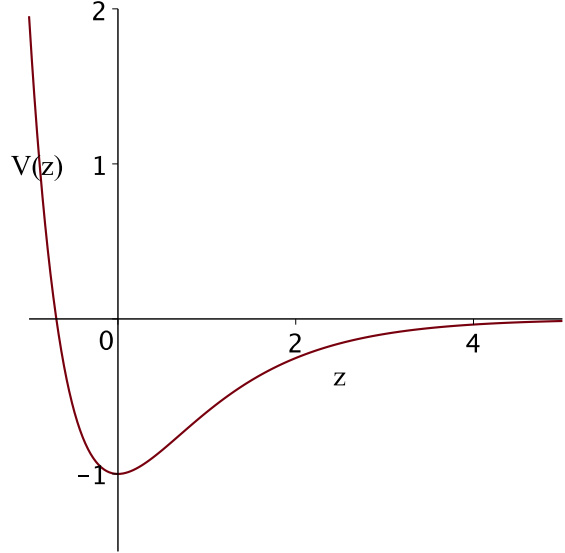}}
     \caption{Morse potential (\ref{e1.3.2-morse}) at $A = \al = 1$. It has a minimum at
     $z=0$, $V_{min}=-1$\,. The discrete spectra consists of a single bound state (the ground state)
     with energy $E_0=-\frac{1}{4}$, and eigenfunction $\Psi_0 = e^{-\frac{z}{2}-e^{-z}}$ .
      }
    \label{Morse-E}
\end{center}
\end{figure}
which is called the {\it Morse} potential, see Fig. \ref{Morse-E}. In general, this equation is
characterized by a finite number of bound states,
\[
  E_k = - [1 + \frac{\al}{2}(1-2k)]^2 ,\quad k=0,1 \ldots k_{max}\ ,
\]
its number $k_{max}=[\frac{1}{\al}+\frac{3}{2}]$ depends on the parameter $\al$ only, and
where $[a]$ denotes the integer part of $a$.
The potential $V(z)$ is periodic with the imaginary period $i\, \frac{2\pi}{\al}$ and
the Hamiltonian is translation-invariant, $T_c: z \rar z + i\, \frac{2\pi}{\al}$.
Taking the invariant $t_{\mp}=e^{\mp\al z}$ as a new variable (note, that $t_- t_+ =1$ ), we get the Morse potential
in polynomial form
\begin{equation}
\label{e1.3.2-morse-t}
     V(t_-) = A^2 t_-^2 - 2 A t_- \ ,\ V(t_+) =  \frac{A^2}{t_+^2} - \frac{2 A}{t_+} \ ,
\end{equation}
and the second derivative ($1D$ Laplacian) becomes the Laplace-Beltrami operator
with metric $g^{11}= \al^2 t_{\pm}^2$ ,
\[
    \dzz\ =\ \al^2 t^2 \dtt + \al^2 t \dt \ ,\quad t=t_{\pm}\ .
\]
Hence, the Laplace-Beltrami operator does not depend on the sign of $\al$ or, saying differently,
it is invariant under the transformation $t \rar 1/t$.

The Hamiltonian of the Morse oscillator in $t_-$-variable
\begin{equation}
\label{e1.3.2-morse-tHam}
  {\mathcal H}(t_-)\ =\ -\al^2 t_-^2 \,\frac{d^2}{dt_-^2} - \al^2 t_- \frac{d}{dt_-} + A^2 t_-^2 - 2 A t_- \ =\ {\mathcal H}(1/t_+)\ ,
\end{equation}
is the algebraic operator. It has the form of the Heun operator.

The Morse oscillator is widely used in molecular physics to model the interaction
of the atoms in diatomic molecules.

\newpage

{\bf Case I.}

\vskip .3cm
Consider the following bilinear combination of the $\mathfrak{sl}(2,\bf{R})$
generators (\ref{sl2r}) with index $n$ (see \cite{Turbiner:1988qes})
\renewcommand{\theequation}{1.3.16-{\arabic{equation}}}
\setcounter{equation}{0}
\begin{equation}
\label{e1.3.25}
 T_2 = -\al ^2J^+_n J^-_n + 2\al a J^+_n -
 \al [\al (n+1)- 2b] J^0_n - 2\al c J^-_n
\end{equation}
\[
- \frac{\al n }{ 2} [\al (n+1) - 2b]\ ,
\]
where $\al \neq 0, a, b, c$ are real parameters.
If $a=0$ the term of positive grading in (\ref{e1.3.25}) disappears and $T_2$
becomes exactly-solvable. Since $T_2$ is proportional to $\al$ it is convenient
to divide it by $\al$ and measure also the spectral parameter $\veps$ in
units of $\al$:
\[
  \veps \rar \al \veps\ .
\]
After the substitution of the generators (\ref{sl2r}) into (\ref{e1.3.25})
and division on $\al$ we get an algebraic operator
\begin{equation}
\label{e1.3.25.1}
 T_2(x,d_x) = -\al x^2d_x^2 + [2ax^2+(2b-\al)x-2c]d_x -
 2 a n x \ .
\end{equation}
In the basis of monomials $\{1, x, x^2, \ldots, x^k, \ldots x^n, \ldots \}$ this operator has the form of tri-diagonal, Jacobi matrix,
\begin{equation}
\label{e1.3.25.2}
 t_{k,k-1} = -2 c\, k\ ,\ t_{k,k} = (2 b -\al k)\, k\ ,\ t_{k,k+1} = 2 a (k - n)\ .
\end{equation}
If $k=n$, the matrix element $t_{n,n+1}=0$ and the Jacobi matrix $T_2$ becomes block-triangular. In this case
the characteristic polynomial is factorized, $\det (T_2 - \veps) = P_n(\veps) P_{\infty}(\veps)$.

\renewcommand{\theequation}{1.3.{\arabic{equation}}}
\setcounter{equation}{16}
\noindent
In general, this operator has $(n+1)$ polynomial eigenfunctions. However,
at $a=0$ the operator $T_2(x,d_x)$ becomes exactly-solvable: it has infinitely-many
polynomial eigenfunctions. In this case the upper subdiagonal of $T_2$ vanishes, this matrix becomes
lower-triangular.

As an illustration let us consider $n=1$ in (\ref{e1.3.25.1}). In this case
there are two polynomial eigenfunctions,
\[
 \vphi_{\pm}=4 a x + (2b-\al) \mp \sqrt{(2b-\al)^2
 +16 a c} \ ,
\]
with eigenvalues
\begin{equation}
\label{e1.3.25E}
 \veps_{\pm}=\frac{(2b-\al) \pm \sqrt{(2b-\al)^2
 +16 a c}}{2} \ ,
\end{equation}
(cf. (\ref{esl28E})) .
Both eigenfunctions at fixed $x$ as well as both eigenvalues form a two-sheeted Riemann surface
in any of the parameters $\al,a,b,c$ if the others are kept fixed with square-root branch points at
$(2b-\al)= \pm 4 i (ac)^{1/2}$.  In order for the spectra of (\ref{e1.3.25}) to be real
\[
(2b-\al)^2 +16 a c > 0 \ ,
\]
(cf. (\ref{esl28ES})), where the equality is excluded as a consequence of
Hermiticity of (\ref{e1.3.25.1}) which prohibits the degeneracy of energy. Since the domain for
(\ref{e1.3.25.1}) is $x \in [0, +\infty)$ and also $a c > 0$ (see below), $\vphi_-$ describes the
ground state (it has no zeroes (nodes) at $x>0$) with energy $\veps_-$.

In the universal enveloping algebra $\mathfrak{sl}(2,\bf{R})$ there exists an element
which is equivalent to (\ref{e1.3.25}). It can be obtained from (\ref{e1.3.25})
by conjugation (\ref{e1.2.2})
\renewcommand{\theequation}{1.3.18-{\arabic{equation}}}
\setcounter{equation}{0}
\begin{equation}
\label{e1.3.25equiv}
 T_2 = -\al ^2 J^+_n J^-_n + 2\al c J^+_n +
 \al [\al (n-1)- 2b] J^0_n - 2\al a J^-_n
\end{equation}
\[
- \frac{\al n }{ 2} [\al (n+1) - 2b]\ .
\]
In fact, the operator (\ref{e1.3.25equiv}) coincides with (\ref{e1.3.25}) with renamed
parameters (see below, (\ref{e1.3.25.1param})). So, this conjugation does
not bring anything new when we are making a mapping (\ref{e1.3.25}) to itself.
The operator (\ref{e1.3.25equiv}) has $(n+1)$ polynomial eigenfunctions with
the same eigenvalues as (\ref{e1.3.25}).
The polynomial eigenfunctions of the operators (\ref{e1.3.25}) and
(\ref{e1.3.25equiv}) are related with each other through the invariance
condition (\ref{e1.1.6}). However, if $c=0$ a single term of positive
grading in (\ref{e1.3.25equiv}) disappears and the operator becomes exactly-solvable.
Naturally, this exactly-solvable form does not occur for (\ref{e1.3.25}).

The algebraic operator, which corresponds to (\ref{e1.3.25equiv}), after
division by $\al$ takes the form
\begin{equation}
\label{e1.3.25.1equiv}
 T_2(x,d_x)= -\al x^2d_x^2 + [2cx^2+(\al (2n-1)-2b)x - 2a]d_x -
 2 c n x - n (\al n -2b)\ ,
\end{equation}
(cf. (\ref{e1.3.25.1})). This operator can be obtained from (\ref{e1.3.25.1})
through the following change of parameters
\[
 a \lrar c \ ,
\]
\[
 \al \rar \al \ ,
\]
\begin{equation}
\label{e1.3.25.1param}
 b \rar -b + \al n \ ,
\end{equation}
if we neglect constant terms in the operator, these terms can always be adjusted
through a change of the reference point for eigenvalues.
\renewcommand{\theequation}{1.3.{\arabic{equation}}}
\setcounter{equation}{18}

\begin{figure}[tb]
\begin{center}
     {\includegraphics*[width=3in]{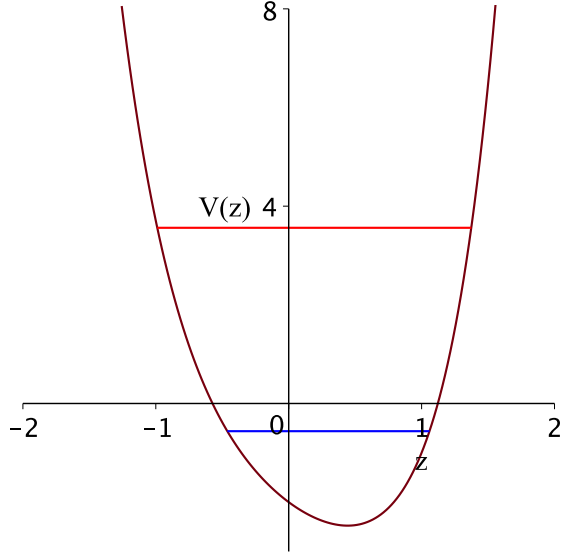}}
     \caption{QES Morse potential I (\ref{e1.3.26}) at $a = b = c = \al = 1$ and
     $n=1$:\ $V_I(z) = e^{-2 z} - e^{- z} - 3 e^{ z} + e^{2 z}$. It has minimum at
     $z \sim 1/2$, $V_{min} \sim -2.2$\,. Algebraic discrete spectra consists
     of two bound states (ground state and the first excited state) with energies
     $E_{\pm}=\frac{3}{2} \pm \frac{\sqrt{17}}{2} $, and eigenfunctions
     $\Psi_{\pm} = (4 e^{-z} + 1 \mp \frac{\sqrt{17}}{2}) e^{-e^{-z} + z - e^{z}}$ .
      }
    \label{Morse-I}
\end{center}
\end{figure}

The spectral problem for the operator (\ref{e1.3.25}) is reduced to the Schr\"odinger
equation (\ref{e1.3.21}) by the procedure (\ref{e1.3.22})-(\ref{e1.3.23.1}).
As a result the spectral problem (\ref{e1.3.24}) with the potential
\begin{equation}
\label{e1.3.26}
 V_I(z) = a^2e^{-2\al z} + a[2b - \al (2n +1)]e^{-\al z} -
 c(2b+\al)e^{\al z} + c^2e^{2\al z}
\end{equation}
\[
+b^2 - 2ac \ ,
\]
occurs, where
\[
 x=e^{-\al z}\ .
\]
see Fig. \ref{Morse-I}.
The constant part of the potential, $(b^2 - 2ac)$, can be absorbed into the definition
of energy $E=\veps - (b^2 - 2ac)$. Of course, a similar potential will occur if the operator
(\ref{e1.3.25.1}) is reduced to the Schr\"odinger operator (\ref{e1.3.21})
by the procedure (\ref{e1.3.22})-(\ref{e1.3.23.1}). This potential will correspond to
a change of parameters in (\ref{e1.3.26}) by (\ref{e1.3.25.1param}).
The derivative of the gauge factor $y=-A'$ is
\begin{equation}
\label{e1.3.26.log}
 y(z) \ = \ -a e^{-\al z} - b + c e^{\al z} \ .
\end{equation}
and has a meaning of the logarithmic derivative of the ground state eigenfunction at $n=0$
with negative sign, see (\ref{e1.3.23.1}).

The Schr\"odinger equation is defined on the real line $z \in
(-\infty, +\infty)$, while the spectral problem for $T_2$ is
restricted to $x \in [0, +\infty)$. For non-vanishing values $a,c$ the potential
(\ref{e1.3.26}) grows at $|z| \rar \infty$ and always has infinite discrete spectra.
For $n=1$ it implies that one of the functions $\vphi_{\pm}$ in (\ref{e1.3.25E})
has either no nodes or one node, while another one has either one node or no nodes,
respectively, in agreement with Sturm (oscillation) theorem ($a c \geq 0$, see below).
Thus, the Schr\"odinger equation with the potential (\ref{e1.3.26}) is a
quasi-exactly-solvable Schr\"odinger equation of the first type. A finite number of
eigenfunctions and eigen-energies can be found through a linear algebra procedure.

The potential $V(z)$ (\ref{e1.3.26}) is periodic with the imaginary period
$i\,\frac{2\pi}{\al}$ and the Hamiltonian is translationally-invariant,
$z \rar z + i\,\frac{2\pi}{\al}$.
The variable $x$ has the meaning of the invariant $t=x (= e^{-\al z})$. Taking $t$
as the new variable, we get the quasi-exactly-solvable potential (\ref{e1.3.26})
in rational form
\begin{equation}
\label{Vqes1}
 V(t) = a^2 t^2 + a[2b - \al (2n +1)] t - \frac{c(2b+\al)}{t} + \frac{c^2}{t^2}\ .
\end{equation}
The quasi-exactly-solvable Hamiltonian in $t$-variable is
\begin{equation}
\label{Hqes1}
  {\mathcal H}\ =\ -\al^2 t^2 \dtt + \al^2 t \dt + a^2 t^2 + a[2b - \al (2n +1)] t
  - \frac{c(2b+\al)}{t} + \frac{c^2}{t^2}\ .
\end{equation}

\hskip 0.5cm
The $(n+1)$ polynomial solutions (up to a multiplicative factor) of the Schr\"odinger equation (\ref{e1.3.21}) define the \textit{algebraic} part of the spectra; they have the form
\begin{equation}
\label{e1.3.26.eif}
 \Psi_{n}^{(k)} (z) \ = \  p_{n}^{(k)} (e^{-\al z})
 \exp{\bigg(-\frac{a}{\al} e^{-\al z} + b z - \frac{c}{\al} e^{ \al z}\bigg)} \ ,\ k=0,1,\ldots n
\end{equation}
where $p_{n}^{(k)}(x)$ is the polynomial eigenfunction of the operator $T_2$. Multiplicative factor in (\ref{e1.3.26.eif})
\begin{equation}
\label{e1.3.26.eif-0}
 \Psi_0 (z) \ = \  \exp{\bigg(-\frac{a}{\al} e^{-\al z} + b z - \frac{c}{\al} e^{ \al z}\bigg)} \ ,
\end{equation}
if normalizable, defines the ground state. So far, these functions (\ref{e1.3.26.eif}) are formal solutions of the Schr\"odinger equation. In order to answer the question whether these solutions have the meaning of eigenfunctions, we should check their normalizability.
This will be performed later.

In the invariant variable $t$, the eigenfunction (\ref{e1.3.26.eif}) is
\begin{equation}
\label{Pqes1}
 \Psi (t) \ = \  p_n (t)\ t^{-\frac{b}{\al}}\ \exp{\bigg(-\frac{a}{\al}\, t - \frac{c}{\al\, t}\bigg)} \ .
\end{equation}

\hskip 0.5cm
In order to find the algebraic eigenfunctions (\ref{e1.3.26.eif}) we have to diagonalize
the Jacobi matrix $T_2$ of size $(n+1)$ by $(n+1)$ with the matrix elements
\[
  t_{k,k}=  k (2b - \al k)\ ,\ t_{k, k+1} = -2 a(n-k)\ , \
  t_{k+1,k}= -2 c(k+1)\ ,
\]
(cf. (\ref{e1.3.25.2})),
where $k=0,1,2,\ldots n$. Its eigenvalues are the energies of the corresponding algebraic
eigenfunctions, $E=\veps/\al$ (hence, they are measured in units of $\al$) while its
eigenfunctions define the coefficients of polynomials $p_n$. Reality of the eigenvalues of $J$
is guaranteed if $t_{k+1,k}\, t_{k,k+1} >0$ (see e.g. \cite{Mishina:1962}, p. 28).
The characteristic equations for energies $E$ (measured in the units of $\al$) at $n=1,2$
have the form
\[
  E^2 + (\al - 2b) E - 4 a c =0\ ,
\]
\[
 E^3 + (5 \al - 6 b) E^2 + 4 [(\al-b)(\al-2b)- 4ac] E - 32 ac (\al-b) = 0\ ,
\]
respectively. For a given $n$ the characteristic equation has order $(n+1)$,
its eigenvalues (as well as eigenfunctions at fixed $x$ or $z$) form a $(n+1)$-sheeted Riemann
surface in any parameter when all the other parameters are kept fixed. One can easily see that
when $n$ tends to infinity, the parameter $a$ tends to zero. One subdiagonal in the matrix $t$
vanishes, the eigenvalues coincide with diagonal matrix elements and the Riemann surface splits
into separate (disconnected) sheets.

Now we return to the question about normalizability of (\ref{e1.3.26.eif}).
If $\al >0$, the eigenfunctions (\ref{e1.3.26.eif}) are, in
general, normalizable for any $a, c > 0$ and real $b$, the potential has either
one or two minima depending on the values of parameters~
\footnote{It is worth noting that a rather amusing situation occurs at $2b=-\al < 0$,
$a>0$ and $c\equiv i c$ pure imaginary. The potential (\ref{e1.3.26}) is real and unbounded
from below (bottomless), $V(z) = a^2 e^{-2\al z} - 2 a \al (n +1) e^{-\al z} - c^2 e^{2\al z}$.
All eigenfunctions (\ref{e1.3.26.eif}) are normalizable. In such a potential,
which is sometimes called {\it inverted} the time for a particle to travel to $+\infty$
is finite (see for a discussion the book by Titchmarsh \cite{Titchmarsh}}.

If $c=0$, then for $a>0,\ b < 0$ the normalizability of (\ref{e1.3.26.eif}) is guaranteed
for any $n$; the potential has a single minimum.

On the other hand, if $a=0$ (which corresponds to the
exactly-solvable Morse problem~
\footnote{The form in which this potential usually appears
in the textbooks assumes that the parameter $\alpha$ is negative}), the parameters
$c > 0$ and $b > 0$. If the integer part $[b/\al] = n$, the first $(n+1)$ eigenstates of the
potential (\ref{e1.3.26}) are characterized by normalizable eigenfunctions (\ref{e1.3.26.eif}).
Their energies are
\begin{equation}
\label{e1.3.26.eiv}
 E_k\ =\ \al k (2b - \al k)\ ,\ k=0,1,\ldots, n \ .
\end{equation}
It is worth noting that the operator $T_2$ (\ref{e1.3.25}) has infinitely many polynomial
eigenfunctions with eigenvalues which are given by (\ref{e1.3.26.eiv}) with integer $k$ running
from 0 up to infinity. However, for $k>n$ the corresponding eigenfunctions (\ref{e1.3.26.eif})
are non-normalizable and, hence, they are irrelevant physically. It is known
(see, for example, the textbook by Landau-Lifschitz
\cite{LL-QM:1977}) that the above eigenstates with $k \leq n$
exhaust all bound states in the potential (\ref{e1.3.26}). If $\al
< 0$, the eigenfunctions (\ref{e1.3.26.eif}) are, in general,
normalizable at $a, c < 0$ and real $b$, though if $c=0$, then
$a<0,\ b > 0$. Now, let us take $a=0$ (which again corresponds to the
exactly-solvable problem); for normalizability the parameters $c > 0$ and $b/\al > 0$.
If the integer part $[b/\al] = n$, where $n=0,1,2,\ldots$, the first $(n+1)$ eigenstates
of the potential (\ref{e1.3.26}) have normalizable eigenfunctions
(\ref{e1.3.26.eif}) and describe bound states with energies
(\ref{e1.3.26.eiv}).

It is worth to emphasize how the limit to the exactly-solvable problem, $a \rar 0$, is taken.
The procedure for energies is quite obvious but not at all for the eigenfunctions. Let
us consider the case $n=1$ (and thus $\vphi_{\pm}$) as an example.
At $a \rar 0$ the function $\vphi_{-}$ tends to a constant, which is the
lowest eigenfunction, and $E_- \rar 0$. As for the state $\vphi_{+}$
one can see immediately that the energy $E_+ \rar (2b - \al)$ at $a \rar 0$.
In order to get a meaningful eigenfunction $\vphi_{+}$ at $a \rar 0$ we should
introduce a normalization factor $\propto 1/a$ and then take the limit $a \rar 0$.
Eventually,
\[
\frac{1}{a}\ \vphi_{+} (x) \rar 4x - \frac{8 c}{(2b - \al)^2} \ ,
\]
or, in $z$-representation,
\[
\frac{1}{a}\ \vphi_{+} (z) \rar 4 e^{-\al z} - \frac{8 c}{(2b - \al)^2} \ .
\]


{\bf Case II}.

\vskip .3cm
Let us take a bilinear combination of the $\mathfrak{sl}(2,\bf{R})$
generators with index $n$ (\ref{sl2r}) (see  \cite{Turbiner:1988qes})
\renewcommand{\theequation}{1.3.27-{\arabic{equation}}}
\setcounter{equation}{0}
\begin{equation}
\label{e1.3.27}
 T_2 = -\al J^0_n J^-_n + 2a J^+_n + 2c J^0_n -
 \bigg(\frac{n+2}{2}\al + 2b\bigg) J^-_n + c\, n \ ,
\end{equation}
(cf. (\ref{e1.3.25})), where $\al \neq 0, a, b, c$ are real parameters.
As for the corresponding algebraic operator it looks like
\begin{equation}
\label{e1.3.27.1}
 T_2(x,d_x)= -\al x d_x^2 - (2ax^2 + 2cx - 2b - \al)d_x + 2 a n x
 \ .
\end{equation}

In the basis of monomials $\{1, x, x^2, \ldots, x^k, \ldots x^n, \ldots \}$ this operator
has the form of a tri-diagonal, Jacobi matrix,
\begin{equation}
\label{e1.3.27.2}
 t_{k,k-1} = [2 (b +\al )-\al k]\, k \ ,\quad t_{k,k} = 2 c\, k\ ,\quad t_{k,k+1} = -2 a (k - n)\ .
\end{equation}
If $k=n$, the matrix element $t_{n,n+1}=0$ and the Jacobi matrix $T_2$ becomes block-triangular.
In this case, the characteristic polynomial can be factorized,
$\det (T_2 - \veps) = P_n(\veps) P_{\infty}(\veps)$.

Evidently, this operator has $(n+1)$ polynomial eigenfunctions, which
form the algebraic sector of eigenstates. For example, at $n=1$ there
exist two polynomial eigenfunctions
\renewcommand{\theequation}{1.3.{\arabic{equation}}}
\setcounter{equation}{27}
\[
 \vphi_{\pm}= (2 b + \al) x - c \mp \sqrt{c^2 +2 a (2b+\al)} \ ,
\]
with eigenvalues
\begin{equation}
\label{e1.3.27E}
 \veps_{\pm}=-c \pm \sqrt{c^2 + 2 a (2b+\al)} \ .
\end{equation}
Both eigenfunctions and both eigenvalues form a two-sheeted Riemann
surface in any of the parameters $\al,a,b,c$ if the other parameters are fixed.
Reality of the spectra of (\ref{e1.3.27}) requires, in particular,
\[
 c^2 + 2 a (2b+\al) > 0 \ ,
\]
(cf. (\ref{esl28ES})). Equality is excluded as a consequence of
requirement of Hermiticity of (\ref{e1.3.27.1}) which prohibits degeneracy
in energy. Since the domain for  (\ref{e1.3.27.1}) is $[0, +\infty)$ and
$a (2b+\al) > 0$ (see below), $\vphi_-$ describes the ground state (it has
no zeroes at $x>0$).

It is worth mentioning that in the universal enveloping algebra
$U_{\mathfrak{sl}(2,\bf{R})}$ there exists an element which is
equivalent to (\ref{e1.3.27}). It can be obtained from
(\ref{e1.3.27}) by conjugation (\ref{e1.2.2})
\renewcommand{\theequation}{1.3.28-{\arabic{equation}}}
\setcounter{equation}{0}
\begin{equation}
\label{e1.3.27equiv}
 T_2 = -\al J^+_n J^0_n + \bigg(\frac{n+2}{2}\al + 2b\bigg) J^+_n - 2c J^0_n - 2a J^-_n
 + c\,n \ ,
\end{equation}
(see Case III, cf. (\ref{e1.3.29})).
It has $(n+1)$ polynomial eigenfunctions with the same eigenvalues as (\ref{e1.3.27}).
However, since $\al \neq 0$ the terms of positive grading in (\ref{e1.3.27equiv})
never disappear; this operator never becomes exactly-solvable. It does not preserve
the flag of polynomials. The limit $\al \rar 0$ is singular, the operators
(\ref{e1.3.27}) and (\ref{e1.3.27equiv}) loose their bilinearity in generators
becoming linear in generators. The polynomial eigenfunctions of the operators
(\ref{e1.3.27}) and (\ref{e1.3.27equiv}) are related to each other through the
invariance condition (\ref{e1.1.6}). The algebraic operator which corresponds
to (\ref{e1.3.27equiv}) has the form
\[
 T_2(x,d_x)= -\al x^3 d_x^2 + [2(\al n + b)x^2 - 2cx - 2a - \al]d_x
\]
\begin{equation}
\label{e1.3.27.1equiv}
 -[\al \frac{(n+3)}{2} +2b]n x + c (n+1) \ ,
\end{equation}
(cf. (\ref{e1.3.27.1})).
\renewcommand{\theequation}{1.3.{\arabic{equation}}}
\setcounter{equation}{28}

\begin{figure}[tb]
\begin{center}
     {\includegraphics*[width=3in]{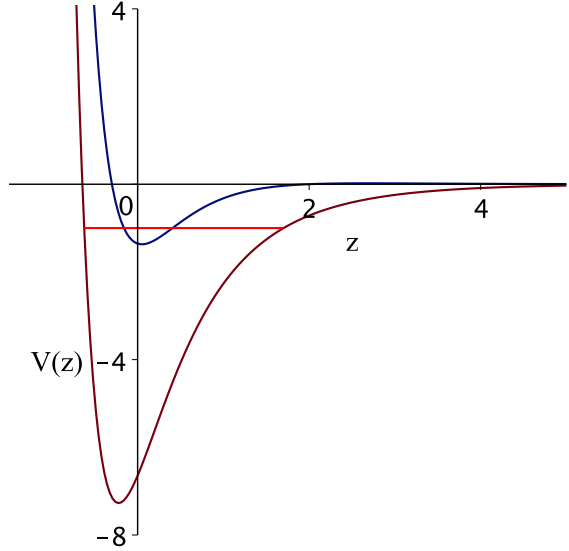}}
     \caption{QES Morse potential II (\ref{e1.3.28}) at $a = b = c = \al = 1$ and $n=1$.\ Two potentials $V_{II}^{(\pm)}(z) = e^{-4 z} + 2 e^{- 3z} + 6 e^{-2z} -(3+\veps_{\pm}) e^{ -z}$,  where the ground state energy for $\veps_{+}$ (brown line)
     is equal to the 1st excited state energy for $\veps_{-}$ (blue line) and is equal to $E=-1$; $\veps_{\pm}=\pm {\sqrt{7}}$.
     Its eigenfunctions are $\Psi_{\pm} = (3 e^{-z} - 1 \mp {\sqrt{7}}) e^{ - \frac{1}{2} e^{-2 z} - e^{-z} - z}$\,.
      }
    \label{Morse-II}
\end{center}
\end{figure}

\hskip 0.5cm
Rewriting the operator (\ref{e1.3.27}) to the form of the Schr\"odinger operator,
we arrive at a generalized spectral problem (\ref{e1.3.24}) with the potential
\begin{equation}
\label{e1.3.28}
 \tilde V(z)_{II}\ =\ a^2e^{-4\al z} + 2 a c e^{-3\al z}
 + [c^2 - 2a(b +\al a n + \al)]e^{-2\al z} - c(2b+\al )e^{-\al z}
 + b^2\ ,
\end{equation}
where
\[
 x=e^{-\al z}\ ,
\]
and the weight factor is given by
\[
 \vrho=\frac{1}{\al} e^{-\al z} \ .
\]

see Fig. \ref{Morse-II}.
The constant part of the potential, $b^2$, can be absorbed into the definition of energy $E=\veps - b^2$.

This potential (\ref{e1.3.28}) grows in one direction and tends to a constant $b^2$ in another direction.
Since this constant can always be chosen as a reference point for the energy $E_{ref}=-b^2$,
the new potential will vanish in that direction.
The derivative of the gauge factor $y=-A'$ (logarithmic derivative of the ground state
eigenfunction at $n=0$ with sign minus, see (\ref{e1.3.23.1})) is
\begin{equation}
\label{e1.3.28log}
 y(z) \ = \ -a e^{-2\al z} - c e^{-\al z} + b \ .
\end{equation}
Similar to Case I, the spectral problem (\ref{e1.3.24}) is
defined again on the real line $z \in (-\infty, +\infty)$, while the
spectral problem for $T_2$ is actually restricted to $x \in [0, +\infty)$. Thus, the
quasi-exactly-solvable Schr\"odinger equation with potential (\ref{e1.3.28}) belongs
to the second type. A finite number of eigenfunctions, each of them within its own
potential, at a fixed energy, can be found through an algebraic procedure.

The potential $\tilde V(z)$ (\ref{e1.3.28}) is periodic with the imaginary period
$i \frac{2\pi}{\al}$, hence, the Hamiltonian is translationally-invariant, $z \rar z + i \frac{2\pi}{\al}$.
The variable $x$ has a meaning of the invariant with respect to translations,
$t=x (= e^{-\al z})$. Taking $t$ as a new variable, we get the quasi-exactly-solvable potential (\ref{e1.3.28}) in polynomial form
\begin{equation}
\label{Vqes2}
 {\tilde V}(t) = a^2 t^4 + 2 a c t^{3}
 + [c^2 - 2 a (b + \al a n + \al)]t^{2} - c (2b + \al )t\ .
\end{equation}
The quasi-exactly-solvable Hamiltonian in $t$-variable is
\begin{equation}
\label{Hqes2}
  {\mathcal H}\ =\ -\al^2 t^2 \dtt + \al^2 t \dt + a^2 t^4 + 2 a c t^{3}
 + [c^2 - 2 a (b + \al a n + \al)]t^{2} - c (2b + \al )t\ .
\end{equation}

The algebraic eigenfunctions in the potential (\ref{e1.3.28}) become
\begin{equation}
\label{e1.3.28eif}
 \Psi (z) \ = \  p_n (e^{-\al z}) \exp{(-\frac{a}{2\al}e^{-2\al z}
 - \frac{c}{\al}e^{-\al z} - b z)} \ .
\end{equation}
and written in the invariant variable $t$, the eigenfunction (\ref{e1.3.28eif}) is
\begin{equation}
\label{Pqes2}
 \Psi (z) \ = \  p_n (t)\ t^{\frac{b}{\al}}\ \exp{\bigg(-\frac{a}{2\al}\, t^2
                          - \frac{c}{\al}\, t \bigg)} \ .
\end{equation}

\hskip 0.5cm
For $\al > 0$ the eigenfunctions (\ref{e1.3.28eif}) are normalizable, if $a,\ b >0$ and
$c$ is arbitrary. If $a=0$, which corresponds to an exactly-solvable situation (the Morse oscillator), the parameter $c$ should be positive, $c > 0$. For this case, the eigenvalues are
\begin{equation}
\label{e1.3.28eiv}
 \veps_k\ =\ -2 c k \ ,\ k=0,1,\ldots \ ,
\end{equation}
and, thus, the Morse oscillator, treated in the Sturm representation, possesses an
infinite discrete spectra. For $\al < 0$, normalizability of (\ref{e1.3.28eif}) occurs
if both $a,\ b < 0$ and $c$ is arbitrary. If $a=0$, the exactly-solvable situation
appears again with the condition $c < 0$.

In order to find the algebraic eigenfunctions (\ref{e1.3.28eif}) we have to
diagonalize the Jacobi matrix $T_2$ of size $(n+1)$ by $(n+1)$ with the matrix elements
\[
  t_{k, k}= -2 c k\ ,\ t_{k, k+1} = (k+1)[2b - \al (k-1)]\ , \
  t_{k+1,k}= \ 2a (n-k),
\]
where $k=0,1,2,\ldots n$. Its eigenvalues do not have a meaning of the energies
of the algebraic eigenfunctions. Instead they have a meaning of the coefficient in front
of the term $e^{-\al z}$ in the potential. Its eigenfunctions define the coefficients of
polynomials $p_n$. Reality of the eigenvalues of $T_2$ is guaranteed if $t_{k+1,k}\, t_{k,k+1} >0$
(see e.g. \cite{Mishina:1962}, p. 28). The characteristic equations for $n=1,2$ have the form
\[
  \veps^2 + 2 c \veps - 2 a (2b + \al) =0\ ,
\]
\[
 \veps^3 - 6 c \veps^2 + 4[a(\al+4b)-2c^2]\veps
 + 16 ac (\al+2b)=0\ ,
\]
respectively. For arbitrary $n$ the characteristic equation has the order $(n+1)$, its eigenvalues
(as well as eigenfunctions) form a $(n+1)$-sheeted Riemann surface in any parameter when all others
are kept fixed. One can easily see that when $n$ tends to infinity, the parameter $a$ tends to
zero. One subdiagonal in the matrix $T_2$ vanishes, the eigenvalues coincide with diagonal matrix
elements and the Riemann surface splits into separate (disconnected) sheets.

The spectral problem (\ref{e1.3.27}) has a quite outstanding property: once the parameter $c$ vanishes,
$c=0$. In this case, in the matrix $T_2$ all the diagonal matrix elements $J_{kk}$ vanish.
Thus, the $(n+1)$ eigenvalues $\veps$ from the algebraic sector (where eigenfunctions are polynomials)
are distributed in such a way that they have a center of symmetry (see Lemma 1.2.4). The energy reflection
symmetry occurs, see \cite{Shifman:1999}. If the center of symmetry is chosen as the
reference point for energy such that $\veps_k=-\veps_{n+1-k}$, it becomes evident that the characteristic
equation of the algebraic sector actually depends on $\veps^2$, when a possible zero eigenvalue is not
taken into account. In particular,

\[
n=1 \quad,\quad \veps^2 = 2a(2b+\al)\ ,
\]

\[
n=2 \quad,\quad \veps^2=  4a(4b+\al)\ ,\ \veps_0=0 \ ,
\]

\[
n=3 \quad,\quad (\veps^2)_{\pm} = 20ab \pm 2a\sqrt{64 b^2 + 9
\al^2}\ ,
\]

\[
n=4 \quad,\quad (\veps^2)_{\pm} = 10a(4b-\al) \pm 6a\sqrt{16
b^2 -8 b \al + 9 \al^2}\ ,\ \veps_0=0 \ .
\]

\newpage

{\bf Case III}.

Let us take a bilinear combination of the $\mathfrak{sl}(2,\bf{R})$ generators
with index $n$ (\ref{sl2r})
(see \cite{Turbiner:1988qes})
\begin{equation}
\label{e1.3.29}
 T_2 = -\al J^+_n J^0_n + (2 b - 3 \al \frac{n}{2})\, J^+_n - 2 a\, J^0_n -
 2 c\, J^-_n - a\, n \ .
\end{equation}
(cf. (\ref{e1.3.25}), (\ref{e1.3.27})), where $\al\neq 0, a, b, c$ are real parameters.
It can be immediately seen that after renaming the parameters
\[
 a \lrar c \ ,
\]
\[
 \al \rar \al \ ,
\]
\begin{equation}
\label{e1.3.29param}
 2b \rar 2b + \al (2n +1) \ ,
\end{equation}
the operator (\ref{e1.3.29}) coincides with the operator (\ref{e1.3.27equiv}). Therefore, if both
parameters $a,c \neq 0$, the quasi-exactly-solvable problems of Case II and Case III are related.
As an algebraic operator the operator (\ref{e1.3.29}) looks like,
\[
 T_2(x,d_x)= -\al x^3 d_x^2 + [(2b-\al)x^2-2ax-2c]d_x + (\al n -2b)nx
 \ ,
\]
(cf.(\ref{e1.3.27.1equiv})). This operator has $(n+1)$ polynomial eigenfunctions which can be found by
linear algebra means. They form the algebraic sector of eigenstates.

In the $L^2$-space, the spectral problem for (\ref{e1.3.29}) leads to a generalized spectral problem
(\ref{e1.3.24}) with the potential
\begin{equation}
\label{e1.3.30}
 \tilde V(z)_{III}\ =\ c^2e^{4\al z} +2ac e^{3\al z}+
 [a^2 - 2c(b +\al)]e^{2\al z}-
 a(2b+\al) e^{\al z}\ +b^2+ \al n(\al n -2b) \ ,
\end{equation}
where
\[
x=e^{-\al z}\ ,
\]
and the weight factor
\[
 \vrho= \frac{1}{\al} e^{\al z} \ .
\]
This potential grows in one direction and tends to a constant in another direction. This constant
can always be chosen as a reference point for the energy. Hence, a new potential will vanish in that
direction. The derivative of the gauge factor $y=-A'$ (minus logarithmic derivative of the ground
state eigenfunction at $n=0$, see (\ref{e1.3.23.1})) is
\begin{equation}
\label{e1.3.30log}
 y(z) \ = \ c e^{2\al z} + a e^{\al z} - b \ .
\end{equation}
Similar to Cases I, II the spectral problem (\ref{e1.3.24})
with the potential (\ref{e1.3.30}) is defined on the real line $z \in (-\infty, +\infty)$, while the spectral problem for $T_2$ is actually restricted to the half-line $x \in [0, +\infty)$. Thus, this is the quasi-exactly-solvable Schr\"odinger equation with potential (\ref{e1.3.30}) of the second type. A finite number of eigenfunctions, each of them appears in its own potential with the same energy, can be found through an algebraic procedure.

The algebraic eigenfunctions become
\begin{equation}
\label{e1.3.30eif}
 \Psi (z) \ = \  p_n (e^{-\al z}) \exp{(-\frac{c}{ 2\al} e^{2\al z} -
 \frac{a}{\al} e^{\al z} + bz )}\ .
\end{equation}
For $\al > 0$, the eigenfunctions (\ref{e1.3.30eif}) are normalizable if $b >0, c \geq 0$ and $a$ is arbitrary.

In order to find the algebraic eigenfunctions (\ref{e1.3.30eif}) we have to diagonalize the Jacobi matrix $T_2$
of size $(n+1)$ by $(n+1)$ with the matrix elements
\[
  t_{k, k}= - 2 a k\ ,\ t_{k, k+1} = -2 c (k+1)\ , \
  t_{k+1,k}= \ (k-n)[2b - \al (n+k)],
\]
where $k=0,1,2,\ldots n$. Its eigenvalues do not have a meaning of the energies of the algebraic eigenfunctions.
Instead they have a meaning of the coefficient in front of the term $e^{\al z}$ in the potential.
Its eigenfunctions define the coefficients of polynomials $p_n$. Reality of the eigenvalues of $T_2$ is guaranteed
if $t_{k+1,k}\,t_{k,k+1} > 0$ (see e.g. \cite{Mishina:1962}, p. 28).
The characteristic equations for $n=1,2$ have the form
\[
  \veps^2 + 2a \veps - 2c (2b - \al) =0\ ,
\]
\[
 \veps^3 + 6 a \veps^2 + 4[c(5\al-4b)+2a^2]\veps
 + 32 ac (\al+b)=0\ ,
\]
respectively. For arbitrary $n$, the characteristic equation has order $(n+1)$, its eigenvalues (as well
as eigenfunctions) form $(n+1)$-sheeted Riemann surface in any parameter when all others are kept fixed.
One can easily see that when $n$ tends to infinity, the parameter $a$ tends to zero. One subdiagonal
in the matrix $J$ vanishes, the eigenvalues coincide with diagonal matrix elements and the Riemann surface
splits into separate (disconnected) sheets.

The spectral problem (\ref{e1.3.29}) has a quite outstanding
property: if the parameter $a$ takes zero value, $a=0$, in the matrix $T_2$ all the diagonal matrix
elements $t_{k, k}$ vanish. Thus, the $(n+1)$ eigenvalues $\veps$ from the algebraic sector
(where eigenfunctions are polynomials) are distributed in such a way that they have a center of symmetry
(see Lemma 1.2.4). The energy reflection symmetry occurs \cite{Shifman:1999}.
If the center of symmetry is chosen as the reference point for energy such that $\veps_k=-\veps_{n+1-k}$,
where $k=0, 1, \ldots , [\frac{n+1}{2}]$ the characteristic equation of the algebraic sector takes the form
$\veps^p P_{[\frac{n+1}{2}]}(\veps^2)=0$, where $p=\frac{1+(-1)^n}{2}$. If $n$ is even, a
zero eigenvalue occurs. For the particular cases,
\[
 n=1 \quad,\quad \veps^2 - 2c(2b-\al)\ =\ 0 \ ,
\]

\[
 n=2 \quad,\quad \veps^2 - 4[c(4b-5\al)-2a^2]\ =\ 0 \ ,
 \ \veps_0=0 \ ,
\]

\[
 n=3 \quad,\quad (\veps^2)_{\pm} = 20c(b-2\al) \pm 2c\sqrt{64 b^2
 - 256 \al b + 265 \al^2}\ ,
\]

\[
 n=4 \quad,\quad (\veps^2)_{\pm} = 10c(4b-11\al) \pm 6c\sqrt{16
 b^2 - 88 b \al + 129 \al^2}\ ,\ \veps_0=0 \ .
\]

\newpage

Next two quasi-exactly-solvable potentials are associated to the
(hyperbolic) P\"{o}schl-Teller or one-soliton potential.

\vskip .3cm

\subsection{P\"{o}schl-Teller-type potentials}

\begin{figure}[tb]
\begin{center}
     {\includegraphics*[width=3in]{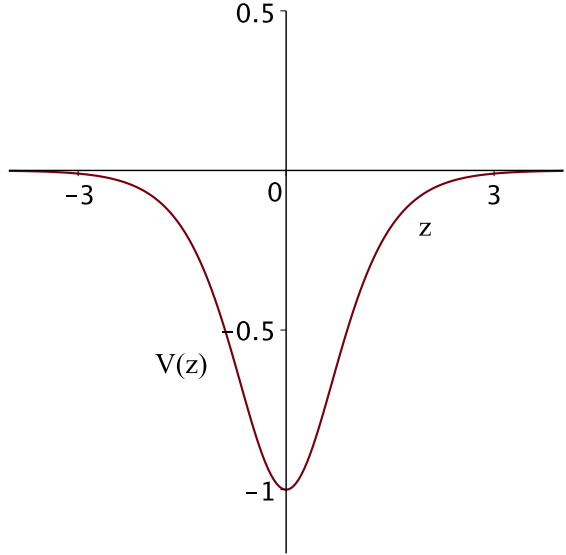}}
     \caption{P\"{o}schl-Teller potential (\ref{e1.3.41-Poschl-Teller}) at $A = \al = 1$:\ $V(z) = -\frac{1}{\cosh^2 {z}}$\ ,
     it has minimum at $z=0$, $V_{min}=-1$\,.
     Discrete spectra consists of a single bound state (the ground state) with energy $E_0=-\frac{1}{4}$, its eigenfunction
     $\Psi_0 = e^{-\frac{z}{2}-e^{-z}}$ .
      }
    \label{PT-e}
\end{center}
\end{figure}

\vskip .5truecm

\ The P\"{o}schl-Teller potential or, in other words, the one-soliton potential
describes a well-known exactly-solvable quantum-mechanical problem
(see e.g. Landau and Lifschitz \cite{LL-QM:1977})
\begin{equation}
\label{e1.3.41-Poschl-Teller}
V(z) = -  \frac{A^2 }{\cosh^2 {\al z}} \ ,\ \al \neq 0\ ,
\end{equation}
see Fig. \ref{PT-e}.

This potential is defined on the real line $z \in {\bf R}$. The potential
$V(z)$ is periodic with the imaginary period $i\, \frac{\pi}{\al}$ and has
infinitely many second order poles distributed uniformly along the imaginary axis.
The Hamiltonian is translationally-invariant, $T_c: z \rar z + i\, \frac{\pi}{\al}$.
Taking $T_c$-invariant
\[
    \ta\ =\ \cosh^2{\al z}\ ,
\]
as a new variable, we get the P\"{o}schl-Teller potential in a very simple, rational form,
\begin{equation}
\label{e1.3.41-Poschl-Teller-t}
  V(\ta) = - \frac{A^2}{\ta} \ ,
\end{equation}
and the second derivative ($1D$ Laplacian) becomes the $1D$ Laplace-Beltrami operator
with metric $g^{11} = 4 \al^2 \ta(\ta-1)$ ,
\[
    \dzz\ =\ 4 \al^2 \bigg(\ta(\ta-1) \frac{d^2}{d {\ta}^2}\ + (\ta-\frac{1}{2}) \frac{d}{d {\ta}}\bigg)\ ,
\]
Hence, the Laplace-Beltrami operator does not depend on the sign of $\al$, it changes the overall sign when $\al \rar i\al$\,.

Depending on the value of $A$ the potential (\ref{e1.3.41-Poschl-Teller}) has the finite number of bound states which can be found by linear algebra means.
This potential has the unique property of reflectionless scattering.
It is probably the simplest solution of the so called
Korteweg-de Vries equation playing an important role in the {\it
inverse problem method} (for detailed discussion see e.g. the book
by V.E. Zakharov et al \cite{Zakharov:1980}).

The potential (\ref{e1.3.41-Poschl-Teller}) can be generalized to
\begin{equation}
\label{e1.3.41-PT-G}
V(z) = - \frac{A^2 }{\cosh^2 {\al z}} + \frac{B}{\sinh^2 {\al z}} \ ,\ \al \neq 0\ ,
\end{equation}
with $z \in [0, \infty)$ preserving the property of exact-solvability.
It is called a modified P\"{o}schl-Teller potential. The potential
(\ref{e1.3.41-PT-G}) can be written in the form of $BC_1$-hyperbolic potential
\begin{equation}
\label{e1.3.41-PT-BC1}
V(z) = \frac{a}{\sinh^2 { 2 \al z}} + \frac{b}{\sinh^2 {\al z}} \ ,\ \al \neq 0\ ,
\end{equation}
which occurs in the Hamiltonian Reduction method, see Olshanetsky-Perelomov
\cite{Olshanetsky:1983}.

Sometimes, trigonometric versions of (\ref{e1.3.41-Poschl-Teller}) and
(\ref{e1.3.41-PT-G}) are considered by changing $\al \rar i \al$ , for instance,
\begin{equation}
\label{e1.3.41-PT-G-trig}
V(z) = -   \frac{A^2}{\cos^2 {\al z}} + \frac{B}{\sin^2 {\al z}} \ ,\ \al \neq 0\ ,
\end{equation}
at $z \in [0, \frac{\pi}{\al}]$.
This potential has infinite discrete spectra which can be found by algebraic means.
The potential (\ref{e1.3.41-PT-G-trig}) can be also written in the form of $BC_1$-trigonometric
potential
\begin{equation}
\label{e1.3.41-PT-BC1-trig}
V(z) = \frac{a}{\sinh^2 { 2 \al z}} + \frac{b}{\sinh^2 {\al z}} \ ,\ \al \neq 0\ ,
\end{equation}
which occurs in the Hamiltonian Reduction method, see Olshanetsky-Perelomov
\cite{Olshanetsky:1983}.


\pagebreak

{\bf Case IV}.

\vskip .3cm

Let us take the following bilinear combination of the $\mathfrak{sl}(2,\bf{R})$
generators with index $n$ (\ref{sl2r}) (see \cite{Turbiner:1988qes})
\renewcommand{\theequation}{1.3.30-{\arabic{equation}}}
\setcounter{equation}{0}
\[
 T_2 = - 4 \al ^2 J^+_n J^0_n  + 4 \al ^2 J^+_n J^-_n
 -2 \al [(3n+2p+1)\al + 2c] J^+_n
\]
\begin{equation}
\label{e1.3.31}
 + 4 \al [(n+1)\al  + c - a] J^0_n + 4\al a J^-_n +
 2\al n[(n+1)\al + c - a]\ ,
\end{equation}
where $\al, a, c$ are real parameters, $p=0,1$. Since $T_2$ is proportional
to $\al$ it is convenient to divide it by $\al$ and to measure the spectral
parameter $\veps$ in units of $\al$:
\[
  \veps \rar \al \veps\ .
\]
Since $\al \neq 0$ the terms of positive grading in (\ref{e1.3.31}) never
disappear, this operator never becomes exactly-solvable. It never preserves
the flag of polynomials.

After the substitution of the generators (\ref{sl2r}) into (\ref{e1.3.31}) and
division by $\al$ we get an algebraic differential operator,
\[
 T_2(x,d_x)\ =\ -4 \al x^2(x-1) d_x^2 - 2 \{[(2 p +3)\al + 2c] x^2 - 2 (\al - a + c) x - 2a \}d_x
\]
\begin{equation}
\label{e1.3.31.1}
 + 2n [(2n+2p+1)\al + 2c] x \ .
\end{equation}

In the basis of monomials $\{1, x, x^2, \ldots, x^k, \ldots x^n, \ldots \}$ this operator has the form of tri-diagonal, Jacobi matrix,
\begin{equation}
\label{e1.3.31.2}
 t_{k,k-1} = 4 a\, k\ ,\ t_{k,k} = 4(-a + c +\al k)\, k\ ,\ t_{k,k+1} =2 (n-k)[(2n+2p+2k+1)\al +2 c]\ .
\end{equation}
If $k=n$, the matrix element $t_{n,n+1}=0$ and the Jacobi matrix $T_2$ becomes block-triangular.
The characteristic polynomial is factorized, $\det (T_2 - \veps) = P_n(\veps) P_{\infty}(\veps)$.

\renewcommand{\theequation}{1.3.{\arabic{equation}}}
\setcounter{equation}{30}

This operator has $(n+1)$ polynomial eigenfunctions which can be found by
linear algebra means. They form the algebraic sector of eigenstates. As illustration
of general situation we take $n=1$ in (\ref{e1.3.31.1}). There are two polynomial
eigenfunctions,
\[
 \vphi_{\pm}= [(2p + 3\al)+2c] x -\al +a - c \mp \sqrt{(\al+a+c)^2 + 2 a (2p+1) \al} \ ,
\]
with eigenvalues
\begin{equation}
\label{e1.3.31E}
 \frac{\veps_{\pm}}{2}\ =\ \al-a+c \pm \sqrt{(\al+a+c)^2 + 2 a (2p+1) \al} \ .
\end{equation}
Both eigenfunctions and both eigenvalues form two-sheeted Riemann
surface in any of the parameters $\al,a,c$ if others are fixed.
Reality of the spectra of (\ref{e1.3.31}) requires, in particular,
\[
 (\al+a+c)^2 + 2 a (2p+1) \al > 0 \ ,
\]
(cf. (\ref{esl28ES})) and equality is also excluded as a consequence of
requirement of Hermiticity of (\ref{e1.3.31.1}) which prohibits degeneracy
of energy. Since the domain of (\ref{e1.3.31.1}) is $[0, +\infty)$ (see below),
$\vphi_-$ describes the ground state (it does not vanish at $x>0$, hence, no nodes).

It is worth mentioning that in the universal enveloping algebra
$\mathfrak{sl}(2,\bf{R})$ there exists an element which is almost equivalent
to (\ref{e1.3.31}). It can be obtained from (\ref{e1.3.31}) by conjugation (\ref{e1.2.2})
\renewcommand{\theequation}{1.3.32-{\arabic{equation}}}
\setcounter{equation}{0}
\[
 T_2 = 4 \al ^2 J^+_n J^-_n - 4 \al ^2 J^0_n J^-_n - 4\al a J^+_n
 - 4 \al [(n-1)\al + c - a] J^0_n
\]
\begin{equation}
\label{e1.3.31equiv}
 + 2 \al [(3n+2p+3)\al + 2c] J^-_n + 2 \al n[(n+1)\al + c - a]\ ,
\end{equation}
(cf. (\ref{e1.3.34}) of Case V).
It has $(n+1)$ polynomial eigenfunctions with the same eigenvalues as (\ref{e1.3.31}).
The limit $\al \rar 0$ is singular, the operators (\ref{e1.3.31}) and
(\ref{e1.3.31equiv}) loose their bilinearity in generators becoming linear in generators.
Unlike (\ref{e1.3.31equiv}) the term of positive grading in (\ref{e1.3.31equiv}) can be
easily vanished at $a=0$, this operator becomes exactly-solvable. It preserves the flag of
polynomials.

The polynomial eigenfunctions of the operators (\ref{e1.3.31}) and (\ref{e1.3.31equiv})
are related to each other through the invariance condition (\ref{e1.1.6}). The algebraic
differential operator which corresponds to (\ref{e1.3.31equiv})
has the form (after division by $\al$)
\[
 T_2(x,d_x)= 4 \al x (x-1) d_x^2 -2\{2 a x^2 + 2(n \al - c + a)x + [(2n +2p +3)\al + 2c]\}d_x
\]
\begin{equation}
\label{e1.3.31.1equiv}
 + 4 n \al \ ,
\end{equation}
(cf. (\ref{e1.3.31.1})).
\renewcommand{\theequation}{1.3.{\arabic{equation}}}
\setcounter{equation}{32}

\begin{figure}[tb]
\begin{center}
     {\includegraphics*[width=3in]{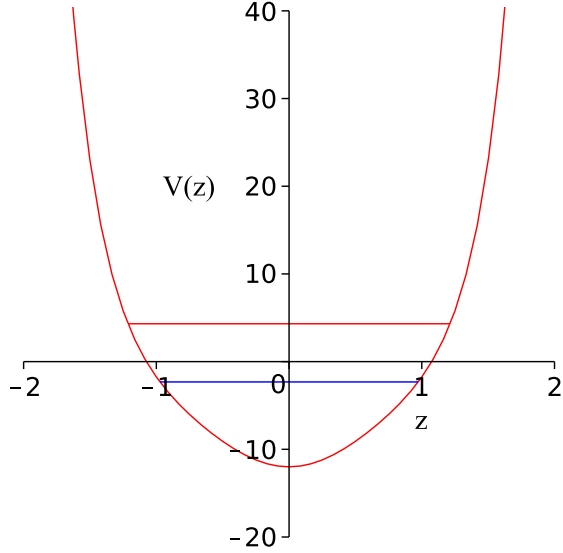}}
     \caption{P\"{o}schl-Teller potential (\ref{e1.3.32}) at $a = c = \al = n = 1$:
     \ $V(z) = \cosh^4 {z}-\cosh^2 {z}-\frac{12}{\cosh^2 {z}} $\ ,
     it has minimum at $z=0$, $V_{min}=-12$\,. Discrete spectra contains infinitely-many bound states,
     the ground state and the 2nd excited state are found algebraically, their energies $E_{\pm}= 2 \mp 2 {\sqrt {11}}$,
     their eigenfunctions
     $\Psi_{\pm} = (\frac{5}{\cosh^2 {z}} - 1 \mp {\sqrt {11}}) \cosh{z} \exp{(-\frac{1}{{4}} \cosh{2z})}$\, .
      }
    \label{PT-I}
\end{center}
\end{figure}

The spectral problem for the operator (\ref{e1.3.31}) is reduced to the Schr\"odinger equation
(\ref{e1.3.21}) by the above-described procedure (\ref{e1.3.22})-(\ref{e1.3.23.1}), where a new variable
\[
x=\cosh^{-2}{\al z}\ ,
\]
is introduced. Finally, the spectral problem (\ref{e1.3.24}) with the potential
\begin{equation}
\label{e1.3.32}
 V(z) = a^2 \cosh^4 {\al z} - a (a + 2\al - 2c) \cosh^2 {\al z}
\end{equation}
\[
- \{c(c+\al) + \al (2n+p)[\al (2n+p+1) + 2c]\} \cosh^{-2}  {\al z}
+ c^2 + a \al - 2ac \ ,
\]
occurs. This is the quasi-exactly-solvable modification of the P\"oschl-Teller potential of the first type, see Fig. \ref{PT-I}.
The derivative of the gauge factor $y=-A'$ is
\begin{equation}
\label{e1.3.31log}
 y(z) \ = \ -c \tanh {\al z} + \frac{a}{{2}} \cosh{ 2 \al z} \ .
\end{equation}
It has a meaning of logarithmic derivative of the ground state eigenfunction at $n=0$
with negative sign, see (\ref{e1.3.23.1}).

The Schr\"odinger equation with the potential (\ref{e1.3.32}) is defined on the real line
$z \in (-\infty, +\infty)$, while the spectral problem for $T_2$ is
restricted to $x \in [0, +\infty)$. For non-vanishing $a$ the potential
(\ref{e1.3.32}) grows at $|z| \rar \infty$ and always has infinite discrete spectra.
Since the potential is even there are two families of eigenstates, one of positive and
another one is of negative parity, $m=(-1)^p ,\ p=0,1\,$, respectively.
Thus, the Schr\"odinger equation with the potential (\ref{e1.3.32}) is the
quasi-exactly-solvable Schr\"odinger equation of the first type. A finite number of
eigenfunctions and eigen-energies of definite parity $m$ can be found through
linear algebraic procedure.

The $(n+1)$ solutions of the Schr\"odinger equation
(\ref{e1.3.21}) with the potential (\ref{e1.3.32}), possibly giving rise to an
``algebraized" part of the spectra, have the form
\begin{equation}
\label{e1.3.33eif}
 \Psi (z) \ = \  (\tanh {\al z})^p p_n (\tanh^2 {\al z}) ( \cosh{\al z})^{-c/\al}
\exp{(-\frac{a}{{4\al}} \cosh{ 2 \al z})}\ ,
\end{equation}
where the polynomials $\tilde p_n(x) = p_n(1-x)$ are the eigenfunctions of the
operator $T_2(x,d_x)$ (\ref{e1.3.31.1}). In order to answer the question whether these
solutions have a meaning of the eigenfunctions we should check their normalizability.
It is evident if $\frac{a}{\al} > 0$ for any real $c$ and integer $n, p$ the function
(\ref{e1.3.33eif}) is normalizable. If $a=0$ the potential (\ref{e1.3.32}) becomes
exactly-solvable. A question of normalizability in this case which occurs at
$\frac{c}{\al} + p >0$ needs a certain clarification. It will be done later.

In order to find the algebraic eigenfunctions (\ref{e1.3.33eif}) we have to
diagonalize the Jacobi matrix $T_2$ of the size $(n+1)$ by $(n+1)$ with the matrix elements
\[
  t_{k, k}= 4 k (k \al - a +c)\ ,\ t_{k, k+1} = 4a (k+1)\ , \
  t_{k+1,k}= \ 2 (n-k)[(2n+2p+2k+1)\al +2 c],
\]
where $k=0,1,2,\ldots n$, see (\ref{e1.3.31.2}). Its eigenvalues have a meaning
of the energies of the algebraic eigenfunctions. Reality of the eigenvalues of $T_2$
is guaranteed if $t_{k+1,k}\, t_{k,k+1} > 0$ is fulfilled (see e.g. \cite{Mishina:1962}, p. 28).
In particular, the characteristic equation for $n=1$ has the form
\[
  \veps^2 - 4(\al - a + c) \veps - 8a[(2p+3)\al + 2c]\ =\ 0\ .
\]
For arbitrary $n$ the characteristic equation has the order $(n+1)$, its
eigenvalues (as well as eigenfunctions) form $(n+1)$-sheeted Riemann surface
in any parameter when all others are kept fixed. If parameter $a$ tends to zero
one subdiagonal in the matrix $T_2$ vanishes, the eigenvalues coincide with
diagonal matrix elements and the Riemann surface splits into separate
(disconnected) sheets.

\vskip .5truecm

{\bf Case V}.
\vskip .3truecm

Let us take the following bilinear combination of the $\mathfrak{sl}(2,\bf{R})$
generators with the mark $n$ (\ref{sl2r}) (see \cite{Turbiner:1988qes})
\renewcommand{\theequation}{1.3.36-{\arabic{equation}}}
\setcounter{equation}{0}
\[
 T_2 = -4 \al^2 J^+_n J^-_n + 4 \al ^2 J^0_n J^-_n + 4\al b J^+_n
 -2\al [\al (2n + 2p + 3)+2a+ 4b] J^0_n
\]
\begin{equation}
\label{e1.3.34}
 + 2\al [\al (n+2) + 2a + 2b] J^-_n - \al  [\al n (2n + 2p + 3) + 2an - 2 b k] \ ,
\end{equation}
where $\al, a, b$ are real parameters, $p=0,1$. Since $T_2$ is proportional
to $\al$ it is convenient to divide it by $\al$ and to measure also the spectral
parameter $\veps$ in units of $\al$:
\[
  \veps \rar \al \veps\ .
\]
If $b = 0$ the term of positive grading in (\ref{e1.3.34}) disappears,
this operator becomes exactly-solvable.

After the substitution of the generators (\ref{sl2r}) into (\ref{e1.3.34}) and
division by $(2\al)$ we get an algebraic differential operator,
\[
 T_2(x,d_x)= -2\al x(x-1) d_x^2 + [2 b x^2 - (2a + 4b + 2p \al+ 3\al) x+ 2 (\al + a + b) ]d_x
\]
\begin{equation}
\label{e1.3.34.1}
 - 2 b n x + b (2n + p) \ .
\end{equation}

In the basis of monomials $\{1, x, x^2, \ldots, x^k, \ldots x^n, \ldots \}$ this operator has the form of tri-diagonal, Jacobi matrix,
\[
 t_{k,k-1} = 2(a + b + \al k)\,k \ ,\ t_{k,k} = - 2 \al k^2 - (2a + 4b +\al (2p+1))\,k + b (2n + p)\ ,
\]
\begin{equation}
\label{e1.3.34.2}
 t_{k,k+1} = 2b\,(k - n)\ .
\end{equation}
If $k=n$, the matrix element $t_{n,n+1}=0$ and the Jacobi matrix $T_2$ becomes block-triangular.
The characteristic polynomial is factorized, $\det (T_2 - \veps) = P_n(\veps) P_{\infty}(\veps)$.

\renewcommand{\theequation}{1.3.{\arabic{equation}}}
\setcounter{equation}{36}

The operator (\ref{e1.3.34.1}) has $(n+1)$ polynomial eigenfunctions which can be found
by linear algebra means. They form the algebraic sector of eigenstates. For instance, at $n=0$ two polynomial eigenfunctions occur for different $p$,
\[
 \vphi_{p} = \mbox{const} \ ,
\]
with eigenvalue
\begin{equation}
\label{e1.3.34E}
 \veps_{p}\ =\ b (p+2) \ ,\ p=0,1\ .
\end{equation}

The spectral problem for the operator (\ref{e1.3.34}) is reduced to the Schr\"odinger equation
(\ref{e1.3.21}) by the above-described procedure (\ref{e1.3.22})-(\ref{e1.3.23.1}).
Finally, the spectral problem (\ref{e1.3.24}) with the potential
\begin{equation}
\label{e1.3.35}
 V(z) = -b^2 \cosh^{-6} {\al z} + b[2a+ 3b + \al (4n+2p +3)] \cosh^{-4}{\al z}
\end{equation}
\[
-[(a+3b)(a+b + \al) + 2(2n+p) \al b ] \cosh^{-2} {\al z} + (a+b)^2\ ,
\]
occurs, where
\[
 x=\cosh^{-2}{\al z}\ ,\ \vrho= \cosh^{-2}{\al z}\ .
\]
It is quasi-exactly-solvable Schr\"odinger equation of the second type.
The derivative of the gauge factor $y=-A'$ is
\begin{equation}
\label{e1.3.34log}
 y(z) \ = \ -c \tanh {\al z} + \frac{a}{2} \cosh{2 \al z} \ .
\end{equation}
has a meaning of logarithmic derivative of the ground state eigenfunction at $n=0$
with negative sign, see (\ref{e1.3.23.1}).

The Schr\"odinger equation with the potential (\ref{e1.3.35}) is defined on the real line
$z \in (-\infty, +\infty)$, while the spectral problem for $T_2$ is
restricted to $x \in [0, +\infty)$. For non-vanishing $a$ the potential
(\ref{e1.3.35}) grows at $|z| \rar \infty$ and always has infinite discrete spectra.
Since the potential is even there are two families of eigenstates of positive and
negative parity, $m=(-1)^p, p=0,1\,$, respectively.
Thus, the Schr\"odinger equation with the potential (\ref{e1.3.35}) is the
quasi-exactly-solvable Schr\"odinger equation of the second type. Thus, a finite number
of eigenfunctions and eigen-energies of definite parity $m$ can be found through
linear algebraic procedure.

The $(n+1)$ solutions of the Schr\"odinger equation (\ref{e1.3.21}) with the potential
(\ref{e1.3.35}), giving rise to an {\it algebraized} part of the spectra, have the form
\begin{equation}
\label{e1.3.34eif}
 \Psi (z) \ = \  (\tanh {\al z})^p\,  p_n (\tanh^2 {\al z})
 (\cosh{\al z})^{-\frac{(a+b)}{\al}} \exp{(\frac{b}{2\al}
 \tanh^2 {2\al z} )}\ ,
\end{equation}
at $\al > 0, (a+b) > 0$ and $p=0,1$.

In order to find the algebraic eigenfunctions (\ref{e1.3.34eif}) explicitly we have to
diagonalize the Jacobi matrix $T_2$ of the size $(n+1)$ by $(n+1)$, see (\ref{e1.3.34.2}).
Its eigenvalues have a meaning of the energies of the algebraic eigenfunctions.
Reality of the eigenvalues of $T_2$ is guaranteed if $t_{k+1,k}\, t_{k,k+1} > 0$ is fulfilled
(see e.g. \cite{Mishina:1962}, p. 28). It can be easily checked that for $n=0$ in the potential
\begin{equation}
\label{e1.3.35-0}
 V(z) = -b^2 \cosh^{-6} {\al z} + b[2a+ 3b +\al (2p +3)] \cosh^{-4}{\al z}
\end{equation}
\[
     -[(a+3b)(a + b + \al) + 2 p \al b + b (2+p)] \cosh^{-2} {\al z} \ ,
\]
the lowest eigenstate of parity $(-)^p$ is known
\begin{equation}
\label{e1.3.34eif-0}
 \Psi_0 (z) \ = \  (\tanh {\al z})^p\,(\cosh{\al z})^{-\frac{(a+b)}{\al}} \exp{(\frac{b}{2\al}
 \tanh^2 {2\al z} )}\ ,
\end{equation}
with the energy
\[
     E_0 = - (a+b)^2\ .
\]

\newpage

\subsection{Harmonic oscillator-type potentials}

The next two quasi-exactly-solvable potentials are associated with the harmonic
oscillator potential. Both of them are a particular kind of anharmonic oscillators.

\vskip .5truecm

{\bf Case VI}.

\vskip .3truecm

The sextic polynomial potential was the first example of a potential for which the
quasi-exactly-solvable Schr\"odinger operator has been found. As it usually happens this
potential appeared in a number of different articles in various occasions at more or less
the same time, for an (incomplete) history see \cite{Turbiner:1987}. It turns out that is the
{\it unique} quasi-exactly-solvable problem with polynomial potential \cite{Turbiner:1988qes}.
It provides an enormous wealth of non-trivial properties, it can be used as a paradigm where
different theoretical ideas can be tested. A separate Chapter 2 will be dedicated to
this potential.

Let us take the following non-linear combination in the generators
(\ref{sl2r}), see \cite{Turbiner:1988qes}
\begin{equation}
\label{e1.3.35a}
 T_2 = -4 J^0_n J^-_n + 4a J^+_n + 4b J^0_n - 2(n+1+2p) J^-_n + 2bn
\end{equation}
or as a differential operator,
\begin{equation}
\label{e1.3.35b}
T_2(x,d_x)= -4xd_x^2 + 2(2ax^2+2bx-1-2p)d_x - 4anx  \ ,
\end{equation}
where $x \in {\bf R^+}$ assuming $a>0, \forall b$, or $a \geq 0, b>0$ and $p=0,1$.
In the basis of monomials $\{1, x, x^2, \ldots, x^k, \ldots x^n, \ldots \}$
this operator has the form of tri-diagonal, Jacobi matrix,
\begin{equation}
\label{e1.3.35a.1}
 t_{k,k-1} = -2\,k(2k - 1 + 2 p) \ ,\ t_{k,k} =  4b \,k \ ,\ t_{k,k+1} = 4a\,(k - n)\ .
\end{equation}
If $k=n$, the matrix element $t_{n,n+1}=0$ and the Jacobi matrix $T_2$ becomes block-triangular.
The characteristic polynomial is factorized, $\det (T_2 - \veps) = P_n(\veps) P_{\infty}(\veps)$.
If $b=0$ the principal diagonal of the Jacobi matrix vanishes, $t_{k,k} = 0$, and the energy-reflection symmetry occurs, see \cite{Shifman:1999}, and Lemma \ref{Lemma124},
\[
     P_{n}(\veps)= \veps^p {\tilde P}_{[\frac{n}{2}]} (\veps^2) \ ,
\]
where $p=0$ or 1 depending on $n$ being even or odd, respectively.
For $a=0$ this matrix becomes triangular.

\begin{figure}
\begin{center}
\begin{tabular}{cc}
    \includegraphics[width=140pt,angle=0]{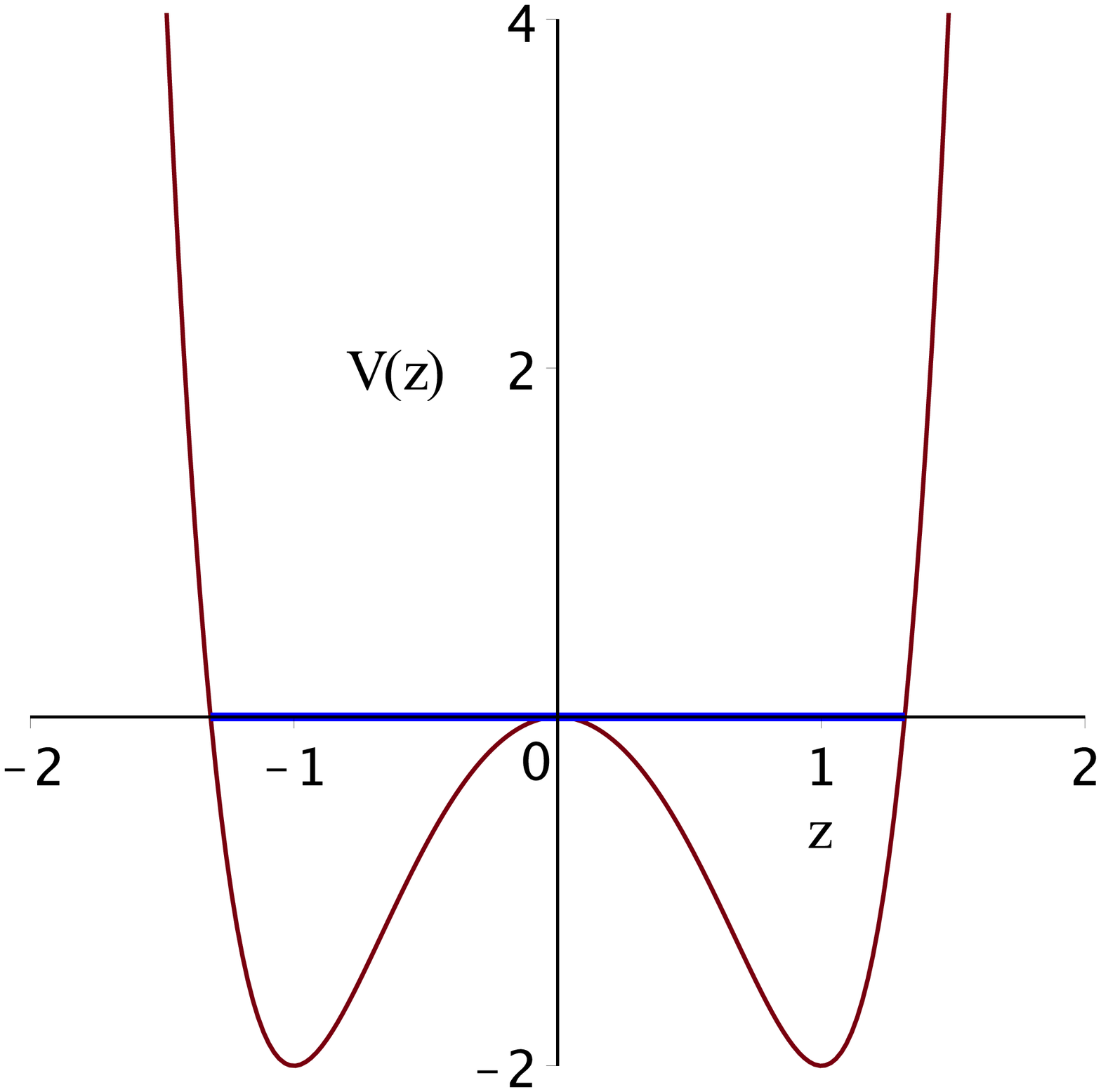} &
\includegraphics[width=140pt,angle=0]{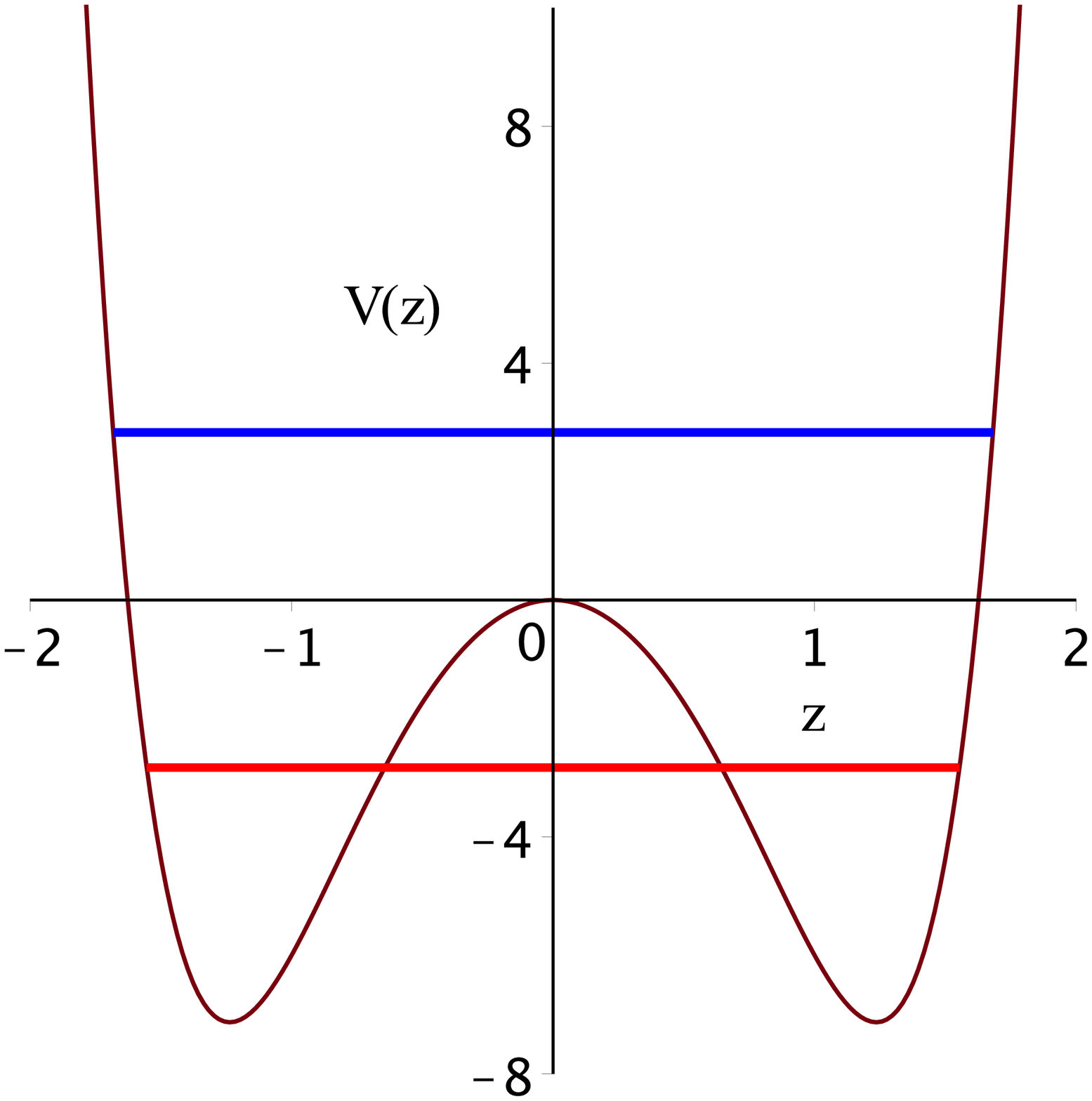} \\
    (a) & (b)
\end{tabular}
\end{center}
 \caption{
 \label{figVa}
 Sextic QES potential (\ref{e1.3.36}) with $a=1, b=0$ at $n=0, p=0$ (a) $V_a = x^6 - 3 x^2$ with the known
 ground state (blue line), $E_0=0, \Psi_0 = e^{ - \frac{z^4}{4}}$ and at $n=2, p=0$ (b) $V_b = x^6 - 7 x^2$
 with two known states (red and blue lines),
 $E_{\mp}=\pm 2 {\sqrt 2}, \Psi_{\mp} = (2 z^2 \pm {\sqrt 2}) e^{ - \frac{z^4}{4}}$.}
\end{figure}

Putting $x=z^2$ and choosing the gauge phase
$$A = \frac{ax^2}{4} + \frac{bx}{2} - \frac{p}{2} \ln{x}\ ,$$
we arrive at the spectral problem (\ref{esl29}), following the above-described procedure (\ref{e1.3.22})-(\ref{e1.3.23.1}), with the singular potential
\renewcommand{\theequation}{1.3.45-{\arabic{equation}}}
\setcounter{equation}{0}
\begin{equation}
\label{e1.3.36sing}
   V(z) = a^2 z^6 + 2ab z^4 + [b^2 - (4n+3+2p)a] z^2 - b(1+2p) + \frac{p(p-1)}{z^2}\ ,
   \ z \in {\bf R^+}\ .
\end{equation}
If $p=0,1$ the singular term disappears and the symmetric polynomial potential occurs,
see \cite{Turbiner:1987},
\begin{equation}
\label{e1.3.36}
   V(z) = a^2 z^6 + 2ab z^4 + [b^2 - (4n+3+2p)a] z^2\ ,
   \ z \in {\bf R}\ .
\end{equation}
\renewcommand{\theequation}{1.3.{\arabic{equation}}}
\setcounter{equation}{45}
\hskip -0.25cm
For $a \geq 0$ and $p=0\ (p=1)$ the first $(n+1)$ eigenfunctions, even
(odd) with respect to reflection $z \lrar -z$, can be found algebraically,
by linear algebra means. Hence, the Schr\"odinger operator with the potential
(\ref{e1.3.36}) is quasi-exactly-solvable.
The number of those {\it algebraized} eigenfunctions is nothing
but the dimension of the irreducible representation of the algebra (\ref{sl2r}).
Therefore, the $(n+1)$ {\it algebraized} eigenfunctions of (\ref{esl29})
have the form
\begin{equation}
\label{e1.3.37}
 \Psi^{(p)} (z) \ =\ z^p\, p_n (z^2)\,  e^{ - \frac{az^4}{4} - \frac{bz^2}{2}},
\end{equation}
where $p_n(y)$ is a polynomial of the $n$th degree. It is worth
noting that if the parameter $a$ goes to 0, the potential
(\ref{e1.3.35a}) becomes the harmonic oscillator potential and the
polynomial $z^p p_n (z^2)$ reconciles to the Hermite polynomial
$H_{2n+p} (z)$ (see discussion below). For illustration, two double-well potentials at $b=0$ and $n=0,1$, respectively,
are shown on Fig.~\ref{figVa}.

If the parameter $n$ in (\ref{e1.3.36}) is a real number, the potential becomes a generic sextic symmetric polynomial potential. Its energies are branches of an infinitely-valued analytic function in $b$ (if $a$ and $n$
are fixed) with infinitely-many square-root branch points on every Riemann sheet. Due to reflection symmetry $z \rar -z$, two separate infinitely-sheeted Riemann surfaces occur: $E^{(+)}(b)$ describes even states while $E^{(-)}(b)$ describes odd ones. The branch points form pairs with complex-conjugated locations. If $n$ takes integer value, the $(n+1)$ -sheeted Riemann surface splits away into an infinitely-sheeted one. The $(n+1)$ -sheeted Riemann surface is determined by the characteristic equation of $(n+1)$ degree of the Jacobi matrix (\ref{e1.3.35a.1}). This matrix depends on parameters $a,b$ linearly, while
the coefficients of the characteristic equation depend on them polynomially.

It is worth presenting several particular cases explicitly:

(i) \ Ground state (the lowest energy state of positive parity)

at~$n=0,p=0$\ ,
\[
 E_0^{(+)} = b\ ,\ \Psi_0^{(+)} = e^{ - \frac{a z^4}{4} - \frac{b z^2}{2}}\ ,
\]
in potential
\[
   V(z) = a^2 z^6 + 2ab z^4 + (b^2 - 3 a) z^2 \ .
\]
The Riemann sheet $E_0^{(+)}(b)$ of the ground energy splits away from the infinitely-sheeted Riemann surface $E^{(+)}(b)$.

(ii) \ Ground state (the lowest energy state of negative parity)

at~$n=0,p=1$\ ,
\[
 E_0^{(-)} = 3 b\ ,\ \Psi_0^{(-)} = z\,e^{ - \frac{a z^4}{4} - \frac{b z^2}{2}}\ ,
\]
in potential
\[
   V(z) = a^2 z^6 + 2ab z^4 + (b^2 - 5 a) z^2 \ .
\]
The Riemann sheet $E_0^{(-)}(b)$ of the ground energy splits away from the infinitely-sheeted Riemann surface $E(^{(-)}b)$.

\begin{figure}[tb]
\begin{center}
     {\includegraphics*[width=3in]{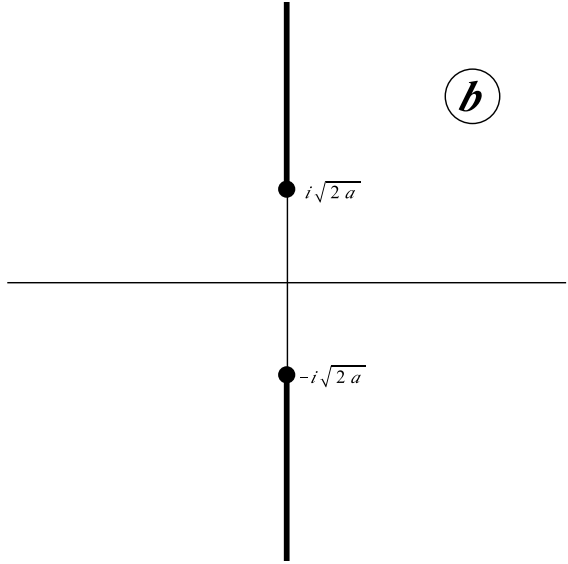}}
     \caption{$b$-Complex plane of the ground state energy at $n = 1$,\ $p=0$ in the QES potential
     \ $V(z) = a^2 z^6 + 2 a b z^4 + (b^2 - 7 a) z^2 $\ . The square-root branch points are marked by bullets, branch cuts go along the imaginary axis to $\pm i \infty$ infinity, respectively.
      }
    \label{bcomp-V6-2}
\end{center}
\end{figure}

(iii) \ Ground state and the second excited state, both of positive parity,

at~$n=1,p=0$\ ,
\[
E_1^{(0/2)} = 3 b \pm 2 (b^2 + 2a)^{\half}\ ,\ \Psi_1^{(0/2)} = (2 a z^2 + b \mp (b^2 + 2a)^{\half})
\, e^{ - \frac{a z^4}{4} - \frac{b z^2}{2}}\ ,
\]
in potential
\[
   V(z) = a^2 z^6 + 2ab z^4 + (b^2 - 7 a) z^2 \ .
\]
The two-sheeted Riemann surface $E_1^{(0/2)}(b)$ which describes the ground state energy and the second excited state splits away from the infinitely-sheeted Riemann surface $E^{(+)}(b)$. These levels intersect at $b = \pm i {\sqrt {2 a}}$ where the square-root branch points occur, see Fig. \ref{bcomp-V6-2}. These are the Landau-Zener singularities: making the analytic continuation around any of them, starting from ground state energy at real $b$ (and fixed
$a > 0$), we arrive at the energy of the 2nd excited state at the same $b$.

\begin{figure}[tb]
\begin{center}
     {\includegraphics*[width=3in]{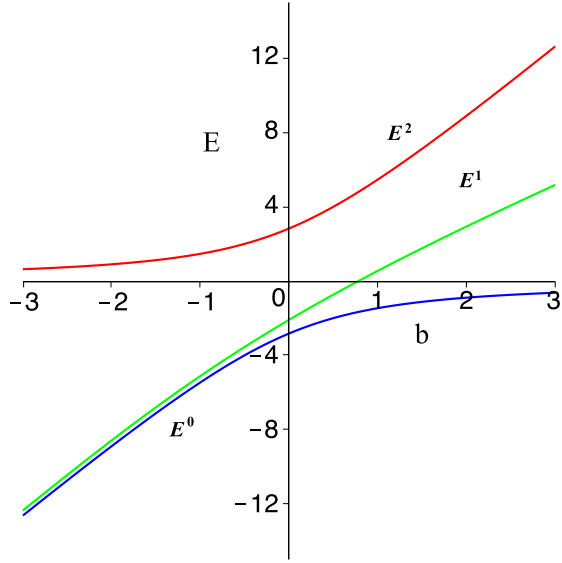}}
     \caption{Sextic QES potential (\ref{e1.3.36}) at $a = 1$,\ $n = 1$,\ $p=0$:
     \ $V(z) = z^6 + 2 b z^4 + (b^2 - 7 ) z^2 - b$\ ,
     it has maximum at $z=0$, $V_{max}=0$\,. Discrete spectra contains infinitely-many bound states,
     the ground state and the 2nd excited state are found algebraically, their energies
     $E^{0,2} \equiv E_1^{(0/2)} = 2 b \pm 2 {\sqrt {b^2 +2}}$, their eigenfunctions
     $\Psi_{\pm} = (2 z^2 + b \mp (b^2 + 2)^{\half}) \, e^{ - \frac{z^4}{4} - \frac{b z^2}{2}}$\,,
     the first excited state energy $E^1$ vs $b$ is shown, in particular,
     $E^1(b=0) = -2 {\sqrt 2}\, \frac{\Gamma(\frac{5}{4})}{\Gamma(\frac{3}{4})}$ (see text).
      }
    \label{V6-2}
\end{center}
\end{figure}

As an illustration, plots of the energy versus $b$ for the first three states are shown on Fig. \ref{V6-2}; the energy of the first excited state is found in convergent perturbation theory \cite{Turbiner:1984},
\[
     E^1\ \approx \ 3\, b - 2\ \frac{\int_0^{\infty} z^4\,e^{ - \frac{z^4}{2} - b z^2}\,dz}
     {\int_0^{\infty} z^2\,e^{ - \frac{z^4}{2} - b z^2}\,dz}\ .
\]
The behavior of $E_1^{(0/2)}(b)$ demonstrates the effect of avoiding singularities: a minimal distance between these levels occurs at $b=0$,
\[
    \De E_1 = E_1^{(2)} - E_1^{(0)} = 4 (2a)^{\half}\ .
\]

(iv) \ The first and the third excited states, both of negative parity,

at~$n=1,p=1$\ (or, in other words, the ground state and the second excited state of negative parity),
\[
  E_1^{(1/3)} = 5 b \pm 2 (b^2 + 6a)^{\half}\ ,\ \Psi_1^{(1/3)} =
z\,(2 a z^2 + b \mp (b^2 + 6a)^{\half}) \, e^{ - \frac{a z^4}{4} - \frac{b z^2}{2}}\ ,
\]
in potential
\[
   V(z) = a^2 z^6 + 2ab z^4 + (b^2 - 9 a) z^2 \ .
\]
The two-sheeted Riemann surface $E_1^{(1/3)}(b)$ which describes the first and the third excited states splits away from the infinitely-sheeted Riemann surface $E^{(-)}(b)$. These levels intersect at $b = \pm i {\sqrt {6 a}}$ where the square-root branch points occur, see e.g. Fig. \ref{bcomp-V6-2}.
These are the Landau-Zener singularities: making the analytic continuation around any of this branch point, starting from the 1st excited state energy at real $b$ (and fixed
$a > 0$), we arrive at the energy of the 3rd excited state at the same $b$.

The behavior of $E_1^{(1/3)}(b)$ demonstrates the effect of avoiding singularities: a minimal distance between these levels occurs at $b=0$,
\[
    \De E_1 = E_1^{(3)} - E_1^{(1)} = 4 (6a)^{\half}\ .
\]

\hskip 1cm
Naively, the analytic continuation can be made from positive $a$
to negative $a$. The potential (\ref{e1.3.36}) at negative $a$ remains
confined and contains infinitely-many bound states, the operator (\ref{e1.3.35a})
has $(n+1)$ polynomial eigenfunctions, the only thing that happens is that, the
eigenfunctions (\ref{e1.3.37}) become non-normalizable. A non-trivial
fact is that for negative $a$ a linear combination of (\ref{e1.3.37})
with the second, linearly-independent solution (obtained by keeping the Wronskian
equal to constant) can never be made normalizable! It implies the absence of analytical
continuation from positive $a$ to negative $a$. Thus, there exist
two well-defined but analytically disconnected spectral problems
with infinite discrete spectra:
one for positive $a$, another one - for negative $a$ (see \cite{Bender-Turbiner:1993}).

\vskip .5truecm

{\it Comment 3.3}\ . The $d$-dimensional Laplacian
\[
   \De\ =\ \sum_1^d \frac{\pa^2}{{\pa x_i}^2}\ ,
\]
written in spherical coordinates $(r, \Om)$ has the well-known form
\[
   \De\ =\ \frac{\pa^2}{{\pa r}^2} + \frac{d-1}{r} \frac{\pa}{\pa r} + \frac{\De_{S^d}}{r^2}\ ,
\]
where $\De_{S^d}$ is Laplacian on the sphere ${S^d}$. The eigenfunctions of $\De_{S^d}$ are the $d$-dimensional spherical harmonics $Y_{\{ l \}}(\Om)$ with a certain total angular momentum $l$,
\[
   \De_{S^d} Y_{\{ l \}}(\Om)\ =\ l(l+d-2) Y_{\{ l \}}(\Om)\ .
\]
Separating out the angular variables leads to a spectral problem for finding the radial function of the operator
\begin{equation}
\label{e1.3.38R}
  \De_r\ =\ \frac{\pa^2}{{\pa r}^2} + \frac{d-1}{r} \frac{\pa}{\pa r} + \frac{l(l+d-2)}{r^2}\ ,
\end{equation}
which is sometimes called the {\it radial Laplacian}. Making different gauge rotations we arrive either to the spectral problem (\ref{esl29}) on half-line $r \in {\bf R^+}$ with an effective singular potential,
\[
  r^{\frac{(d-1)}{2}} \De_r r^{\frac{-(d-1)}{2}}\ =\ \frac{\pa^2}{{\pa r}^2} + \frac{C}{r^2}\ ,
\]
where $C$ is a constant, or at the spectral problem for the radial operator
\[
  r^{-l} \De_r r^{l}\ =\ \frac{\pa^2}{{\pa r}^2}+ \frac{d + 2 l - 1}{r} \frac{\pa}{\pa r}\ ,
\]
defined also on the half-line, $r \in {\bf R^+}$. If the Hamiltonian with spherically symmetric potential $V(r)$ is considered,
\[
     {\mathcal H}\ =\ -\De\ +\ V(r)\ ,
\]
the notion of the {\it radial} Hamiltonian can be introduced,
\begin{equation}
\label{e1.3.38H}
  {\mathcal H}_r\ =\ -\frac{\pa^2}{{\pa r}^2} - \frac{d + 2 l -1}{r} \frac{\pa}{\pa r} +  V(r)\ .
\end{equation}
This Hamiltonian is Hermitian with measure $\sim r^{d + 2 l - 1}$.

\vskip .5truecm

{\bf Case VII}.

\vskip .3truecm

Let us take the following non-linear combination in the generators
(\ref{sl2r}), see\cite{Turbiner:1988qes},
\begin{equation}
\label{e1.3.38}
 T_2 =
 -4 J^0_n J^-_n + 4a J^+_n + 4b J^0_n - 2(n + d + 2 l - 2 c) J^-_n + 2 b n\ ,
\end{equation}
(c.f. (\ref{e1.3.35a}) at $2 p = d + 2 l - 2 c -1$),
or as the differential operator,
\begin{equation}
\label{e1.3.38a}
T_2(x,d_x)= -4 x d_x^2 + 2(2 a x^2 + 2 b x - d - 2 l + 2 c)d_x - 4 a n x\ ,
\end{equation}
where $a > 0, b, c, d, l, n$ are parameters. If the mark of representation (\ref{sl2r}) $n$ is a non-negative integer, the operator (\ref{e1.3.38}) has
finite-dimensional invariant subspace in polynomials in $x$ of degree not higher than $n$.

In the basis of monomials $\{1, x, x^2, \ldots, x^k, \ldots x^n, \ldots \}$
this operator has the form of a tri-diagonal, Jacobi matrix,
\begin{equation}
\label{e1.3.38.1}
 t_{k,k-1} = -2\,k(2k + d + 2 l - 2 c - 2) \ ,\ t_{k,k} =  4b \,k \ ,\ t_{k,k+1} = 4a\,(k - n)\ ,
\end{equation}
(c.f. (\ref{e1.3.35a.1})).
If $k=n$, the matrix element $t_{n,n+1}=0$ and the Jacobi matrix $T_2$ becomes block-triangular.
The characteristic polynomial is factorized, $\det (T_2 - \veps) = P_n(\veps) P_{\infty}(\veps)$.
If $b=0$ the principal diagonal of the Jacobi matrix vanishes and the energy-reflection symmetry occurs, see \cite{Shifman:1999}, and Lemma \ref{Lemma124},
\[
     P_{n}(\veps)= \veps^p {\tilde P}_{[\frac{n}{2}]} (\veps^2) \ ,
\]
where the parameter $p=0$ or $1$ depending on whether the integer $n$ is even or odd, respectively.
For $a=0$ this matrix becomes triangular. Its eigenvalues can be found explicitly, $\veps_k = 4 b k, k=0,1,\ldots$\ .

Choosing the gauge phase
\[
 A\ =\ \frac{ax^2}{4} + \frac{bx}{2} - \frac{D_c - 1}{4}\, \ln{x} \ ,
\]
where $D_c \equiv (d + 2 l - 2 c)$ and following the above-described procedure (\ref{e1.3.22})-(\ref{e1.3.23.1}) at
\[
     x=r^2\ ,\ \vrho=1  \ ,
\]
we arrive at the spectral problem for the radial Hamiltonian (\ref{e1.3.38H}) with the potential, see \cite{Turbiner:1987},
\begin{equation}
\label{e1.3.39}
      V(z) = a^2 r^6 + 2ab r^4 + [b^2 - (4n+D_c + 2)a] r^2 - \frac{c(c + D_c - 2)}{r^{2}}\ ,
\end{equation}
where
\[
   \Psi (r) \ = \  p_n (r^2) r^{l-c}\,  e ^ {-\frac{ar^4}{4} - \frac{br^2}{2}},
\]
at $a>0, \forall b$ or $a \geq 0, b>0$ and $(d+l-c)>1$ at $z \in [0, \infty)$.

It corresponds to the radial part of a $d$-dimensional Schr\"odinger
equation with angular momentum $l$\,, see (\ref{e1.3.38H}).
At $d=1$ and $l=0$ the radial operator coincides with the ordinary Schr\"odinger
equation (\ref{e1.3.21}). The potential (\ref{e1.3.39}) becomes a
generalization of the potential (\ref{e1.3.36}), with an additional singular
term proportional to $ r^{-2}$, see (\ref{e1.3.36sing}). Note that the operator
(\ref{e1.3.38H}) with potential (\ref{e1.3.39}) depends on a combination $D \equiv (d+2l)$.
It implies that the spectra for different $d, l$ but the same $D$ coincide
(but not the multiplicities). In particular, for $d>3$ the spectra of $S$-states in the
$d$-dimensional case coincide with the spectra of $P$-states in the $(d-2)$-dimensional case.
There is no single feature which makes the physical dimension $d=3$ special.

It is worth presenting two particular cases explicitly,

(i) \ Ground state, \ $n=0$\ ,
\[
  E^{(l)}_0 = b D_c \ ,\ \Psi_0 = r^{l-c}\,e^{ - \frac{a r^4}{4} - \frac{b r^2}{2}}\ ,
\]
in potential
\[
   V(r) = a^2 r^6 + 2ab r^4 + [b^2 - (D_c+2) a] r^2 - \frac{c(c+D_c-2)}{r^{2}} \ .
\]
The Riemann sheet $E^{(l)}_0(b)$ of the ground energy for given $l$ splits away from the
infinitely-sheeted Riemann surface $E^{(l)}(b)$.

(ii) \ Ground state (marked by superscript -) and the first excited state (marked by  superscript +), \ $n=1$\ ,
\[
E_1^{(\pm)} = b D_c + 2 b \pm 2 (b^2 + 2aD_c)^{\half}\ ,\ \Psi_1^{(\pm)} = (2 a r^2 + b \mp (b^2 + 2 a D_c)^{\half})
\,r^{l-c}\, e^{ - \frac{a r^4}{4} - \frac{b r^2}{2}}\ ,
\]
in potential
\[
   V(r) = a^2 r^6 + 2ab r^4 + (b^2 - (D_c+6) a) r^2 - \frac{c(c+D_c-2)}{r^{2}}\ .
\]
The two-sheeted Riemann surface $E_1^{(\pm)}(b)$ which describes the ground state energy and the first excited state splits away from the infinitely-sheeted Riemann surface $E(b)$. These levels intersect at $b = \pm i {\sqrt {2 a D_c}}$\,, where the square-root branch points occur, see e.g. Fig. \ref{bcomp-V6-2}.

\newpage

\subsection{Coulomb-type potentials}

\vskip .3truecm

Next we consider two quasi-exactly-solvable potentials which are associated with the two-body
Coulomb problem in $d$-dimensional space,
\[
        V_c(r)\ =\ -\frac{\al}{r}\ ,\ x \in {\bf R}^d \ .
\]
where $r$ is the distance between two charged particles, $\al$ is the parameter. Separating out the center-of-mass motion,
we arrive at the Hamiltonian of relative motion of the form,
\[
      {\mathcal H}(r)\ =\ -\De^{(d)}\ +\ V_c(r)\ .
\]
Introducing the Euler coordinates $(r, \Om)$ in ${\mathcal H}(r)$ and separating out the angular
degrees of freedom $\{\Om\}$, and then making a gauge rotation with $r^l$ as a gauge factor
we arrive at the radial Hamiltonian ${\mathcal H}_r$ (\ref{e1.3.38H}) and the eigenvalue problem for the radial motion
\begin{equation}
\label{e1.3.38Hr}
  \bigg(-\frac{\pa^2}{{\pa r}^2} - \frac{d + 2 l -1}{r} \frac{\pa}{\pa r} +  V_c(r)\bigg) \Psi(r)\
  =\ E \Psi(r)\ ,
\end{equation}
see {\it Comment 3.3}\,.
If we assume that the energy is fixed and negative, $$-E\ =\ E'\ \equiv\ k^2$$, and we look for the spectra
of $\al$, the equation for radial motion is changed,
\begin{equation*}
  \bigg(-\frac{\pa^2}{{\pa r}^2} - \frac{d + 2 l -1}{r} \frac{\pa}{\pa r} + E' \bigg) \Psi(r)\
  =\ \frac{\al}{r} \Psi(r)\ .
\end{equation*}
Multiplying both sides by $r$, we arrive at the spectral problem for radial motion
\begin{equation}
\label{e1.3.38Hc}
  H_r \Psi(r) \equiv \bigg(-r \frac{\pa^2}{{\pa r}^2} - (d + 2 l -1) \frac{\pa}{\pa r} + E' r \bigg) \Psi(r)\
  =\ \al \Psi(r)\ ,\ \Psi(r) \in L^2 ({\bf R}^+)\ .
\end{equation}
This eigenvalue problem has infinite discrete spectra. Such an approach to the Coulomb problem where
we quantize the $\al$-parameter keeping the energy fixed is called the {\it Sturm} approach. The corresponding representation of the spectral problem  is called the {\it Sturm} representation.

The operator $H_r$ can be gauge-rotated with the gauge factor $e^{-k r}$ to obtain
\[
    h_r \equiv e^{k r}\, H_r\, e^{-k r}\ =\ -r \frac{\pa^2}{{\pa r}^2} + (2 k r - d - 2 l +1) \frac{\pa}{\pa r}
    + k (d + 2 l -1)\ .
\]
The resulting operator has infinitely-many finite-dimensional invariant subspaces in polynomials. Those
subspaces form an infinite flag. In action on monomials this operator is triangular, its spectra is linear
in radial quantum number $n_r$,
\begin{equation}
\label{aln}
      \al_n\ =\ 2 k n_r + k (d + 2 l -1)\ \equiv 2k n,\ n_r=0,1,\ldots
\end{equation}
and its eigenfunctions are the Laguerre polynomials. At $d=3$ the parameter $n$ coincides with the principal
quantum number, $n=n_r+l+1$, see e.g. \cite{LL-QM:1977}. The operator $h_r$ can be rewritten in terms
of the generators $J^{0,-} \equiv J^{0,-}_0$ of the Borel subalgebra, $\mathfrak{b}_2 \subset \mathfrak{sl}(2,\bf{R})$, see (\ref{sl2r}),
\[
     h_r\ =\ -J^{0} J^{-} + 2 k J^{0} - (d + 2 l -1) J^{-} + k (d + 2 l -1)\ .
\]
To summarize, we state that the existence of the Sturm representation implies that the two-body Coulomb problem can be considered as a (quasi)-exactly-solvable problem of the second type where the parameter in front of the Coulomb term is quantized while the energy plays the role of the external parameter.

Note that using (\ref{aln}) and assuming the parameter $\al$ is kept fixed in (\ref{e1.3.38Hc}), one can obtain the energy quantization
\begin{equation}
\label{kn}
      k \ =\ \frac{\al}{2n}\ ,\ E\ =\ -\frac{\al^2}{4 n^2}\ .
\end{equation}

The Sturm representation can be constructed for a general spherical symmetrical potential $V(r)$,
\begin{equation}
\label{e1.3.38Hrg}
  \bigg(-\frac{\pa^2}{{\pa r}^2} - \frac{d + 2 l -1}{r} \frac{\pa}{\pa r} +  V \bigg) \Psi(r)\
  =\ E \Psi(r)\ .
\end{equation}
which contains the singular, Coulomb-type potential term $V_c(r)=-\al/r$.  The analogue of the radial equation (\ref{e1.3.38Hc})
\begin{equation}
\label{e1.3.38Hcg}
  \bigg(-r \frac{\pa^2}{{\pa r}^2} - (d + 2 l -1) \frac{\pa}{\pa r} + E' r + r(V - V_c) \bigg)
  \Psi(r)\ =\ \al \Psi(r)\ ,\ \Psi(r) \in L^2 ({\bf R}^+)\ ,
\end{equation}
defines a QES problem of the second type, c.f. (\ref{e1.3.24.2}). Also the Sturm representation can be constructed for the case when the potential $V(r)$ contains the singular term $V_s(r)=-\la/r^2$. The analogous radial equation in this case is
\begin{equation}
\label{e1.3.38Hsg}
  \bigg(-r^2 \frac{\pa^2}{{\pa r}^2} - (d + 2 l -1) r \frac{\pa}{\pa r} + E' r^2 + r^2 (V - V_s) \bigg) \Psi(r)\ =\ \la \Psi(r)\ ,\ \Psi(r) \in L^2 ({\bf R}^+)\ .
\end{equation}


\vskip .5truecm

{\bf Case VIII}.

\vskip .3truecm

Let us take the following bilinear combination of the $\mathfrak{sl}(2,\bf{R})$
generators with the mark $n$ (\ref{sl2r}) (see \cite{Turbiner:1988qes})
\begin{equation}
\label{e1.3.40}
 T_2 =
 -J^0_n J^-_n + 2 a J^+_n + 2b J^0_n - (\frac{n}{2}+d+2l-2c-1) J^-_n - bn\ ,
\end{equation}
or, as the differential operator,
\begin{equation}
\label{e1.3.40a}
T_2(x,d_x)= -xd_x^2 + (2ax^2+2bx+2c-d-2l+1)d_x - 2anx\ .
\end{equation}
where $a > 0, b, c, d, l, n$ are parameters. If $n$ is a non-negative integer,
which plays the role of the mark of representation (\ref{sl2r}), the operator (\ref{e1.3.40}) has a
finite-dimensional invariant subspace in polynomials of degree not higher than $n$.

In the basis of monomials $\{1, x, x^2, \ldots, x^k, \ldots x^n, \ldots \}$
this operator has the form of a tri-diagonal, Jacobi matrix,
\begin{equation}
\label{e1.3.40.1}
 t_{k,k-1} = -k( k + d + 2 l - 2 c - 2) \ ,\ t_{k,k} =  2 b \,k \ ,\ t_{k,k+1} = 2a\,(k - n)\ ,
\end{equation}
(c.f. (\ref{e1.3.38.1})).
If $k=n$, the matrix element $t_{n,n+1}=0$ and the Jacobi matrix $T_2$ becomes block-triangular.
The characteristic polynomial is factorizable, $\det (T_2 - \al) = P_n(\al) P_{\infty}(\al)$.
If $b=0$ the principal diagonal of the Jacobi matrix vanishes and the energy-reflection symmetry occurs, see \cite{Shifman:1999}, and Lemma \ref{Lemma124},
\[
     P_{n}(\al)= \al^p p_{[\frac{n}{2}]} (\al^2) \ ,
\]
where the parameter $p=0$ or 1 depending on whether the integer $n$ is even or odd, respectively.
For $a=0$ this matrix becomes triangular. Its eigenvalues can be found explicitly,
$\al_k = 2 b k\ , k=0,1,\ldots$\ .

Making a gauge rotation of (\ref{e1.3.40}) to (\ref{e1.3.38Hcg})
and a change of variables we arrive at the potential
\[
   V(r) - V_c(r)\ =\ a^2 r^2 + 2 a b r - \frac{b (D_c - 1)}{r} - \frac{c(D_c + c - 2)}{r^2}
\]
\begin{equation}
\label{e1.3.41}
  + b^2 - a (2n  + D_c) \ ,
\end{equation}
c.f. (\ref{e1.3.24.2}), (\ref{e1.3.24.3}), where
\[
                 x=r\ ,\  V_c = -\frac{\al}{r}   \ ,
\]
and $D_c \equiv (d + 2 l - 2 c)$, and the reference point for energy is chosen to be equal to zero.
The eigenfunctions of (\ref{e1.3.38Hcg}) in the algebraic sector are
\begin{equation}
\label{e1.3.41e}
         \Psi (r) \ = \  p_n (r) r^{l-c}  e^{-\frac{a}{2}\, r^2  - b\, r}\ ,
\end{equation}
at $ a \geq 0, b>0$ and $D_c > 2$, where $p_n(r)$ are eigenpolynomials of the operator (\ref{e1.3.40}).
Eventually, we arrive at a family of Schr\"odinger equations,
\begin{equation}
\label{e1.3.41S}
   (-\De^{(d)} + V(r) )\Psi \ =\ E \Psi
\end{equation}
where the $(n+1)$ radial excited states (the ground state of the 1st potential, the 1st excited state
in the 2nd potential, $k$ excited state in the $(k+1)$ potential, ... , the $n$th excited state in the
$(n+1)$ potential) can be found algebraically, solving the spectral problem (\ref{e1.3.40}).
For a fixed $n$,  the family of potentials is given by
\begin{equation}
\label{e1.3.41V}
 V(r)\ =\ a^2 r^2 + 2 a b r - \frac{b (D_c - 1) + \al}{r} - \frac{c(D_c + c - 2)}{r^2}\ .
\end{equation}
The parameter $\al$ takes the values of the eigenvalues of the finite-size matrix (\ref{e1.3.40.1})
ordered in the following manner:

$\al_0 > \al_1 > \ldots >  \al_n$. The energies of all these states are equal to
\[
  E_n\ =\ a (2n  + D_c) - b^2 \ ,
\]
and the eigenfunctions are (\ref{e1.3.41e}).

\begin{figure}
\begin{center}
\begin{tabular}{cc}
    \includegraphics[width=140pt,angle=0]{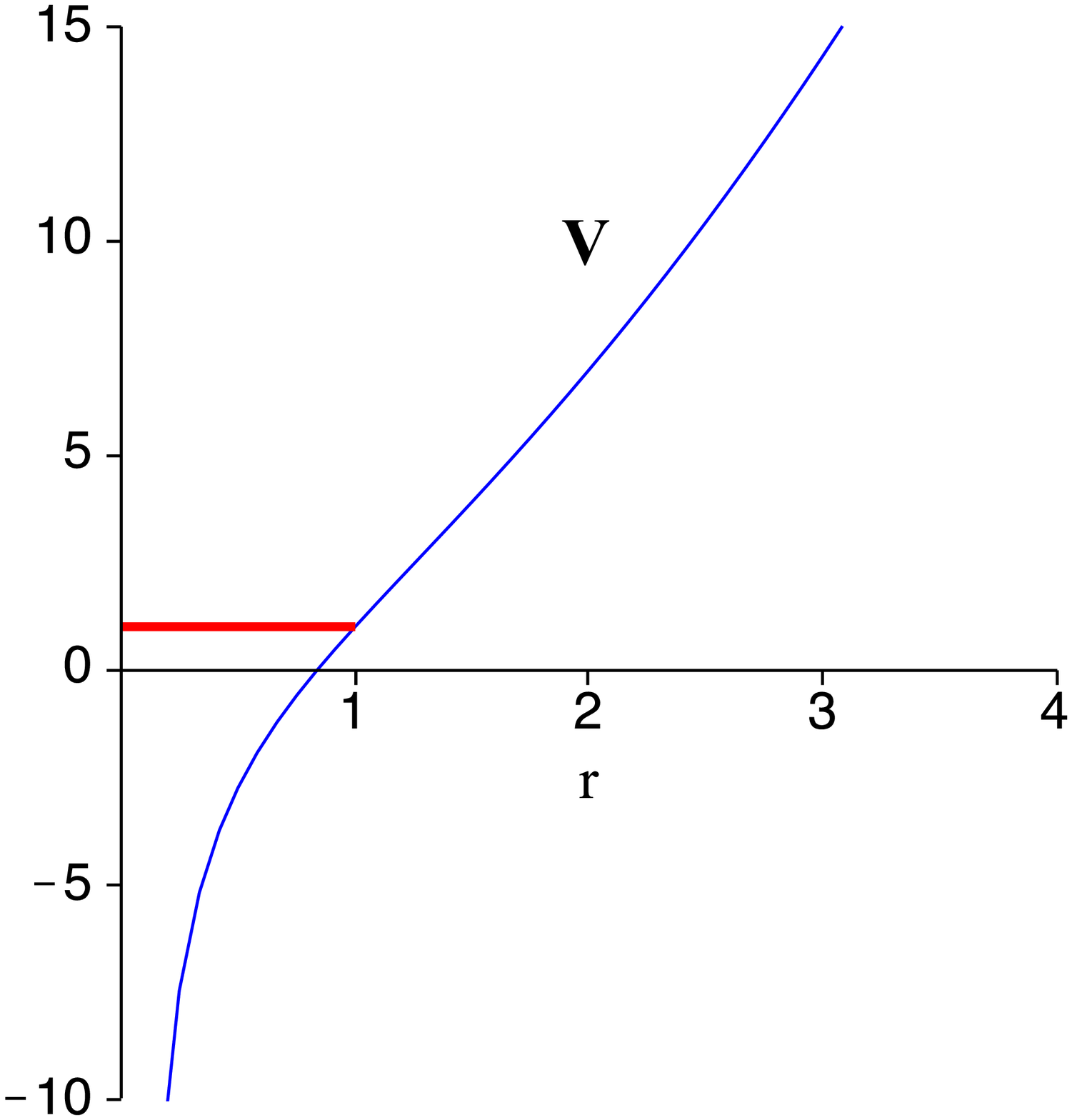} &
\includegraphics[width=140pt,angle=0]{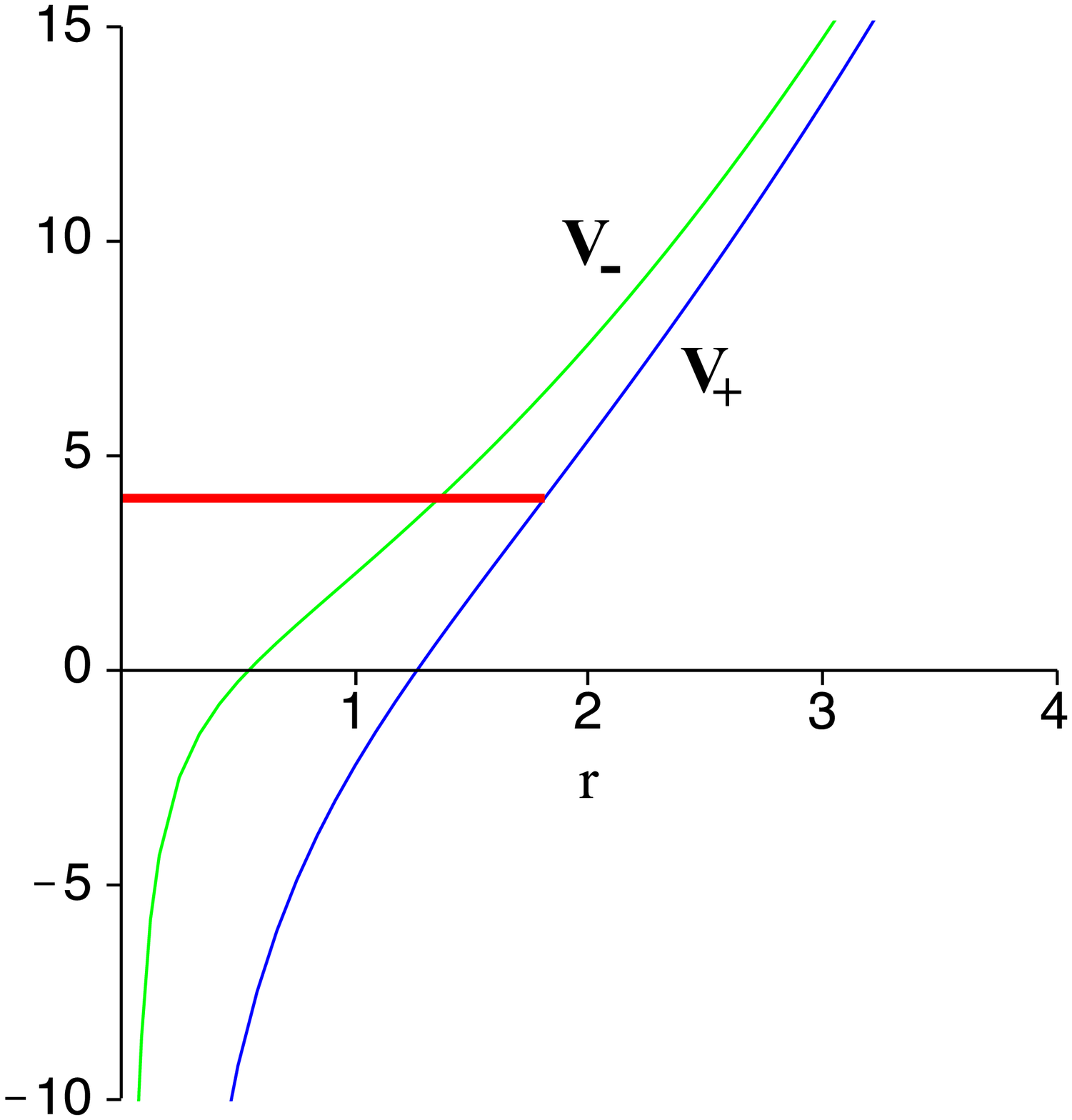} \\
    (a) & (b)
\end{tabular}
\end{center}
 \caption{
 \label{figVA}
 QES potential (\ref{e1.3.41V}) in 3-dimensional space at $l=0$:
 (a) Ground state in the potential $V(r) = r^2 + 2 r - \frac{2}{r}$ (blue line) with energy
      $E_0=1$ (red line) and the eigenfunction $\Psi_0\ =\ e^{ - \frac{r^2}{2} - r}$;
 (b) Ground state (with superscript -, green line) and the first excited state (with +, blue line)
     in the potentials $V_\pm = r^2 + 2 r - \frac{3 \pm {\sqrt 5}}{r}$, respectively, with energy
     $E=4$ (red line) and eigenfunctions $\Psi^{\pm}\ =\ \bigg(2 r + 1 \pm {\sqrt {5}}\bigg)
     \,e^{ - \frac{r^2}{2} - r}$.}
\end{figure}

The potential (\ref{e1.3.41V}) is a funnel-type potential which interpolates between the Coulomb
potential at small distances and the harmonic oscillator potential at large distances, see Fig. \ref{figVA}. This QES potential appears in a number of applications: (i) the Hooke's ``pseudo-atom" - two electrons in the external harmonic oscillator potential \cite{Turbiner:1994fe}, (ii) two electrons on a hypersphere \cite{Gill:2009}, (iii) two charges on a plane in a constant magnetic field \cite{Turbiner:2013} - to mention a few.

Let us present two particular cases,

(i) \ Ground state, \ $n=0$\ ,
\[
  E^{(l)}_0 = a D_c - b^2 \ ,\ \Psi_0 = r^{l-c}\,e^{ - \frac{a r^2}{2} - b r}\ ,
\]
in potential
\[
   V(r) = a^2 r^2 + 2 a b r - \frac{b (D_c - 1)}{r} - \frac{c(D_c + c - 2)}{r^2} \ ,
\]
see Fig. \ref{figVA}a.
The Riemann sheet $E^{(l)}_0(b)$ of the ground energy for a given $l$ and fixed $a,c$ splits away
from the infinitely-sheeted Riemann surface $E^{(l)}(b)$.

(ii) \ Ground state (marked by superscript -) and the first excited state (marked by  superscript +), \ $n=1$\ .

At energy
\[
   E^{(l)}_1\ =\  a (D_c+2) - b^2 \ ,
\]
in the potentials
\[
   V_{\pm}(r)\ =\ a^2 r^2 + 2 a b r - \frac{b (\al^{\pm} + D_c - 1) }{r}
   - \frac{c(D_c + c - 2)}{r^2} \ ,
\]
see Fig. \ref{figVA}b, where
\[
        \al^{\pm} = b \pm {\sqrt {b^2 + 2a (D_c-1)}}\ ,
\]
the eigenfunctions are
\[
  \Psi^{\pm}_1\ =\ \bigg(2 a r + b \pm {\sqrt {b^2 + 2a (D_c-1)}}\bigg)  \,
  r^{l-c}\,e^{ - \frac{a r^2}{2} - b r}\ .
\]

\vskip .5truecm

{\bf Case IX}.

\vskip .3truecm

Let us take the following bilinear combination of the $\mathfrak{sl}(2,\bf{R})$
generators with the mark $n$ (\ref{sl2r}) (see \cite{Turbiner:1988qes})
\begin{equation}
\label{e1.3.42}
 T_2 =
 -J^+_n J^-_n + 2a J^+_n - (n+d-1+2l-2c) J^0_n + 2b J^-_n
 - n (d+2l-1-2c)\ ,
\end{equation}
or as the differential operator,
\begin{equation}
\label{e1.3.42-1}
T_2(x,d_x)= -x^2d_x^2 - [2ax^2+(2c-d-2l+1)x -2b]d_x - 2anx\ .
\end{equation}
If $n$ is a non-negative integer, which plays the role of a mark of representation (\ref{sl2r}),
the operator (\ref{e1.3.38}) has the finite-dimensional invariant subspace in polynomials of degree
not higher than $n$.

In the basis of monomials $\{1, x, x^2, \ldots, x^k, \ldots x^n, \ldots \}$
this operator has the form of a tri-diagonal, Jacobi matrix,
\begin{equation}
\label{e1.3.42.1}
 t_{k,k-1} = 2 k b \ ,\ t_{k,k} =  -k(k - d - 2l + 2c) \ ,\ t_{k,k+1} = -2a\,(k - n)\ ,
\end{equation}
(c.f. (\ref{e1.3.38.1})).
If $k=n$, the matrix element $t_{n,n+1}=0$ and the Jacobi matrix $T_2$ becomes block-triangular.
The characteristic polynomial is factorizable, $\det (T_2 - \la) = P_n(\la) P_{\infty}(\la)$.
If $b=0$ this matrix becomes triangular. Its eigenvalues can be found explicitly,
$\la_k =-k(k - d - 2l + 2c) \ ,\ k=0,1, \ldots$\ .

\begin{figure}[tb]
\begin{center}
     {\includegraphics*[width=3in]{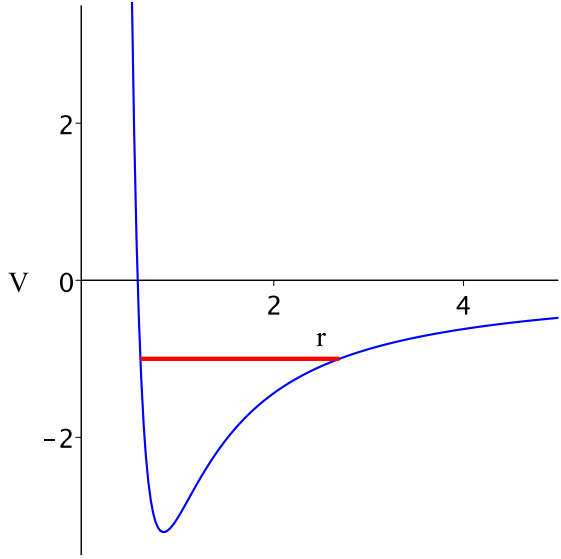}}
     \caption{
     $3D$-Potential (\ref{e1.3.43V}) at $a = b = 1, c=0, D_c = 3$ and $n=0, \la=0$:
     \ $V(r) = 1/r^4 - 2/r^2 - \frac{2}{r}$\,;
     the ground state energy $E=-1$ and the eigenfunction
     $\Psi_0 = e^{ - \frac{r}{2} - \frac{1}{r}}$\, .
      }
    \label{PT-II}
\end{center}
\end{figure}

Making a gauge rotation of (\ref{e1.3.42-1}) to (\ref{e1.3.38Hsg}) and a change
of variables we arrive at the potential
\begin{equation}
\label{e1.3.43}
  V(r) - V_s(r)\ =\ \frac{b^2}{r^4} + \frac{b(D_c - 3)}{r^3} - \frac{c (D_c + c - 2) + 2ab}{r^2}
\end{equation}
\[
  -\ \frac{a(2n+D_c-1)}{r}\ +\ a^2\ ,
\]
c.f. (\ref{e1.3.24.2}), (\ref{e1.3.24.3}), where
\[
     x=r\ ,\  V_s = -\frac{\la}{r^2} \ ,
\]
and $D_c \equiv (d + 2 l - 2 c)$, and the reference point for energy is chosen to be equal to zero.
The eigenfunctions of (\ref{e1.3.38Hsg}) in the algebraic sector are of the form
\begin{equation}
\label{e1.3.43e}
      \Psi (r) \ = \  p_n (r) r^{l-c}  e ^ { - a r - br^{-1}},
\end{equation}
at $a >0, b \geq 0$ and  $c \in R$, and $l, d > 0$ are non-negative integers; if $a=0$, the parameter
$b > 0$ and $d + l + n - c < 0$. Here $p_n(r)$ are eigenpolynomials of the operator (\ref{e1.3.42-1}).
Eventually, we arrive at a family of Schr\"odinger equations (\ref{e1.3.41S})
\[
   (-\De^{(d)} + V(r) )\Psi \ =\ E \Psi\ ,
\]
where the $(n+1)$ radial excited states (the ground state of the 1st potential, the 1st radial excited state in the 2nd potential, $k$th radial excited state in the $(k+1)$ potential, ... , the $n$th radial excited state in the $(n+1)$ potential) can be found algebraically, solving the spectral problem (\ref{e1.3.42}).
For fixed $n$,  the family of potentials is given by
\begin{equation}
\label{e1.3.43V}
 V(r)\ =\ \frac{b^2}{r^4} + \frac{b(D_c - 3)}{r^3} - \frac{c (D_c + c - 2) + 2ab + \la}{r^2}
  -\ \frac{a(2n+D_c-1)}{r}\ ,
\end{equation}
see Fig. \ref{PT-II}.
The parameter $\la$ takes the value of the eigenvalues of the finite-size matrix (\ref{e1.3.42.1}) ordered in the following manner:\\
$\la_0 > \la_1 > \ldots >  \la_n$. The energies of all these states are equal to the same
\[
  E_n\ =\ -a^2 \ ,
\]
and their eigenfunctions are (\ref{e1.3.43e}). For $a=0$ at $d=3$ the potential (\ref{e1.3.43V}) was discovered by Korol \cite{Korol:1973}.

Let us present a particular case,

(i) \ Ground state, \ $n=0$\ ,
\[
  E^{(l)}_0 = -a^2 \ ,\ \Psi_0 = r^{l-c}\,e^{ - \frac{a r}{2} - \frac{b}{r}}\ ,\ \la=0\ ,
\]
in potential
\[
   V(r)\ =\ \frac{b^2}{r^4} + \frac{b(D_c - 3)}{r^3} - \frac{c (D_c + c - 2) + 2ab}{r^2}
  -\ \frac{a(D_c-1)}{r} \ ,
\]
see Fig. \ref{PT-II}.
The Riemann sheet $E^{(l)}_0(b)$ of the ground energy for given $l$ and fixed $a,c$ splits away from the infinitely-sheeted Riemann surface $E^{(l)}(b)$.

\newpage

\subsection{Non-singular Periodic potential}

\vskip .3truecm

Now let us show the unique example of the non-singular periodic
quasi-exactly-solvable potential associated with the Mathieu
potential.

\vskip .5truecm

\noindent
 {\it Comment 3.4} \ The Mathieu potential
\[
V(z)\ =\ A \al^2 \cos{\al z}\ ,
\]
which is sometimes called the sine-Gordon potential,
is among the most important potentials in many branches of
physics and engineering. Detailed description of the properties
of the corresponding Schr\"odinger equation, which is called the Mathieu equation,
can be found in Bateman-Erd\'elyi \cite{Bateman:1953}, Vol.3, also in
Kamke \cite{Kamke:1959}, Equation 2.22 and in Whittaker-Watson \cite{WW:1927}.

The potential $V(z)$ is periodic with period $\frac{2\pi}{\al}$ and, hence, it has
infinitely-many degenerate minima distributed uniformly along the real axis. It implies
the existence of non-analytic, exponentially-small terms at $\al=0$, see e.g.
Dunne-Unsal \cite{DU:2014}. The eigenvalues form four infinitely-sheeted Riemann surfaces
in $\al$, every sheet contains infinitely-many square-root branch points \cite{HG:1981}.

The Hamiltonian is translationally-invariant,

$T: z \rar z + \frac{2\pi}{\al}$.
Taking the simplest $T$-invariant
\[
    \ta\ =\ \cos{\al z}\ ,
\]
as the new variable, we get the Mathieu potential in a very simple form,
\begin{equation}
\label{e1.3.44M}
  V(\ta) = A \al^2\, {\ta} \ ,
\end{equation}
and the second derivative ($1D$ Laplacian) becomes the one-dimensional Laplace-Beltrami operator with metric $g^{11} = \al^2 (1 - \ta^2)$\,,
\renewcommand{\theequation}{1.3.72-{\arabic{equation}}}
\setcounter{equation}{0}
\begin{equation}
\label{e1.3.44M-1}
    \dzz |_{\ta=\cos{\al z}}\ =\ \al^2 \bigg((1-\ta^2) \frac{d^2}{d {\ta}^2}\ -\
    \ta \frac{d}{d {\ta}}\bigg)\ ,
\end{equation}
which is an algebraic operator. This operator preserves the infinite flag of the spaces
of polynomials  ${\mathcal P}$ (\ref{e1.1.7}); it is exactly solvable and can be rewritten in terms of the generators $J^{0,-} \equiv J^{0,-}_0$ of the Borel subalgebra,
$\mathfrak{b}_2 \subset \mathfrak{sl}(2,\bf{R})$, see (\ref{sl2r}),
\[
    -\al^2 (J^{0} J^{0} - J^{-} J^{-})\ ,
\]
hence, it has infinitely many polynomial eigenfunctions with eigenvalues
\[
     \veps_k\ =\ -\al^2 k^2\ ,\ k=0,1,\ldots\ .
\]
Note that there exist three gauge factors such that the gauge rotated Laplacian in $\ta$-variable
remains an algebraic operator,
\[
   \sin^{-1}{\al z}\ \dzz\ \sin{\al z} |_{\ta=\cos{\al z}} \ =\
   \al^2 (1-\ta^2)^{-\half} \bigg((1-\ta^2) \frac{d^2}{d {\ta}^2}\ -\
   \ta \frac{d}{d {\ta}}\bigg) (1-\ta^2)^{\half}\ =\
\]
\begin{equation}
\label{e1.3.44M-2}
  \al^2 \bigg((1-\ta^2) \frac{d^2}{d {\ta}^2}\ - 3 \ta \frac{d}{d {\ta}}\ -\ 1 \bigg)\ ,
\end{equation}

\[
   \sin^{-1}{\bigg(\frac{\al}{2} z\bigg)}\ \dzz\ \sin{\bigg(\frac{\al}{2} z\bigg)} |_{\ta=\cos{\al z}} \ =\
   \al^2 (1-\ta)^{-\half} \bigg((1-\ta^2) \frac{d^2}{d {\ta}^2}\ -\
   \ta \frac{d}{d {\ta}}\bigg) (1-\ta)^{\half}\ =\
\]
\begin{equation}
\label{e1.3.44M-3}
   \al^2 \bigg((1-\ta^2) \frac{d^2}{d {\ta}^2}\ - (1 + 2\ta) \frac{d}{d {\ta}}\ -\ \frac{1}{4}\bigg)\ ,
\end{equation}

\[
   \cos^{-1}{\bigg(\frac{\al}{2} z\bigg)}\ \dzz\ \cos{\bigg(\frac{\al}{2} z\bigg)} |_{\ta=\cos{\al z}} \ =\
      \al^2 (1+\ta)^{-\half} \bigg((1-\ta^2) \frac{d^2}{d {\ta}^2}\ -\
      \ta \frac{d}{d {\ta}}\bigg) (1+\ta)^{\half}\ =\
\]
\begin{equation}
\label{e1.3.44M-4}
   \al^2 \bigg((1-\ta^2) \frac{d^2}{d {\ta}^2}\ + (1 - 2\ta) \frac{d}{d {\ta}}\ -\ \frac{1}{4}\bigg)\ .
\end{equation}

\renewcommand{\theequation}{1.3.{\arabic{equation}}}
\setcounter{equation}{72}

Every operator (\ref{e1.3.44M-2}) - (\ref{e1.3.44M-4}) preserves the infinite flag of the spaces
of polynomials ${\mathcal P}$ (\ref{e1.1.7}); it is exactly solvable and can be rewritten in terms
of the generators $J^{0,-} \equiv J^{0,-}_0$ of the Borel subalgebra,
$\mathfrak{b}_2 \subset \mathfrak{sl}(2,\bf{R})$, see (\ref{sl2r}),
\[
    -\al^2 (J^{0} J^{0} - J^{-} J^{-} + 2 J^0 + 1)\ ,
\]
\[
    -\al^2 (J^{0} J^{0} - J^{-} J^{-} + J^0 + J^- + \frac{1}{4})\ ,
\]
\[
    -\al^2 (J^{0} J^{0} - J^{-} J^{-} + J^0 - J^- + \frac{1}{4})\ ,
\]
respectively, hence, it has infinitely many polynomial eigenfunctions with eigenvalues
\[
     \veps_k\ =\ -\al^2 (k+1)^2\ ,\ k=0,1,\ldots\ ,
\]
\[
     \veps_k\ =\ -\al^2 (k+\frac{1}{2})^2\ ,\ k=0,1,\ldots\ ,
\]
\[
     \veps_k\ =\ -\al^2 (k+\frac{1}{2})^2\ ,\ k=0,1,\ldots\ ,
\]
respectively.

Finally, the Schr\"odinger operator in $\ta$-variable,
\[
  {\mathcal H}\ =\ -\dzz\ +\ A \al^2 \cos{\al z} = -\al^2 \bigg((1-\ta^2) \frac{d^2}{d {\ta}^2}
  + \ta \frac{d}{d {\ta}}\ +\ A\ {\ta}\bigg)\ ,
\]
becomes an algebraic operator, which is the algebraic form of the Mathieu operator, see e.g. \cite{Bateman:1953}. This algebraic operator is a particular form of the Heun operator.

\vskip .5truecm

 {\bf Case X}.

\vskip .3truecm

Let us take the following bilinear combination of the $\mathfrak{sl}(2,\bf{R})$
generators with mark $(n - \mu)$ (\ref{sl2r}) (see \cite{Turbiner:1988qes})
\begin{equation}
\label{e1.3.44}
 T_2\ =\
 \al^2 \bigg( J^+_{n-\mu} J^-_{n-\mu}   - J^-_{n-\mu}  J^-_{n-\mu} +
\end{equation}
\[
2a J^+_{n-\mu} + (n+\mu+1) J^0_{n-\mu} - 2 a J^-_{n-\mu} + \frac{(n-\mu)(n+\mu+1)}{2} \bigg)\ ,
\]
and rewrite it as a differential operator,
\begin{equation}
\label{e1.3.44d}
T_2(x,d_x) = \al^2 \bigg((x^2-1) d_x^2 + [2ax^2+ (1+2\mu)x - 2 a]d_x - 2 a(n-\mu)x \bigg) \equiv \al^2 t_2 (x,d_x)\ ,
\end{equation}
where $a \neq 0, \al, \mu, n$ are parameters. If $(n-\mu)$ is a non-negative integer, the differential
operator (\ref{e1.3.44d}) has a finite-dimensional invariant subspace in polynomials of
degree not higher than $n$.

In the basis of monomials $\{1, x, x^2, \ldots, x^k, \ldots x^n, \ldots \}$
the operator $t_2 (x,d_x)$ has the form of a four-diagonal matrix,
\begin{equation}
\label{e1.3.44.1}
  t_{k,k-2} = - k(k-1) \ ,\ t_{k,k-1} = -2 a k \ ,\ t_{k,k} =\,k(k+2\mu) \ ,\ t_{k,k+1} = 2a\,(k - n + \mu)\ .
\end{equation}
If $k=n-\mu$, the matrix element $t_{n-\mu,n-\mu+1}=0$ and the matrix $t_2$ becomes block-triangular.
The characteristic polynomial is factorizable, $\det (T_2 - \la) = P_{n-\mu}(\la) P_{\infty}(\la)$.

The transformation (\ref{e1.3.22})-(\ref{e1.3.23.1}) leads eventually to the
periodic potential
\begin{equation}
\label{e1.3.45}
  V(z) =  \al^2 \bigg(a ^2  \sin^2{\al z} - a (2n+1) \cos{\al z} + \mu\bigg)\ ,
\end{equation}
where
\[
x=\cos{\al z} \ ,\ \vrho=1  \ ,
\]
\[
\Psi (z) \ = \ (\sin {\al z})^{\mu}   p_{n-\mu} (\cos{\al z} ) \
e ^ {a \cos{\al z} },
\]
at $ a \neq 0,  \al \geq 0$. Here $\mu = 0,1$ and $n-\mu=0,1,\ldots$\,. For the (fixed) integer $n$,
the $(2n+1)$ eigenstates have the meaning of the edges of bands (zones) and can be
found algebraically. The quasi-exactly-solvable potential (\ref{e1.3.45}) was, perhaps,
the first QES potential which was discovered in full generality, it has a number of
different names: the Whittaker-Hill potential \cite{WW:1927}, the Magnus-Winkler potential
\cite{MW:1966}, the non-singular periodic QES potential \cite{Turbiner:1988z},
and the two term (trigonometric) potential \cite{DM:2007}. The polynomials
$p_{n-\mu} (\cos{\al z} )$ are sometimes called the Ince polynomials. The Hamiltonian corresponding to the QES potential (\ref{e1.3.45}) has the form
\begin{equation}
\label{e1.3.45H}
  {\mathcal H} (z)\ =\ -\dzz\ +\  \al^2 \bigg(a ^2  \sin^2{\al z} - a (2n+1) \cos{\al z} + \mu\bigg) \ .
\end{equation}

Now let us take the following operator
\begin{equation}
\label{e1.3.46}
 T_2 =
 \al^2 \bigg(J^+_{n-1} J^-_{n-1} - J^-_{n-1}  J^-_{n-1} +
\end{equation}
\[
  2a J^+_{n-1} + (n+1) J^0_{n-1}  - [2a+ (\nu_1-\nu_2)]  J^-_{n-1}
+ \frac{(n^2-1)}{2}\bigg)\ ,
\]
(c.f. (\ref{e1.3.44})), or in the form of a differential operator,
\begin{equation}
\label{e1.3.46d}
   T_2(x,d_x)= \al^2 \bigg((x^2-1) d_x^2 + [2ax^2+ 2 x-2a
(\nu_1-\nu_2) ]d_x - 2 a(n-1)x\bigg) \ ,
\end{equation}
(c.f. (\ref{e1.3.46d})).

The transformation (\ref{e1.3.22})-(\ref{e1.3.23.1}) leads to the
periodic potential
\begin{equation}
\label{e1.3.47}
  V(z) =  \al^2 \bigg(a^2  \sin^2{\al z} - 2 n a \cos{\al z} + a
(\nu_1-\nu_2) - \frac{1}{4} \bigg)\ ,
\end{equation}
where
\[
x=\cos{\al z} \ ,\ \vrho=1  \ ,
\]
\[
\Psi (z) \ = \ \cos^{\nu_1} {\frac{\al}{2} z} \ \sin^{\nu_2} {\frac{\al}{2} z}\
P_{n-1} (\cos{\al z} ) \ e^{ a \cos{\al z} },
\]
at $ a \neq 0,  \al \geq 0$. Here $\nu_{1,2} = 0,1$, but
$\nu_1 + \nu_2 =1$ . For the fixed $n$, $(2n+1)$ eigenstates, which
have the meaning of the edges of bands can be found algebraically. The
polynomials $P_{n-1} (\cos{\al z} )$ are called the Ince polynomials.
The Hamiltonian corresponding to the QES potential (\ref{e1.3.47}) has the form
\begin{equation}
\label{e1.3.47H}
  {\mathcal H} (z)\ =\ -\dzz\ +\  \al^2 \bigg(a^2  \sin^2{\al z} - 2 n a \cos{\al z} + a
(\nu_1-\nu_2) - \frac{1}{4} \bigg)\ ,
\end{equation}
cf. (\ref{e1.3.45H}).

\newpage

\subsection{Double-Periodic potentials}

\vskip .3truecm

The last two one-dimensional quasi-exactly-solvable potentials we are going to present are
double-periodic potentials, given by elliptic functions. They are written in terms of
the Weierstrass function $\wp (z) \equiv \wp (z|g_2,g_3)$ (see e.g. \cite{WW:1927}),
\begin{equation}
\label{wp}
    (\wp'(z))^2\ =\ 4\ \wp^3 (z) - g_2\ \wp (z)\ -\ g_3\ =\ 4(\wp (z)-e_1)(\wp (z)-e_2)(\wp (z)-e_3)\ ,
\end{equation}
where $g_{2,3}$ are its invariants and $e_{1,2,3}$ are its roots, $e_1+e_2+e_3=0$. Both potentials
appear in relation with the integrable quantum Calogero-Sutherland models with Weyl symmetry
$A_1$ and $BC_1$, thus, respectively, they emerge in the Hamiltonian Reduction Method, for review
see e.g. Olshanetsky-Perelomov \cite{Olshanetsky:1983}.

The $A_1$-quantum elliptic Calogero-Sutherland model describes a two-body system with pairwise interaction
given by the elliptic potential. Its Hamiltonian is defined on the plane $(z_1, z_2)$ as,
\[
    {\mathcal H}\ =\ -\half \,\De^{(2)} \ +\ \frac{\ka}{2} \wp (z_1-z_2)\ ,
\]
where $\ka = m (m+1)$ is the coupling constant.
After center-of-mass separation $Z=z_1+z_2,\ z=z_1-z_2$ the relative motion is described by the Hamiltonian
\begin{equation}
\label{e1.4.0A}
    {\mathcal H}_A\ =\ - d_z^2 \ +\ m(m + 1) {\wp}(z)\ .
\end{equation}
It coincides with the celebrated Lam\'e operator, see e.g. \cite{WW:1927}. Thus, it can be naturally called the
$A_1$-Lam\'e Hamiltonian or, simply, the Lam\'e Hamiltonian. In turn, the Hamiltonian of $BC_1$-quantum
elliptic Calogero-Sutherland model is defined as
\begin{equation}
\label{e1.4.0B}
    {\mathcal H}_B\ =\ - \half d_z^2 \ +\ \ka_2\ \wp (2z)\ +\ \ka_3\ \wp (z)\ ,
\end{equation}
see e.g. Olshanetsky-Perelomov \cite{Olshanetsky:1983}, where $\ka_{2}, \ka_{3}$ are coupling constants. Certainly, it can be considered as the Hamiltonian of two-body relative motion.
If one of the coupling constants vanishes, $\ka_{2}(\ka_{3})=0$, the Hamiltonian becomes $A_1$-Lam\'e Hamiltonian. We will show that in variable $\ta \ =\ \wp(z)$\ both Hamiltonians take algebraic form becoming a particular case of the Heun operator (\ref{QES-2expe}).
Each Hamiltonian can be rewritten in terms of the $\mathfrak{sl}(2,\bf{R})$ generators (\ref{sl2r}), hence, they are quadratic elements $T_2$ of the universal enveloping algebra $U_{\mathfrak{sl}(2,\bf{R})}$. If the mark $n$ of $\mathfrak{sl}(2,\bf{R})$ representation is an integer, the Hamiltonian has a finite-dimensional functional invariant subspace in polynomials. Furthermore, in the algebra $\mathfrak{sl}(2,\bf{R})$ both Hamiltonians have
the remarkable property of factorization
\[
      T_2\ =\ T_1^a \ T_1^b\ ,
\]
where $T_1^{a(b)} = A^{a(b)} J^+_{n} + B^{a(b)} J^0_{n} + C^{a(b)} J^-_{n} + D^{a(b)}$ are linear elements of $U_{\mathfrak{sl}(2,\bf{R})}$. Note that such a factorization occurs for some other exactly-solvable problems as well (see below).

Both potentials we are going to study can be written in terms of the Weierstrass functions
$\wp (z|g_2,g_3)$, $\wp (2z|g_2,g_3)$. The discrete symmetry of the Hamiltonian is $\mathbb{Z}_2 \oplus T_r \oplus T_c$: It consists of reflection $\mathbb{Z}_2 (x \rar -x)$
and two translations $T_r:\ z \rar z + P_r$ and $T_c:\ z \rar z + P_c$, where $P_{r,c}$
are periods. Let us take the simplest invariant with respect to the above discrete symmetry group,
\begin{equation}
\label{tau}
 \ta \ =\ \wp(z)\ ,
\end{equation}
as the new variable. The second derivative ($1D$ Laplacian) becomes the one-dimensional Laplace-Beltrami operator $\De_g$,
\begin{equation}
\label{e1.4.0M-1}
    \dzz |_{\ta=\wp(z)}\equiv g^{-1/2} \ \frac{\pa}{\pa \ta} g^{1/2} g^{11} \frac{\pa}{\pa \ta}\ =\
   (4 \ta^3 - g_2 \ta\ -\ g_3) \pa^2_{\ta} \ +\ (6\ta^2-\frac{g_2}{2}) \pa_{\ta}\ ,
\end{equation}
with flat metric
\[
g^{1 1}\ =\ \ (4 \ta^3 - g_2 \ta\ -\ g_3) \ =\ \frac{1}{g}\ ,
\]
here $g$ is its determinant with lower indices. It is the algebraic operator.

The operator (\ref{e1.4.0M-1}) preserves the one-dimensional space of polynomials
${\mathcal P}_0$ (\ref{e1.1.5}); it is a primitive quasi-exactly solvable operator and
can be rewritten in terms of the generators $J^{+,0,-} \equiv J^{+,0,-}_0$ of the algebra
$\mathfrak{sl}(2,\bf{R})$ (\ref{sl2r}) in one-dimensional representation,
\[
    4J^{+} J^{0} - g_2 J^{0} J^{-} - g_3 J^{-} J^{-} + 2J^{+} - \frac{g_2}{2} J^{-}\ ,
\]
hence, it has a constant as the eigenfunction with eigenvalue
\[
     \veps_0\ =\ 0\ .
\]

\vskip .5truecm

{\bf Case XI.} {\it Lam\'e equation (or $A_1$-quantum elliptic Calogero-Sutherland model)}

\vskip .3truecm

In this Section we consider one of the so-called $m$-zone Lam\'e
equations
\begin{equation}
\label{e1.4.1}
 - d_z^2 \Psi + m(m + 1) {\wp}(z) \Psi \ =\ \veps \Psi \ ,
\end{equation}
where ${\wp}(z)$ is the Weierstrass function in standard notation (\ref{wp}), which
depends on two free parameters, assuming that $m=1,2,\ldots$ . It will be shown that
it is a quasi-exactly-solvable Schr\"odinger equation \cite{Turbiner:1989}.

Introducing the new variable $\xi = {\wp}(z)+ \frac{1}{3} \sum a_i$ in
(\ref{e1.4.1}) (see, e.g. Kamke \cite{Kamke:1959}), a new
equation emerges
\begin{equation}
\label{e1.4.2}
        \eta^{\prime\prime} + {1\over 2} \bigg ({1\over \xi - a_1} +
        {1\over \xi - a_2} + {1\over \xi - a_3}\bigg ) \eta^\prime -
        {m(m+1) \xi + \veps \over 4(\xi - a_1)(\xi - a_2)(\xi -
        a_3)} \eta = 0\ ,
\end{equation}
where $\eta(\xi)\equiv\Psi(z)$. Without loss of generality $a_1=0$. Here the new parameters
$a_i$ satisfy the system of linear equations $a_i - \frac{1}{3} \sum a_i=e_i$.
Equation (\ref{e1.4.2}) is called the {\it algebraic form} for the Lam\'e equation.
It is known that the equation (\ref{e1.4.2}) can have four types of solutions:
\renewcommand{\theequation}{1.3.85-{\arabic{equation}}}
\setcounter{equation}{0}
\begin{equation}
\label{e1.4.3.1}
        \eta^{(1)} \ = \ p_k(\xi) \ ,
\end{equation}
\begin{equation}
\label{e1.4.3.2}
        \eta^{(2)}_i \ =\  (\xi - a_i)^{1/2} p_k(\xi)\quad, \quad i =1,2,3
\end{equation}
\begin{equation}
\label{e1.4.3.3}
        \eta^{(3)}_i \ = (\xi - a_{l_1})^{1/2} (\xi - a_{l_2})^{1/2}
        p_{k-1}(\xi)\ , \quad l_1 \ne l_2; i \ne l_{1,2}; i,l_{1,2} = 1,2,3
\end{equation}
\begin{equation}
\label{e1.4.3.4}
        \eta^{(4)} \ = (\xi - a_1)^{1/2} (\xi - a_2)^{1/2} (\xi - a_3)^{1/2}
 p_{k-1}(\xi)
\end{equation}

\renewcommand{\theequation}{1.3.{\arabic{equation}}}
\setcounter{equation}{85}

\noindent where $p_r (\xi)$ are polynomials in $\xi$ of degree $r$.
If the value of the parameter $m$ is fixed, there are $(2m + 1)$
linear independent solutions of the following form: if $m = 2k$ is
even, then the $\eta^{(1)}(\xi)$ and $\eta^{(3)}(\xi)$ solutions
arise, if $m = 2k +1$ is odd we get solutions of the
$\eta^{(2)}(\xi)$ and $\eta^{(4)}(\xi)$ types. Those eigenvalues
have the meaning of the edges of the zones in the potential
(\ref{e1.4.1}).

\begin{THEOREM} \cite{Turbiner:1989}
\label{Theorem:1.3}
 \  The spectral problem (\ref{e1.4.1})  at $m=1,2,\ldots$
with polynomial solutions (\ref{e1.4.3.1}), (\ref{e1.4.3.2}),
(\ref{e1.4.3.3}), (\ref{e1.4.3.4}) is equivalent to the spectral
problem (\ref{e1.2.5}) for the operator $T_2$ (\ref{QES-2})
belonging to the universal enveloping $\mathfrak{sl}(2,\bf{R})$-algebra in
the representation (\ref{sl2r}) with the coefficients
\begin{equation}
\label{e1.4.4}
        c_{+0} = 4 \quad ,\quad 2 c_{+-} = -4 \sum a_i \quad ,\quad c_{0-}
        = 4 \sum a_i a_j \quad ,\quad c_{--} = a_1 a_2 a_3
\end{equation}
before the terms quadratic in the generators and  the following
coefficients before the linear terms in the generators $J ^{\pm,0}_r$ :
\begin{enumerate}
\item[(1)] \ For $\eta^{(1)}(\xi)$-type solutions at $m=2k, r=k$
\renewcommand{\theequation}{1.3.87-{\arabic{equation}}}
\setcounter{equation}{0}
\begin{equation}
\label{e1.4.5.1}
 c_+ \ = - 6k -2\ , \  c_0\ = \ 4(k + 1)\sum a_i \ ,
 \ c_- = -2(k + 1)\sum a_i a_j
\end{equation}

\item[(2)] \ For $\eta^{(2)}_i (\xi)$-type solutions at $m=2k+1, r=k$
\[
c_+\ =\ -6k -6\ ,\ c_0 \ = 4(k + 2)\sum a_i  - a_i\ , \
\]
\begin{equation}
\label{e1.4.5.2}
 c_-\ =\ -2(k + 1)\sum a_i a_j - 4a_{l_1} a_{l_2} \ ,\quad
 i \ne l_{1,2} , l_1 \ne l_2
\end{equation}

\item[(3)] \ For $\eta^{(3)}_i (\xi)$-type solutions at $m=2k, r=k-1$
\[
c_+\ =\ -6k -4 \ ,\   c_0\ = 4(k + 1)\sum a_i  + 4a_i\ ,
\]
\begin{equation}
\label{e1.4.5.3}
 \ c_-\ =\ -2(k + 2)\sum a_i a_j + 4a_{l_1} a_{l_2} \ ,
 \quad i \ne l_{1,2} , l_1 \ne l_2
\end{equation}

\item[(4)]  \ For $\eta^{(4)}_i (\xi)$-type solutions at $m=2k+1,
r=k-1$
\begin{equation}
\label{e1.4.5.4} c_+ = -6k - 8 \ ,\ c_0 = 4(k + 2)\sum a_i  \ , \
c_- = -2(k + 2)\sum a_i a_j
\end{equation}
\end{enumerate}
\renewcommand{\theequation}{1.3.{\arabic{equation}}}
\setcounter{equation}{87}
\end{THEOREM}
Thus, each type of solution (\ref{e1.4.3.1}), (\ref{e1.4.3.2}),
(\ref{e1.4.3.3}), (\ref{e1.4.3.4})  corresponds to the particular
spectral problem (\ref{e1.4.1})  with a special set of parameters
(\ref{e1.4.4}) plus (\ref{e1.4.5.1}), (\ref{e1.4.5.2}),
(\ref{e1.4.5.3}), (\ref{e1.4.5.4}), respectively.

It can be easily shown that the calculation of eigenvalues $\veps$
of (\ref{e1.4.1}) corresponds to the solution of the characteristic
equation for the four-diagonal matrix:
\[
         C_{l,l-1} \ =\ (l - 1 -2j) [4(j + 1 - l) + c_+] \ ,
\]
\[
        C_{l,l} \ = \ [l(2j + 1 - l)2c_{+-} + (l - j) c_0] \ ,
\]
\[
        C_{l,l+1} \ = \ (l + 1)(j - l) c_{0-} + (l + 1) c_- \ ,
\]
\begin{equation}
\label{e1.4.6}
        C_{l,l+2} \ =\  - (l + 1)(l + 2) c_{--}.
\end{equation}
where the size of this matrix is $(k + 1) \times (k + 1)$ and
$2j=k$ for (\ref{e1.4.3.1}), (\ref{e1.4.3.2}), and $k \times k$
and $2j=k-1$ for (\ref{e1.4.3.3}), (\ref{e1.4.3.4}), respectively.
Since one of $a$'s can always be placed equal to zero, say, $a_1=0$, the coefficient
$c_{--}$ vanishes and the matrix (\ref{e1.4.6}) becomes tri-diagonal, Jacobi matrix.

In connection to Theorem~\ref{Theorem:1.3} one can prove the following
theorem \cite{Turbiner:1989},

\begin{THEOREM}
\label{Theorem:1.4}
 \  Let us fix all the parameters $e'$s ($a'$s) in
(\ref{e1.4.1}) (or (\ref{e1.4.2})) except for one, e.g. $e_1 (a_1)$.
The first  $(2m + 1)$ eigenvalues of (\ref{e1.4.1}) (or
(\ref{e1.4.2})) form  a $(2m + 1)$-sheeted Riemann surface in
parameter $e_1 (a_1)$. This surface contains four disconnected
pieces: one of them corresponds to $\eta^{(1)} (\eta^{(4)})$
solutions and the others correspond to $\eta^{(3)} (\eta^{(2)})$.
At $m = 2k$ the Riemann subsurface for $\eta^{(1)}$ has $(k + 1)$
sheets and the number of sheets in each of the others is equal to
$k$. At $m = 2k + 1$ the number of sheets for $\eta^{(4)}$ is
equal to $k$ and for $\eta^{(2)}$ each subsurface contains $(k +
1)$ sheets.
\end{THEOREM}

It is worth emphasizing that we cannot find a relation between
the spectral problem for the two-zone potential
\begin{equation}
\label{e1.4.7}
 V = -2 \sum^3_{k=1} {\wp} (z - z_i) \ ,\quad
 \sum^3_{i=1} z_i = 0 \ ,
\end{equation}
(see \cite{Dubrovin:1974}) \footnote{The potential (\ref{e1.4.7})
and the original Lam\'e potential (\ref{e1.4.1}) at $m=2$ are
related via the isospectral deformation.} and the spectral
problem (\ref{e1.2.5}) for $T_2$ with the parameters
(\ref{e1.4.4}) and (\ref{e1.4.5.1}) or  (\ref{e1.4.5.3})  at
$k=1$.  In this case the eigenvalues $\veps$ and also the eigenfunctions
(\ref{e1.4.2}) but not (\ref{e1.4.1}) do not depend on parameters
$c_{--}$.

{\it Comment 3.5}  .\ One can generalize the meaning of an isospectral
deformation by saying we want to study a variety of potentials with
the first several coinciding eigenvalues. They can be named {\it
quasi-isospectral} (see below).

Now let us consider such a quasi-isospectral deformation of
(\ref{e1.4.1}) at $m=2$. It arises from the fact that the addition
of the term $c_{++} J^+_r J^+_r$ to the operator $T_2$ with the
parameters (\ref{e1.4.4}) and (\ref{e1.4.5.1}) or (\ref{e1.4.5.3})
at $k=1$ with appropriate $r$does not change the characteristic matrix
(\ref{e1.4.6}). Making the reduction (\ref{e1.3.22})-(\ref{e1.3.23})
from the equation (\ref{e1.2.5}) to the Schr\"odinger equation
(\ref{e1.2.21}), we obtain
\begin{equation}
\label{e1.4.8}
 V(x) = c_{++} {2 c_{++} \xi^6 - 2 c_{+-} \xi^4 - 2 c_{0-}
 \xi^3 -3 c_{--} \xi^2 \over P^2_4 (\xi)} +  P_2 (\xi) \ ,
\end{equation}
where
\begin{equation}
\label{e1.4.9}
 P_4 (\xi) =c_{++} \xi^4 + c_{+0} \xi^3 + 2 c_{+-} \xi^2 +
 c_{0-} \xi + c_{--} \ ,\
 P_2 (\xi) = - m (m + 1) \xi  + {c_0\over 2}\ .
\end{equation}
and $\xi$ is defined via the equation
\begin{equation}
\label{e1.4.10}
        z\ =\ \int {d\xi \over \sqrt{P_4 (\xi)}} \ ,
\end{equation}
In general, the potential (\ref{e1.4.8}) contains four double
poles in $x$ and does not reduce to (\ref{e1.4.7}).  It is worth
noting that the first five eigenfunctions in the potential
(\ref{e1.4.8}) have the form
\begin{equation}
\label{e1.4.11}
        \Psi (z)\ =\
 \left\{ \begin{array}{c}
 A\xi + B  \\ (\xi - a_i)^{1/2} (\xi - a_j)^{1/2}
 \end{array}  \right\}
 \exp { \left( -c_{++} \int {\xi^3 d\xi\over P_4 (\xi)}  \right) } \ ,
 \ i\ne j,\ i,j =1,2,3 .
\end{equation}
Here $\xi$ is given by (\ref{e1.4.10}). These first five eigenvalues
of the potential (\ref{e1.4.8}) do not depend on the parameters
$c_{--}, c_{++}$.

\newpage

\vskip .5truecm

{\bf Case XII.} {\it $BC_1$ Lam\'e equation (or $BC_1$-quantum elliptic Calogero-Sutherland model)}

\vskip .3truecm

The goal of this Section is to show that the Schr\"odinger equation, which describes the
$BC_1$-quantum elliptic Calogero-Sutherland model, is a quasi-exactly-solvable
Schr\"odinger equation. This remarkable observation was made by Gomez-Ullate et al, \cite{Gomez-Ullate:2001} for the $BC_1$-quantum elliptic Calogero-Sutherland model,
which was later extended by Brihaye-Hartmann \cite{Brihaye:2003}. In this presentation
we mostly follow the paper \cite{Turbiner:2015}.

The potential of the $BC_1$ elliptic model, see (\ref{e1.4.0B}), has the following
interesting property:
in $\tau$-variable (\ref{tau}) it is a simple rational function,
\begin{equation}
\label{e1.4.11a}
   V(z)=(\ka_2\ \wp (2z)\ +\ \ka_3\ \wp (z))|_{\ta=\wp (z)}\ =\  \frac{\ka_2 + 4 \ka_3}{4}\ \ta\ +\
 \frac{\ka_2}{16}\ \frac{12 g_2 \ta^2 +36 g_3 \ta + g_2^2}{4 \ta^3 - g_2 \ta\ -\ g_3}\ ,
\end{equation}
which is a superposition of the linear function with three simple poles situated at the roots
of the Weierstrass function. The kinetic energy represented by the Laplace-Beltrami operator is
the algebraic operator (\ref{e1.4.0M-1}). Hence, the $BC_1$ Hamiltonian (\ref{e1.4.0B}) in $\tau$-variable
\[
    {\mathcal H}_B  \ =\ \bigg(- \half\, \dzz + V(z)\bigg)|_{\ta=\wp(z)}
\]
has the very simple form of an algebraic operator with polynomial coefficient functions
in front of the 1st and 2nd derivatives and with a rational function as the no-derivative term.

{\bf (I).}\ It can be checked that the eigenvalue problem for the Hamiltonian (\ref{e1.4.0B}) has a formal exact solution
\begin{equation}
\label{gs-0}
             \Psi_{0}\ =\ [\ \wp'(x)\ ]^{\mu}\ ,
\end{equation}
for coupling constants
\begin{equation}
\label{par0}
      \ka_2\ =\ 2\mu (\mu-1)\ ,\ \ka_3\ =\ 2\mu (1+2\mu)\ ,
\end{equation}
where $\mu$ is an arbitrary parameter, for which the eigenvalue is
\[
   E_{0}\ =\ 0\ .
\]
Let us parametrize the coupling constants as follows
\begin{equation}
\label{par}
      \ka_2\ =\ 2\mu(\mu-1)\ ,\ \ka_3\ =\ (2n+1+2\mu)(n+2\mu)\ ,
\end{equation}
where $\mu$ and $n$ are parameters. Making the gauge rotation
\[
  h_B\ =\ -2(\Psi_{0})^{-1}\,{\mathcal H}_{\rm B}\,\Psi_{0}\ ,\
  \Psi_{0}\ =\ [\ \wp'(x)\ ]^{\mu}\ ,
\]
see (\ref{gs-0}), we arrive at the algebraic operator in variable $\ta$,
\begin{equation}
\label{htauB}
    h_B(\ta)\ =\ \ (4 \ta^3 - g_2 \ta\ -\ g_3) \pa^2_{\ta} \ +\
    (1+2\mu) (6\ta^2-\frac{g_2}{2}) \pa_{\ta}\ -\ 2n(2n+1+6\mu) \ta\ .
\end{equation}
It can be easily checked that if the parameter $n$ is a non-negative integer, the operator
$h_B(\ta)$ (\ref{htauB}) has the invariant subspace
\[
     {\mathcal P}_{n}\ =\ \langle \ta^{p} \vert \ 0 \le p \le n \rangle\ ,
\]
of dimension
\[
      \dim {\mathcal P}_{n}\ =\ (n+1)\ ,
\]
namely,
\[
  h_B\ :\ {\mathcal P}_{n} \ \rar \ {\mathcal P}_{n}\ .
\]
The operator (\ref{htauB}) can be rewritten in terms of $\mathfrak{sl}(2,\bf{R})$-generators
\[
 h_B\ =\ 4\ { J}^+(n)\, { J}^0(n)\ -\ g_2\ { J}^0(n)\, { J}^-\ - \ g_3\ {\ J}^-\, { J}^-
\]
\begin{equation}
\label{htau-sl2}
 +\ 2\ \big(4n + 1 + 6\mu \big)\, { J}^+(n)\ -\ {g_2}\ \big(n+\frac{1}{2}+\mu \big)\,{ J}^- \ .
\end{equation}

Thus, it is the Hamiltonian of the $\mathfrak{sl}(2,\bf{R})$ quantum top in a constant magnetic field. This representation holds for any value of the parameter $n$. Thus, the algebra $\mathfrak{sl}(2,\bf{R})$ is the hidden algebra of the $BC_1$ elliptic model with arbitrary coupling constants $\ka_{2,3}$ parameterized via (\ref{par}).

If we parameterize in (\ref{wp}) the invariants $g_2, g_3$ as follows
\begin{equation}
\label{wp-inv}
     g_2 = 12 (\la^2 - \de)\ ,\qquad  g_3=4 \la (2\la^2-3\de)\ ,
\end{equation}
where $\la, \de$ are parameters, then $e=-\la$ becomes the root of the $\wp-$Weierstrass function, $\wp'(-\la)=0$, see \cite{Sokolov-Turbiner:2015}. The defining equation for
$$\tilde \wp(z) \equiv \wp(z|\la,\de)+\la\ =\ \wp(z|g_2,g_3)+\la$$ takes the form,
\begin{equation}
\label{ff}
   (\tilde \wp(z))'^2=4 (\tilde \wp(z))^3\ -\ 12 \, \la \tilde \wp(z)^2\ +\ 12 \de \tilde \wp(z)\ .
\end{equation}
Now let us shift the variable $\ta$ in (\ref{htauB}),
\[
    \tilde \ta \ =\ \ta + \la\ ,
\]
and we arrive to the following operator,
\[
   h_B({\tilde \ta})\ =\
\]
\begin{equation}
\label{htauBtilde}
    \ 4({\tilde \ta}^3 - 3 \, \la {\tilde \ta}^2\ +\ 3 \de {\tilde \ta}) \pa^2_{{\tilde \ta}} \ +\
    6 (1+2\mu) ({\tilde \ta}^2-2 \la {\tilde \ta}+\de) \pa_{{\tilde \ta}}\ -\ 2n(2n+1+6\mu) ({\tilde \ta} - \la)\ ,
\end{equation}
c.f. (\ref{htauB}). In space of monomials $\ta^k,\ k=0,1,2,\ldots$ the operator
$h_B({\tilde \ta})$ has the form of tri-diagonal, Jacobi matrix. In terms of $\mathfrak{sl}(2,\bf{R})$-generators it reads
\[
 {\tilde h}_B\ =\ 4\ { J}^+(n)\, { J}^0(n)\ -\ 12\la {J}^0(n)\, {J}^0(n)\ + \ 12\de {\ J}^0(n)\, { J}^-
\]
\begin{equation}
\label{httau-sl2}
 +\ 2\,(4n + 1 + 6\mu)\,{J}^+(n)\ -\ 12\,\la (n+2\mu) J^0(n)\ +\ 6\,\de\,(n+1+ 2\mu)\,{J}^-
 \ +\ \la n(n+2)     \ .
\end{equation}
This is different Lie-algebraic form than (\ref{htau-sl2}) but equivalent. Again the operator ${\tilde h}_B$ has the meaning of the Hamiltonian of the $\mathfrak{sl}(2,\bf{R})$ quantum top in a constant magnetic field.

If $n$ takes integer value, the hidden algebra $\mathfrak{sl}(2,\bf{R})$ (\ref{sl2r}) appears
in a finite-dimensional representation, and the operator (\ref{htauB}) has finite-dimensional invariant
subspace ${\mathcal P}_{n}$, which is the finite-dimensional representation space of the
$\mathfrak{sl}(2,\bf{R})$-algebra. It possesses a number of polynomial eigenfunctions
\[
    \vphi_{n,i}\ =\ P_{n,i}(\ta; \mu)\ , \ i=1,\ldots (n+1)\ .
\]
These polynomials are {\it called $BC_1$ Lam\'e polynomials of the first kind} \cite{Turbiner:2015}.
Those polynomials degenerate either to the Lam\'e polynomials of the first kind at $\mu = 0$, see
(\ref{e1.4.3.1}), or to the Lam\'e polynomials of the fourth kind at $\mu = 1$, see (\ref{e1.4.3.4}).

Few examples are presented below.

(i) For $n=0$ at coupling constants
\[
   \ka_2\ =\ 2\mu(\mu-1)\ ,\ \ka_3\ =\ 2\mu(1+2\mu)\ ,
\]
thus, for the potential
\[
    V_0\ =\ \ 2\mu(\mu-1)\wp (2z)\ +\ 2\mu(1+2\mu)\ \wp (z)\ ,
\]
the single eigenstate is known
\[
    E_{0,1}\ =\ 0\ ,\ P_{0,1} = 1\ .
\]

(ii) For $n=1$ at coupling constants
\[
    \ka_2\ =\ 2\mu\ (\mu-1)\ ,\ \ka_3\ =\ 2(1+2\mu)\,(1+\mu)\ ,
\]
thus, for the potential
\[
    V_1\ =\ \ 2\mu(\mu-1)\wp (2z)\ +\ 2\mu(1+2\mu)\,(1+\mu)\ \wp (z)\ ,
\]
two eigenstates are known
\[
   E_{\mp}\ =\ \pm (1+2\mu) \sqrt {3 g_2}\ ,\ P_{1,\mp}\ =\ 2\ta \mp  \sqrt{\frac{g_2}{3}}\ .
\]
As a function of $g_2$ both eigenvalues are branches of a double-sheeted Riemann surface. Note that
if $\mu=-\frac{1}{2}$ the degeneracy occurs: both eigenvalues coincide, they are equal to zero;
any linear combination of $P_{1,\mp}$ is an eigenfunction. If $g_2=0$ but $\mu \neq -\frac{1}{2}$,
the Jordan cell occurs: both eigenvalues are equal to zero but there exists a single eigenfunction, $P=\ta$.

To summarize, it can be stated that for coupling constants (\ref{par}) at integer $n$,
the Hamiltonian (\ref{e1.4.0B}) has $(n+1)$ eigenfunctions of the form
\begin{equation}
\label{eigen-1}
 \Psi_{n,i}\ =\ P_{n,i}(\ta; \mu)\ [\ \wp'(x)\ ]^{\mu}\ , \quad i=1,\ldots (n+1) \ ,
\end{equation}
where $P_{n,i}(\ta; \mu)$ is a polynomial in $\ta$ of degree $n$.

{\bf (II).}\  It can be checked that the eigenvalue problem for the Hamiltonian
(\ref{e1.4.0B}) has a formal exact solution, other than (\ref{gs-0}),
\begin{equation}
\label{gs-k}
 \Psi_{0,k}\ =\ [\ \wp'(x)\ ]^{\mu}\ \big(\wp(x) - e_k\big)^{\frac{1}{2}-\mu}\ ,
\end{equation}
for coupling constants
\begin{equation}
\label{par0-k}
      \ka_2\ =\ 2\mu(\mu-1)\ ,\ \ka_3\ =\ (1+2\mu)(1-\mu)\ ,
\end{equation}
where $\mu$ is an arbitrary parameter, for which the eigenvalue is
\[
   E_{0,k}\ =\ \frac{(4\mu^2-1)}{2} e_k\ ,\ k=1,2,3\ ,
\]
here $e_k$ is the $k$th root of the Weierstrass function (\ref{wp}).
It implies that for parameters (\ref{par0-k}) the Hamiltonian ${\mathcal H}_{B}$ has one-dimensional invariant subspace.

Let us parametrize the coupling constants $\ka_{2,3}$ as follows
\begin{equation}
\label{par-k}
      \ka_2\ =\ 2\mu(\mu-1)\ ,\ \ka_3\ =\ (n+1+2\mu)(2n+1-\mu) - \mu n \ ,
\end{equation}
(cf. (\ref{par})). Making a gauge rotation of the Hamiltonian (\ref{e1.4.0B}) with subtracted $E_{0,k}$ and changing variable to $\ta$,
\[
h_k\ =\ -2(\Psi_{0,k})^{-1}\,({\mathcal H}_{\rm B}-E_{0,k}) \,\Psi_{0,k}|_{\ta=\wp(z)}\ ,
\]
we arrive at the algebraic operator
\[
    h_k(\ta)\ =\ (\ta - e_k)^{-\half + \mu}\ (h_B(\ta)-2E_{0,k})\ (\ta - e_k)^{\half - \mu}
    \ =\ \
\]
\[
(4 \ta^3 - g_2 \ta\ -\ g_3) \pa^2_{\ta}
\ +\
    \big( 2(5+2\mu) \ta^2 + 4(1-2\mu) e_k (\ta + e_k) - (3-2\mu)\frac{g_2}{2}\big) \pa_{\ta}
\]
\begin{equation}
\label{htauB-k}
\ -\ 2n (2n + 3 + 2\mu) \ta\ ,
\end{equation}
(cf. (\ref{htauB})). It can be checked that if the parameter $n$ takes a non-negative integer value, the operator $h_k(\ta)$ has the invariant subspace ${\mathcal P}_{n}$. Furthermore, the operator (\ref{htauB-k}) can be rewritten in terms of $\mathfrak{sl}(2,\bf{R})$-generators (\ref{sl2r}) for any value of $n$,
cf. (\ref{htau-sl2}),
\[
 h_k\ =\ 4\ { J}^+(n)\ {J}^0(n)\ -\ g_2\ { J}^0(n)\ { J}^-\
 - \ g_3\ { J}^-\ {J}^-
\]
\[
 + \ 2(4n + 3 + 2\mu) {J}^+(n)\ + \ 4(1-2\mu)e_k \big({ J}^0(n) + n\big)
\]
\begin{equation}
\label{htau-sl2-k}
 + \ \bigg( 4(1-2\mu)e_k^2 - (2n + 3 - 2\mu)\frac{g_2}{2} \bigg){ J}^- \ .
\end{equation}
Thus, it is $\mathfrak{sl}(2,\bf{R})$ quantum top in a constant magnetic field. This representation holds for any value of the parameter $n$. Thus, the algebra $\mathfrak{sl}(2,\bf{R})$ is the hidden algebra of the $BC_1$ elliptic model with arbitrary coupling constants $\ka_{2,3}$ parameterized via (\ref{par-k}).

If $n$ takes a non-negative integer value, the hidden algebra $\mathfrak{sl}(2,\bf{R})$ (\ref{sl2r}) appears in a finite-dimensional representation, and the operator (\ref{htauB-k}) has finite-dimensional invariant subspace ${\mathcal P}_{n}$, which is the finite-dimensional representation space of the $\mathfrak{sl}(2,\bf{R})$-algebra. It possesses a number of polynomial eigenfunctions
\[
    \vphi_{n,i}\ =\ P_{n,i}(\ta; \mu, e_k )\ , \ i=1,\ldots\ , (n+1)\ ,\ k=1,2,3)\ .
\]
These polynomials are {\it called $BC_1$ Lam\'e polynomials of the second kind} \cite{Turbiner:2015}. Those polynomials degenerate either to the Lam\'e polynomials of the second kind at $\mu = 0$, see (\ref{e1.4.3.2}),
or to the Lam\'e polynomials of the third kind at $\mu = 1$, see (\ref{e1.4.3.3}), to the Lam\'e polynomials of the fourth kind at $\mu = 1/2$, see (\ref{e1.4.3.4}).

For example, for $n=0$ at couplings (\ref{par-k}),
\[
    E_{0,k}\ =\ \frac{(4\mu^2-1)}{2} e_k\ ,\ P_{0,1} = 1\ .
\]

In general, for $n > 0$, the eigenvalues are branches of $(n+1)$-sheeted Riemann surfaces in $g_2$.

To summarize, it can be stated that for coupling constants (\ref{par-k}) with integer $n=0,1,2,\ldots$,
the Hamiltonian (\ref{e1.4.0B}) has $(n+1)$ eigenfunctions of the form
\begin{equation}
\label{eigen-2}
 \Psi_{n,i;k}\ =\ P_{n,i}(\ta; \mu, e_k)\ \Psi_{0,k}\ , \quad i=1,\ldots (n+1) \ ,\ k=1,2,3\ ,
\end{equation}
where $\Psi_{0,k}$ is given by (\ref{gs-k}).

{\bf (III).}\ It can be verified that the eigenvalue problem for the Hamiltonian (\ref{e1.4.0B})
has one more exact solution, other than (\ref{gs-0}) or (\ref{gs-k}),
\begin{equation}
\label{gs-ij}
 \Psi_{0,\tilde k}\ =\ [\ \wp'(x)\ ]^{\nu}\  [(\wp(x) - e_i) (\wp(x) - e_j)]^{\frac{1}{2}-\nu}\ ,
\end{equation}
where $\tilde k$ is complement to $(i,j)$, for coupling constants
\begin{equation}
\label{par0-ij}
      \ka_2\ =\ 2\nu(\nu-1)\ ,\ \ka_3\ =\ \nu(3-2\nu)\ ,
\end{equation}
where $\nu$ is an arbitrary parameter, for which the eigenvalue is
\[
   E_{0,\tilde k}\ =\ \frac{(1-2\nu)(3-2\nu)}{2} e_{\tilde k}\ ,
\]
here $e_{\tilde k}$ is the $\tilde k$th root of the Weierstrass function (\ref{wp}).
It implies that for parameters (\ref{par0-ij}) the Hamiltonian ${\mathcal H}_{B}$ has
one-dimensional invariant subspace. Note that if in (\ref{gs-ij}) we replace $\nu=1-\mu$
we arrive at the function (\ref{gs-k}).

Let us parametrize the coupling constants $\ka_{2,3}$ as follows
\begin{equation}
\label{par-ij}
      \ka_2\ =\ 2\nu(\nu-1)\ ,\ \ka_3\ =\ n(2n + 5)- 2\nu n + \nu (3-2\nu)\ ,
\end{equation}
(cf. (\ref{par}) and (\ref{par-k})), where $\nu$ and $n$ are parameters.

Making a gauge rotation of the Hamiltonian (\ref{e1.4.0B}) with subtracted $E_{0,\tilde k}$,
\[
h_{\tilde k}\ =\ -2(\Psi_{0,\tilde k})^{-1}\,({\mathcal H}_{\rm B}-E_{0,\tilde k}) \,\Psi_{0,\tilde k}
\]
and changing variable $z$ to $\ta$, we arrive at the algebraic operator
\[
    h_{\tilde k}(\ta)\ =\ [(\ta - e_i)(\ta - e_j)]^{-\half + \nu}\
    \big(h^{(e)}(\ta)-2E_{0,\tilde k}\big)\ [(\ta - e_i)(\ta - e_j)]^{\half - \nu} \ =\
\]
\[
    (4 \ta^3 - g_2 \ta\ -\ g_3) \pa^2_{\ta}\ +\ 2\big((7-2\nu) \ta^2 + 2(2\nu-1)
      e_{\tilde k} (\ta + e_{\tilde k}) - (5+2\nu)\frac{g_2}{4}\big) \pa_{\ta}
\]
\begin{equation}
\label{htauB-ij}
\ -\ 2 n (2n + (5-2\nu)) \ta\ ,
\end{equation}
(cf. (\ref{htauB})), where $e_{\tilde k}$ is ${\tilde k}$th root of the Weierstrass function,
see (\ref{wp}).

It can be verified that if the parameter $n$ takes a non-negative integer value, the operator
$h_{\tilde k}(\ta)$ has the invariant subspace ${\mathcal P}_{n}$. Furthermore, the operator
(\ref{htauB-ij}) can be rewritten in terms of $\mathfrak{sl}(2,\bf{R})$-generators (\ref{sl2r})
for any value of $n$, cf. (\ref{htau-sl2}),
\[
 h_{\tilde k}\ =\ 4\ { J}^+(n)\ { J}^0(n) -\ g_2 { J}^0(n)\ { J}^- - \ g_3\ { J}^-\ { J}^-
\]
\[
 + \  2(4n + 5 - 2\nu)\ { J}^+(n)\
 + \ 4(2\nu-1)e_k ({ J}^0(n) + n)
\]
\begin{equation}
\label{htau-sl2-ij}
 + \ 2 \big(2(2\nu-1)e_k^2 - (2n + 5 + 2\nu)\frac{g_2}{4}\big){ J}^- \ .
\end{equation}
Thus, it is $\mathfrak{sl}(2,\bf{R})$ quantum top in constant magnetic field.
This representation holds for any value of the parameter $n$. Thus, the algebra
$\mathfrak{sl}(2,\bf{R})$ is the hidden algebra of the $BC_1$ elliptic model with
arbitrary coupling constants $\ka_{2,3}$ parameterized via
(\ref{par-ij}).

If $n$ takes a non-negative integer value, the hidden algebra $\mathfrak{sl}(2,\bf{R})$ (\ref{sl2r})
appears in a finite-dimensional representation, and the operator (\ref{htauB-ij}) has a
finite-dimensional invariant subspace ${\mathcal P}_{n}$, which is the finite-dimensional
representation space of the $\mathfrak{sl}(2,\bf{R})$-algebra. It possesses a number of
polynomial eigenfunctions
\[
    \vphi_{n,i}\ =\ {\tilde P}_{n,i}(\ta; \nu, e_{\tilde k} )\ , \ i=1,\ldots\ , (n+1)\ ,\ \tilde k = 1,2,3)\ .
\]
These polynomials are {\it called $BC_1$ Lam\'e polynomials of the third kind} \cite{Turbiner:2015}.
The polynomials degenerate either to the Lam\'e polynomials of the third kind at $\nu = 0$, see
(\ref{e1.4.3.3}), or to the Lam\'e polynomials of the second kind at $\nu = 1$, see (\ref{e1.4.3.2}),
or to the Lam\'e polynomials of the fourth kind at $\nu = 1/2$, see (\ref{e1.4.3.4}).
It is important to mention that the $BC_1$ Lam\'e polynomials of the second and third kind are related,
\[
{P}_{n,i}(\ta; \mu, e_{\tilde k} ) =  {\tilde P}_{n,i}(\ta; 1 - \mu, e_{\tilde k} )\ .
\]

For example, for $n=0$ at couplings (\ref{par-k}),
\[
    E_{0,\tilde k}\ =\ \frac{(1-2\nu)(3-2\nu)}{2} e_{\tilde k}\ ,\ P_{0,1} = 1\ .
\]

In general, for $n > 0$, the eigenvalues are branches of $(n+1)$-sheeted Riemann surfaces in $g_2$.

To summarize, it can be stated that for coupling constants (\ref{par-k}) at integer $n=0,1,2,\ldots$,
the Hamiltonian (\ref{e1.4.0B}) has $(n+1)$ eigenfunctions of the form
\begin{equation}
\label{eigen-3}
 \Psi_{n,i;e_{\tilde k}}\ =\ {\tilde P}_{n,i}(\ta; \nu, e_{\tilde k} )\ \Psi_{0,e_{\tilde k} }\ ,
 \quad i=1,\ldots (n+1) \ ,\ e_{\tilde k} = 1,2,3\ ,
\end{equation}
where $\Psi_{0,e_{\tilde k} }$ is given by (\ref{gs-ij}).

The most general one-dimensional elliptic potential which appears in literature is a superposition
of four Weierstrass functions,
\[
    V(z)\ =\ \mu_0 \wp (z) + \mu_1 \wp (z+\om_1)  + \mu_2 \wp (z+\om_2) + \mu_3 \wp (z+\om_3)\ ,
\]
where $\om_{1,2,3}$ are half-periods and $\mu_{0, 1, 2, 3}$ are coupling constants. The quantum model
characterized by this potential is called $BC_1$ elliptic Inozemtsev model \cite{Inozemtsev:1989}.
This is a generalization of $BC_1$-quantum elliptic Calogero-Sutherland model.
Quasi-exact-solvability of this model was shown in \cite{Gomez-Ullate:2001,Takemura:2002},
see also \cite{Brihaye:2003}. Since the analysis of this model is very much similar to one presented above, we will not describe this here referring the reader to the original papers.


\subsection{Exactly-solvable operators and classical orthogonal
polynomials}

For any exactly-solvable operator of the second order $E_2$ (\ref{ES-2})
\[
 E_2\ =\   c_{00} J^0  J^0  + c_{0-} J^0 J^- +
 c_{--} J^-  J^-  + c_0 J^0  + c_- J^-\ ,
\]
where
\[
      J^0=x d_x\ ,\ J^-=d_x\ ,
\]
the $J^{0,-} \equiv J^{0,-}_0$ span the Borel subalgebra,
$\mathfrak{b}_2 \subset \mathfrak{sl}(2,\bf{R})$-generators, the spectra of
polynomial eigenfunctions is the second degree polynomial in quantum number $n$,
\begin{equation}
\label{e1.5.1}
 {\ep}_n =  \ c_{00} n^2 \ + \ c_{0} n \ ,\ n=0,1,2\ldots\ ,
\end{equation}
c.f. (\ref{e1.2.21}).
Here $n$ is a degree of the $n$th polynomial eigenfunction.
It can be shown that the Hermite, Laguerre, Legendre and
Jacobi operators give rise to four families of classical orthogonal
polynomials as eigenfunctions are exactly-solvable operators, respectively.
Hence, the eigenvalue problem (\ref{e1.2.5}) for exactly-solvable operator $E_2$
(see (\ref{ES-2})) leads to the equations having the Hermite, Laguerre, Legendre and
Jacobi polynomials as eigenfunctions \cite{Turbiner:1988f,Turbiner:1988z}.
This is shown below.  In the definition of these polynomials we follow the handbook by Bateman--Erd\'{e}lyi \cite{Bateman:1953} as well as by Koekoek--Swarttouw
\cite{Koekoek:1994}.

{\it 1. Hermite polynomials}

The Hermite polynomials $H_{2n+p}(x),\ n=0,1,2,\ldots , \ p=0,1$
are the polynomial eigenfunctions of the operator
\begin{equation}
\label{e1.5.2}
 E^{(Hermite)} (x)\ =\ -d_x^2 + 2x d_x\ ,
 \quad \ep_{2n+p}= 4n+2p\ ,
\end{equation}
hence, linear in $n$ spectra, c.f. (\ref{e1.5.1}). They can be rewritten in terms
of the generators (\ref{sl2r}) following the Lemma 2.1
\begin{equation}
\label{e1.5.3}
 E^{(Hermite)} = \ -J^- J^-\ +\ 2 J^0  \ .
\end{equation}
Although the operator (\ref{e1.5.2}) can be factorized as a differential operator,
\[
   E^{(Hermite)} (x)  \ =\ -d_x (d_x - 2x) \ ,
\]
it can not be factorized in the $\mathfrak{sl}(2,\bf{R})$-algebra representation,
for a discussion see above eqs.(\ref{e1.4.0A})-(\ref{e1.4.0B}).

Let us note that the parameter $p$ is related to the parity
operator, those eigenvalues are $P=(-1)^p$. The polynomial $H_{2n+p}$ is
characterized by the definite parity $P$ being either even, $P=+1$, or odd, $P=-1$
with respect to reflection, $x \rar -x$. Thus, for a polynomial $H_{2n+p}(x)$ there
exists a remarkable representation
\begin{equation}
\label{e1.5.3h}
 H_{2n+p} (x) = x^p L_n^{(-1/2+p)} (x^2)\ ,
\end{equation}
where $L_n^{(-1/2+p)}$ is the associated Laguerre polynomial (see
below). Here, the factor $x^p$ encodes the information about what representation of reflection
group we are considering, even or odd. The remaining factor is already reflection-invariant.

Making a gauge transformation $x^{-p} E^{(Hermite)} (x,
d_x)x^p$ and then changing the variable $x^2=y$, we arrive at the
operator having the Laguerre polynomial $L_n^{(-1/2+p)}(y)$ as the eigenfunction
\begin{equation}
\label{e1.5.4}
 \bar{E} (y) = -4y d_y^2 + 2(2y-1-2p) d_y + 2p\ ,
\end{equation}
where $p=0,1$. It can be again rewritten in terms of the generators
(\ref{sl2r}) but in variable $y$,
\begin{equation}
\label{e1.5.5}
 \bar{E}\ = \ -4J^0 J^- \ +\ 4 J^0  \ - \ 2(1+2p) J^- + 2p \ ,
\end{equation}
(cf. (\ref{e1.5.3})). Of course, these two representations
(\ref{e1.5.3}) and (\ref{e1.5.5}) are equivalent, however, a
quasi-exactly-solvable generalization can be reached for the
second representation {\it only} (see Cases VI and VII in
Section \ref{qesp}). If the operator (\ref{e1.5.3}) possesses
reflection symmetry $x \rar -x$, the operator (\ref{e1.5.5}) has
no symmetries. Thus, the symmetry of the original operator
$ E^{(Hermite)}$ (\ref{e1.5.2}) is hidden in the variable $y$.
This property is quite general -- usually, a quasi-exactly solvable
generalization exists for operators without a discrete symmetry.

Note that $\bar{E} (y)$ admits factorization as a differential operator
\[
    -4y d_y^2 + 2(2y-1-2p) d_y     = \ -(4y d_y - 2(2y-1-2p)) d_y\ ,
\]
while $(\bar{E} + 2(1+p))$ admits factorization in the algebra
$\mathfrak{sl}(2,\bf{R})$,
\[
    -4J^0 J^- \ +\ 4 J^0  \ - \ 2(1+2p) J^- + 2(1+2p)\ =
    \ -2 (2 J^0 + (1+2p)) (J^- - 1) \ .
\]

{\it 2. Laguerre polynomials}

The associated Laguerre polynomials $L_n^{(a)} (x)$ occur as the
polynomial eigenfunctions of the generalized Laguerre operator
\begin{equation}
\label{e1.5.6}
 E^{(Laguerre)} (x)\ =\ -xd_x^2 + (x-a-1) d_x\ ,\quad \ep_n = n\ ,
\end{equation}
where $a$ is any real, hence, with linear in $n$ spectra, c.f. (\ref{e1.5.1}).
Of course, the operator (\ref{e1.5.6}) can be rewritten as
\begin{equation}
\label{e1.5.7}
 E^{(Laguerre)} = \ - J^0 J^- \ +\  J^0 \ - \ (a+1) J^-\ .
\end{equation}
If $a=-1/2+p$ the operator (\ref{e1.5.7}) coincides with $\bar{E}$
in (\ref{e1.5.4}). Evidently, the operator $E^{(Laguerre)}$ admits factorization
as differential operator as well as in the algebra $\mathfrak{sl}(2,\bf{R})$,
\begin{equation}
\label{e1.5.7-1}
 E^{(Laguerre)}\ = \ - J^0 J^- \ +\  J^0 \ - \ (a+1) J^- + (a+1)\ =
 \ -(J^0 + (a+1)) (J^- - 1) \ .
\end{equation}

{\it 3. Legendre polynomials}

The Legendre polynomials $P_{2n+p} (x),\ n=0,1,2,\ldots , \ p=0,1$
are the polynomial eigenfunctions of the operator
\begin{equation}
\label{e1.5.8}
 E^{(Legendre)} (x, d_x) = (1-x^2)d_x^2 -2x d_x\ ,\quad \ep_{2n+p}=
 -(2n+p)(2n+p+1)\ ,
\end{equation}
or, in $\mathfrak{sl}(2,\bf{R})$-Lie-algebraic form
\begin{equation}
\label{e1.5.9}
 E^{(Legendre)}  = \ - J^0 J^0 \ +\ J^- J^-\ - \ J^0 \ .
\end{equation}
Analogously to the Hermite polynomials, the Legendre polynomials
possess the definite parity and can be represented as
\begin{equation}
\label{e1.5.9p}
 P_{2n+p} (x) = x^p P_n^{(-1/2+p,0)} (x^2)\ ,
\end{equation}
where $P_n^{(a,b)}$ is the Jacobi polynomial (see below).
Making a gauge transformation $x^{-p} E^{(Legendre)} (x, d_x)x^p$ and then
changing the variable $x^2=y$, we arrive at the operator having the Jacobi
polynomial $P_n^{(-1/2+p,0)}(y)$ as the eigenfunction
\begin{equation}
\label{e1.5.10}
 \bar{E}^{(Legendre)} (y) = 4y(1-y)d_y^2 +2[ 1+2p -(3+2p)y] d_y\ ,\ p=0,1\ ,
\end{equation}
and, correspondingly,
\begin{equation}
\label{e1.5.11}
 \bar{E}^{(Legendre)} = \ -4J^+J^-\ +\ 4J^0 J^- \ - \ 2 (3+2p) J^0 \ +\ 2 (1+2p) J^- \ .
\end{equation}
The operator $\bar{E}$ admits factorization as a differential operator as well as
in the algebra $\mathfrak{sl}(2,\bf{R})$,
\[
   \bar{E}^{(Legendre)}\ = \ -4J^0J^0\ +\ 4J^0 J^- \ - \ 2 (1+2p) J^0 \ +\ 2 (1+2p) J^-\ =
\]
\begin{equation}
\label{e1.5.11-1}
 -2 (2J^0 + (1+2p)) (J^0 - J^-) \ .
\end{equation}

{\it 4. Jacobi polynomials}

The Jacobi polynomials $P_n^{(a,b)}$ appear as the polynomial
eigenfunctions of the Jacobi equation taken either in the symmetric form
corresponding to the operator
\begin{equation}
\label{e1.5.12}
 E^{(Jacobi)} (x, d_x) = (1-x^2)d_x^2 +
   [b - a - (a+b+2)x] d_x\ , \quad \ep_n= -n(n+a+b+1) \ ,
\end{equation}
with the $\mathfrak{sl}(2,\bf{R})$-Lie-algebraic representation
\begin{equation}
\label{e1.5.13}
 E^{(Jacobi)}= \ -J^0 J^0 + J^- J^- - (1+a+b) J^0 + (b-a) J^- \ ,
\end{equation}
or in the asymmetric form  (see e.g. the book by Murphy
\cite{Murphy:1960} or by Bateman--Erd\'{e}lyi \cite{Bateman:1953})
\begin{equation}
\label{e1.5.14}
 \bar{E}^{(Jacobi)} (x, d_x) = x(1-x)d_x^2 + [1+a-(a+b+2)x] d_x\ ,
\end{equation}
corresponding to
\begin{equation}
\label{e1.5.15}
 \bar{E}^{(Jacobi)} = \ -J^0 J^0  + J^0 J^- - (1+a+b) J^0 + (a+1) J^-\ .
\end{equation}

It is not surprising that under a special choice of a general polynomial
element of the universal enveloping algebra of Borel subalgebra
$\mathfrak{b}_2 \subset \mathfrak{sl}(2,\mathcal{C})$ one can find
Lie-algebraic forms of {\it all} known fourth-, sixth-, and
eighth-order differential equations giving rise to infinite
sequences of orthogonal polynomials (see e.g. the paper by Littlejohn
\cite{Littlejohn:1986} and other papers in this volume).

\vskip 1.truecm

In \cite{Olver:1993} it was given the complete description of the second-order polynomial elements of $U_{\mathfrak{sl}_2({\bf R})}$ (\ref{QES-2}) at $c_{++}=c_{+0}=0$
in the representation (\ref{sl2r}) leading after transformation
(\ref{e1.3.22})-(\ref{e1.3.23}) to the square-integrable eigenfunctions
of the Sturm-Liouville problem (\ref{e1.2.21}). Similar classification
for non-vanishing $c_{++} \neq 0$ and/or $c_{+0} \neq 0$ is still missing.

Consequently, for second-order ordinary differential equation (\ref{ES-2eq})
a combination of Theorems 2.1, 2.2 leads to the statement that a general
solution of the problem of classification of equations, possessing the finite
number of orthogonal polynomial solutions, is related to a chance to rewrite
these equations in terms of ${\mathfrak{sl}_2({\bf R})}$-generators in
finite-dimensional representation.

\clearpage
%
%
%
%
\newpage
\renewcommand{\thesection}{2.{\arabic{section}}}
\setcounter{section}{0}

\begin{center}
{\bf \large Chapter\ 2.}\\[20pt]
 {\Large Quasi-Exactly-Solvable Anharmonic Oscillator }
\end{center}
\addcontentsline{toc}{toc}{Chapter 2. Quasi-Exactly-Solvable
Anharmonic Oscillator \hfill}
\label{BT}
\vskip 1cm


\renewcommand{\theequation}{2.{\arabic{equation}}}
 \setcounter{equation}{0}

 \renewcommand{\thefigure}{2.{\arabic{figure}}}
 \setcounter{figure}{0}

 \renewcommand{\thetable}{2.{\arabic{table}}}
 \setcounter{table}{0}

In this Chapter we present a detailed description of the unique
polynomial one-dimensional potential, which has the property
of quasi-exact-solvability, see Case VI in Chapter 1:
\begin{equation}
\label{e2.1-1}
   V(x)\ =\ a^2 x^6 + 2 a b x^4 + [b^2 - a (4 n + 2k + 3)]x^2
   - b(1 + 2k)\ ,
\end{equation}
where the parameter $k=0,1$ is introduced for convenience, see below.
If $a, b$ and $n$ are real parameters we have the general sextic, even polynomial
potential with infinite discrete spectra. The property of the Quasi-Exact-Solvability
(QES) takes place when two conditions are fulfilled: $n$ is non-negative
integer and $a$ is a non-negative number, $a \geq 0$. In this case $(n+1)$
eigenstates of parity $(-1)^k$ with square-integrable eigenfunctions
can be found algebraically; we say that those states form the algebraic
sector of eigenstates or, saying differently, those eigenstates are algebraic.
In $b$-space the algebraic eigenvalues (eigenfunctions) form $(n+1)$ sheeted-Riemann
surface with the Landau-Zener square-root branch points. These branch points have
the meaning of the points of level crossing.  If $a < 0$, the algebraic eigenstates
are absent, although the Schr\"odinger equation has $(n+1)$ algebraic,
non-square-integrable eigenfunctions.

A quantum system which is described by this potential can be
called the {\it anharmonic sextic quasi-exactly-solvable oscillator}.
For simplicity we call it the {\it sextic QES oscillator}.
Since the potential (\ref{e2.1-1}) is even, $V(-x)=V(x)$, the eigenfunctions
are characterized by the definite parity: they are either even or odd with
respect to reflection $(x \rar -x)$: $\Psi(-x)=\pm \Psi(x)$. The parameter
$k$ takes value $k=0$ for even eigenstates and $k=1$ for odd eigenstates,
correspondingly. From physical point of view the potential (\ref{e2.1-1})
with the QES property does not seem very much different from generic sextic,
even potential. For arbitrary $n$ in (\ref{e2.1-1}) the even (odd) eigenvalues
in $b$-space are branches of infinitely-sheeted Riemann surface with square-root
branch points (for a discussion see \cite{Bender:1969,Bender:1973,Turbiner:1988z,Turbiner:1988qes,Eremenko:2009}).
If $n$ takes an integer value, $(n+1)$-sheeted Riemann surface in $b$-space of
even (odd) eigenstates splits away from the infinite-sheeted Riemann surface.
It implies that zeroing of the residues of the branch points connecting the sheets
of the $(n+1)$-sheeted Riemann surface with ones from the infinite-sheeted Riemann
surface. Usually, there are infinitely-many such branch points. Thus, if $n$ takes
an integer value, infinitely-many conditions are fulfilled!
In the same time for arbitrary $n$ any eigenfunction is an entire function in $x$.
It has infinitely-many simple zeroes at complex $x$ and finitely-many at real $x$
(for a discussion see \cite{Eremenko:2008}).
If $n$ takes an integer value, for any algebraic eigenfunction infinitely-many zeroes
disappear and the $(n+1)$ simple zeroes remain. Classically, it is not seen anything
specific when $n$ takes an integer value (for a discussion see \cite{Turbiner:1988z,Turbiner:1988qes,Eremenko:2008}).

The QES property gives us a unique chance to see exact
solutions of non-trivial quantum problems. One of the important features of
these solutions is a possibility to test results obtained in some approximation
methods.
To the best of our knowledge the sextic potential (\ref{e2.1-1}) with the
QES property had appeared in an explicit form for the first time in
the paper by Peter Leach \cite{Leach:1984}. It is worth mentioning
that there is no physical system with the QES property among the
most popular quartic anharmonic oscillators
\[
   V(x)\ =\ m^2 x^2 + g x^4 \ ,
\]
for a discussion see \cite{Bender:1998}. Hence, the limit $a$ tends to zero in (\ref{e2.1-1}) leads to destruction of the QES property.  In the presentation of (\ref{e2.1-1}) we mainly
follow \cite{Turbiner:1987,Bender-Turbiner:1993,Turbiner:1992}.

\section{QES sextic potential}

The first $(n+1)$ even or odd eigenfunctions in (\ref{e2.1-1}) are
of the form
\begin{equation}
\label{e2.2}
 \Psi^{(qes)}_{n,i}\ =\
 x^k P_n^{(i)} (x^2) e^{-\frac{b}{2}x^2-\frac{a}{4}x^4}\ ,\ i=0,1,\ldots n\ ,
\end{equation}
where $P_n$ is a polynomial of the degree $n$, it is an element of $(n+1)$-dimensional
representation space of the $sl(2)$-algebra.  These eigenfunctions and corresponding
eigenvalues can be found algebraically, by solving a system of $(n+1)$ linear homogeneous
equations. Remaining eigenfunctions can not be written in a form of polynomial multiplied
by the exponential, they are of non-algebraic nature. They are elements of infinite-dimensional representation space of the $\mathfrak{sl}(2,\bf{R})$-algebra.

\begin{figure}[tb]
\begin{center}
     {\includegraphics*[width=4.2in]{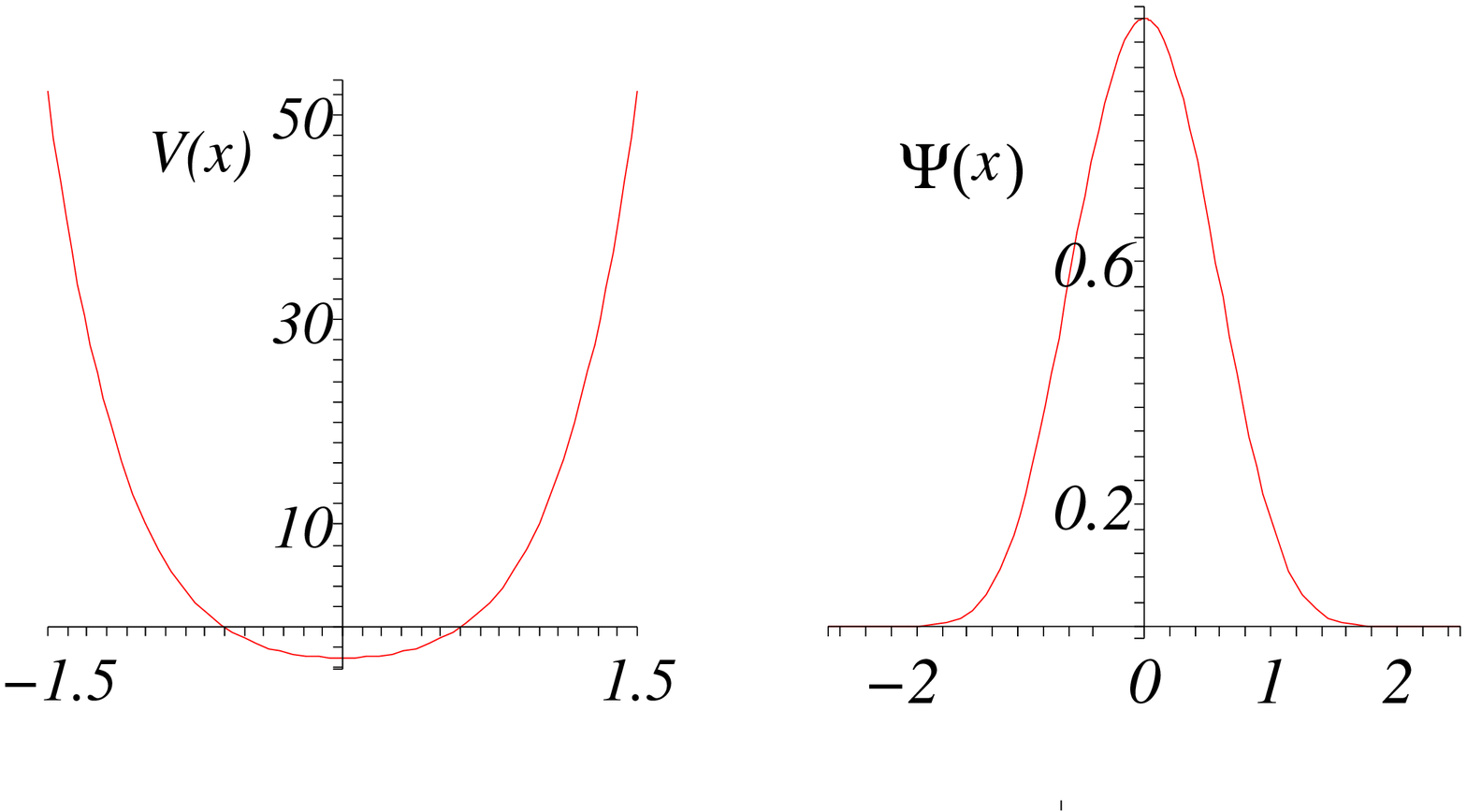}}
     \caption{The potential (\ref{e2.1-1}) at $a=1, n=k=0$ and $b=3$ and the ground state eigenfunction $\Psi = \Psi_0$ with zero eigenvalue, $E_0=0$.}
         \label{fig:2.1.1}
\end{center}
\end{figure}

\begin{figure}[tb]
\begin{center}
     {\includegraphics*[width=4.2in]{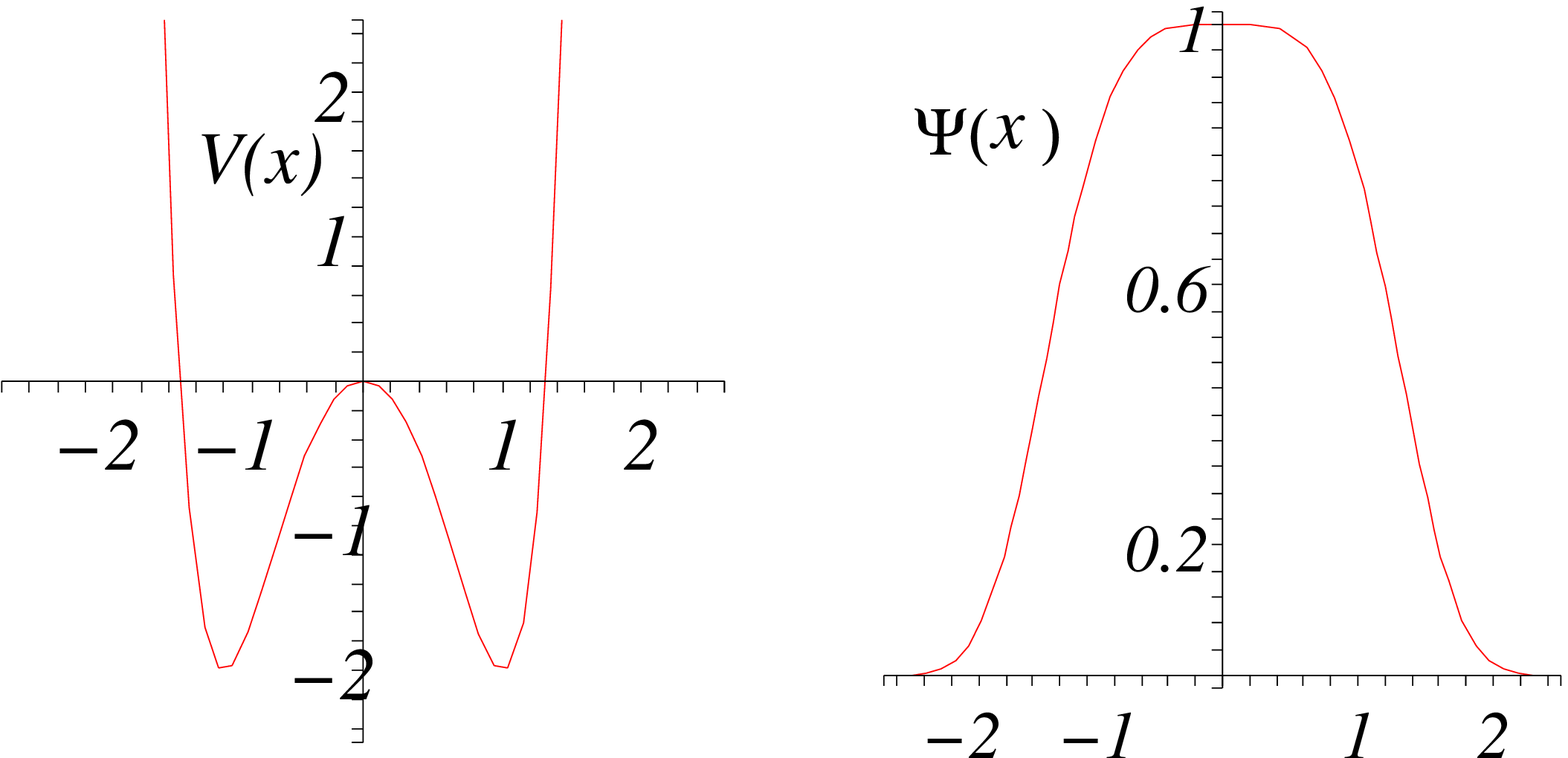}}
     \caption{The potential (\ref{e2.1-1}) at $a=1, n=k=0$ and $b=0$ and the ground state eigenfunction $\Psi = \Psi_0$ with zero eigenvalue, $E_0=0$.}
         \label{fig:2.1.2}
\end{center}
\end{figure}

\begin{figure}[tb]
\begin{center}
     {\includegraphics*[width=4.2in]{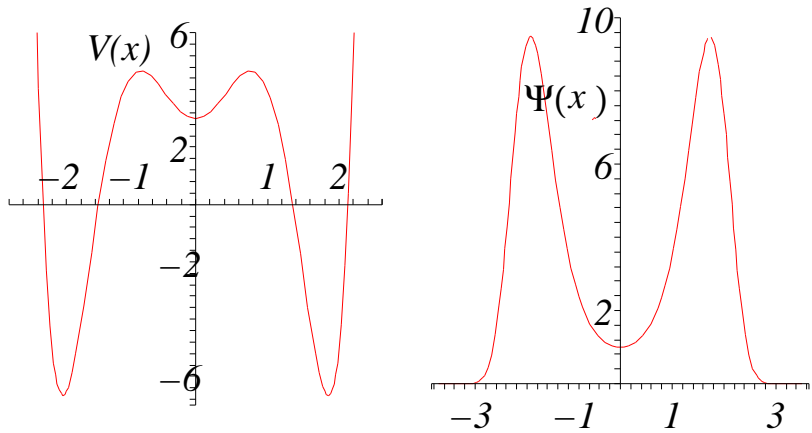}}
     \caption{The potential (\ref{e2.1-1}) at $a=1, n=k=0$ and $b=-3$ and the ground state eigenfunction $\Psi = \Psi_0$ with zero eigenvalue, $E_0=0$.}
         \label{fig:2.1.3}
\end{center}
\end{figure}

A pattern of the potential curve (\ref{e2.1-1}) depends on a
relation between parameters $a,b$ and corresponds to one-, two- and
three-minima one (for illustration, see
Fig.~\ref{fig:2.1.1}-\ref{fig:2.1.3}), respectively. It is worth
mentioning that at $b=0$ and $n=k=0$ the potential (\ref{e2.1-1})
looks like a standard double-well potential but the ground state
eigenfunction is characterized by a single peak (!). This peak appears
at a position of unstable equilibrium (see Fig.~\ref{fig:2.1.2}).
It looks as a striking contradiction to a straightforward (naive)
intuition based on (semi)classical picture. Similar contradiction
occurs for $b=-3$ (see Fig.~\ref{fig:2.1.3}): a triple-well
potential with the two-peaked ground state eigenfunction. These
results demonstrate that a semi-classical treatment of these cases
is not yet applicable: the corresponding wells are not deep enough and
the barriers are not large enough.

Let us consider the Hamiltonian corresponding to the potential
(\ref{e2.1-1})
\begin{equation}
\label{e2.3}
 \h \ =\  -\dxx + a^2 x^6 + 2 a b x^4 +
 [b^2 - a (4 n + 2k + 3)]x^2 - b(1 + 2k)\ ,
\end{equation}
make a gauge rotation with a change of variable,
\[
 h \ =\  (\Psi^{(qes)}_0)^{(-1)}{\h} \Psi^{(qes)}_0\vert_{y = x^2}
\]
\begin{equation}
\label{e2.4}
 =\ -4y\dyy\ +\ 2(2ay^2 + 2by - 1 - 2k)\dy\ -\ 4 a n y \ .
\end{equation}
The first $(n+1)$-eigenfunctions of the eigenvalue problem for the operator
$h$,
\begin{equation}
\label{e2.5}
  h \varphi\ =\ \ep \varphi \ ,
\end{equation}
are polynomials of degree $n$:\ $\varphi=P_n (y)$ (cf.
(\ref{e2.2})).

The gauge rotated Hamiltonian $h$ (\ref{e2.4}) can be immediately
rewritten as a quadratic combination in the
$\mathfrak{sl}(2,\bf{R})$-generators (1.2.3) written in variable $y$,
\[
J^+_n = y^2 \dy - n y \ ,
\]
\begin{equation*}
 J^0_n = y \dy - \frac{n}{2} \  ,
\end{equation*}
\[
J^-_n = \dy \ ,
\]
as follows \cite{Turbiner:1988qes}
\begin{equation}
\label{e2.7}
 h\ =\ -2 \{J^0_n, J^-_n\} \ +\ 4a J^+_n\ +\ 4b J^0_n\ -\ 2(n+2+2k) J^-_n\ +\ 2 b n \ .
\end{equation}
This representation immediately reveals the meaning of the parameter $n$:
$j=\frac{n}{2}$ is the spin of the representation (\ref{sl2r}) of the algebra $\mathfrak{sl}(2,\bf{R})$.
\textit{
It shows that the one-dimensional quantum dynamics in a generic sextic, even
potential is equivalent a quantum top with spin $j=\frac{n}{2}$ in a constant
magnetic field with a constraint
\[
 \{J^+_n,J^-_n\} - 2 J^0_n J^0_n\,+\,n\, \big (\frac{n}{2}+1 \big )=0\ ,
\]
cf. 
(1.2.4), where $\{ , \}$ denotes anti-commutator. If the spin $j$ is integer
or half-integer, it occurs the irreducible finite-dimensional representation.}
In this case the generators (\ref{sl2r}) have a common invariant subspace
${\mathcal P}_n$ (1.1.5), which is a representation space of the finite-dimensional
representation. Correspondingly, the operator (\ref{e2.4}) has this space as a
finite-dimensional invariant subspace. Certainly, the number of those {\it algebraic}
eigenfunctions (whose can be found algebraically) is nothing but the dimension of
the irreducible finite-dimensional representation of the algebra
(\ref{sl2r}).

\vskip 1cm

\vskip 1cm

\vskip 1cm

One can find explicitly the algebraic eigenstates of (\ref{e2.3}) for $n=0,1$ at $a > 0$.

\renewcommand{\theequation}{2.8.{\arabic{equation}}}
\setcounter{equation}{0}
\newpage
\begin{itemize}
    \item $n=0$
\begin{equation}
\label{e2.8.1}
   V(x)\ =\ a^2 x^6\ +\ 2 a b x^4\ +\ [b^2 - a (2k + 3)]x^2
   \ -\ b(1 + 2k)\ ,
\end{equation}

\[
  E_0^{(k)}\ =\ 0\ ,\ \vphi_0^{(k)}\ =\ 1\ ,
\]

\[
    \Psi^{(k)}_0\ =\ x^k\, e^{-\frac{b}{2}x^2-\frac{a}{4}x^4}\ ,
\]

where the algebraic states are the ground state of the positive parity $(k=0)$ and of the negative parity $(k=1)$.

    \item $n=1$
\begin{equation}
\label{e2.8.2}
   V(x)\ =\ a^2 x^6 + 2 a b x^4 + [b^2 - a (2k + 7)]x^2 - b(1 + 2k)\ .
\end{equation}

\[
   E_{\pm}^{(k)}\ =\ 4(1+k) b \pm 2 \big(b^2 + 2(1+2k)a\big)^{\half}\ ,
\]
\[
   \vphi_{\pm}^{(k)}\ =\ x^k\, \bigg(2 a x^2 + b \mp \big(b^2 + 2(1+2k)a\big)^{\half}\bigg)\ ,
\]
\[
   \Psi_{\pm}^{(k)}\ =\ x^k\, \bigg(2 a x^2 + b \mp \big(b^2 + 2(1+2k)a\big)^{\half}\bigg)
     \, e^{-\frac{b}{2}x^2-\frac{a}{4}x^4}\ ,
\]
where the algebraic states are the ground/second-excited states of the positive parity $(k=0)$ and of the negative parity $(k=1)$.

\end{itemize}

\renewcommand{\theequation}{2.{\arabic{equation}}}
\setcounter{equation}{9}

\section{``Phase transitions" in sextic QES potential}

\begin{figure}[tb]
\begin{center}
     {\includegraphics*[width=4.2in]{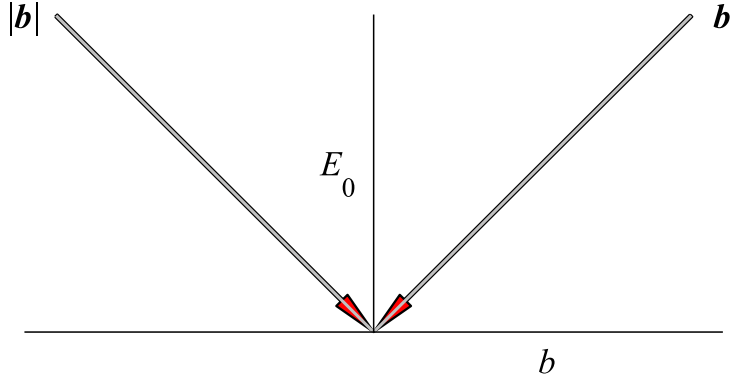}}
     \caption{Harmonic oscillator: the ground state energy $E_0$ {\it vs.} $b$.}
         \label{fig:2.1.4}
\end{center}
\end{figure}

Let us consider the harmonic oscillator
\[
          V(x) \ =\ b^2 x^2\ , \ x \in {\bf R}\ .
\]
For any real $b \neq 0$ it has the infinite discrete spectra, which has a certain non-analytic behavior at $b=0$ as a function of $b$. For instance, the ground state is given by
\[
      E_0\ =\ |b|\ ,\ \Psi_0\ =\ e^{-|b|\frac{x^2}{2}}\ ,
\]
see Fig. \ref{fig:2.1.4}. The ground state energy as a function of $b$ is continuous, but its derivative $\frac{dE_0}{db}$ has a discontinuity at $b=0$: it jumps from $+1$ at positive $b$ to $-1$ at negative $b$. Similar behavior occurs for {\it any} energy level. Of course, if the harmonic oscillator is placed in a box of the size $2 L$, $x \in [-L, L]$, such a non-analytic behavior disappears. Seemingly, at large $L$ the ground state energy behaves like
\[
    E_0\ \sim \bigg( b^2 + \frac{\al}{L^4}\bigg)^{\half}\ ,
\]
where $\al$ is a positive number. Thus, we have two symmetric square-root branch points on the imaginary axis with branch cuts tending to $\pm i \infty$, respectively. In the limit $L \rar \infty$ both branch points
coincide and the non-analytic function $|b|$ occurs. It resembles a behavior typical for a second-order phase transition.

Now we turn to the sextic QES potential at $n=0$ (\ref{e2.8.1}) and $k=0$ (cf. (\ref{e2.8.1})),
\begin{equation}
\label{e2.9}
   V_0 (x; a, b)\ =\ a^2 x^6\ +\ 2 a b x^4\ +\ (b^2 - 3 a) x^2 \ -\ b\ ,
\end{equation}
for which the ground state is known exactly if $a>0$,

\[
  E_0(a, b)\ =\ 0\ ,\
\]

and

\[
    \Psi_0(x)\ =\ e^{-\frac{b}{2}x^2-\frac{a}{4}x^4}\ .
\]

It can be easily shown that for $a < 0$ the lowest eigenvalue can {\it not} be equal to zero, $E_0 \neq 0$. In order to see it let us take $\Psi_0(x)$ which is a solution of the homogeneous Schr\"odinger equation with the potential $V_0(x; a, b)$ for any value $a$. Using the fact that the Wronskian should be a constant, one can construct the second, linearly-independent solution
\[
   \Psi_1(x)\ =\ \Psi_0(x)\, \int_{-\infty}^{x} \Psi_0^{-2}(x') dx'\ ,
\]
and then consider a linear combination
\[
    c_0 \Psi_0(x) + c_1 \Psi_1(x)\ .
\]
In order to fulfill the boundary conditions we are looking for a square-integrable solution of the Schr\"odinger equation. If $a>0$ it corresponds to $c_1=0$. It can be immediately seen that for
$a < 0$, it does not exist $c_0, c_1$ for which a linear combination of $\Psi_{0,1}$ is
square-integrable. A simple analysis shows that for $a < 0$, the lowest eigenvalue must be
positive, $E_0 > 0$. It implies a non-analytic behavior at $a=0$. It turns out that this behavior depends on sign of $b$.

It can be derived \cite{Bender-Turbiner:1993,Turbiner:1992} that at $a \rar -\infty$ the ground state energy behaves asymptotically like
\[
   E_0\ =\ 1.93556\ |a|^{\half}\ + \ldots
\]
independently on the value of $b$. This result is exact for $b=0$.

The QES potential (\ref{e2.9}) can be studied perturbatively, writing it as
\[
    V_0 (x; a, b)\ = V_0 + a V_p + a^2 V_{pp}\ \equiv\ (b^2 x^2 - b) \ +\ a (2 b x^4\ -\ 3 x^2)\
    +\ a^2 x^6\ ,
\]
where $a$ is perturbation (formal) parameter, assuming that $b > 0$. It is evident that for the ground state energy
\begin{equation}
\label{e2.10}
   E_0\ =\ \sum_{i=0}^{\infty} e_i(b) a^i\ ,
\end{equation}
{\it all} perturbative coefficients vanish
\[
   e_i(b)\ =\ 0\ ,
\]
(see \cite{Herbst:1978} and also \cite{Turbiner:1992})
\footnote{This remarkable observation implies the existence of non-trivial identities involving matrix elements of $x^2, x^4$ and $x^6$, if the coefficients $e_i$ are treated as the coefficients in the Rayleigh-Schr\"odinger perturbation theory. On the other hand, the coefficients $e_i$ can be calculated using the vacuum Feynman diagrams \cite{Bender:1969}:
the vanishing of these coefficients implies the existence of non-trivial relations between Feynman integrals. The present author is not aware that those identities and relations are profoundly investigated somewhere. A similar perturbation theory study can be carried out for other $n \neq 0$ sextic QES potentials for different algebraic states. The coefficients $e_i$ are always rational numbers. It is quite evident that one can find many potentials for which a perturbative theory analysis leads to similar conclusions. Perhaps, the most interesting class of such potentials is related to polynomial perturbations of the two-body Coulomb problem.}.
In turn, the ground state eigenfunction has a non-trivial perturbative expansion,
\[
    \Psi_0(x)\ =\
    e^{-\frac{b}{2}x^2}\ \bigg(\sum_{i=0}^{\infty} \frac{(-)^i}{4^i\, i!} a^i\, x^{4i} \bigg) \ .
\]
It is evident that for $a>0$ the sum (\ref{e2.10}) is equal to zero, $E_0=0$. On the other hand, we showed that for $a<0$, the ground state energy is non-zero. Hence, the sum (\ref{e2.10}) is {\it not} equal to zero, $E_0 \neq 0$. It leads to a conclusion about the existence of exponentially-small terms at $a \rar 0-$ for $b>0$.

\begin{figure}[tb]
\begin{center}
     {\includegraphics*[width=4.2in]{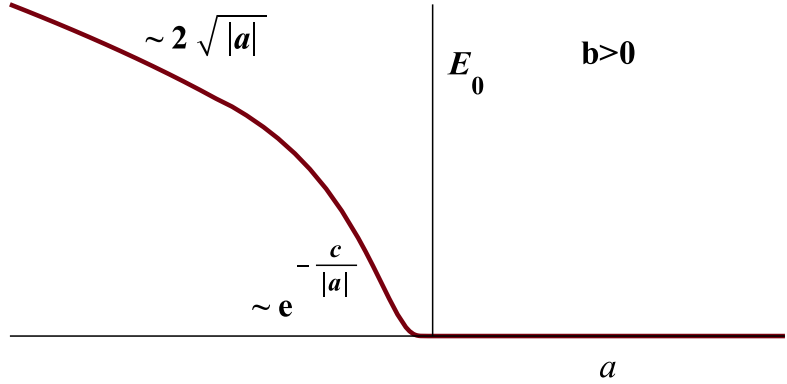}}
     \caption{Sextic QES oscillator at $n=0$: the ground state energy $E_0$ {\it vs.} $a$ for positive $b$.}
         \label{fig:2.1.5}
\end{center}
\end{figure}

\begin{figure}[tb]
\begin{center}
     {\includegraphics*[width=4.2in]{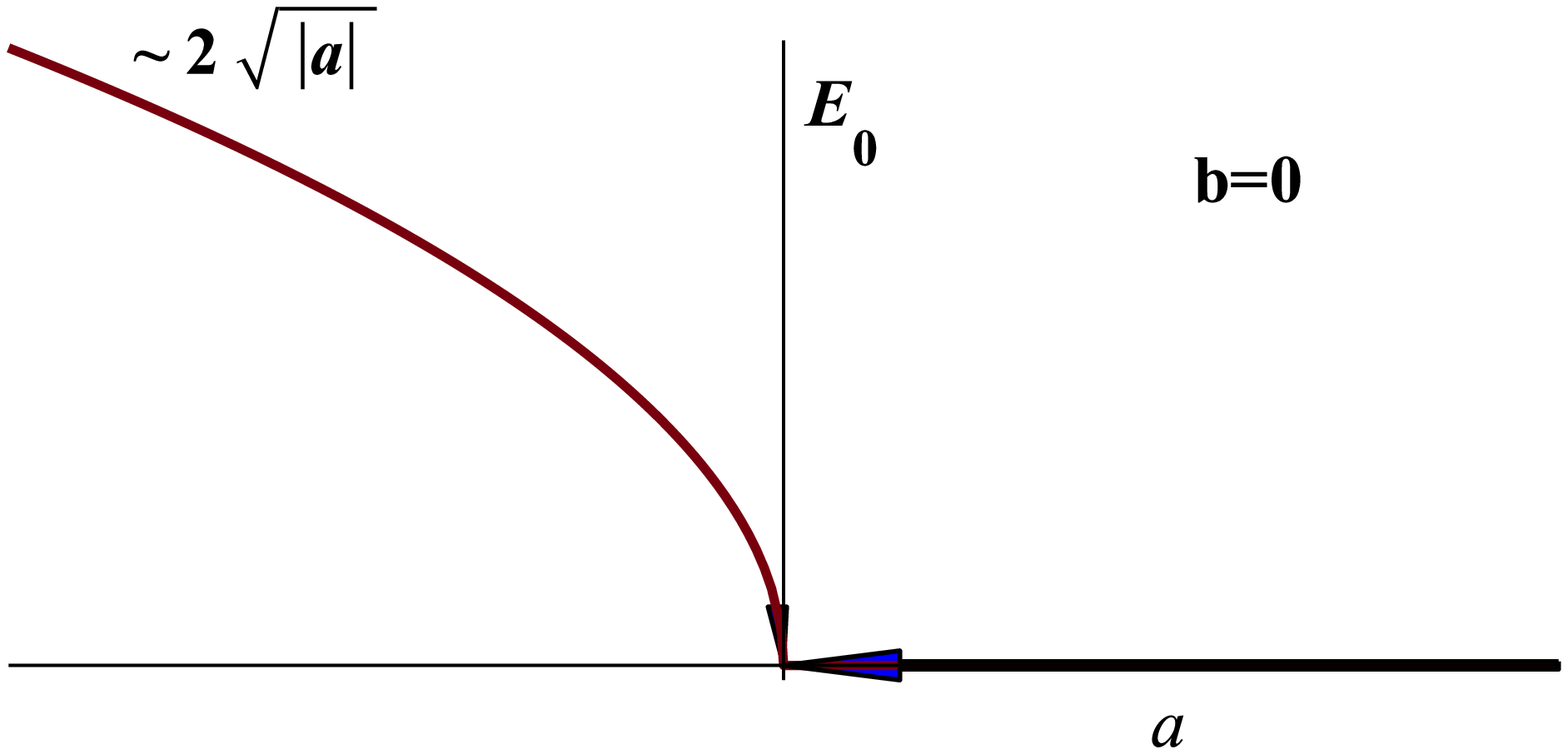}}
     \caption{Sextic QES oscillator at $n=0$: the ground state energy $E_0$ {\it vs.} $a$ for $b=0$.}
         \label{fig:2.1.6}
\end{center}
\end{figure}

\begin{figure}[tb]
\begin{center}
     {\includegraphics*[width=4.2in]{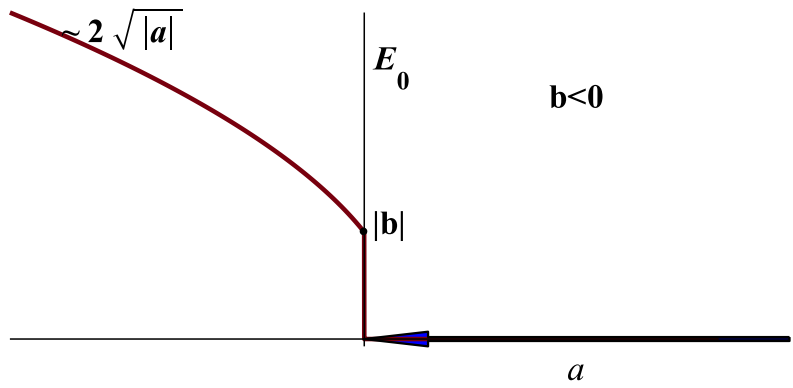}}
     \caption{Sextic QES oscillator at $n=0$: the ground state energy $E_0$ {\it vs.} $a$ for negative $b$.}
         \label{fig:2.1.7}
\end{center}
\end{figure}

A simple analysis shows that there are three types of non-analytic behavior of the ground state energy at $a=0$. If $b<0$ the ground state energy is discontinuous at $a=0$: it jumps from $E_0=0$ at $a \rar 0+$ to $E_0=|b|$ at $a \rar 0-$. It looks like the 1st order phase transition. If $b=0$ the ground state energy is continuous at $a=0$ but the first derivative (and all others) is discontinuous: it jumps from $\frac{dE_0}{da}=0$ at $a \rar 0+$ to $\frac{dE_0}{da}=\infty$ at $a \rar 0-$. It looks like the 2nd order phase transition. If $b>0$ the ground state energy is continuous at $a=0$ as well as all others being equal to zero. It looks like the infinite-order phase transition (the Kosterlitz-Thouless type phase transition).
On Figs. \ref{fig:2.1.5}, \ref{fig:2.1.6}, \ref{fig:2.1.7} the behavior of the ground state energy in the potential $V_0$ {\it vs.} $a$ is illustrated for positive $b$, $b=0$ and negative $b$, respectively. It is evident that similar discontinuities of the energy {\it vs.} $a$ at $a=0$ will occur for any eigenstate in the QES problem (\ref{e2.1-1}).

Of course, if the sextic QES oscillator is placed in the box of the size $2 L$, $x \in [-L, L]$, such a non-analytic behavior disappears: all curves on Figs. \ref{fig:2.1.5}, \ref{fig:2.1.6}, \ref{fig:2.1.7} begin to behave smoothly around $a=0$.


\clearpage
%
%
%
%
\newpage
\renewcommand{\thesection}{3.{\arabic{section}}}
\setcounter{section}{0}

\begin{center}
{\bf \large Chapter\ 3.}\\[20pt]
 {\Large Algebraic Perturbations of Exactly-Solvable Problems}
\end{center}

\addcontentsline{toc}{section}{Chapter 3. Algebraic Perturbations of
Exactly-Solvable Problems \protect
 \hfill}
\vskip 1cm


\renewcommand{\theequation}{3.{\arabic{equation}}}
 \setcounter{equation}{0}

 \renewcommand{\thefigure}{3.{\arabic{figure}}}
 \setcounter{figure}{0}

 \renewcommand{\thetable}{3.{\arabic{table}}}
 \setcounter{table}{0}

Perturbation theory is one of the most developed and powerful approaches in quantum mechanics. This approach is quite universal and can be applied, sometimes easily, to practically
any problem. Many realistic problems of quantum mechanics can be naturally considered as perturbations of exactly-solvable problems. It explains the
theoretical and practical importance of perturbative studies in
quantum mechanics. In practice, the perturbation theory is usually employed in Rayleigh-Schr\"odinger form where construction of perturbative corrections
is reduced to a calculation of matrix elements, given by certain integrals,
and summation over intermediate states. However, there exists a large family
of perturbations of exactly-solvable problems, where the perturbative corrections
can be found by linear algebra means. We call such perturbations {\it algebraic}.
The Lie-algebraic formalism described in Chapter 1 allows us to give a partial
classification of the algebraic perturbations.
In particular, it sheds light to a reason why a perturbation theory in powers $g$ for one-dimensional quartic anharmonic oscillator
\begin{equation}
\label{QAHO}
   V(x)\ =\ m^2 x^2 + g x^{2n} \ ,\ n=2,3,\ldots
\end{equation}
developed in \cite{Bender:1969,Bender:1973} is constructed by
algebraic means through solving the recurrence relations.

In Chapter 1 a classification of the one-dimensional
exactly-solvable linear differential operators was given. In particular,
it was shown that the most general exactly-solvable differential operator
of the second order is given by a general second-order element of the universal
enveloping algebra of the Borel sub-algebra $b_2$ of the algebra
$\mathfrak{sl}(2,\bf{R})$ of the differential operators of
the first order,
\begin{equation}
\label{ES-4}
 E_2\ =\   c_{00} J^0  J^0  + c_{0-} J^0 J^- +
 c_{--} J^-  J^-  + c_0 J^0  + c_- J^-  + c\ ,
\end{equation}
with the number of free parameters equal to $par (E_2) = 6$. Here
\[
 J^0 \equiv J^0_0\ =\ x \dx \  ,
\]
\[
 J^- \equiv J^-_0\ =\ \dx \ ,
\]
(cf. (\ref{sl2r})). Substituting $J^0, J^-$ into (\ref{ES-4}) we
obtain the hypergeometrical operator
\begin{equation}
\label{ES-4op}
 E_2(x)\ =\   -Q_{2}(x) \dxx \ +\ Q_{1}(x) \dx \ +\ Q_{0}(x) \ ,
\end{equation}
where the $Q_{j}(x)$ are polynomials of $j$th order
\[
 - Q_2 (x)\ =\  c_{00}\,x^2 + c_{0-}\,x + c_{--}\ ,
\]
\[
 Q_1 (x)\ = \ (c_{00} + c_{0})\,x + c_{-} \ ,
\]
\begin{equation}
\label{ES-4coef}
 Q_0 (x)\ = \ c \ .
\end{equation}
The parameter $c$ defines the reference point for energy and
without loss of generality it can vanish, $c=0$. For a sake of
simplicity, we denote $c_{00}+c_{0}=\tilde c_0$.
It always can be chosen the reference point for the coordinate $x$ in such
a way that $Q_2(0)=0$. This, without a loss of generality the parameter $c_{--}=0$.

Let us consider the spectral problem for the perturbed
exactly-solvable operator $E_2$,
\begin{equation}
\label{Epert}
 h\ =\ E_2(x) + g V_p(x) \ ,
\end{equation}
where $V_p(x)$ is a perturbation potential and $g$ is a perturbation parameter (the coupling constant),
\begin{equation}
\label{e4.1}
 h \vphi \,= \,\veps\, \vphi\ .
\end{equation}
We develop perturbation theory for the equation (\ref{e4.1}) in
powers of $g$ by following a standard procedure,
\begin{equation}
\label{e4.2}
 \vphi = \sum_{k=0}^{\infty} \vphi_k g^k\ ,\quad
 \veps = \sum_{k=0}^{\infty} \veps_k g^k\ \, ,
\end{equation}
assuming a solution of unperturbed problem
\[
  h_0 \vphi_0 \,= \,\veps_0\, \vphi_0\ ,
\]
is known.
It seems clear that is nothing but the so-called Dalgarno-Lewis form of perturbation theory
in quantum mechanics \cite{Dalgarno:1955}. It is easy to derive an equation for the $k$th correction
\begin{equation}
\label{e4.3}
 (E_2 - \veps_0) \vphi_k \ =\ \sum_{i=1}^{k} \veps_i \,\vphi_{k-i} -
V_p \,\vphi_{k-1} \ ,
\end{equation}
where the solution $\vphi_k(x)$ is looked for in the form of a polynomial. This requirement plays a role of a boundary condition.
The important feature of this perturbation theory is related to a fact that perturbation corrections
can be calculated for each eigenstate separately. It differs from a standard Rayleigh-Schr\"odinger perturbation theory, which is widely accepted, where the correction to eigenfunction is given by an expansion of the eigenfunctions of the unperturbed problem (see e.g. \cite{LL-QM:1977} and a discussion below).

In \cite{Turbiner:2002} it was proved a general theorem which states that {\it if}
\begin{quote}
{\it (i) unperturbed problem can be written as an element of universal enveloping algebra of an algebra of differential operators which preserves an infinite flag of finite-dimensional representation spaces
and,

\noindent
(ii) perturbation is an element of a finite-dimensional invariant subspace(s) in the flag,

{\it then,}

\noindent
the perturbation theory is algebraic. It implies that any perturbation correction to the eigenfunction is an element of a finite-dimensional invariant subspace(s) in the flag. It can be found by linear algebra means.}
\end{quote}

Following this Theorem it is evident if $h$ is exactly-solvable operator associated with the algebra
$\mathfrak{sl}(2,\bf{R})$ and perturbation potential $V_p$ is a polynomial of finite degree in $x$, hence, it belongs a finite-dimensional invariant subspace of the algebra
$\mathfrak{sl}(2,\bf{R})$, the perturbation theory (\ref{e4.2}) should be algebraic. Thus, all perturbation corrections $\vphi_k(x)$, which are found by means of solving (\ref{e4.3}), are polynomials in $x$ of a finite degree. Hence, the construction of perturbation theory is a linear algebraic procedure. In particular, all energy corrections should be rational functions of parameters of $E_2$ (see (\ref{ES-4coef})).

As an illustration let us consider a simplest non-trivial perturbation
\[
   V_p\ =\ x\ ,
\]
of the operator
\[
     h_0\ \equiv E_2\ =\ -(c_{00}\,x^2 + c_{0-}\,x)\,\dxx + ({\tilde c}_{0}\,x + c_{-}) \dx\ ,
\]
cf. (\ref{Epert}), at $c_{00} \neq 0$ \footnote{At $c_{00} = 0$ the perturbed operator (\ref{Epert}) can be transformed to $E_2$ via canonical transformation $\pa_x \rar \pa_x + \al, x \rar x$ with a certain parameter $\al$.} , with spectra of polynomial eigenfunctions of the form
\[
     \veps_0^{(n)}\ =\ -c_{00}\,n(n-1) + {\tilde c}_{0} n\ ,\ n=0,1,2,\ldots \ .
\]

We focus on the ground state. It can be seen immediately, that the lowest eigenstate of the exactly-solvable operator $E_2$ is
\[
   \veps_0 = 0\ ,\ \vphi_0 = 1 \ .
\]
It is a straightforward simple calculation by using (\ref{e4.3}) to find the first several corrections, for example,

\[
    \veps_1\ =\ -\frac{c_-}{\tilde c_0}\ ,\ \vphi_1\ =\ -\frac{x}{\tilde c_0} \ ,
\]
and
\[
    \veps_2\ =\ -  \frac{c_-}{{\tilde c}_0^2}\ \frac{c_{0-}{\tilde c}_0 - c_{00}c_{-}}
    {- c_{00}+{\tilde c}_0}\ ,\
\]

\[
    \vphi_2\ =\  \ \frac{x^2}{2{\tilde c}_0 (- c_{00}+{\tilde c}_0)}
    + \frac{x}{\tilde c_0^2} \ \frac{c_{0-}{\tilde c}_0 - c_{00}c_{-}}
    {- c_{00}+{\tilde c}_0}\ .
\]
In general, the $n$th correction $\vphi_n$ has the form of the $n$th degree polynomial without a constant term.

In a straightforward way the spectral problem for the operator (\ref{Epert}) can be "lifted" to the Fock space,
\[
    (E_2(b,a) + g V_p(b)) \phi(b) |0>\ =\ \veps \phi(b) |0>\ ,
\]
where the boundary conditions are converted to the search for polynomial $\phi(b)$, here $\veps$ is the spectral parameter.
Theorem \cite{Turbiner:2002} (see above) can be modified accordingly - in the case of a polynomial perturbation $V_p(b)$ the corrections $\phi_n(b)$ are polynomials, they can be found by algebraic means. Furthermore, these corrections $\phi_n(b)$ remain the same as in the $x$-space as well as the corrections to energy $\veps_n$. Taking the realization of $b,a$ in finite-difference operators we arrive at the spectral problem for a finite-difference operator, where the exactly-solvable finite-difference operator is perturbed by a finite-difference operator(!). In this case the perturbative corrections $\veps_n$ remain the same as in continuous case, the corrections $\vphi_n=\phi_n(b) |0>$ remain polynomials in $x$ with coefficients changed accordingly. The case of the perturbed harmonic oscillator (\ref{QAHO}) (continuous, on uniform lattice and on exponential lattice) was studied in some details in \cite{Turbiner:2000p,Turbiner:2001}.

\clearpage
%
%
%
%
\newpage
\renewcommand{\thesection}{4.{\arabic{section}}}
\setcounter{section}{0}

\begin{center}
{\bf \large Chapter\ 4.}\\[20pt]
 {\Large Quasi-Exactly-Solvable finite-difference equations in one variable }
\end{center}
\addcontentsline{toc}{toc}{Chapter 4. Quasi-Exactly-Solvable
finite-difference equations in one variable \hfill}

\vskip 1cm


\renewcommand{\theequation}{4.{\arabic{equation}}}
 \setcounter{equation}{0}

 \renewcommand{\thefigure}{4.{\arabic{figure}}}
 \setcounter{figure}{0}

 \renewcommand{\thetable}{4.{\arabic{table}}}
 \setcounter{table}{0}

\section{Finite-difference equations in one variable (exponential lattice)}
\subsection{General consideration}
Let us define a multiplicative finite-difference operator (derivative), or a shift operator,
or the so-called Jackson symbol (derivative) (see e.g. Exton \cite{Exton:1983}, Gasper and Rahman \cite{GR:1990}),
\begin{equation}
\label{e5.1}
{D}_q f(x)\ =\ \frac{f(x) - f(qx)}{(1 - q) x} \ ,
\end{equation}
where $q \in R$ and $f(x)$ is real function $x \in R$\,. All these names are used in literature for (\ref{e5.1}). The Leibnitz rule for the operator $D_q$ is
\[
D_q (f(x) g(x))\ =\ (D_q f(x)) g(x)+ f(qx) (D_q g(x))\ ,
\]
it connects two neighbouring points in the lattice space.
It is easy to check that
\begin{equation}
\label{e5.1-1}
D_q \, x^n\ =\ \frac{1 - q^n}{1 - q}\,x^n \equiv \{ n \}_q \, x^n\ ,
\end{equation}
where $\{ n \}_q$ is the so called $q$-number $n$, at $q \rar 1$, $\{ n \}_q \rar n$. Finite-difference derivative $D_q$ $q$-commutes with coordinate,
\begin{equation}
\label{e5.1-2}
 [x, D_q]_q \equiv x D_q - q D_q x\ =\ 1\ ,
\end{equation}
where $[A, B]_q \equiv A B - q B A$ sometimes called {\it quommutator}, following
David Fairlie suggestion. The expression (\ref{e5.1-2}) together with $[1, x]=[1, D_q]=0$
defines the $q$-deformed Heisenberg algebra. This algebra like non-deformed one naturally
acts in the space of monomials mapping monomial to monomial.

Now one can easily introduce a finite-difference analogue of the $\mathfrak{sl}(2,\bf{R})$-algebra of the differential operators  (\ref{sl2r}),
where the operator $D_q$ replaces the continuous
derivative, see e.g. \cite{Ogievetsky:1991},
\[
  \tilde  J^+_n = x^2 D_q - \{ n \}_q\, x \ ,
\]
\begin{equation}
\label{E5.2}
\tilde  J^0_n = \  x D_q - \hat{n}\ ,
\end{equation}
\[
 \tilde  J^-_n = \ D_q \ ,
\]
where $\hat n \equiv \frac{{\{n\}_q\{n+1\}_q}}{\{2n+2\}_q}$\,.
The operators (\ref{E5.2}) after multiplication by some factors
\[
\tilde  j^0\ =\ \frac{q^{-n}}{q+1} \frac{\{2n+2\}_q}{\{n+1\}_q} \tilde J^0_n\ ,
\]
\[
\tilde  j^{\pm}\ =\ q^{-n/2} \tilde  J^{\pm}_n\ ,
\]
see \cite{Ogievetsky:1991}, span a quantum algebra $\mathfrak{sl}(2,\bf{R})_q$ with
the following commutation (quommutation) relations
\[
q \tilde  j^0\tilde  j^- \ - \ \tilde  j^-\tilde  j^0 \ = \ - \tilde  j^- \ ,
\]
\begin{equation}
\label{E5.3}
 q^2 \tilde  j^+\tilde  j^- \ - \ \tilde  j^-\tilde  j^+ \ = \ - (q+1) \tilde  j^0 \ ,
\end{equation}
\[
 \tilde  j^0\tilde  j^+ \ - \ q\tilde  j^+\tilde  j^0 \ = \  \tilde  j^+ \ .
\]
The parameter $q$ does characterize the deformation of the
commutators of the classical Lie algebra $\mathfrak{sl}(2,\bf{R})$.
If $q \rar 1$, the commutation relations (\ref{E5.3}) reduce to the standard
$\mathfrak{sl}(2,\bf{R})$-algebra ones. A remarkable property of generators (\ref{E5.2})
is that, {\it if $n$ is a non-negative integer, they form the finite-dimensional
representation corresponding the finite-dimensional representation
space ${\p}_{n+1}$ (\ref{e1.1.5}),
\[
   \tilde J^{\pm, 0}_n: {\p}_{n+1} \rar {\p}_{n+1} \ ,
\]
the same as of the non-deformed} $\mathfrak{sl}(2,\bf{R})$
(see (\ref{sl2r})). Note that, in general, for complex $q$ other than prime root of unity,
$|q| \neq 1$, this representation is irreducible.

{\it Comment 4.1}\  The algebra (\ref{E5.3}) is known in the literature as
{\it the second Witten quantum deformation} of the $sl_2$-algebra,
in the classification by C.~Zachos \cite{Zachos:1991}.

Similarly as for differential operators one can introduce
quasi-exactly-solvable $\tilde  T_k(x,D_q)$ and exactly-solvable  $\tilde  E_k(x,D_q)$ finite-difference operators.

\begin{LEMMA} \cite{Turbiner:1994}

{\it (i) Suppose $n > (k-1)$.  Any
quasi-exactly-solvable operator $\tilde T_k$, can be represented by a $k$th
degree polynomial of the operators (\ref{E5.2}). If $n \leq (k-1)$, the part
of the quasi-exactly-solvable operator $\tilde T_k$ containing
derivatives up to order $n$ can be represented by a $n$th
degree polynomial in the generators (\ref{E5.2}).

(ii) Conversely, any polynomial in (\ref{E5.2}) is quasi-exactly solvable.

(iii) Among quasi-exactly-solvable operators
there exist exactly-solvable operators $\tilde E_k$.}
\end{LEMMA}

\noindent {\it Comment 4.2} \ If we define an analogue of the
universal enveloping algebra $U_g$ for the quantum algebra
$\tilde g$ as an algebra of all ordered polynomials in generators, then a
quasi-exactly-solvable operator $\tilde T_k$ at $k < n+1$ is
simply an element of the universal enveloping algebra
$U_{\mathfrak{sl}(2,\bf{R})_q}$ of the algebra $\mathfrak{sl}(2,\bf{R})_q$ taken in
representation (\ref{E5.2}). If $k \geq n+1$, then $\tilde T_k$ is
represented as an element of $U_{\mathfrak{sl}(2,\bf{R})_q}$ plus $B D_q^{n+1}
$, where $B$ is any linear difference operator of order not higher
than $(k-n-1)$.

Similar to $\mathfrak{sl}(2,\bf{R})$ (see Definition 2.3), one can introduce
the grading of generators (\ref{E5.2}) of $\mathfrak{sl}(2,\bf{R})_q$ (cf. (\ref{e1.2.3}))
and, hence, the grading of monomials of the universal enveloping
$U_{\mathfrak{sl}(2,\bf{R})_q}$ (cf. (\ref{e1.2.4})).

\begin{LEMMA} {\it A quasi-exactly-solvable operator
$\tilde T_k \subset U_{\mathfrak{sl}(2,\bf{R})_q}$ has no terms of positive
grading, iff it is an exactly-solvable operator.}
\end{LEMMA}
\begin{THEOREM} \cite{Turbiner:1994}

Let $n$ be a non-negative integer.
Take the eigenvalue problem for a linear difference operator of the $k$-th order
in one variable
\begin{equation}
\label{E5.4}
 {\tilde T}_k (x,D_q)\, \vphi (x) \ = \ \veps\, \vphi (x)\ ,
\end{equation}
where ${\tilde T}_k$ is symmetric. The problem (\ref{E5.4}) has $(n+1)$
linearly independent eigenfunctions in the form of a polynomial in
variable $x$ of order not higher than $n$, if and only if ${\tilde T}_k$ is
quasi-exactly-solvable. The problem (\ref{E5.4}) has an infinite sequence
of polynomial eigenfunctions, if and only if the operator is
exactly-solvable $\tilde E_k$.
\end{THEOREM}

{\it Comment 4.3} \ Saying the operator ${\tilde T}_k$ is symmetric, we imply
that, considering the action of this operator on a space of polynomials
of degree not higher than $n$, one can introduce a positively-defined
scalar product, and the operator $\tilde T_k$ is symmetric with respect
to it.

This theorem gives a general classification of finite-difference equations
\begin{equation}
\label{e5.5}
 \sum_{j=0}^{k} {\tilde a}_j (x) D_q^j \vphi (x) \ = \ \veps \vphi(x)
\end{equation}
having polynomial solutions in $x$.  The coefficient functions must
have the form
\begin{equation}
\label{e5.6}
\tilde a_j (x) \ = \ \sum_{i=0}^{k+j} \tilde a_{j,i} x^i .
\end{equation}
In particular, this form occurs after substitution (\ref{E5.2}) into a
general $k$th degree polynomial element of the universal
enveloping algebra $U_{\mathfrak{sl}(2,\bf{R})_q}$. It guarantees the
existence of at least a finite number of polynomial solutions. The
coefficients $\tilde a_{j,i}$ are related to the coefficients of
the $k$th degree polynomial element of the universal enveloping
algebra $U_{\mathfrak{sl}(2,\bf{R})_q}$. The number of free parameters of
the polynomial solutions is defined by the number of free
parameters of a general $k$-th order polynomial element of the
universal enveloping algebra $U_{\mathfrak{sl}(2,\bf{R})_q}$\,~
\footnote{
For quantum $\mathfrak{sl}(2,\bf{R})_q$ algebra there are no polynomial Casimir
operators (see, e.g. Zachos \cite{Zachos:1991}). However, in the
representation (\ref{E5.2}) the relationship between generators
analogous to the quadratic Casimir operator
\[
q\tilde J^+_n\tilde J^-_n - \tilde J^0_n \tilde J^0_n + (\{ n+1 \}_q
- 2 \hat{n}) \tilde J^0_n = \hat{n} (\hat{n} - \{ n+1 \}_q)
\]
appears. It reduces the number of independent parameters of the
second-order polynomial element of  $U_{\mathfrak{sl}(2,\bf{R})_q}$. It
becomes the standard Casimir operator at $q \rar 1$. }\,.
A rather straightforward calculation leads to the following formula
\[
par (\tilde T_k) = (k+1)^2+1
\]
(for the second-order finite-difference equation $par(\tilde T_2) = 10$).
For the case of an infinite sequence of polynomial solutions the formula
(\ref{e5.6}) simplifies to
\begin{equation}
\label{e5.7}
\tilde a_j (x) \ = \ \sum_{i=0}^{j} \tilde a_{j,i} x^i
\end{equation}
and the number of free parameters is given by
\[ par (\tilde E_k) = {(k+1)(k+2) \over 2} + 1 \]
(for $k=2$, $par(\tilde E^2) = 7$).
The increase in the number of free parameters
compared to ordinary differential equations is due to the presence of the
deformation parameter $q$.

\subsection{Second-order finite-difference exactly-solvable equations.}

In \cite{Turbiner:1992p} it is implemented a description
in the present approach of the $q$-deformed Hermite, Laguerre, Legendre
and Jacobi polynomials (for definitions of these polynomials, see Exton
\cite{Exton:1983}, Gasper-Rahman \cite{GR:1990}).
In order to reproduce the known $q$-deformed classical Hermite, Laguerre,
Legendre and Jacobi polynomials
(for the latter, there exists the $q$-deformation of the asymmetric form
(\ref{e1.5.14}) only, see e.g. \cite{Exton:1983} and \cite{GR:1990}),
one should modify the spectral problem (\ref{E5.4}):
\begin{equation}
\label{e5.8}
 \tilde T_k (x,D_q)\, \vphi (x) \ = \ \tilde \veps\, \vphi (qx)\ ,
\end{equation}
by introducing the r.h.s. function the dependence on the argument
$qx$ (cf. (2.5) and (3.4)) as it follows from the book
\cite{Exton:1983} (see also \cite{GR:1990}).

{\it Comment 4.4}
The spectral problem (\ref{e5.8}) can be considered as a generalized spectral
problem
\[
   \tilde T_k (x,D_q)\, \vphi (x) \ = \
   \tilde \veps\, \big[1 - (q-1)x D_q \big] \vphi (x)\ ,
\]
It is evident that the operator in rhs is the element of
$U_{\mathfrak{sl}(2,\bf{R})_q}$ in $(n+1)$-dimensional representation,
\[
   \big[1 - (q-1)x D_q \big]\ =\ 1 - (q-1) (\tilde  J^0_n + \hat{n})\ .
\]

The corresponding $q$-difference operators having $q$-deformed
classical Hermite, Laguerre, Legendre and Jacobi polynomials as
eigenfunctions (see the equations (5.6.2), (5.5.7.1), (5.7.2.1),
(5.8.3) in the book by Exton \cite{Exton:1983}, respectively) are
given by the combinations in the generators:
\renewcommand{\theequation}{4.11.{\arabic{equation}}}
\setcounter{equation}{0}
\begin{equation}
\label{e5.9.1}
      \tilde E_2 =  \tilde  J^-_0 \tilde  J^-_0 \ -\ \{ 2 \}_q \tilde  J^0_0\ ,\quad
        {\tilde \veps}_n\ =\ -\{2\}_{1/q} \{n\}_{1/q}\ ,
\end{equation}

\begin{equation}
\label{e5.9.2}
\tilde E_2 =  \tilde  J^0_0 \tilde  J^-_0 \ -\ q^{-a-1} \tilde  J^0_0 \ +
\ (q^{-a-1}\{ a+1\}_q) \tilde  J^-_0\ ,\ {\tilde \veps}_n\ =\ -q^{-a-2}\,\{n\}_{1/q}\ ,
\end{equation}

\begin{equation}
\label{e5.9.3}
\tilde E_2 = - q\tilde  J^0_0 \tilde  J^0_0  + \tilde  J^- \tilde  J^-\ +
\ (q -\{ 2 \}_q) \tilde  J^0_0\ ,
\end{equation}

\begin{equation}
\label{e5.9.4}
\tilde E_2= - q^{a+b-1}\tilde  J^0_0 \tilde  J^0_0  + q^a \tilde  J^0_0
\tilde  J^-_0 + [q^{a+b-1}-\{a\}_q\, q^b -\{b\}_q] \tilde  J^0_0 + \{a\}_q \tilde  J^-_0 \ ,
\end{equation}
respectively.
\renewcommand{\theequation}{4.{\arabic{equation}}}
\setcounter{equation}{11}
\begin{LEMMA}
{\it If the operator ${\tilde T}_2$ (for the definition, see e.g. (\ref{QES-2}))
is such that
\begin{equation}
\label{e5.10}
{\tilde c}_{++}=0 \quad and \quad {\tilde c}_{+} =({\hat n} - \{ m \})
{\tilde c}_{+0} \ , \ at \ some\ m=0,1,2,\dots
\end{equation}
 then the operator ${\tilde T}_2$ preserves both ${\p}_{n+1}$ and
${\p}_{m+1}$, and polynomial solutions in $x$ with 8 free
parameters occur.}
\end{LEMMA}
(cf. Lemma \ref{Lemma123}).

As usual in quantum algebras, a rather outstanding situation occurs
if the \linebreak deformation parameter $q$ is equal to a primitive root of unity.
For instance, the following statement holds.
\begin{LEMMA} {\it If a quasi-exactly-solvable operator $\tilde T_k$
preserves the space ${\p}_{n+1}$ and the parameter $q$ satisfies
to the equation
\begin{equation}
\label{e5.11}
q^n\ = \ 1 \ ,
\end{equation}
then the operator  $\tilde T_k$ preserves an infinite flag of
polynomial spaces $ {\p}_0 \subset  {\p}_{n+1} \subset
{\p}_{2(n+1)} \subset \dots \subset {\p}_{k(n+1)} \subset \dots
$.}
\end{LEMMA}

It is worth emphasizing that, in the limit as $q$ tends to one,
Lemmas 4.1.1, 4.1.2, 4.1.3 and Theorem 4.1.1
coincide with Lemmas 2.1, 2.2, 2.3 and Theorem 2.1, respectively.
Thus, the case of differential equations in one variable can be treated
as a limiting case of finite-difference ones, of course, if $q$ is
not a prime root of unity.

\section{$\mathfrak{sl}(2,\bf{R})$-algebra in the Fock space}

Take two operators (letters) $a$ and $b$ obeying the commutation relations
\begin{equation}
\label{e5.1.1}
             [a,b] \equiv ab  -  ba \ =\ 1\ ,\ [a,1]=[b,1]=0\ ,
\end{equation}
with the identity operator on the r.h.s. -- they span the
three-dimensional Heisenberg algebra. In other words, $a$ and $b$
form canonical pair. By definition the universal enveloping algebra
of the Heisenberg algebra is the algebra of
all ordered polynomials in $a,b$: any monomial is taken to
be of the form $b^k a^m$ \footnote{Sometimes this is called the
Heisenberg-Weyl algebra}. If, besides the polynomials, all entire
functions in $a,b$ are considered, then the {\it extended}
universal enveloping algebra of the Heisenberg algebra appears
or, in other words, the extended Heisenberg-Weyl algebra. In the
(extended) Heisenberg-Weyl algebra one can find the non-trivial
embeddings of the Heisenberg algebra
\footnote{This means that there exists a family of pairs of the
non-trivial elements of the Heisenberg-Weyl algebra obeying the
commutation relations (\ref{e5.1.1})} ,
whose can be treated as a certain type of quantum canonical transformations.
We say that the (extended) Fock space appears if we take the (extended)
universal enveloping algebra of the Heisenberg algebra and add to it the
vacuum state $|0>$ defined as follows
\begin{equation}
\label{e5.1.2}
a|0>\  = \ 0\ .
\end{equation}
One can take a polynomial element of the Heisenberg-Weyl algebra $L(b,a)$
and define the eigenvalue problem in the Fock space \cite{Smirnov:1995a}
\[
    L(b,a) \phi (b)\, |0>\  = \ \veps \phi (b)\, |0>\ .
\]
Here $\veps$ is spectral parameter. As for boundary conditions: we are looking for
eigenpolynomials in $b$, $\phi (b)$. Thus, technically the problem of finding 
eigenpolynomial $\phi (b)$ is reduced to reordering (the normal ordering) of 
$L(b,a) \phi (b)$ to superposition of monomials $b^p a^q$.

{(i)}\ One of the most important realizations of (\ref{e5.1.1}) is the
coordinate-momentum representation:
\begin{equation}
\label{e5.1.3}
a\ =\ \frac{d}{dx} \equiv \pa_x\ ,\ b\ =\ x\ ,
\end{equation}
where $x$ stands for the multiplication operator in a
space of functions $f(x)$. In this case the vacuum is a constant,
without a loss of generality we put $|0>\  = \ 1$. In this representation
\begin{equation}
\label{e5.1.3n}
    \phi_n \equiv b^n |0>\ =\ x^n\ .
\end{equation}

\begin{figure}[tb]
\begin{center}
     {\includegraphics*[width=6in]{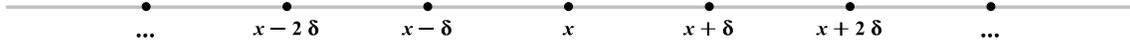}}
     \caption{
     One-dimensional infinite uniform lattice $(-\infty, +\infty)$ with cental point $x$ and spacing $\de$.
      }
    \label{Uni-Latt}
\end{center}
\end{figure}

{(ii)}\ Infinite uniform lattice
\begin{equation}
\label{e5.1.uni}
 \{ \ldots,\ x-2\de\ ,\ x -\de\ ,\ x\ ,\ x+\de\ ,\ x+2\de\ ,\ \ldots  \}
\end{equation}
is marked by $x \in {\bf R}$ - a position of a central or reference point of the lattice and spacing $\de$, see for illustration Fig.~\ref{Uni-Latt}.
Finite-difference analogue of (\ref{e5.1.3}) on uniform lattice
has been found since long time ago (see e.g. \cite{Smirnov:1995a,Smirnov:1995hj}) and
studied profoundly in e.g. \cite{Gorski:1998,Gorski:2000},
\begin{equation}
\label{e5.1.4}
a_{\de}\ =\ {\mathcal D}_+\ ,\ b_{\de}\ =\ x(1-\de{\mathcal D}_-) \equiv x_{\de}\ ,
\end{equation}
where
\[
        {\mathcal D}_{\pm} f(x) = \frac{f(x {\pm}\de) - f(x)}{{\pm}\de}\ ,
\]
is the finite-difference (shift) operator, $\de \in {\bf C}$ and
${\mathcal D}_+\rightarrow {\mathcal D}_-$, if $\de \rightarrow -\de$.
${\mathcal D}_+$ makes sense of finite-difference derivative. Sometimes, it is
called the Norlund derivative, it connects two neighbouring points in the lattice space.
Its canonical partner $x_{\de}$ is a finite-difference analogue of a position operator.
It also connects two neighbouring points in the lattice space.
Interestingly, the Euler operator
\[
    (b_{\de}\, a_{\de})\ =\ x_{\de} {\mathcal D}_+ \ =\ x {\mathcal D}_-\ ,
\]
depends on the lattice spacing $\de$. We define as "vacuum" $|0>\  = \ 1$.
All that gives rise to the so-called {\it umbral calculus} due to G.-C.~Rota,
see e.g. \cite{Rota:1978}. In this representation
\begin{equation}
\label{e5.1.4n}
    \phi_{n+1} \equiv b^{n+1} |0> = x(x-\de)\ldots (x - n \de) \equiv x^{(n+1)}(\de)\ ,
\end{equation}
where $x^{(n)}$ is the Pochhammer symbol which is called in this context
the {\it quasi-monomial}. It is worth noting that the commutator
\[
     [{\mathcal D}_+, x]\ =\ 1 + \de {\mathcal D}_+ \ ,
\]
depends on ${\mathcal D}_+ $ only.

\begin{figure}[tb]
\begin{center}
     {\includegraphics*[width=6in]{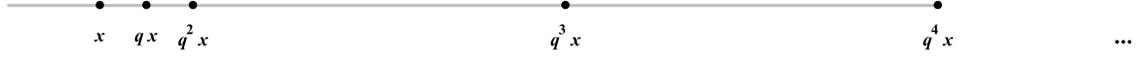}}
     \caption{
     One-dimensional semi-infinite exponential lattice $[x, +\infty)$ with reference point $x$ and dilation $q$.
      }
    \label{Exp-Latt}
\end{center}
\end{figure}

{(iii)}\ Semi-Infinite exponential lattice
\begin{equation}
\label{e5.1.exp}
     \{ \ x\ ,\ x q\ ,\ x q^{2}\ ,\ \ldots  \}\ ,
\end{equation}
is marked by $x \in {\bf R}$ - a position of a reference point of the lattice, see Fig.~\ref{Exp-Latt}.
Finite-difference analogue of (\ref{e5.1.3}) on exponential lattice
has been found in \cite{Chryssomalakos:2001}
\begin{equation}
\label{e5.1.5}
a_{q}\ =\ {\mathcal D}_q\ ,\ b_{q}\ =\ x_q \ ,
\end{equation}
where
\[
   {\mathcal D}_q\ =\ \frac{1}{1-q} x^{-1} (1 - q^A)\ =\ x^{-1} \{ A\}_q\ =\
                       \frac{\{ A+1\}_q}{A+1} \pa_x \ ,\
      A\ =\ x\,\pa_x \ ,
\]
cf.(\ref{e5.1}), $q \in {\bf C}$, is the Jackson derivative and $\{ A\}_q = \frac{1 - q^A}{1-q}$ is the so-called $q$-operator $A$ (cf. with $q$-number (\ref{e5.1-1})) 
\footnote{Since we move from quantum calculus associated with quantum algebra $\mathfrak{sl}(2,\bf{R})_q$ to the same one but associated the Lie algebra $\mathfrak{sl}(2,\bf{R})$ (see below), hereafter we change the notation for the Jackson derivative from $D_q$ (see (\ref{e5.1})) to ${\mathcal D}_q$ (see (\ref{e5.1.5})). We do not expect any confusion due to that.}.
It connects two neighbouring points in the lattice space, see (\ref{e5.1}). The canonical partner of ${\mathcal D}_q$ is
\[
   x_q\ =\ \frac{A}{\{ A\}_q}\,x\ =\ x\,\frac{A+1}{\{ A+1\}_q}\ .
\]
Both ${\mathcal D}_q$ and $x_q$ are pseudodifferential operators. It seems important to mention
that ${\mathcal D}_q$, $x_q$ can be  related to $\pa_x$, $x$ via a similarity transformation \cite{Chryssomalakos:2001}
\[
   {\mathcal D}_q\ =\ U(A)^{-1}\, \pa_x\, U(A)\ ,\ x_q\ =\ U(A)^{-1}\, x\, U(A)\ ,\ A\equiv x \pa_x\ ,
\]
where
\[
   U(A)=\Gamma_q(A+1)/\Gamma(A+1)\ ,
\]
and $\Gamma(x+1) = x\, \Gamma_q(x)$ is the gamma function, and $\Gamma_q(x+1) = \{ x \} \Gamma_q(x)$ the $q$-deformed gamma function (see, e.g.~\cite{Exton:1983}).

The Euler operator $(b_q\, a_q)$ does not depend on $q$,
\begin{equation}
\label{e5.1.6}
  b_{q} a_{q}\ =\ x_q\,{\mathcal D}_q\ =\ x\,\pa_x \ =\ A \ ,
\end{equation}
hence, remains non-deformed. Hence, it is the local operator. Interestingly,
\[
   x_q x^n\ =\ \frac{n+1}{\{ n+1\}_q}\,x^{n+1}\ .
\]
The "vacuum" remains constant $|0>\  = \ 1$. In this representation
\begin{equation}
\label{e5.1.6n}
    \phi_{n} \equiv b_q^{n} |0>\ =\ \frac{n!}{\{ n \}_q!}  x^{n}\ ,
\end{equation}
where $n!=1 \cdot 2 \cdot \ldots n$ is factorial and
$\{ n \}_q! = \{ 1 \}_q \cdot \{ 2 \}_q \cdot \ldots \{ n \}_q$ is
the so-called $q$-factorial.

It is interesting to look at the classical limit of (\ref{e5.1.5}) when
the commutators are replaced by the Poisson brackets.
Indeed, with $p, q$ satisfying $\{p, q\} = 1$, where $\{ \cdot, \cdot \}$ is
the Poisson bracket, one can easily verify that
\[
\{f (A)\,p, q\,f^{-1} (A)\} = 1\ ,
\]
where $A = q p$ and $f(A)$ is a holomorphic function, giving rise to a wide
class of classical canonical transformations. To the best of our knowledge
such canonical transformations were introduced for the first time in
\cite{Chryssomalakos:2001}.

{\bf (a).}
It is easy to check that if the operators $a,b$ obey (\ref{e5.1.1}), then for any
$n \in {\bf C}$ the following three operators
\[
\hat J^+_n = b^2 a - n b\ ,
\]
\begin{equation}
\label{e5.a.1}
\hat J^0_n = ba - \frac{n}{2}\ ,
\end{equation}
\[
\hat J^-_n=a\ ,
\]
span the $\mathfrak{sl}(2,\bf{R})$-algebra with the commutation relations:
\[
[\hat J^0,\hat J^{\pm}]=\pm \hat J^{\pm}\ ,\  [\hat J^+,\hat J^-]=-2 \hat J^0\ .
\]

For the representation (\ref{e5.a.1}) the quadratic Casimir operator
is equal to
\begin{equation}
\label{e5.a.2}
C_2 \equiv \frac{1}{2}\{\hat J^+_n,\hat J^-_n\} - \hat J^0_n \hat J^0_n\ =\
-\frac{n}{2} \bigg(\frac{n}{2} + 1\bigg)\ ,
\end{equation}
where $\{\ ,\ \}$ denotes the anticommutator. If $n$ is a
non-negative integer, then (\ref{e5.a.1}) possesses a
finite-dimensional, irreducible representation in the Fock space
leaving invariant the space
\begin{equation}
\label{e5.a.3}
{\mathcal P}_{n}(b) \ = \ \langle 1, b, b^2, \dots , b^n \rangle
|0\rangle,
\end{equation}
of dimension $\dim{\mathcal P}_{n}=(n+1)$.

Substituting of (\ref{e5.1.3}) into (\ref{e5.a.1}) leads to a
well-known realization of the $\mathfrak{sl}(2,\bf{R})$-algebra as an algebra of
differential operators of the first order
\[
J^+_n\ =\ x^2 \pa_x - n x \ ,
\]
\begin{equation*}
J^0_n\ =\ x \pa_x - \frac{n}{2}\ ,
\end{equation*}
\[
J^-\ =\ \pa_x\ ,
\]
see (\ref{sl2r}), where the finite-dimensional representation space
(\ref{e5.a.3}) becomes the space of polynomials of degree not higher
than $n$
\begin{equation}
\label{e5.a.5}
{\mathcal P}_{n}(x) \ = \ \langle 1, x, x^2, \dots , x^n \rangle \ .
\end{equation}

{\bf (b-1).}\ (Uniform lattice)\, The existence of a non-trivial embedding of the
Heisenberg algebra into its extended universal enveloping
algebra, namely, $[\hat a (a,b),\hat b(a,b)]=[a,b]=1$ allows to
construct different representations of the algebra $\mathfrak{sl}(2,\bf{R})$ by
$a\rightarrow \hat a, b \rightarrow \hat b$ in (\ref{e5.a.1}).
In particular, such an embedding of the Heisenberg algebra into its extended
universal enveloping algebra is realized by the following two
operators \cite{Turbiner:1999},
\[
    \hat a\ =\ \frac{(e^{\de a} -1)}{\de}\ ,
\]
\begin{equation}
\label{e5.b.1}
   \hat b\ =\ b e^{ - \de a}\ ,
\end{equation}
where $\de$ can be any real (complex) number. If $\de$ goes to zero then
$\hat a \rightarrow a, \hat b \rightarrow b$. In other words,
(\ref{e5.b.1}) is a one-parameter quantum canonical transformation
of the deformation type of the Heisenberg algebra (\ref{e5.1.1}).
It is one of the (several) possible quantum analogies of a
point-to-point canonical transformation. The substitution of the
representation (\ref{e5.b.1}) into (\ref{e5.a.1}) results in the
following representation of the $\mathfrak{sl}(2,\bf{R})$-algebra
\[
J_n^+\ =\ b\bigg((\frac{b}{\de}-1) (1 - e^{ - \de a}) - n \bigg)
e^{ - \de a}\ ,
\]
\begin{equation}
\label{e5.b.2}
J_n^0\ =\ \frac{b}{\de} (1 - e^{ - \de a}) -\frac{n}{2}\ ,
\ J^-\ =\ \frac{1}{\de} ( e^{\de a}-1)\ .
\end{equation}
If $n$ is a non-negative integer, then (\ref{e5.b.2}) possesses a
finite-dimensional irreducible representation of dimension
$\dim{\mathcal P}_{n}=(n+1)$ coinciding with (\ref{e5.a.3}). It is
worth noting that the vacuum for (\ref{e5.b.1}) remains the same,
for instance (\ref{e5.1.2}). Also the value of the quadratic
Casimir operator for (\ref{e5.b.2}) coincides with that given by
(\ref{e5.a.2}).

After substitution of (\ref{e5.1.3}) into (\ref{e5.b.2}) we arrive at
a representation of the $\mathfrak{sl}(2,\bf{R})$-algebra by finite-difference operators
\cite{Smirnov:1995hj,Smirnov:1995a}, 
\[
J^+_n\ =\ x\bigg((\frac{x}{\de}-1) (1 - e^{ - \de \pa_x}) - n \bigg)
e^{ - \de \pa_x}\ ,
\]
\begin{equation}
\label{e5.b.5}
J^0_n\ =\ \frac{x}{\de} (1 - e^{ - \de \pa_x})-\frac{n}{2}\ ,
\ J^-\ =\ \frac{1}{\de} ( e^{\de \pa_x}-1)\ ,
\end{equation}
or, equivalently,
\[
J^+_n\ =\ -x\bigg((x-\de) {\mathcal D}_- - n \bigg)(1- \de {\mathcal D}_-)\ ,
\]
\begin{equation}
\label{e5.b.6}
J^0_n\ =\ x {\mathcal D}_- - \frac{n}{2}\ ,
\ J^-= \ {\mathcal D}_+ .
\end{equation}
The finite-dimensional representation space for
(\ref{e5.b.5})--(\ref{e5.b.6}) for integer values of $n$ is again
given by the space of polynomials of degree not
higher than $n$ ${\mathcal P}_{n}(x)$ (\ref{e5.a.5}).

{\bf (b-2).}\ (Exponential lattice)\, Another non-trivial embedding
of the Heisenberg algebra into its extended universal enveloping
algebra has been found (proved) in \cite{Chryssomalakos:2001}
\[
    \hat a\ =\ \frac{\{ ba + 1\}_q}{ba +1} a \ ,\
\]
\begin{equation}
\label{e5.b.7}
   \hat b\ =\ b\,\frac{ba+1}{\{ ba+1\}_q}\ =\ \frac{ba}{\{ ba\}_q}\,b\ ,
\end{equation}
(cf. (\ref{e5.b.1})), where $q$ is any real (complex) number and
$\{ ba+1\}_q = \frac{1-q^{ba+1}}{1-q}$ is the $q$ operator.
If $\de$ goes to zero, then $\hat a \rightarrow a, \hat b \rightarrow b$.
In other words, (\ref{e5.b.7}) is a one-parameter quantum canonical
transformation of the deformation type of the Heisenberg algebra (\ref{e5.1.1}).
It is one of the (several) possible quantum analogies of a
point-to-point canonical transformation. The substitution of the
representation (\ref{e5.b.7}) into (\ref{e5.a.1}) results in the
following representation of the $\mathfrak{sl}(2,\bf{R})$-algebra
\[
J_n^+\ =\ b\,\frac{ba+1}{\{ ba+1\}_q}\,(b a - n)\ =\ \frac{ba}{\{ ba\}_q}\,b\, (b a - n)   \ ,
\]
\begin{equation}
\label{e5.b.8}
J_n^0\ =\ b a  -\frac{n}{2}\ \ ,\
\ J^-\ =\ \frac{\{ ba + 1\}_q}{ba +1} a\ .
\end{equation}
If $n$ is a non-negative integer, then (\ref{e5.b.2}) possesses a
finite-dimensional irreducible representation of dimension
$\dim{\mathcal P}_{n}=(n+1)$ coinciding with (\ref{e5.a.3}). It is
worth noting that the vacuum for (\ref{e5.b.1}) remains the same,
for instance (\ref{e5.1.2}). Also the value of the quadratic
Casimir operator for (\ref{e5.b.2}) coincides with that given by
(\ref{e5.a.2}).

After substitution of (\ref{e5.1.3}) into (\ref{e5.b.8}) we arrive at
a representation of the $\mathfrak{sl}(2,\bf{R})$-algebra by finite-difference operators on
exponential lattice, \cite{Chryssomalakos:2001}
\[
J^+_n\ =\ \frac{(1-q)}{1- q^{x\pa_x}}\, x\,(x \pa_x +1)\,(x \pa_x - n)\ ,
\]
\begin{equation}
\label{e5.b.9}
J^0_n\ =\ x \pa_x - \frac{n}{2}\ ,
\end{equation}
\[
\ J^-\ =\ \frac{1- q^{x\pa_x+1}}{1-q}\,\frac{1}{1 + x\pa_x} \pa_x \ ,
\]
or, equivalently,
\[
J^+_n\ =\ x_q (x \pa_x - n)\ ,
\]
\begin{equation}
\label{e5.b.10}
J^0_n\ =\ x \pa_x - \frac{n}{2}\ ,\ J^-\ = \ {\mathcal D}_q\ .
\end{equation}
The finite-dimensional representation space for
(\ref{e5.b.9})--(\ref{e5.b.10}) for integer values of $n$ is again
given by the space of polynomials of degree not higher than $n$
${\mathcal P}_{n}(x)$ (\ref{e5.a.5}).

Taking into account that
\[
    {\mathcal D}_q\, x^k\ =\ \{k\}_q\, x^{k-1}\ ,\ x_q\, x^k\ =\ \frac{k+1}{\{k+1\}_q}\, x^{k+1}\ ,
\]
one find the action of generators on monomial $x^k$,
\[
J^+_n\,x^k\ =\ x_q (x \pa_x - n)\, x^k\ =\ \frac{k+1}{\{k+1\}_q} (k - n) \, x^{k+1}\ ,
\]
\begin{equation}
\label{e5.b.11}
J^0_n \,x^k\ =\ (x \pa_x - \frac{n}{2})\,x^k\ =\ \,(k - \frac{n}{2})\,x^k\ ,\
\end{equation}
\[
  J^-\,x^k\ = \ {\mathcal D}_q \,x^k\ =\ \{k\}_q \,x^{k-1}\ .
\]

{\bf (c).}\
Another example of quantum canonical transformation is given by the
oscillator representation
\[
\hat a = \frac{b+a}{\sqrt{2}}\ ,
\]
\begin{equation}
\label{e5.c.1}
\hat b = \frac{b-a}{\sqrt{2}}\ .
\end{equation}
Inserting (\ref{e5.c.1}) into (\ref{e5.a.1}) it is easy to check that
the following three generators form a representation of the
$\mathfrak{sl}(2,\bf{R})$-algebra,
\[
J^+_n = \frac{1}{2^{3/2}}[b^3 + a^3 -b(b+a)a -(2n+1)(b-a) - 2 b]\ ,
\]
\begin{equation}
\label{e5.c.2}
J^0_n = \frac{1}{2} (b^2-a^2 - n-1)\ ,
\end{equation}
\[
J^-= \frac{b+a}{\sqrt{2}}\ ,
\]
where $n$ is any real (complex) number. In this case the vacuum state
\begin{equation}
\label{e5.c.3}
(b+a)|0>\  = \ 0 ,
\end{equation}
differs from (\ref{e5.1.2}).
If $n$ is a non-negative integer, then (\ref{e5.c.2}) possesses
a finite-dimensional irreducible representation in a subspace
of the Fock space
\begin{equation}
\label{e5.c.4}
{\mathcal P}_{n}(b) \ = \ \langle 1, (b-a), (b-a)^2, \dots , (b-a)^n
\rangle |0\rangle \ ,
\end{equation}
of dimension $\dim{\mathcal P}_{n}=(n+1)$.

Taking $a,b$ in the realization (\ref{e5.1.3}) and substituting
them into (\ref{e5.c.2}), we obtain
\[
J^+_n = \frac{1}{2^{3/2}}[x^3 + \pa_x^3 -x(x+\pa_x)\pa_x
-(2n+1)(x-\pa_x) - 2 \pa_x]\ ,
\]
\begin{equation}
\label{e5.c.5}
J^0_n = \frac{1}{2} (x^2-\pa_x^2 - n-1)\ ,
\end{equation}
\[
J^-= \frac{x+\pa_x}{\sqrt{2}}\ ,
\]
which represents the $\mathfrak{sl}(2,\bf{R})$-algebra by means of differential
operators of finite order (but not of first order as in
(\ref{sl2r})). The operator $J^0_n$ coincides with the
Hamiltonian of the harmonic oscillator (with the reference point
for eigenvalues changed). The vacuum state is
\begin{equation}
\label{e5.c.6}
|0>\  = \ e^{-\frac{x^2}{2}}\ ,
\end{equation}
and the representation space is
\begin{equation}
\label{e5.c.7}
{\mathcal P}_{n}(x) \ = \ \langle 1, x, x^2, \dots , x^n \rangle\ 
e^{-\frac{x^2}{2}}\ ,
\end{equation}
(cf.(\ref{e5.c.4})).

{\bf (d).}
The following three operators
\[
J^+ = \frac{a^2}{2}\ ,
\]
\begin{equation}
\label{e5.d.1}
J^0 = - \frac{\{ a,b\}}{4}\ ,
\end{equation}
\[
J^-= \frac{b^2}{2}\ ,
\]
are generators of the $\mathfrak{sl}(2,\bf{R})$-algebra and the quadratic Casimir
operator for this representation is
\[
C_2 = \frac{3}{16}\ .
\]
This is the so-called metaplectic representation of the $sl_2$-algebra (see,
for example, \cite{Perelomov:1986}). This representation is
infinite-dimensional. Taking the realization (\ref{e5.1.2}) or
(\ref{e5.1.4}) of the Heisenberg algebra we get the well-known
representation
\begin{equation}
\label{e5.d.2}
J^+\ =\ \frac{1}{2}\pa_x^2\ ,\ J^0\ =\ - \frac{1}{2}(x\pa_x\ +\ \frac{1}{2})
\ ,
\ J^-\ =\ \frac{1}{2}x^2\
\end{equation}
in terms of differential operators, sometimes it is called the {oscillator representation},
or
\[
J^+\ =\ \frac{1}{2}{\mathcal D}_+^2\ ,\ J^0\ =\ - \frac{1}{2}
(x{\mathcal D}_-\ +\ \frac{1}{2})\ ,
\]
\begin{equation}
\label{e5.d.3}
J^-\ =\ \frac{1}{2}x(x\ -\ \de)(1-2\de{\mathcal D}_-\ -\ \de^2{\mathcal D}_-^2)\ ,
\end{equation}
in terms of finite-difference operators, correspondingly.

{\bf (e).}
For the sake of completeness let us take two operators $a$ and $b$ from
the Clifford algebra $s_2$,
\begin{equation}
\label{e5.e.1}
             \{a,b\} \equiv ab  +  ba \ =\ 0\ ,\ a^2=b^2=1\ .
\end{equation}
It can be shown that the operators
\begin{equation}
\label{e5.e.2}
J^1= a\ ,\ J^2= b\ , \ J^3= ab\ ,
\end{equation}
span the $sl_2$-algebra.

\section{Spectral problem in the Fock space}

Let $L(b,a)$ is a polynomial in $a,b$ - the generators of the Heisenberg algebra, $[a,b]=1$.
Define the eigenvalue problem in the Fock space, see e.g. \cite{Smirnov:1995a}, as
\begin{equation}
\label{e5.2.eigen}
    L(b,a) \phi (b)\, |0>\  = \ \veps \phi (b)\, |0>\ ,
\end{equation}
where $\veps$ is spectral parameter. Instead of boundary conditions we impose a condition that
we are looking for eigenpolynomials in $b$.
Technically, the problem of finding eigenpolynomial is reduced to reordering (the normal ordering)
of $L(b,a) \phi (b)$ to superposition of monomials $b^p a^q$,
\[
    L(b,a) \phi (b)\ =\ \sum_0 L_i (b) a^i \ ,
\]
and then making a study of $L_0(b)$. In addition to standard relation,
\[
     [a, b^k]\ =\ k b^{k-1}\ ,\ [a^k, b]\ =\ k a^{k-1}\ ,
\]
several identities can be useful \cite{Fleury:1994,Turbiner:1995},
\[
      (aba)^k\ =\ a^k b^k a^k\ , \  (bab)^k\ =\ b^k a^k b^k\ ,
\]
\[
       [a^k b^k , a^m b^m]\ =\ 0\ ,
\]
where $k, m$ are integer.

In general, if the operator $L(b,a)$ can be represented in terms of the $\mathfrak{sl}(2,\bf{R})$-algebra generators (\ref{e5.a.1}) at integer $n$,
\[
    L(b,a)\ =\ L_{qes}(\hat J^+_n, \hat J^0_n, \hat J^-_n)\ ,
\]
it is evident there exists $(n+1)$ polynomial solutions of (\ref{e5.2.eigen}), thus, the problem is quasi-exactly-solvable. If the generator $\hat J^+_n$ is not present,
\[
    L(b,a)\ =\ L_{es}(\hat J^0_n, \hat J^-_n)\ ,
\]
it is evident there exists infinitely-many polynomial solutions of (\ref{e5.2.eigen}), thus, the problem is
exactly-solvable. As an illustration let us calculate the spectra of polynomial solutions for the Cartan
generator $\hat J^0_0 \equiv \hat J^0$.

{\it Example.} Let

\[
     L(b,a)\ =\ \hat J^0 = ba\ ,
\]
It is easy to check that
\[
     \phi_k\ =\ b^k\ ,\ \veps_k\ =\ k\ .
\]

It can be drawn the immediate conclusion: changing the operator $L$ by adding $J^-_n F(J^0_n, J^-_n)$, where $F$ is any polynomial, is isospectral. The eigenpolynomial $\phi_k$ is modified getting extra monomials of degrees less than $k$.

Concrete realizations of the Heisenberg algebra in terms of differential or finite-difference operators leads to the
spectral problem for differential or finite-difference operators with the same spectra!

1) in coordinate-momentum representation (\ref{e5.1.3}) the operator $J^0$ becomes the Euler operator

\[
    L(x, \pa_x)\ =\ x \pa_x\ ,
\]
and
\[
     \vphi_k(x) \ =\ \phi_k(b)\,|0>\ =\ x^k\ ,\ \veps_k\ =\ k\ .
\]

2) in $\de-$representation (\ref{e5.1.4}) on the uniform lattice the operator $J^0$ becomes

\[
   L(x_{\de}, D_{\de})\ =\  x {\mathcal D}_-\ ,
\]
while the spectral problem is two-point finite-difference equation,
\[
     \frac{x}{\de} \vphi(x) - \frac{x}{\de} \vphi(x-\de) \ =\ \veps \vphi(x)
\]
and with infinitely-many polynomial eigenfunctions,
\[
    \vphi_k(x)\ =\ \phi_k(b)\,|0>\ =\ x^{(k)}(\de) \equiv x(x-\de)(x -2\de)\ldots \big(x - (k-1)\de\big) {}\ ,\ \veps_k\ =\ k\ .
\]

3) in $q-$representation (\ref{e5.1.5}) on the exponential lattice the operator $J^0$ becomes

\[
    L(x, \pa_x)\ =\ x \pa_x\ ,
\]
and
\[
    \vphi_k(x)\ =\ \phi_k(b)\,|0>\ =\ x^k\ ,\ \veps_k\ =\ k\ .
\]

Concluding one can state the property of isospectrality of the spectral problem in the Fock space with respect to
quantum canonical transformations, $[a,b] = [\hat a (a,b), \hat b (a,b)]=1$:
\begin{quote}
 {\it If vacuum $|0>$ remains unchanged under quantum canonical transformations
\[
      a |0>\ =\ \hat a |0>\ =\ 0\ ,
\]
the spectra of the operator $L(b,a)$ with polynomial eigenfunctions $\phi(b)$ coincides with the spectra
of the operator $L(\hat b, \hat a)$ with polynomial eigenfunctions $\phi(\hat b)$.}
\end{quote}
Among known quantum canonical transformations realized in action on functions there are two special:
one associated with shift operator,
\[
      T_{\de} f(x)\ =\ f(x + \de)\ ,
\]
see (\ref{e5.b.1}), and another one associated with dilatation operator,
\[
      T_{q} f(x)\ =\ f(q x)\ ,
\]
see (\ref{e5.b.7}), respectively. It gives a chance to make use of these transformation for
{\it isospectral discretization}, connecting linear differential operators with finite-difference
linear operators. It also allows us to connect, preserving isospectrality, different finite-difference operators,
\[
    L(b,a)\ \Leftrightarrow \ L(b_{\de},a_{\de})\ \Leftrightarrow \ L(b_{q},a_{q})\ .
\]
It can be proved \cite{Chryssomalakos:2001} that if the eigenvalue problem
\[
   L(x, \pa_x)\,\vphi(x)\ =\ \la \ \vphi(x)\ ,
\]
has a polynomial eigenfunction
\[
  \vphi(x)\ =\ \sum_k^N  a_k x^k\ ,
\]
at a certain $\la_N$, then the eigenvalue problem
\[
   L(x_{\de}, {\mathcal D}_+)\, \vphi_{\de}(x)\ =\ \la \ \vphi_{\de} (x)\ ,
\]
has a polynomial eigenfunction
\[
  \vphi_{\de}(x)\ =\ \sum_k^N  a_k x^{(k)}\ ,
\]
for the same $\la_N$, then the eigenvalue problem
\[
   L(x_{q}, {\mathcal D}_q) \,\vphi_{q}(x)\ =\ \la \ \vphi_{q}(x)\ ,
\]
has a polynomial eigenfunction
\[
  \vphi_{q}(x)\ =\ \sum_k^N \ a_k\, \frac{k!}{\{ k \}_q!} \, x^k\ ,
\]
for the same $\la_N$. Assume the operator $L(b,a)$ can be rewritten in terms of the
$\mathfrak{sl}(2,\bf{R})$-algebra generators (\ref{e5.a.1}), hence, it is the $\mathfrak{sl}(2,\bf{R})$-Lie-algebraic operator.
Then $L(x, \pa_x)$ is the $\mathfrak{sl}(2,\bf{R})$-Lie-algebraic differential operator as well as the operators
$L(x_{\de}, {\mathcal D}_+)$ and $L(x_{q}, {\mathcal D}_q)$ which are the $\mathfrak{sl}(2,\bf{R})$-Lie-algebraic finite-difference operators. Transition from differential operator $L(x, \pa_x)$ to $L(x_{\de}, {\mathcal D}_+)$ or to $L(x_{q}, {\mathcal D}_q)$ is nothing but the {\it $\mathfrak{sl}(2,\bf{R})$-Lie-algebraic discretization} for which polynomial eigenfunctions remains polynomial ones with the property isospectrality of the corresponding eigenvalues. Thus, for all (quasi)-exactly-solvable problems, see Cases I-XII, it can be constructed ``polynomially-isospectral", {\it quasi-exactly-solvable discrete}
systems on uniform or exponential lattice:
they have $(n+1)$ polynomial eigenfunctions in a form of polynomial in $x$ of the degree $n$
with the same eigenvalues.
It will be presented in the next Section.

\section{Quasi-exactly-solvable finite-difference operators}

In this Section we give a list of the quasi-exactly-solvable operators of Cases I-X, XII
in the Fock space; for some cases their representations on the uniform and exponential lattices
are also given.

{\bf Case I.}

\vskip .2cm

Let us take the $\mathfrak{sl}(2,\bf{R})$-Lie algebraic operator $T_2/\al$ (\ref{e1.3.25}). Then
substitute the generators (\ref{e5.a.1}) into (\ref{e1.3.25}). We get the
$\mathfrak{sl}(2,\bf{R})$-Lie algebraic operator in the Fock space of Case I,
\begin{equation}
\label{e5.I.1}
     T_2(\hat b, \hat a)\ =\ -\al\, {\hat b}^2 {\hat a}^2\ +\
 [2a\,{\hat b}^2 + (2b-\al)\,{\hat b} - 2c]\, {\hat a} - 2 a\, n\, {\hat b} \ ,
\end{equation}
cf. (\ref{e1.3.25.1}),
where $[{\hat a},\,{\hat b}]=1$. In $\de-$representation (\ref{e5.1.4}) on the uniform lattice, see Fig.~\ref{Uni-Latt}, the operator (\ref{e5.I.1}) becomes,
\[
T_2(x_{\de}, D_{\de})\ =\ -(\al + 2\de a) x(x-\de) {\mathcal D}_- {\mathcal D}_- \ +\
\]
\begin{equation}
\label{e5.I.2}
    [2ax + 2a \de (n-1) + 2b - \al] x {\mathcal D}_-\ -\ 2c{\mathcal D}_+ - 2 a n x\ ,
\end{equation}
It is the four-point, quasi-exactly-solvable, finite-difference operator, which is polynomially-isospectral to (\ref{e1.3.25}). In $q-$representation (\ref{e5.1.5}) on the exponential lattice, see Fig.~\ref{Exp-Latt}, the operator (\ref{e5.I.1}) becomes,
\begin{equation}
\label{e5.I.3}
    T_2(x_{q}, {\mathcal D}_{q})\ =\ -\al x \pa_x x \pa_x\ +\
    2ax_q x \pa_x +2b x \pa_x -\ 2c{\mathcal D}_q - 2 a n x_q\ .
\end{equation}
It is the quasi-exactly-solvable differential-difference operator, which is polynomially-isospectral to (\ref{e1.3.25}) and (\ref{e5.I.2}).

{\bf Case II.}

\vskip .2cm

Let us take the $\mathfrak{sl}(2,\bf{R})$-Lie algebraic operator $T_2$ (\ref{e1.3.27}). Then
substitute again the generators (\ref{e5.a.1}) into (\ref{e1.3.27}). We get the
$\mathfrak{sl}(2,\bf{R})$-Lie algebraic operator in the Fock space of Case II,
\begin{equation}
\label{e5.II.1}
 T_2(\hat b, \hat a)\ =\ -\al\, {\hat b} {\hat a}^2\ -\
 (2a\,{\hat b}^2 + 2c\,{\hat b} - 2b - \al)\, {\hat a} + 2 a\, n\, {\hat b} \ ,
\end{equation}
cf. (\ref{e1.3.27.1}).

{\bf Case III.}

\vskip .2cm

Let us take the $\mathfrak{sl}(2,\bf{R})$-Lie algebraic operator $T_2$ (\ref{e1.3.29}) and
substitute the generators (\ref{e5.a.1}) into (\ref{e1.3.29}). We get the
$\mathfrak{sl}(2,\bf{R})$-Lie algebraic operator in the Fock space of Case III,
\begin{equation}
\label{e5.III.1}
 T_2(\hat b, \hat a)\ =\ -\al\, {\hat b}^3 {\hat a}^2\ +\
 [(2b-\al)\,{\hat b}^2 - 2a\,{\hat b} - 2c]\, {\hat a} + (\al\, n -2 b)\, {\hat b} \ ,
\end{equation}
cf. (\ref{e1.3.27.1}).

{\bf Case IV.}

\vskip .2cm

Let us take the $\mathfrak{sl}(2,\bf{R})$-Lie algebraic operator $T_2$ (\ref{e1.3.31}) and then
substitute the generators (\ref{e5.a.1}) into it. We get the $\mathfrak{sl}(2,\bf{R})$-Lie algebraic operator in the Fock space of Case IV,
\begin{equation}
\label{e5.IV.1}
 T_2(\hat b, \hat a)\ =\ -\al\, {\hat b}^3 {\hat a}^2\ +\
 [(2b-\al)\,{\hat b}^2 - 2a\,{\hat b} - 2c]\, {\hat a} + (\al\, n -2 b)\, {\hat b} \ ,
\end{equation}
cf. (\ref{e1.3.31.1}).

{\bf Case V.}

\vskip .2cm

Let us take the $\mathfrak{sl}(2,\bf{R})$-Lie algebraic operator $T_2$ (\ref{e1.3.34}). After
substitution of the generators (\ref{e5.a.1}) into (\ref{e1.3.34}), we get the
$\mathfrak{sl}(2,\bf{R})$-Lie algebraic operator in the Fock space of Case V,
\[
 T_2(\hat b, \hat a)\ =\ - 2 \al\, {\hat b}({\hat b}-1) {\hat a}^2\ +\
 [2b\,{\hat b}^2 -(2a + 4b + 2p \al+ 3\al)\,{\hat b} + 2 (\al + a + b)]\, {\hat a}
\]
\begin{equation}
\label{e5.V.1}
 -\, 2 b n\, {\hat b} + b (2n + p) \ ,
\end{equation}
cf. (\ref{e1.3.34.1}).

{\bf Case VI.}

\vskip .2cm

Let us take the $\mathfrak{sl}(2,\bf{R})$-Lie algebraic operator $T_2$ (\ref{e1.3.35a}). Then
substitute the generators (\ref{e5.a.1}) into (\ref{e1.3.35a}). We get the
$\mathfrak{sl}(2,\bf{R})$-Lie algebraic operator in the Fock space of Case VI,
\begin{equation}
\label{e5.VI.1}
 T_2(\hat b, \hat a)\ =\ - 4\, {\hat b} {\hat a}^2\ +\
 2(2a\,{\hat b}^2\,+\,2b\,{\hat b} - 1-2p) {\hat a}\ -\ 4 a n \,{\hat b}
 \ ,
\end{equation}
cf. (\ref{e1.3.35b}), where $[{\hat a},\,{\hat b}]=1$. In $\de-$representation (\ref{e5.1.4}) on the uniform lattice, see Fig.~\ref{Uni-Latt}, the operator (\ref{e5.VI.1}) becomes,
\[
    T_2(x_{\de}, D_{\de})\ =\ -4 a \de\, x(x-\de)\, {\mathcal D}_- {\mathcal D}_- \ +\
\]
\begin{equation}
\label{e5.VI.2}
    4 x [a x + a \de (n-1) + b + \frac{1}{\de}] {\mathcal D}_-\ -\ 2(\frac{2 x }{\de}-1 - 2p){\mathcal D}_+ - 4 a n \,x\ .
\end{equation}
It is the four-point, quasi-exactly-solvable, finite-difference operator, which is polynomially-isospectral to (\ref{e1.3.35a}). In $q-$representation (\ref{e5.1.5}) on the exponential lattice, see Fig.~\ref{Exp-Latt}, the operator (\ref{e5.VI.1}) becomes,
\begin{equation}
\label{e5.VI.3}
    T_2(x_{q}, {\mathcal D}_{q})\ =\ -4 x \pa_x {\mathcal D}_q\ +\
    4ax_q x \pa_x +4b x \pa_x -\ 2(1+2p) {\mathcal D}_q - 4 a n \,x_q\ .
\end{equation}
It is the quasi-exactly-solvable differential-difference operator, which is polynomially-isospectral to (\ref{e1.3.25}) and (\ref{e5.VI.2}). It is also polynomially-isospectral to the Hamiltonian (\ref{e2.3}) at $a \geq 0$.

{\bf Case VII.}

\vskip .2cm

Let us take the $\mathfrak{sl}(2,\bf{R})$-Lie algebraic operator $T_2$ (\ref{e1.3.38}) and
substitute the generators (\ref{e5.a.1}) into it. We get the $\mathfrak{sl}(2,\bf{R})$-Lie algebraic operator in the Fock space of Case VII,
\begin{equation}
\label{e5.VII.1}
 T_2(\hat b, \hat a)\ =\ - 4\, {\hat b} {\hat a}^2\ +\
 2(2a\,{\hat b}^2\,+\,2b\,{\hat b} - d - 2 l + 2 c) {\hat a}\ -\ 4 a n \,{\hat b}
 \ ,
\end{equation}
cf. (\ref{e1.3.38a}).

{\bf Case VIII.}

\vskip .2cm

Let us take the $\mathfrak{sl}(2,\bf{R})$-Lie algebraic operator $T_2$ (\ref{e1.3.40}). After
substitution the generators (\ref{e5.a.1}) into it we get the
$\mathfrak{sl}(2,\bf{R})$-Lie algebraic operator in the Fock space of Case VIII,
\begin{equation}
\label{e5.VIII.1}
 T_2(\hat b, \hat a)\ =\ - {\hat b} {\hat a}^2\ +\
 (2a\,{\hat b}^2\,+\,2b\,{\hat b} - d - 2 l + 2 c +1) {\hat a}\ -\ 2 a n \,{\hat b}
 \ ,
\end{equation}
cf. (\ref{e1.3.40a}).

{\bf Case IX.}

\vskip .2cm

Let us take the $\mathfrak{sl}(2,\bf{R})$-Lie algebraic operator $T_2$ (\ref{e1.3.42}). Then
substitute the generators (\ref{e5.a.1}) into (\ref{e1.3.42}). We get the
$\mathfrak{sl}(2,\bf{R})$-Lie algebraic operator in the Fock space of Case IX,
\begin{equation}
\label{e5.IX.1}
 T_2(\hat b, \hat a)\ =\ - {\hat b} {\hat a}^2\ -\
 [2a\,{\hat b}^2\,+\,(2c-d-2l+1)\,{\hat b} -2b] {\hat a}\ -\ 2 a n \,{\hat b}
 \ ,
\end{equation}
cf. (\ref{e1.3.42-1}).

{\bf Case X.}

\vskip .2cm

Let us take the $\mathfrak{sl}(2,\bf{R})$-Lie algebraic operator $T_2/\al^2$ (\ref{e1.3.44}). Then substitute the generators (\ref{e5.a.1}) into (\ref{e1.3.44}). We get the
$\mathfrak{sl}(2,\bf{R})$-Lie algebraic operator in the Fock space of Case X,
\begin{equation}
\label{e5.X.1}
 T_2(\hat b, \hat a)\ =\ ({\hat b}^2-1)\, {\hat a}^2\ +\
 [2a\,{\hat b}^2\,+\,(1+2\mu)\,{\hat b} -2 a] {\hat a}\ -\ 2 a (n-\mu) \,{\hat b}
 \ ,
\end{equation}
cf. (\ref{e1.3.44d}). It is polynomially-isospectral to the Hamiltonian (\ref{e1.3.45H}).

{\bf Case XII.}

\vskip .2cm

{\it I.}

Let us take the $\mathfrak{sl}(2,\bf{R})$-Lie algebraic operator $h_B$ (\ref{htau-sl2}) and
substitute the generators (\ref{e5.a.1}) into (\ref{htau-sl2}). We get the
$\mathfrak{sl}(2,\bf{R})$-Lie algebraic operator in the Fock space of Case XII (I),
\begin{equation}
\label{e5.XII-I.1}
 h_B(\hat b, \hat a)\ =\ (4 {\hat b}^3 - g_2 {\hat b} -\ g_3)\, {\hat a}^2\ +\
(1+2\mu) (6\,{\hat b}^2\,-\,\frac{g_2}{2})] {\hat a}\ -\ 2 n\,(2n+1+6\mu) \,{\hat b}
 \ ,
\end{equation}
cf. (\ref{htauB}). At $\mu=0$ the operator (\ref{e5.XII-I.1}) degenerates to the L\'ame case, see Case XI.

{\it II.}

Let us take the $\mathfrak{sl}(2,\bf{R})$-Lie algebraic operator $h_B$ (\ref{htau-sl2-k}). Then
substitute the generators (\ref{e5.a.1}) into (\ref{htau-sl2-k}). We get the
$\mathfrak{sl}(2,\bf{R})$-Lie algebraic operator in the Fock space of Case XII (II),
\[
  h_{B,k}(\hat b, \hat a)\ =\ (4 {\hat b}^3 - g_2 {\hat b} -\ g_3)\,{\hat a}^2\ +
\]
\begin{equation}
\label{e5.XII-II.1}
  \big(2(5+2\mu)\,{\hat b}^2\,+\, 4(1-2\mu)\,e_k\,({\hat b} + e_k) \
  - \ (3-2\mu)\,\frac{g_2}{2} \big) {\hat a}\ -\ 2 n\,(2n + 3 + 2\mu) \,{\hat b} \ ,
\end{equation}
cf. (\ref{htauB-k}). At $\mu=0$ the operator (\ref{e5.XII-II.1}) degenerates to the L\'ame case, see Case XI.

\section{Quasi-exactly-solvable 3-point finite-difference operators (Generalized Hahn Polynomials)}

Typical second-order finite-difference equation relates an unknown function
at three lattice points. In a form of the eigenvalue problem it is written as
\begin{equation}
\label{e5.5.1}
 A(x) \varphi (x+\de) - B(x)\varphi (x)+ C(x) \varphi (x-\de)\ =\ \la \varphi (x)\ ,
\end{equation}
where $A(x), B(x), C(x)$ are some functions and $\la$ is spectral parameter.
The equation (\ref{e5.5.1}) can also be rewritten in operator form
\begin{equation}
\label{e5.5.2}
 \bigg(A(x) e^{\de \pa_x}\ -\ B(x)\ +\ C(x) e^{-\de \pa_x}\bigg) \varphi (x)\ =\
 \la \varphi (x)\ .
\end{equation}
The celebrated Harper equation, which appears, in particular, in the Azbel-Hofstadter problem 
(see e.g. \cite{Wiegmann:1992,Wiegmann:1994}) is of this type.

One can pose a natural question: {\it what are the coefficient
functions $A(x), B(x)$, $C(x)$ for which the equation (\ref{e5.5.1}) or (\ref{e5.5.2})
admits a finite number of polynomial eigenfunctions ?} It is evident if the operator
\begin{equation}
\label{e5.5.2o}
  H_{\de}\ =\  A(x) e^{\de \pa_x}\ -\ B(x)\ +\ C(x) e^{-\de \pa_x}\ ,
\end{equation}
in the r.h.s. of (\ref{e5.5.2}) can be rewritten in terms of generators of $\mathfrak{sl}(2,\bf{R})$-algebra, realized as finite-difference operators (\ref{e5.b.5}), in finite-dimensional representation (hence, the $n$ is an integer number) polynomial eigenfunctions can occur. 

Instead of developing the general theory of polynomial eigenfunctions of
the finite-difference operators we limit ourselves by a consideration of the (generalized)
Hahn polynomials as the eigenfunctions of a certain operator which we will call the {\it Hahn operator}. As for potential applications of quantum canonical transformations
discussed in previous Sections, we present here various deformations
of the Hahn operator and its eigenfunctions, the deformed Hahn polynomials.
Needless to say that the Hahn polynomials are among the most important
(and the most complicated) polynomials of discrete variable which appear
in numerous applications (see e.g. \cite{Askey:1985,Atakishiev:1985,Nikiforov:1991}
and references therein). Their degenerations contain the celebrated Meixner, Charlier,
Tschebyschov and Krawtchouk polynomials.

Take the spectral problem
\[
\frac{1}{\de^2}(A_1 x^2 + A_2 x + A_4 \de) \, f(x+\de)\ -
\]
\[
\frac{1}{\de^2}[2 A_1 x^2 - (\de A_1 - 2 A_2 + A_3 \de + A_4\de^2)x] \, f(x) \ +
\]
\begin{equation}
\label{hahn}
 \frac{1}{\de^2}[A_1 x^2 - (\de A_1 - A_2 + A_3\de )x ] \, f(x-\de)\ =\ \la f(x) \ .
\end{equation}
where $A_{1,2,3,4}$ and $\de$ are parameters, and $\la$ is spectral parameter. 
It can be easily verified that the equation (\ref{hahn}) has infinitely-many polynomial
eigenfunctions with the eigenvalues
\begin{equation}
\label{hahn-spec}
 \la_k\ =\ A_1\, k^2 + A_3\, k\quad ,\quad k=0,1,2,\ldots\ ,
\end{equation}
which depend on $k$ quadratically at large $k$. It was shown in \cite{Smirnov:1995a} 
that the operator (\ref{hahn}) is the most general three-point finite-difference possessing infinitely-many polynomial eigenfunctions. The operator in the rhs of (\ref{hahn})
\[
  H_{\de}\ =\  \frac{1}{\de^2}(A_1 x^2 + A_2 x + A_4 \de) \,  e^{ \pa_x}\ -\
  \frac{1}{\de^2}[2 \de A_1 x^2 - (\de A_1 - 2 A_2 + A_3 \de + A_4\de^2)x] \,
\]
\begin{equation}
\label{e5.5.3}
  \ +\ \frac{1}{\de^2}[A_1 x^2 - (\de A_1 - A_2 + A_3\de )x ]\, e^{- \pa_x}\ ,
\end{equation}
is called {\it the Hahn operator}.

Without a loss of generality we put spacing $\de = 1$ and $ A_1=-1$, and parametrize other
parameters as
\begin{equation}
\label{hahn-par}
 A_2= N - 2 - \beta\ ,\ A_3= -\al-\beta-1\ ,\ A_4=(\beta+1)(N-1) \ ,
\end{equation}
where $N$ and $\al, \beta$ are new parameters. We call the corresponding eigenfunctions
the {\it Hahn polynomials of continuous argument} $h_k^{(\al,\beta;\de)} (x, N)$
\footnote{
It must be emphasized that the Hahn polynomials of the {\it continuous} argument  {\it do not} coincide to the so-called continuous Hahn polynomials known in literature \cite{Askey:1985}}.
Besides that, if we choose
\[
 A_1\,=\,1\ ,\ A_2\,=\, 2-2N-\nu\ ,\ A_3\,=\, 1-2N-\mu-\nu\ ,\ A_4\,=\,(N+\nu-1)(N-1)\ ,
\]
where $N$ and $\nu, \mu$ are new parameters, the so-called analytically-continued Hanh polynomials $\bar h_k^{(\mu,\nu;\de)} (x, N)\ ,\ k=0,1,2\ldots$ as the eigenfunctions appear, 
see e.g. \cite{Nikiforov:1991}.

Taking in (\ref{e5.5.3})
\[
A_1=0\ ,\ A_2= -\mu \ ,\ A_3= \mu -1, A_4= \gamma \mu \ ,
\]
we reproduce the operator having the Meixner polynomials as the eigenfunctions.
Furthermore, if
\[
A_1=0\ ,\ A_2= 0\ ,\ A_3 =  -1\ ,\ A_4=  \mu \ ,
\]
the operator (\ref{e5.5.3}) has the Charlier polynomials as
the eigenfunctions  (for the definition of the Meixner and Charlier polynomials
see e.g. \cite{Nikiforov:1991}). For a certain particular choice of the parameters,
one can reproduce the equations having Tschebyschov and Krawtchouk
polynomials as the solutions. It must be emphasized that if $x$ is the
continuous argument, we will arrive at {\it continuous} analogues of all the 
above-mentioned polynomials. These polynomials are poorly studied in literature, 
for discussion see \cite{Smirnov:1995hj}.

Since the operator (\ref{e5.5.3}) is the most general exactly-solvable
finite-difference three-point operator, one can construct corresponding
\begin{equation}
\label{3-points-Js}
      H_{\de}\ =\ A_1 J^0 J^0 (\de J^- + 1)\ +\ A_2 J^0J^- +
                 A_3 J^0 + A_4 J^- \ ,
\end{equation}
which is the cubic polynomial in generators from the universal enveloping $\mathfrak{sl}(2,\bf{R})$-algebra leading to (\ref{hahn}). Hence, the Hahn polynomials 
are related to the finite-dimensional representation of a cubic element of the universal enveloping $\mathfrak{sl}(2,\bf{R})$-algebra.

If $N$ is a positive integer number and the variable $x$ is restricted
to a lattice $x=0,1,2\ldots, (N-1)$, these Hahn polynomials of
continuous argument coincide to the standard Hahn polynomials
$h_k^{(\al,\beta)} (x, N)$ of the {\it discrete} argument (we
use the notation of \cite{Nikiforov:1991}). The Hahn polynomial of
continuous argument can be represented as
\begin{equation}
\label{hahn-pol}
  h_k^{(\al,\beta;\de)} (x, N) = \sum_{i=0}^{k} \gamma_i x^{(i)}\ ,
\end{equation}
where $x^{(i+1)}\equiv x (x-\de)\ldots (x-i\, \de)$ is the
so-called {\it $\de-$quasi-monomial} (see (\ref{e5.1.4n})) and
$\gamma_i$ are known coefficients. If the parameter $N$ is integer
and $x$ is continuous argument, then for the higher Hahn
polynomials $k\geq N$ there is the property,
\begin{equation}
\label{hahn-pol-deg}
  h_k^{(\al,\beta;\de)} (x, N)=x^{(N)}p_{k-N}(x)\  ,
\end{equation}
where $p_{k-N}(x)$ is a Hahn polynomial of degree $(k-N)$.
It explains why a finite number of the Hahn polynomials solely
exists if continuous argument $x$ is restricted to the finite
lattice $x=0,1,2\ldots, (N-1)$.

In \cite{Smirnov:1995a,Smirnov:1995hj} it was proven that
the operator $H_{\de}$ in the r.h.s. (\ref{hahn}) belongs to the
Heisenberg-Weyl universal enveloping algebra,
\begin{equation}
\label{adbd-hahn}
 H_{\de}=A_1\, b_{\de}a_{\de}b_{\de}a_{\de}(\de a_{\de} + 1)
 + A_2 b_{\de}a_{\de}^2 + A_3 b_{\de}a_{\de} + A_4 a_{\de} \ ,
\end{equation}
cf. (\ref{3-points-Js}), where
\[
 a_\de = \de^{-1}
(e^{\de \pa} -1) \, , \qquad \qquad b_\de = x
 e^{-\de \pa} \, ,
\]
where $[a_\de, \, b_\de] =1$ -- this is the explicit $\de$-realization of
the quantum canonical commutation relations (\ref{e5.1.4}), (\ref{e5.b.1}).

Replacing in (\ref{adbd-hahn}), $a_\de, \, b_\de \rightarrow
a,b$ with $a=\pa, b=x$ we arrive at a linear differential
operator isospectral to (\ref{adbd-hahn}) and thus to (\ref{hahn}),
\begin{equation}
\label{ab-hahn}
 H\ =\
 \de A_1 x^2 {\pa}^3\ +\
 [(\de A_1+A_2)+ A_1 x] x {\pa}^2\ +\ [A_4 + (A_1 + A_3)x] {\pa} \ .
\end{equation}
This operator has infinitely-many polynomial eigenfunctions: they are
simply related to the Hahn polynomials,
\begin{equation}
\label{hahn-pol-dif}
 {\tilde h}_k^{(\al,\beta)} (x, N) = \sum_{i=0}^{k} \gamma_i x^i\ ,
\end{equation}
and, in particular,
\begin{equation}
\label{hahn-pol-dif-deg}
 {\tilde h}_N^{(\al,\beta)} (x, N) = \gamma_N x^N\ ,
\end{equation}
cf. (\ref{hahn-pol}). Explicitly, their eigenvalues are given by (\ref{hahn-spec}). 
Note that since the operator (\ref{ab-hahn}) is of the third order there
is no continuous measure with respect of which the polynomials (\ref{hahn-pol-dif}) are orthogonal.

Taking $q-$realization of the Heisenberg algebra, see (\ref{e5.1.5}) and (\ref{e5.b.7}), in the Fock space representation,
\[
    \hat a\ =\ \frac{\{ ba + 1\}_q}{ba +1}\, a \ ,
\]
\[
  \hat b\ =\ b\,\frac{ba+1}{\{ ba+1\}_q}\ =\ \frac{ba}{\{ ba\}_q}\,b\ ,
\]
where $[\hat a, \hat b]=1$, and replacing in (\ref{adbd-hahn}), 
$(a_\de, \, b_\de) \rightarrow (\hat a, \hat b)$ and then by $(\pa_q, x_q)$ 
we get the remarkable, non-local, differential-difference operator
\begin{equation}
\label{aqbq-hahn}
 H_{q}\ =\  A_1 x\, \pa x\, \pa\,(\de\, \pa_q\, +\, 1)\ +\ A_2 x\,\pa\, \pa_q\ +\ A_3 x\, \pa\ +\  A_4 \pa_q \ ,
\end{equation}
which is (polynomially)-isospectral to both (\ref{adbd-hahn}) and (\ref{ab-hahn}) 
for any $q$. The operator (\ref{aqbq-hahn}) has infinitely-many polynomial eigenfunctions 
and they are closely related to the Hahn polynomials,
\begin{equation}
\label{hahn-pol-q}
 h_k^{(\al,\beta;q)} (x, N) = \sum_{i=0}^{k} \gamma_i
 \ \frac{i!}{\{ i \}_q!}\  x^i\ ,
\end{equation}
and
\begin{equation}
\label{hahn-pol-q-deg}
 h_N^{(\al,\beta;q)} (x, N) = \gamma_N \ \frac{N!}{\{ N \}_q!}\ x^N \ ,
\end{equation}
(c.f.(\ref{hahn-pol-dif}), (\ref{hahn-pol-dif-deg})), where the corresponding eigenvalues 
are given by (\ref{hahn-spec}).

It seems the presence of the eigenfunction of the type (\ref{hahn-pol-q-deg})
indicates that there is no continuous measure with respect of which the polynomials (\ref{hahn-pol-q}) are orthogonal. It also hints that the operator (\ref{aqbq-hahn}) 
is not self-adjoint.

It is evident that the differential operator (\ref{ab-hahn}) continues to have a finite-dimensional invariant subspace in polynomials even though the replacement
$(\pa, \, x) \rightarrow (\pa_q,\, x)$ is made, which {\it not} a canonical transformation. 
Thus, the underlying Heisenberg algebra is replaced by $q$-Heisenberg algebra (\ref{e5.1-2}). 
In this case we get a non-local, finite-difference
operator
\begin{equation}
\label{aqb-hahn}
 H\ =\
 \de A_1 x^2\, {\pa_q}^3\ +\ [(\de A_1+A_2)\ +\ A_1 x]\, x\, {\pa_q}^2\ +
 \ [A_4 + (A_1 + A_3)\,x]\, {\pa_q} \ ,
\end{equation}
(c.f.(\ref{aqbq-hahn})). Certainly, this operator is not isospectral to (\ref{aqbq-hahn}), 
but still has infinitely-many polynomial eigenfunctions. Its eigenvalues are
\begin{equation}
\label{hahn-spec-q}
 \tilde\la_k\ =\ A_1\, \{k\}_q\,(\{k-1\}_q+1)\ +\ A_3\, \{k\}_q\ ,\quad k=0,1,2,\ldots\ .
\end{equation}
The operator (\ref{aqb-hahn}) does not support a special property of the eigenstate 
at $k=N$ (see (\ref{hahn-pol-dif-deg}), (\ref{hahn-pol-q-deg})).
This can be easily arranged modifying the coefficients $A's$:
\begin{equation}
\label{hahn-par-q}
 A_1=-1\ ,\ A_2= \{N - 2\}_q - \beta\ ,\ A_3= -\al -\beta-1\ ,\
A_4=(\beta+1)(N-1) \ ,
\end{equation}
(cf. (\ref{hahn-par})). The operator (\ref{aqb-hahn}) with coefficients (\ref{hahn-par-q}) 
still possesses infinitely-many polynomial eigenfunctions and
\begin{equation}
\label{hahn-pol-qt-deg}
 {\tilde h}_N^{(\al,\beta;q)} (x, N)\ =\ \gamma_N x^N \ ,
\end{equation}
(c.f. (\ref{hahn-pol-dif-deg}), (\ref{hahn-pol-q-deg})). The eigenvalues are continued 
to be given by the formula (\ref{hahn-spec-q}). So, by changing the $A's$-coefficients 
we make an isospectral deformation (\ref{aqb-hahn}), which now support an exceptional 
nature of the $Nth$ eigenstate.

\bigskip

Among the three-point equations (\ref{e5.5.1}) there also exist quasi-exactly-solvable
equations possessing a finite number of polynomial eigenfunctions \cite{Smirnov:1995a}.
Corresponding operators in the r.h.s. of these equations are classified via the cubic
polynomial element of the universal enveloping $\mathfrak{sl}(2,\bf{R})$-algebra taken in the representation (\ref{e5.b.5})-(\ref{e5.b.6}), which is the explicit $\de$-realization of
the quantum canonical commutation relations (\ref{e5.1.4}), (\ref{e5.b.1}).
\begin{equation}
\label{e5.5.1q}
 \tilde T\ =\ A_+\, (J^+_n + \de J^0_n J^0_n)\ +\ A_1 J^0_n J^0_n
(1 + \de J^-_n)\ +\ A_2 J^0_n J^-_n\ +\ A_3 J^0_n\ +\ A_4 J^-_n \ ,
\end{equation}
(cf. (\ref{3-points-Js})), where the $A$'s are free parameters and $n$ takes integer value.
This is nothing but the quasi-exactly-solvable generalization of the Hahn operator. In terms of the
Heisenberg algebra generators
\[
 \tilde T\ =\  A_+\, (b_{\de}^2 a_{\de} - n b_{\de} + \de
 b_{\de}a_{\de}b_{\de}a_{\de} - n \de b_{\de}a_{\de} + \frac{n^2}{4}\de)\ +\
\]
\begin{equation}
\label{e5.5.1h}
 A_1\, b_{\de}a_{\de}b_{\de}a_{\de}(\de a_{\de}\,+\,1)\ +\ A_2 b_{\de}a_{\de}^2\ +\ A_3 b_{\de}a_{\de}\ +\ A_4 a_{\de} \ ,
\end{equation}
where $[a_{\de}, b_{\de}]=1$\,.
Replacing in (\ref{e5.5.1h}), $a_\de, \, b_\de \rightarrow a,b$ with $a=\pa, b=x$ we arrive at
a linear differential operator, which is polynomially-isospectral to (\ref{e5.5.1q}) or to (\ref{e5.5.1h})\,,
\[
 \tilde T (x, \pa_x)\ =\
 \de A_1 x^2 {\pa}^3\ +\ [(\de A_1+A_2)+ (A_1 + \de A_+) x] x {\pa}^2\ +\
\]
\begin{equation}
\label{e5.5.1d}
    [A_4 + (A_1 + \de A_+ + A_3)x + \de A_+ x^2] {\pa}\ - \ \de A_+ n x \ .
\end{equation}
It has $(n+1)$ polynomial eigenfunctions in a form of polynomial of the degree $n$,
\[
     \phi^{(n)}_{k}\ =\ \sum_{i=0}^{n} a_{k,i}\, x^i\ ,\ k=0,1,\ldots n\ .
\]
Replacing in (\ref{e5.5.1h}), $(a_\de, \, b_\de) \rightarrow (a_q, b_q)$ with $a_q={\mathcal D}_q, b=x_q$ we arrive at a linear differential-difference operator, which is polynomially-isospectral to (\ref{e5.5.1q}), (\ref{e5.5.1h}) or to (\ref{e5.5.1d})\,. It has $(n+1)$ polynomial eigenfunctions in the form of polynomial of the degree $n$.

\clearpage
\newpage

\begin{center}
{\it CONCLUSIONS }
\end{center}

\addcontentsline{toc}{section}{Conclusions \protect \hfill}

\vskip 1.5cm

\renewcommand{\theequation}{C.{\arabic{equation}}}
 \setcounter{equation}{0}

 \renewcommand{\thefigure}{C.{\arabic{figure}}}
 \setcounter{figure}{0}

 \renewcommand{\thetable}{C.{\arabic{table}}}
 \setcounter{table}{0}

We have presented 12 families of quasi-exactly-solvable (QES) one-dimensional Schr\"odinger
equations. Each family depends on a number of free parameters. Usually,
QES problem represents a type of anharmonic oscillator with a specific anharmonicity: it is
a perturbation of one of well-known exactly-solvable problems (the Harmonic oscillator, the Coulomb potential, the Morse oscillator, the P\"oschl-Teller potential).
The most prominent QES example is the sextic anharmonic oscillator.
An outstanding feature of each QES family is that if one parameter takes an integer value,
a finite number of eigenstates (eigenfunctions and eigenvalues) can be found algebraically,
by linear algebra means. It implies that the corresponding QES Schr\"odinger operator
has a finite-dimensional invariant subspace.

In general, any one-dimensional QES Hamiltonian can be transformed to a Heun operator,
\[
     h_e(x)\ =\ P_3(x) \pa_x^2 + P_2(x) \pa_x + P_1(x)\ ,
\]
where $P_{3,2,1}(x)$ are polynomials of degrees 3,2,1\,, respectively. This operator can always be rewritten in terms of the generators of the algebra $sl_2$ in a representation marked by spin $\nu$ and realized by first order differential operators in one variable,
\[
J^+_{\nu} = x^2 \pa_x - 2\nu x \ ,
\]
\begin{equation}
\label{C.1}
 J^0_{\nu} = x \pa_x - \nu \  ,
\end{equation}
\[
J^-_{\nu} = \pa_x \ .
\]

It reveals a surprising connection between one-dimensional quantum dynamics and the $sl_2$-algebra quantum top in a constant magnetic field \cite{Turbiner:1988qes} - their Hamiltonian is
\[
 h_e\ =\ g_{+0}\,\{ J_{\nu}^+, J_{\nu}^0\}\ +\  g_{+-}\,\{ J_{\nu}^+, J_{\nu}^-\}\ +\  g_{00}\, J_{\nu}^0 J_{\nu}^0\ +\  g_{0-}\,\{ J_{\nu}^0, J_{\nu}^-\}\
\]
\begin{equation}
\label{C.2}
 +\ b_+\, J^+_{\nu}\ +\ b_0\, J^0_{\nu}\ +\ b_-\, J^-_{\nu}\ ,
\end{equation}
with a constraint
\[
 \{J^+_{\nu},J^-_{\nu}\} - 2 J^0_{\nu} J^0_{\nu}\,+\,2{\nu}\, \big ({\nu}+1 \big )=0\ ,
\]
where $\{ , \}$ denotes the anti-commutator; $g_{i,j}$ is the tensor of inertia and
$\vec{b}=(b_+,b_0,b_-)$ is a magnetic field.
If the spin $\nu$ is (half)-integer, $\nu=\frac{n}{2},\ n=0,1,\ldots$, the irreducible finite-dimensional representation occurs.
In this case the generators $J^{\pm,0}_{\nu}$ have a common invariant subspace
${\mathcal P}_n$ (in polynomials), which is a representation space of the finite-dimensional
representation. Correspondingly, the operator (\ref{C.2}) has the space ${\mathcal P}_n$ as a
finite-dimensional invariant subspace. Certainly, the number of those {\it algebraic}
eigenfunctions (those which can be found by linear algebra means) is nothing but
the dimension of the irreducible finite-dimensional representation of the algebra $sl_2$.
If $g_{+0} \neq 0$ the quantum top corresponds to the $A_1, BC_1$-Calogero-Moser-Sutherland
(Cases XI, XII in Chapter 1) and $BC_1$-Inozemtsev models.
If $g_{+0} = 0$ the quantum top corresponds to all other QES models (Cases I-X).
If in addition to it the parameter $b_+ = 0$, the quantum top corresponds to
exactly-solvable problems (the Harmonic oscillator, the Coulomb potential, the Morse oscillator,
the P\"oschl-Teller potential).

The $sl_2$-algebra can be realized by finite-difference operators by replacing in (\ref{C.1})
the derivative by the Norlund (Jackson) derivative and the position operator by the canonical conjugate. In this case the $sl_2$-algebra quantum top in a constant magnetic field corresponds to a discrete system on uniform (exponential) lattice. If the spin $\nu$ is (half)-integer, the discrete system becomes quasi-exactly-solvable: it has a finite-dimensional invariant subspace in polynomials.
The QES discrete operator is polynomially-isospectral to the QES Schr\"odinger operator: the spectra of polynomial eigenfunctions coincides, they do not depend on spacing.

A natural extension of the idea of quasi-exact-solvability to multi-dimensional Schr\"odinger equations leads to the question: on what finite-dimensional space of multi-variate inhomogeneous polynomials can the differential operators act? Certainly, a natural candidate might be the space of finite-dimensional representation of a Lie algebra of differential operators of the first order. One of such algebras is the $gl(d+1)$-algebra acting on functions in ${\bf R}^d$.
\begin{eqnarray}
\label{gln}
   {\mathcal J}_i^- &=& \frac{\pa}{\pa \ta_i},\qquad \quad
                i=1,2\ldots d \ , \non \\
   {{\mathcal J}_{ij}}^0 &=&
    \ta_i \frac{\pa}{\pa \ta_j}, \qquad i,j=1,2\ldots d \ ,
 \non \\
   {\mathcal J}^0 &=& \sum_{i=1}^{d} \ta_i\frac{\pa}{\pa \ta_i}-n\, ,
 \\
   {\mathcal J}_i^+ &=& \ta_i {\mathcal J}^0 =
    \ta_i\, \left( \sum_{j=1}^{d} \ta_j\frac{\pa}{\pa \ta_j}-n \right),
           \quad i=1,2\ldots d \ . \non
\end{eqnarray}
where $n$ is an arbitrary number. The total number of generators is $(d+1)^2$.
If $n$ takes the integer values, $n=0,1,2\ldots$,  the finite-dimensional irreps occur
\[
 {\mathcal P}_n^{(d)}\ =\ \langle {\ta_{ 1}}^{p_1}
 {\ta_{2}}^{p_2}\ldots
 {\ta_{ d}}^{p_{d}}
 \vert \ 0 \le \Sigma p_i \le n \rangle\ .
\]
It is a common invariant subspace for (\ref{gln}). It is shown that any quantum $A_d, BC_d, D_d$ rational, trigonometric, elliptic Calogero-Moser-Sutherland Hamiltonian has the finite-dimensional invariant subspace ${\mathcal P}_n^{(d)}$, see \cite{Ruhl:1995}, \cite{Brink:1997}, \cite{Turbiner:2011rat}, \cite{Turbiner:2013trig}, \cite{Gomez-Ullate:2001}, \cite{Sokolov-Turbiner:2015} \footnote{For the $A_d$ elliptic Calogero-Moser-Sutherland Hamiltonian, this is demonstrated for $d=2$ \cite{Sokolov-Turbiner:2015} and is conjectured for $d>2$}. For the rational and trigonometric cases the invariant subspace exists with any integer $n=0,1,\ldots$. All of these Hamiltonians are equivalent to $gl(d+1)$ quantum top (or, saying differently, $sl_{d+1}$ quantum top in a constant magnetic field). The quantum $G_2$ (rational, trigonometric, elliptic) Calogero-Moser-Sutherland Hamiltonian has the finite-dimensional invariant subspace
\[
 {\tilde {\mathcal P}}_n^{(2)}\ =\ \langle {\ta_{1}}^{p_1}
 {\ta_{2}}^{p_2}
 \vert \ 0 \le p_1 + 2 p_2 \le n \rangle\ ,
\]
see \cite{Rosenbaum:1998,Turbiner:1998}, \cite{Boreskov:2004}, \cite{Sokolov-Turbiner:2015}. For the rational and trigonometric cases there exist infinitely-many finite-dimensional invariant subspaces, marked by integer $n=0,1,\ldots$. They form the infinite flag. The hidden algebra is not $gl(3)$ anymore: it is infinite dimensional, 10-generated algebra of differential operators in two variables with generalized Gauss decomposition property (for discussion see the reviews \cite{Turbiner:2011rat,Turbiner:2013trig}).
For any quantum $F_4, E_{6,7,8}$ rational, trigonometric Calogero-Moser-Sutherland Hamiltonian there exist  infinitely-many finite-dimensional invariant subspaces in polynomials of a special form \cite{Boreskov:2004}, this is summarized in \cite{Boreskov:2011}. Analysis of the respectful hidden algebras is still incomplete. Quantum $F_4, E_{6,7,8}$ elliptic Calogero-Moser-Sutherland models were never studied.

It is known that the quantum $A_d, BC_d, D_d$ rational and trigonometric Calogero-Moser-Sutherland models admit the super-symmetric extension \cite{Freedman:1990,Shastry:1993}, see for discussion \cite{Brink:1997} and references therein. All of these models are equivalent to the supersymmetric $gl(d+1,d)$-superalgebra quantum top in a constant magnetic field \cite{Brink:1997}.

\clearpage

\newpage

\centerline{\large Acknowledgement}

\addcontentsline{toc}{section}{Acknowledgement \protect \hfill}

\vskip 1cm

{\bf Historical reminiscence}

There were three people who contributed significantly to
establishing the subject. For many years I worked on different
aspects of perturbation theory in quantum mechanics. One of
essential elements of the approach I used was a construction of
the potential by an arbitrary chosen, square-integrable function.
For many years it was a puzzle how to choose several
square-integrable functions which lead to sensible
eigenfunctions in the same potential. When, finally, I succeeded to
find non-trivial examples, A.B.(`Sasha')~Zamolodchikov, who was
the first to whom I told about my findings, immediately
conjectured that the corresponding Schr\"odinger operators can be
related to the algebra $sl_2$. From the very
beginning I.M.~Gel'fand expressed interest in the finding and postulated
that a relation to the Lam\'e operator should exist -- a
study of this relation shed the light on the general
construction. Later V.I.~Arnold asked me to write for him a one
page description of quasi-exact-solvability -- after about two
years of working I was able to do so, clarifying
the general nature of the subject. I would like to express my
deep gratitude to them.

{\bf Thanks}

I want to thank my two teachers K.A.~Ter-Martirosyan and Ya.B.~Zel'dovich,
and also A.B.~Zamolodchikov and Al.B.~Zamolodchikov, I.M.~Gel'fand,
\linebreak S.P.~Novikov and V.I.~Arnold, A.G.~Ushveridze, W.~R\"uhl,
J.C.~L\'opez Vieyra and K.G.~Boreskov, L.~Brink, M.A.~Shifman, and N.~Wyllard,
Y.~Brihaye, J.~Ellis, N.A.~Nekrasov, P.~Olver, M.A.~Shubin, G.~Veneziano,
for generous interest to the subject and useful discussions in different stages
of the work.
With some of them I had a real pleasure to work together being a co-author.
The text is partly based on a course of lectures given by the author at CERN, Geneva
and University of Minnesota, Minneapolis at 1991, and ETH, Zurich at 1992 and
at Instituto de Fisica de la UNAM, Mexico City at 1994 and numerous talks given
worldwide, in particular, at the Simons Center for Geometry and Physics (Stony Brook).

The story of this review paper comes back to 1990 when R.C.~Slansky
(Los Alamos) as the editor of Physics Reports has invited me to write a review
for Physics Reports. I conditionally accepted saying that the subject is
underdeveloped and it may take several years to write it up. Several times
he reminded me on my promise. As a result it took 25 years. I thank Dick for
the proposal and feel sorry that he had no chance to see the review.

This work was supported in part by PAPIIT grant {\bf IN108815} and CONACyT
grant {\bf 166189} (Mexico) as well as PASPA grant (UNAM, Mexico).
The author thanks Physics Department of Stony Brook University, and the Simons Center for Geometry and Physics where this review was completed during its sabbatical leave. 
\clearpage
\bibliography{refs-QES}
\end{document}